\documentclass[11pt,letterpaper,oneside,openany]{book}
\usepackage[left=1.2in,right=1.2in,top=1in,bottom=1in]{geometry}

\title{Phenomenology of baryon dynamics with directed flow in relativistic heavy-ion collisions}
\author{Tribhuban Parida}

\usepackage[utf8]{inputenc}
\DeclareUnicodeCharacter{00AF}{\= }

\usepackage[T1]{fontenc}
\usepackage{lmodern}
\usepackage[black]{raleway} 
\usepackage{lato}
\usepackage{comment}
\usepackage{subcaption}

\usepackage{mathtools}
\usepackage{amssymb}

\usepackage{graphicx}
\graphicspath{{figures/}}

\usepackage{wrapfig}
\usepackage{tikz}
\usetikzlibrary{shapes,calc,matrix}
\usepackage{xcolor}
\definecolor{veryblack}{RGB}{0,0,0}
\color{veryblack} 

\definecolor{theblue}{RGB}{22,79,149}
\usepackage[pdfusetitle,colorlinks=true,allcolors=theblue,bookmarksdepth=2,linktoc=section,bookmarksopen=false]{hyperref}

\usepackage{fancyhdr}
\usepackage{titlesec}
\titleformat{\chapter}[display]{\Huge\sffamily\color[RGB]{83,94,113}\filcenter}{\thechapter}{1ex}{}
\titleformat*{\section}{\Large\bfseries\color{theblue}}
\titleformat*{\subsection}{\large\bfseries\color{theblue}}
\titleformat*{\subsubsection}{\bfseries\color{theblue}}
\titleformat*{\paragraph}{\bfseries\color[gray]{0.1}}
\titlespacing{\paragraph}{0pt}{*1.4}{*2.5}

\usepackage[font=small,labelfont={bf,color=theblue},labelsep=quad]{caption}

\usepackage{lettrine}

\usepackage{enumitem}

\setlist[itemize,1]{label=\raisebox{.25ex}{\color{theblue}\tiny\textbullet}}

\usepackage{booktabs}
\usepackage{multirow}

\usepackage[style=phys,biblabel=brackets,eprint=true,url=true,maxcitenames=3]{biblatex}
\def \beq{\begin{equation}}
\def \eeq{\end{equation}}
\def \beqa{\begin{eqnarray}}
\def \eeqa{\end{eqnarray}}

\newcommand{\sNN}{\sqrt{s_{\textrm{NN}}}}

\begin{document}

\frontmatter

\makeatletter
\begin{titlepage}
    \centering
    \sffamily
    \color{veryblack} 
    
    {\Huge Phenomenology of Baryon Dynamics with Directed Flow in Relativistic Heavy-Ion Collisions \\[2cm]}
    
    {\Large A thesis submitted in partial fulfillment\\
    of the requirements for the degree of\\
    Doctor of Philosophy \\[2cm]}

    {\Large by \\[0.2cm]
    Tribhuban Parida\\
    Registration Number- 1820502 \\[2cm]}

    {\Large to the \\[0.2cm]
    Department of Physical Sciences \\
    Indian Institute of Science Education and Research Berhampur \\[0.3cm]}

    \includegraphics[width=60mm]{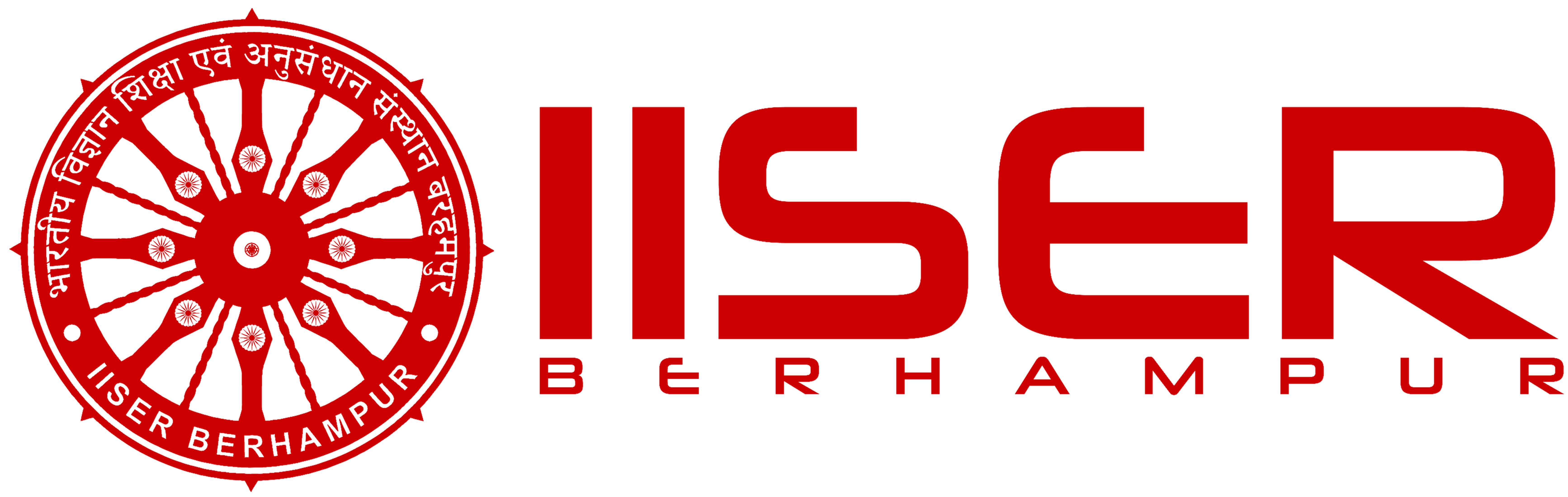} \\[2cm]

    {\Large December, 2024 \\[2cm]}
    
\end{titlepage}
\makeatother

\chapter*{CERTIFICATE}

The undersigned have examined the Ph.D. thesis entitled:
\vspace{1mm}
\\
\textbf{Phenomenology of baryon dynamics with directed flow in relativistic heavy-ion collisions}
presented by \textbf{Tribhuban Parida}, a candidate for the degree of Doctor of Philosophy in \textbf{Department of Physical Sciences}, and hereby certify that it is worthy of acceptance.

\vspace{2cm} 

\noindent
\textbf{Thesis Supervisor:} \underline{Dr. Sandeep Chatterjee} \hfill \textbf{Signature:} \underline{\hspace{3cm}}

\vspace{1.0cm}

\noindent
\textbf{Thesis Examiner:} \underline{} \hfill \textbf{Signature:} \underline{\hspace{3cm}}

\vspace{1.0cm}

\noindent
\textbf{Chairperson:} \underline{} \hfill \textbf{Signature:} \underline{\hspace{3cm}}

\vspace{1.0cm}

\clearpage

\chapter*{ACADEMIC INTEGRITY AND COPYRIGHT DISCLAIMER}
\begin{quote}
I hereby declare that this thesis is my own work and, to the best of my
knowledge, it contains no materials previously published or written by any
other person, or substantial proportions of material which have been accepted
for the award of any other degree or diploma at IISER Berhampur or any
other educational institution, except where due acknowledgment is made in
the thesis.

I certify that all copyrighted material incorporated into this thesis is in
compliance with the Indian Copyright (Amendment) Act, 2012 and that I have
received written permission from the copyright owners for my use of their
work, which is beyond the scope of the law. I agree to indemnify and save
harmless IISER Berhampur from any and all claims that may be asserted or
that may arise from any copyright violation.
%

\noindent

\vspace{1.5cm}
\noindent

\noindent 
\hspace{0.00cm}\textsc{Signature: }  \\
\noindent 
\hspace{0.00cm}\textsc{Date: }  \\
\noindent 
\hspace{0.00cm}\textsc{Place: } {}

\end{quote}
\clearpage

\chapter*{List of publications}
\section*{Journals}
\begin{enumerate}
\item   Tribhuban Parida, Rupam Samanta and Jean-Yves Ollitrault, Probing collectivity in heavy-ion collisions with fluctuations of the $p_T$ spectrum, \href{https://www.sciencedirect.com/science/article/pii/S0370269324005434?via%3Dihub}{Phys. Lett. B 857 (2024), 138985}.
\item Tribhuban Parida, Sandeep Chatterjee and Md. Nasim, Effect of hadronic interaction on the flow of $K^{*0}$, \href{https://journals.aps.org/prc/abstract/10.1103/PhysRevC.109.044905}{Phys. Rev. C 109 (2024) 4, 044905}. ({$\star$})
\item   Tribhuban Parida, Piotr Bozek and Sandeep Chatterjee, Charm balance function in relativistic heavy-ion collisions, \href{https://journals.aps.org/prc/abstract/10.1103/PhysRevC.109.014903}{Phys. Rev. C 109 (2024) 1, 1}.
\item Tribhuban Parida and Sandeep Chatterjee, Splitting of elliptic flow in a tilted fireball, \href{https://journals.aps.org/prc/abstract/10.1103/PhysRevC.106.044907}{Phys. Rev. C 106 (2022) 4, 044907}. ({$\star$})  
\end{enumerate}

\section*{arXiv submissions}
\begin{enumerate}
\item Tribhuban Parida and Sandeep Chatterjee, Baryon diffusion coefficient of the strongly interacting medium, \href{https://arxiv.org/abs/2305.10371}{2305.10371 [nucl-th]}. ({$\star$})  
\item Tribhuban Parida and Sandeep Chatterjee, Baryon inhomogeneities driven charge dependent directed flow in heavy ion collisions, \href{https://arxiv.org/abs/2305.08806}{ 2305.08806 [nucl-th]}. ({$\star$})  
\item Tribhuban Parida and Sandeep Chatterjee, Directed flow of light flavor hadrons for Au+Au collisions at $\sqrt{s_{NN}}$ 7.7-200 GeV, \href{https://arxiv.org/abs/2211.15659}{ 2211.15659 [nucl-th]}. ({$\star$})  
\item Tribhuban Parida and Sandeep Chatterjee, Directed flow in a baryonic fireball, \href{https://arxiv.org/abs/2211.15729}{ 2211.15729 [nucl-th]}. ({$\star$})   
\end{enumerate}

\section*{Conference proceedings}
\begin{enumerate}
\item Tribhuban Parida and Sandeep Chatterjee, Charged particle elliptic flow splitting at non-zero rapidity in heavy ion collisions. \href{https://inspirehep.net/literature/2632363}{DAE Symp.Nucl.Phys. 66 (2023) 946-947}. ({$\star$})   
\item   Tribhuban Parida and Sandeep Chatterjee, Flavour dependent freezeout scenarios within relativistic hydrodynamics framework. \href{https://inspirehep.net/literature/2027294}{65th DAE BRNS Symposium on nuclear physics, 658-659}.
\end{enumerate}

($\star$) denotes the articles that are included in this thesis.

\chapter*{Conferences}
\section*{Oral presentations}
\begin{enumerate}
\item \href{https://indico.ihep.ac.cn/event/22462/contributions/170766/}{Role of Stopping and Diffusion of Baryons in BES Phenomenology.}\\ The 1st International Workshop on Physics at High Baryon Density, PHD 2024. 1-4 November 2024, CCNU, Wuhan, China. 

\item \href{https://theory.tifr.res.in/~saumen/PhaseDg/}{Estimating baryon diffusion coefficient of the strongly interacting medium.}\\ Aspects of the QCD Phase Diagram. A discussion meeting on the phases of strongly interacting matter. 18-20 November 2023, IISER Bhopal, India.

\item \href{https://www.iitgoa.ac.in/HeavyFlavourMeet2023/index.html}{Charm balance function in heavy ion collsions.}\\ $4^{th}$ heavy flavour meet 2023. 2-4 Nov 2023, Goa, India. 

\item \href{https://events.vecc.gov.in/event/19/contributions/812/}{Model study of directed flow of light flavor hadrons, resonances and light nuclei.}\\International Conference on Physics and Astrophysics of Quark Gluon Plasma (ICPAQGP-2023). 7-10 February 2023, Puri, Odisha, India.

\item \href{https://www.niser.ac.in/events/et-hcvm/}{Discriminating the models of initial matter deposition using transverse momentum differential directed flow.}
\\Emergent topics in relativistic hydrodynamics, chirality, vorticity and magnetic field. 2-5 February 2023, National Institute of Science Education and Research, Bhubaneswar, India. 

\item \href{https://events.vecc.gov.in/event/18/abstracts/467/}{Constraining the initial baryon profile from baryon and anti-baryon directed flow split.}\\
DAE-BRNS symposium on Contemporary and Emerging Topics in High Energy Nuclear Physics (CETHENP 2022). 15-17 November 2022, Variable Energy Cyclotron Centre, Kolkata, India.

\item \href{http://sympnp.org/proceedings/} {Flavor dependent freezeout scenarios within relativistic hydrodynamics framework.}\\
65th DAE symposium on Nuclear Physics, 2021 (online).

\end{enumerate}

\section*{Poster presentations}
\begin{enumerate}
\item \href{https://indico.cern.ch/event/895086/contributions/4721578/} {Signature of early freeze-out of strangeness in relativistic heavy ion collisions.} \\
XXIXth International Conference on Ultra-relativistic Nucleus-Nucleus Collisions, Quark Matter 2022. April 4-10, 2022, Krakow, Poland (online).
\item \href{https://indico.cern.ch/event/792436/contributions/3549042/}{Study of possible fluctuations of thermal
parameters at freezeout using HRG model.} \\
XXVIIIth International Conference on Ultra-relativistic Nucleus-Nucleus Collisions, Quark Matter 2019. November 3-9, 2019, Wuhan, China.

\end{enumerate}

\chapter*{Dedication}
\thispagestyle{empty}


\begin{center}
    This thesis is dedicated to my grandparents, \\[1ex]
    Shri Bishnu Charan Lenka \\[1ex]
    and \\[1ex]
    Smt. Sulochana Lenka, \\[1ex]
    whose love is a priceless source of inspiration for me.
\end{center}
\vspace{.2cm}
\hspace{12cm}{-- Tribhuban  }

\vfill


\vspace*{\fill}

\chapter{Abstract}
This thesis aims to elucidate the role of initial baryon stopping and its diffusion in heavy-ion collisions (HIC) using hydrodynamic model. In this regard, we have studied the observable-directed flow of identified hadrons, particularly the directed flow of baryons and antibaryons, as well as the splitting observed between them in detail. 

We propose a new ansatz for the initial baryon distribution. By employing this initial baryon deposition model alongside a tilted energy distribution as inputs to a hybrid framework (combining hydrodynamics and hadronic transport), we successfully describe the rapidity-odd directed flow ($v_1$) of identified hadrons, including the elusive baryon-antibaryon splitting of $v_1$ across a wide range of $\sqrt{s_{NN}}$. The model reproduces the observed double sign change in the slope of the directed flow for net protons and net lambdas in the collision energy range $\sqrt{s_{NN}} = 7.7 - 39$ GeV. 

Our model, incorporating baryon stopping and it's subsequent diffusion within a relativistic hydrodynamic framework and employing a crossover equation of state derived from lattice QCD calculations, establishes a non-critical baryonic baseline. This baseline is instrumental in ongoing efforts to identify the QCD critical point.

Moreover, we demonstrate that recent STAR measurements of the centrality and system-size dependence of $v_1$ splitting between oppositely charged hadrons—attributed to electromagnetic field effects—are significantly influenced by background contributions from baryon stopping and its diffusion. 

Furthermore, we show that the rapidity dependence of the splitting of the rapidity-even component of $v_1$ between protons and anti-protons is highly sensitive to the initial baryon deposition scheme. If measured experimentally, this could constraint the rapidity dependence of the initial baryon depostion profile. Moreover, it could offer valuable phenomenological insights into the baryon junction picture and help refine constraints on the baryon diffusion coefficient of the medium.

Notably, utilizing this phenomenologically successful baryon deposition model, we present the first estimation of the baryon diffusion coefficient for the strongly interacting QCD matter created in HIC.

\titlespacing{\chapter}{0pt}{-15pt}{30pt}
\setcounter{tocdepth}{2}
\tableofcontents
\titlespacing{\chapter}{0pt}{50pt}{40pt}

\mainmatter
\chapter{Introduction}
\label{ch:intro}
\def \la{\langle}
\def \ra{\rangle}

Significant progress in particle physics has been made over the past century, starting with the discovery of subatomic particles like electron and proton. Presently, it is established that all visible matter in the universe is fundamentally composed of six types of quarks (up, down, strange, charm, bottom, top) and six types of leptons (electron, muon, tau, and their corresponding neutrinos) along with their anti-particles \cite{ParticleDataGroup:2022pth,Burgess:2006hbd,Langacker:2017uah,Griffiths:2008zz,Veltman:2018nzg,Cottingham:2007zz}. The interactions among these particles are governed by three fundamental forces: electromagnetic, strong, and weak forces, mediated by 12 gauge bosons, including 8 color-charged gluons, photons, and three weak bosons ($W^+, W^-, Z^0$) \cite{Griffiths:2008zz,Veltman:2018nzg,Cottingham:2007zz,Glashow:1961tr,Peskin:1995ev,Weinberg:1967tq,Gell-Mann:1964ewy}. Alongside these particles, the recently discovered Higgs boson, responsible for providing mass to all elementary particles, completes the current standard model of particle physics \cite{Higgs:1964pj,ATLAS:2012yve,CMS:2012qbp}. However, gravity, the most familiar fundamental force, remains unincorporated in the standard model.

Among the four fundamental forces of nature, the strongest one binds quarks and gluons together within protons and neutrons in the atomic nucleus \cite{Wilson:1974sk,Griffiths:2008zz,Veltman:2018nzg,Glashow:1961tr,Burgess:2006hbd}. Quantum Chromodynamics (QCD) is the quantum field theory that governs this strong interactions \cite{Burgess:2006hbd,Langacker:2017uah,Peskin:1995ev}. The QCD Lagrangian density is given by \cite{Gross:1973id,Politzer:1973fx,Peskin:1995ev,Greiner:2007}: 
\beq
\mathcal{L}_{\text{QCD}} = \sum_{f=1}^{N_f} \bar{\psi}^{\alpha}_{f} ( i \gamma^{\mu} D_{\mu,\alpha \beta} - m_f \delta_{\alpha \beta}) \psi_f^{\beta} - \frac{1}{4} G_{\mu \nu}^a G^{\mu \nu a} 
\label{eq:QCDLag}
\eeq
where $\alpha$ and $\beta$ are color indices running from 1 to 3, and $f$ is the flavor index ranging from 1 to $N_f$. The index $a=1,…,8$ labels the eight gluon fields corresponding to the generators of the SU(3) color gauge group. While one could explicitly include spinor indices for the quark fields $\psi_f$ , they are suppressed here to make the expression simple.

In the above expression, $\gamma^{\mu}$ are the Dirac gamma matrices, and the covariant derivative is defined as
\beq
D_{\mu,\alpha \beta} \equiv \partial_{\mu} \delta_{\alpha \beta} + \frac{i}{2} g_s T_{\alpha \beta}^{a} A_{\mu}^{a}
\eeq
where $g_s$ is the dimensionless strong coupling constant, $A_{\mu}^{a}$ are the gluon gauge fields, and $T^a$ are the Gell-Mann matrices (the generators of SU(3) in the fundamental representation).

The gluon field strength tensor $G_{\mu \nu}^{a}$ is given by: 
\beq
G_{\mu \nu}^{a} = \partial_{\mu} A_{\nu}^{a} - \partial_{\nu} A_{\mu}^{a} - g_s f^{abc} A_{\mu}^{b} A_{\mu}^{c}
\eeq  
where $f^{abc}$ are the structure constants of the SU(3) Lie algebra.

The QCD Lagrangian is invariant under local SU(3) gauge transformations, which form the basis of Quantum Chromodynamics \cite{Peskin:1995ev,Greiner:2007,Greiner:2007}. Quark fields transform under the fundamental representation of SU(3), corresponding to the group’s natural action on a three-dimensional complex vector space. The basis vectors of this space define the three color states—red, green, and blue—which together form a color triplet. These states are represented as components of the quark color wavefunction. In contrast, gluon fields transform under the adjoint representation of SU(3). Since the SU(3) Lie algebra has eight independent generators, gluons span an eight-dimensional vector space and are said to transform under the adjoint (octet) representation. Consequently, there are eight distinct types of gluons, each associated with one of the eight generators $T^a$ of SU(3), which satisfy the commutation relation: \beq [T^a, T^b] = 2 i f^{abc} T^c, \eeq. These non-trivial commutation relations reflect the non-Abelian nature of QCD, implying that gluons themselves carry color charge and can interact with one another—unlike the photon in QED, which is electrically neutral.

The momentum dependence, or running, of the QCD coupling constant $g_s$ is governed by the QCD beta-function. At leading order in perturbation theory, the beta-function at a renormalization scale $Q$ is given by \cite{Peskin:1995ev}: 
\beq
\beta(g_s) \equiv - \frac{\partial g_s}{\partial \ln Q} = - \frac{g_s^3}{(4\pi)^2} \left( \frac{11}{3} N_c - \frac{2}{3} N_f \right), 
\eeq
where $N_c=3$ is the number of colors, and $N_f$ is the number of quark flavors. Solving this differential equation yields the scale dependence of the coupling \cite{Peskin:1995ev}: 
\beq
g_s^2(Q) = \frac{g_s^2}{1 + \beta_0 g_s^2 \ln(Q/M)}, \quad \text{with} \quad \beta_0 \equiv \frac{1}{8\pi^2} \left(11 - \frac{2}{3} N_f \right), 
\eeq where $M$ is an arbitrary reference scale. This expression shows that $g_s^2(Q)$ decreases logarithmically with increasing energy scale $Q$, implying that quarks and gluons interact more weakly at high energies. In the limit $Q \rightarrow \infty$, the coupling approaches zero: $g_s^{2} (Q) \rightarrow 0  $, a phenomenon known as asymptotic freedom.

By introducing a new energy scale $\Lambda$, defined via the relation $g_s^2 \beta_0 \ln{(M/\Lambda)}=1$, and defining the strong coupling constant $\alpha_s(Q)= g_S^2(Q)/4 \pi$, one obtains the following compact form \cite{Peskin:1995ev}: 
\beq 
\alpha_s(Q) = \frac{1}{\beta_0 \ln(Q/\Lambda)}, \label{eq:alphas} 
\eeq
which explicitly shows the logarithmic decrease of the strong coupling with energy scale. The $\Lambda$ is a free parameter known as $\Lambda_{\text{QCD}}$, representing an intrinsic energy scale of QCD. Its value is not predicted by theory but is instead fixed by comparison of Eq.~\ref{eq:alphas} with experimental data \cite{Bethke:2006ac} which provides $\Lambda_{\text{QCD}} \approx 200$ MeV \cite{Peskin:1995ev,ParticleDataGroup:2004fcd,Ellis:1996mzs}.  At higher energy scales $(Q \gg \Lambda_{\text{QCD}})$, perturbative methods become applicable in QCD calculations.
Conversely, at lower energy scales, the value of $\alpha_s$ increases, making perturbative series expansions invalid for determining QCD contributions to physical observables. Thus, $\Lambda_{\text{QCD}}$ offers a crucial estimate of the energy scale below which perturbation theory in QCD loses validity.

Lattice QCD provides a first-principles, non-perturbative framework for studying Quantum Chromodynamics (QCD) by discretizing spacetime into a finite grid or lattice. In this formulation, quark fields are defined on the lattice sites, while gluon fields are represented as SU(3) link variables that connect neighboring sites. This discretization preserves local gauge invariance and allows the evaluation of QCD observables from the path integral using Monte Carlo techniques \cite{Karsch:2022nci,Soltz:2015ula,Soltz:2015ula}.

Using the QCD Lagrangian, the grand canonical partition function for a thermal ensemble of quarks, antiquarks, and gluons in equilibrium at temperature $T$ is given by \cite{Karsch:2022nci,Soltz:2015ula,Soltz:2015ula,Heinz:2004qz}:
\beq
\mathcal{Z} = \int \mathcal{D}A^{a}_{\mu} (x) \mathcal{D} \bar{\psi}(x) \mathcal{D} \psi(x) e^{ -\int_0^{\frac{1}{T}} d\tau \int d^3x \mathcal{L}_{\text{QCD}} (A^{\mu}_a, \bar{\psi}, \psi) }
\label{eq:LatZ}
\eeq
From this partition function, one can extract thermodynamic quantities such as pressure, energy density, and entropy density by calculating the integral numerically.

\begin{figure}
  \centering
  \includegraphics[width=0.6\textwidth]{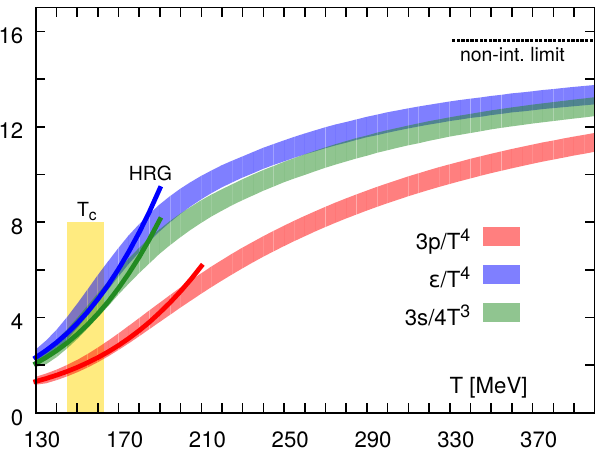} 
  \caption{Temperature dependence of the energy density ($\epsilon$), pressure ($p$), and entropy density ($s$), each normalized by appropriate powers of temperature, from lattice QCD calculations by the HotQCD collaboration~\cite{HotQCD:2014kol}. Results are shown after continuum extrapolation. The lattice results are compared with the Hadron Resonance Gas (HRG) model at low temperatures, showing good agreement for $T \lesssim 150$ MeV, and with the Stefan-Boltzmann limit at high temperatures. A rapid rise in the thermodynamic quantities around $T_c =( 154 \pm 9)$ MeV marks the crossover transition from a hadronic gas to a quark-gluon plasma.}

 \label{fig:Lat_eT4}
\end{figure}

Fig.\ref{fig:Lat_eT4} shows the continuum-extrapolated results for the energy density ($\epsilon$), pressure ($p$), and entropy density ($s$), each normalized appropriately by powers of temperature ($T$), as functions of $T$, obtained from lattice QCD calculations by the HotQCD collaboration\cite{HotQCD:2014kol}. These results were obtained for a system with 2+1 quark flavors using physical values for the quark masses. The convergence of the numerical integration of Eq.~\ref{eq:LatZ} is verified by comparing results across different temporal lattice extents ($N_t = 6, 8, 10, 12$) \cite{HotQCD:2014kol}.

The HotQCD results show excellent agreement with previous lattice QCD calculations by the Wuppertal-Budapest (WB) collaboration~\cite{Borsanyi:2013bia}. Despite differences in lattice spacing and computational techniques between the two collaborations, their results coincide in the continuum limit, providing a fundamental cross-check of the reliability of lattice QCD in describing the thermodynamics of strongly interacting matter. A comparison between the equations of state obtained from these two collaborations can be found in Fig.1 of Ref. \cite{Ratti:2018ksb}.

For comparison, Fig.~\ref{fig:Lat_eT4} also displays the Hadron Resonance Gas (HRG) model predictions at low temperatures. In the region $T \lesssim 150$ MeV, lattice QCD results show good agreement with the HRG model, consistent with the expectation that, at low energies, quarks and gluons are confined within hadrons, forming a hadronic gas. At higher temperatures, a horizontal dotted line marks the Stefan-Boltzmann limit corresponding to the energy density of an ideal, non-interacting massless quark gas.

A rapid rise in the energy density is observed around the critical temperature $T_c = (154 \pm 9)$ MeV. The sharp increase in $\epsilon$ around $T_c$ reflects a sudden rise in the number of effective degrees of freedom, indicative of a crossover transition from a hadronic gas to a deconfined quark-gluon plasma (QGP).

Lattice QCD works well at zero net baryon density, which is a good approximation for conditions in the early Universe or for certain regions in the phase space of ultrarelativistic heavy-ion collisions. However, the matter formed in heavy-ion collisions in general involve a finite net baryon number. In the presence of a nonzero baryon chemical potential ($\mu_B$), lattice QCD suffers from the fermion sign problem \cite{deForcrand:2010ys,Aarts:2015tyj}, which severely limits its applicability in studying the properties and phase structure of QCD matter at finite baryon density.

\section{The Quark Gluon Plasma (QGP)}
In our everyday experiences, the energy scale is considerably lower than $\Lambda_{\text{QCD}}$. Consequently, at such energies, the coupling strength between the fundamental constituents—quarks and gluons—is notably high, leading to their confinement within protons and neutrons inside atomic nuclei. Even upon further increment of energy, surpassing the binding energy of nucleons would release them from the nucleus, yet this energy scale remains significantly lower than $\Lambda_{\text{QCD}}$, as the binding energy per nucleon is on the order of 10 MeV. However, it is anticipated that at sufficiently high energies, well beyond $\Lambda_{\text{QCD}}$, quarks and gluons could almost freely exist, giving rise to a state of matter where deconfined color charges serve as the ultimate degrees of freedom. This distinctive state of matter is known as the Quark-Gluon Plasma \cite{Shuryak:1978ij,Shuryak:1980tp,Shuryak:1988ck,Kapusta:1979fh,Kislinger:1975uy,Freedman:1976xs,Chin:1978gj}.

In physics, plasma is generally referred to as the fourth state of matter, following solid, liquid, and gas. As the temperature of a system (for example, water molecules in a container) increases with fixed pressure, it transits from solid to liquid and then to gas, as the mean free path between atoms or molecules grows. With a further increase in temperature, the atoms in the medium become ionized, separating electrons from ions, leading to the formation of a plasma state. In this electron-ion plasma, the constituents are electrically charged. Typically, the ionization energy ($E$) of an atomic nucleus ranges from a few electron volts (eV) to several kilo-electron volts (KeV), requiring a temperature ($T = E / k_B$)\footnote{$k_B$ is the Boltzmann constant.} between $10^4$ and $10^7$ Kelvin to create this plasma \cite{LI20181}. For comparison, the temperature at the core of the sun is about $10^7$ Kelvin, making it extremely difficult to reach such high temperatures in a laboratory setting. However, if the system is heated beyond this temperature, surpassing the binding energy of nucleus (which is on the order of MeV per nucleon), a gas of hadrons containing protons and neutrons is formed. With even higher temperatures, the thermal energy eventually becomes sufficient to dissociate the quarks and gluons that are tightly bound within the nucleons. This process requires a temperature of approximately $10^{12}$ Kelvin \cite{Karsch:2001cy,Borsanyi:2013bia,HotQCD:2014kol,Ding:2015ona}. At this stage, a system of strongly interacting matter composed of quarks and gluons is expected to emerge. The constituents of this state carry color quantum numbers along with electric charge. However, in this system, the relevance of color charge is dominant over electric charge, as the strong interaction is approximately 100 times stronger than electromagnetic interactions. Similar to the electron-ion plasma, which is considered an electrically charged fluid, this new state of matter, consisting of free quarks and gluons, can be thought of as a color-charged fluid, commonly referred in the literature as Quark-Gluon Plasma (QGP) \cite{Shuryak:1978ij,Shuryak:1980tp,Shuryak:1988ck,Kapusta:1979fh,Shuryak:2024zrh,Florkowski:2010zz,Vogt:2007zz,Yagi:2005yb,Wong:1995jf}.

The formation of QGP requires nuclear matter to be under extremely high temperature and density. Such conditions are believed to have existed in the early universe and within the cores of super-dense stars, where the central density exceeds 10 to 15 times the nuclear saturation density ($\rho_{\text{nm}} \approx 0.16$ fm$^{-3}$) \cite{Rafelski:2013qeu,Kapusta:2003qyx,Boeckel:2010bey,Heinz:2013wva,Dolgov:1981hv,Annala:2019puf,Witten:1984rs,Farhi:1984qu,Weinberg:1977ji,Yagi:2005yb,Annala:2019puf,Annala:2023cwx,Lattimer:2012nd,Yagi:2005yb}. Beyond these natural occurrences, QGP can also be created in laboratory settings. Collider experiments, such as those at the Large Hadron Collider (LHC) at CERN and the Relativistic Heavy Ion Collider (RHIC) at Brookhaven National Laboratory (BNL), are designed to recreate the necessary conditions to produce QGP for very brief moments \cite{Baym:2001in,Shuryak:2014zxa,Heinz:2008tv,Niida:2021wut}.

In these experiments, nuclei are accelerated to ultrarelativistic speeds and collided, depositing their kinetic energy into a very small volume. In such relativistic heavy-ion collisions, the center-of-mass energy per nucleon of the colliding nuclei exceeds 100 GeV, concentrated in an area on the order of 1 fm$^2$. This creates a system with an extremely high energy density and temperature, leading to the formation of a deconfined phase of quarks and gluons. Although QGP is created, its lifetime is very short. It expands rapidly and cools down, eventually transitioning back into a phase of hadrons as quarks and gluons recombine. This phase transition from the deconfined QGP phase to a hadronic gas phase occurs within extremely small time interval which is in the order of $10$ fm/c. The resulting hadrons are then detected by detectors, meaning that QGP cannot be observed directly in experiments. However, indirect evidence of QGP formation can be inferred from the characteristics of the emitted hadrons \cite{Bass:1998vz,Gyulassy:2004zy,Teaney:2000cw,Heinz:2000bk,Sorge:1998mk,McLerran:1986zb,Rischke:1996em,Rischke:1995cm,Chandra:2024ron,Stoecker:1986ci,Chin:1978gj,Mohanty:2011fp,Bjorken:1982tu,Appel:1985dq}. This approach is analogous to probing the early universe through remnants like the cosmic microwave background (CMB). Several experimental signatures of QGP formation in relativistic heavy-ion collisions have been observed, such as the enhanced production of hadrons containing non-zero strange and charm quantum numbers, increased production of antiparticles, suppression of high $p_T$ particles due to energy loss in the QGP medium, and the collective behavior of the produced hadrons \cite{ALICE:2016fzo,STAR:2011fbd,PHENIX:2001hpc,PHENIX:2003qdj,STAR:2002ggv,PHENIX:2004vcz,STAR:2005gfr,BRAHMS:2004adc,STAR:2017ieb,Bjorken:1982tu,Aitbayev:2024ikp,BRAHMS:2002efn,NA49:2003njx,STAR:2001ksn}.

\section{Relativistic heavy-ion collisions}
\subsection{Basic terminology, convention and units}
\begin{figure}
  \centering
  \includegraphics[width=1\textwidth]{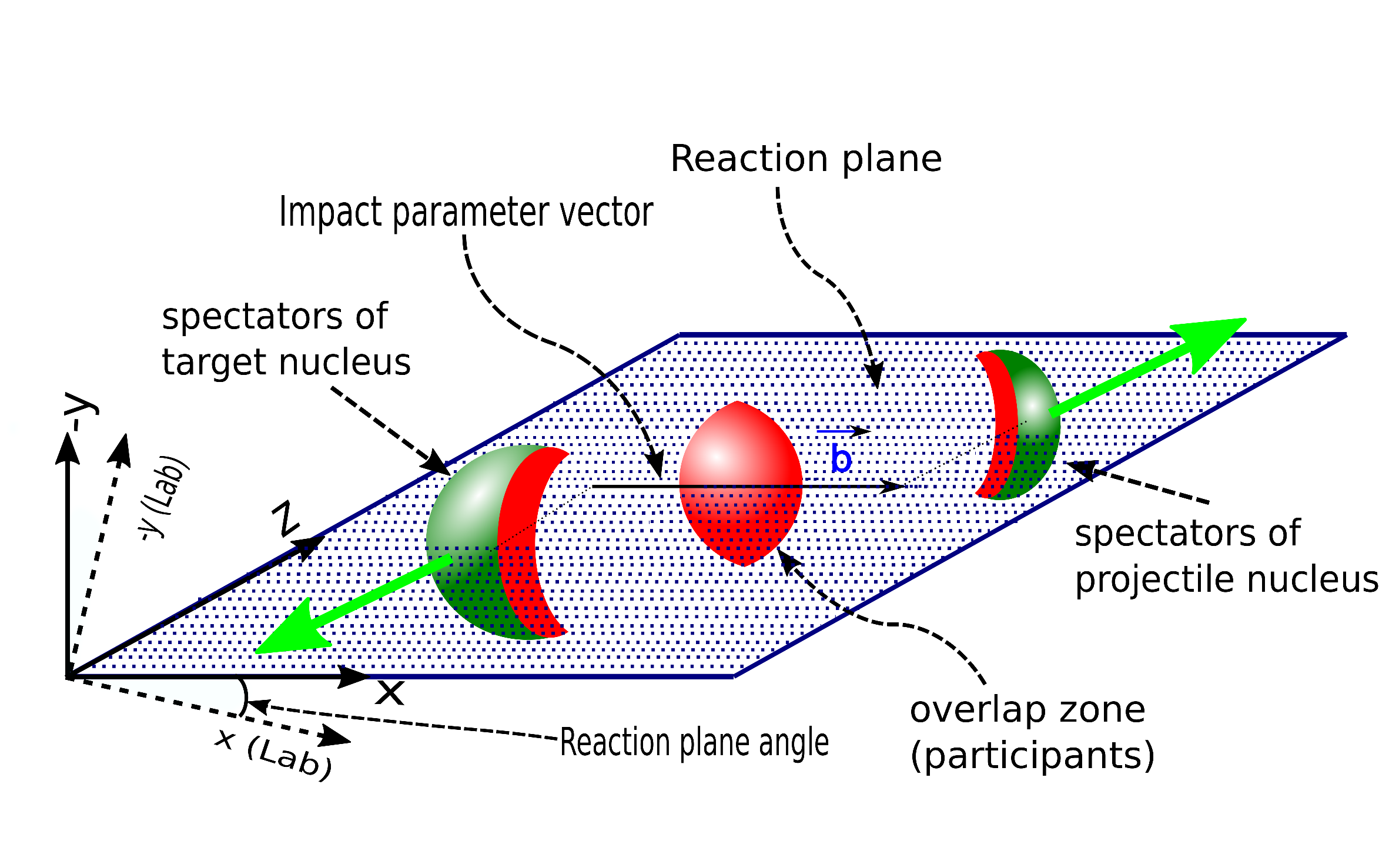} 
  \caption{ A schematic of a typical heavy-ion collision event depicting the convention of the coordinate system. The longitudinal axis, parallel to the beam pipe of the collider experiment, is labeled as the z-axis, while x and y represent the transverse coordinates. The x-axis of the collision event is parallel to the impact parameter vector ($\vec{b}$). The x and y coordinates of the laboratory system are denoted as x (Lab) and y (Lab), respectively. The reaction plane of the collision is the plane formed by $\vec{b}$ and the longitudinal z-axis. The reaction plane forms a non-zero angle with the xz plane of the laboratory coordinate system, known as the reaction plane angle. The almond-shaped interaction region, where the two nuclei overlap, is distinguished from the spectators of the projectile and target nuclei. }
 \label{fig:convention}
\end{figure}

The most commonly used coordinate system used in heavy-ion collision experiments is depicted in the diagram in Fig. \ref{fig:convention}. The direction parallel to the beam pipe is designated as the z-axis of the laboratory frame. The plane perpendicular to the z-axis is called the transverse plane. The nucleus moving in the positive z direction is referred to as the projectile nucleus, while the one moving with a negative $v_z$ (the z-component of velocity) is known as the target nucleus. The vector in the transverse plane that connects the center of the target nucleus to the center of the projectile nucleus is called the impact parameter vector, and its direction is defined as the positive x-axis of the collision event. Once the axes for the positive z and x-directions are established, the positive y-axis of the event can be automatically determined using right hand rule of cross product. The plane formed by the beam axis and the impact parameter is termed the reaction plane of the collision, and the angle between the reaction plane and the lab x-axis is known as the reaction plane angle. In heavy-ion collisions, this reaction plane angle fluctuates from event to event and is crucial for calculating flow-related observables in experiments. Although this angle cannot be directly measured, various proxies are used to estimate it. More details on this will be discussed in later sections.

In a simple geometrical model of heavy-ion collisions, nucleons from the projectile nucleus interact with those of the target nucleus during the collision. Nucleons that engage in the collision or contribute their kinetic energy to the interaction region are termed participants, while those that do not engage are referred to as spectators. The number of participants ($N_{part}$) depends on the magnitude of the impact parameter ($\vert \vec{b} \vert$) of the collision. For smaller values of $\vert \vec{b} \vert$, a larger proportion of nucleons participate in the collision, whereas for larger $\vert \vec{b} \vert$ values, the $N_{part}$ decreases. In heavy-ion collisions, events are classified into centrality classes based on $N_{part}$ values or the magnitude of the impact parameter. However, both the impact parameter magnitude and $N_{part}$ cannot be directly measured in experiments. Therefore, events are categorized into centrality classes based on final state observables such as mid-rapidity charged particle yield (multiplicity) or transverse energy measured in the mid-rapidity region \footnote{Rapidity(y) is defined in Sec. \ref{sec:kinematicvar}. The mid-rapidity corresponds to $y=0$. }. These observables are correlated with the impact parameter and $N_{part}$, although the relationship is not linear due to initial state fluctuations, such as the varying positions of nucleons within the nucleus, and the differing amounts of entropy production in different events during the collision's evolution phase.

\begin{figure}
  \centering
  \includegraphics[width=0.6\textwidth]{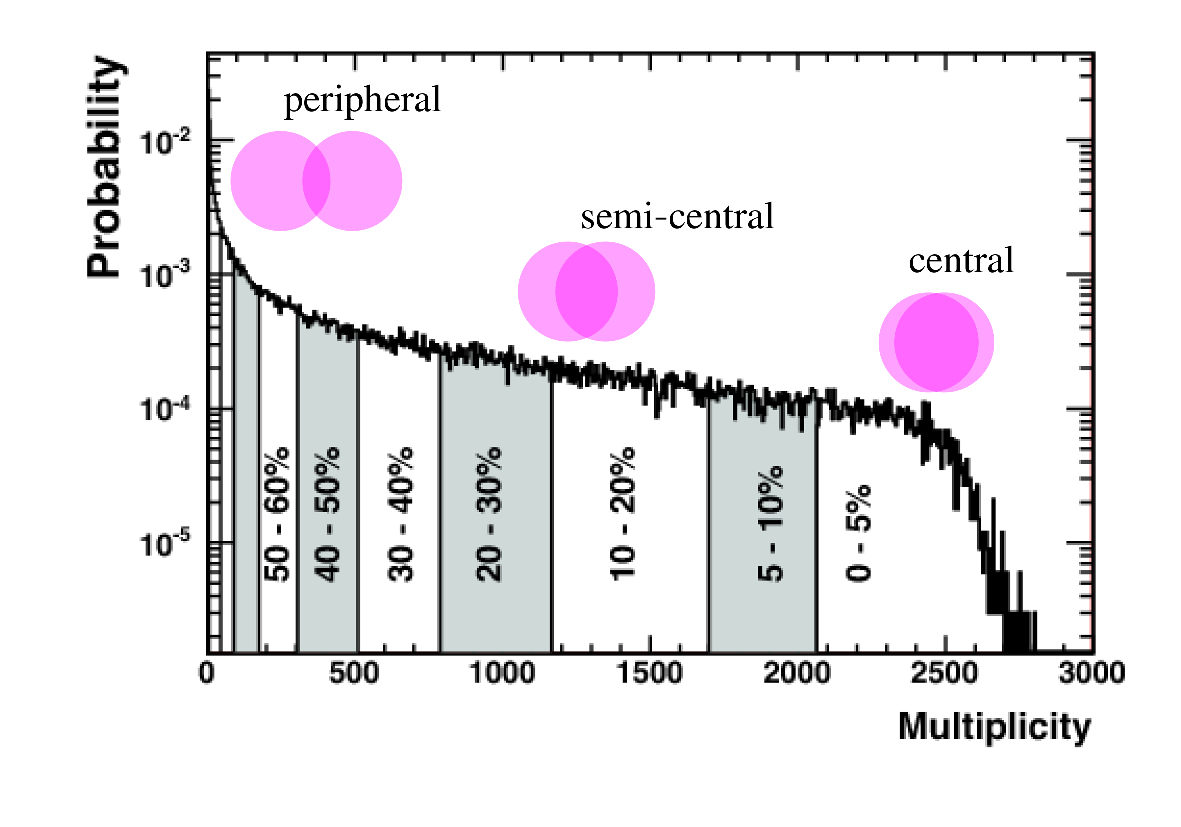} 
  \caption{ The multiplicity distribution of charged particles recorded by the time projection chamber (TPC) detector in the ALICE experiment for Pb+Pb collisions at $\sqrt{s_{NN}} = 2.76$ TeV. The multiplicity is calculated with a pseudorapidity cut of $\vert \eta \vert < 0.8$. Events are categorized into various centrality classes. Above the multiplicity distribution, there are cartoons depicting nuclear collisions at different impact parameters, providing visualizations of central, semi-central, and peripheral collision events. The figure is adapted from \cite{ALICE:2010suc}.  }
 \label{fig:multdist}
\end{figure}

In minimum bias collisions, events are collected and then categorized into centrality classes.  In Fig. \ref{fig:multdist}, the probability distribution of the charged particle multiplicity in the mid-rapidity region is shown, as obtained from minimum bias Pb+Pb collisions at $\sqrt{s_{NN}}=2.76$ TeV, measured in the ALICE experiment \cite{ALICE:2010suc}. The top 5\% of events, which produce the highest yield of charged particles in the mid-rapidity region, are categorized as 0-5\% centrality events. The next 5\% are grouped into the 5-10\% centrality class, and so on. Geometrically, it is apparent that collision events with larger impact parameters are more frequent than those with smaller impact parameters. Therefore, in a minimum bias event set, more events correspond to larger impact parameters, resulting in a higher probability of observing events with lower multiplicities compared to those with higher multiplicities. This trend can be observed in Fig. \ref{fig:multdist}, though the y-axis is presented on a logarithmic scale. Consequently, while the number of events within each centrality class is the same, the range of multiplicity values within a given centrality class varies. From Fig. \ref{fig:multdist}, it can be seen that the multiplicity range for the 0-5\% centrality class is approximately 2000 to 2700, whereas for the 5-10\% centrality class, it ranges from 1700 to 2000, despite both classes containing the same number of events.

In this thesis, we will express the magnitude of any dimensional observable in natural units. In natural units, the Planck constant ($\hbar$), the speed of light ($c$), and the Boltzmann constant ($k_B$) are set to one. Consequently, quantities such as energy, pressure, temperature, mass, length, and time can all be expressed in units of energy. For instance, in natural units, $\hbar c = 0.197$ GeV·fm = 1, which means 1 fm = 5.076 GeV$^{-1}$. Additionally, the Boltzmann constant is $k_B = 8.6173 \times 10^{-14}$ GeV·K$^{-1}=1$, so 1 Kelvin is equivalent to $8.6173 \times 10^{-14}$ GeV. Furthermore, time can also be expressed in GeV, with the relation $1$ fm/c = $0.333 \times 10^{-23}$ sec. $=5.076$ GeV$^{-1}$.

\subsection{Kinematic variables}
\label{sec:kinematicvar}
In heavy-ion collisions, it is more convenient to describe the momenta of the produced particles using transverse momentum ($p_T$), azimuthal angle ($\phi$), and rapidity ($y$). For a particle with four-momentum $p^{\mu} = (E,p_x,p_y,p_z)$, where $E$ denotes the energy and $p_x,p_y,p_z$ the three-momentum of the particle, one can define $p_T,\phi$ and $y$ as follows:
\begin{equation}
    p_T = \sqrt{\left( p_x^2 +p_y^2 \right)}
\end{equation} 
\begin{equation}
    \phi = \arctan \left( \frac{p_y}{p_x} \right).
\end{equation}
and
\begin{equation}
    y = \frac{1}{2} \ln{\left( \frac{E+p_z}{E-p_z} \right)}.
    \label{eq:defny}
\end{equation}
From the rapidity expression, we could derive \cite{Sahoo:2016hln,Florkowski:2010zz}:
\begin{equation}
    E = m_T \cosh{(y)}, \ \     p_z = m_T \sinh{(y)}
    \label{eq:mTcoshy}
\end{equation}
where $m_T = \sqrt{p_T^2 + m^2}$ is the transverse mass, with $m$ being the rest mass of the particle. One advantage of using rapidity over longitudinal momentum ($p_z$) is that rapidity is additive under a longitudinal boost. If a particle has rapidity $y$ in one inertial frame, it will have a rapidity of $y' = y + \Delta y$ in another inertial frame moving with rapidity $\Delta y$ or with a longitudinal velocity $v_z = \tanh(\Delta y)$ relative to the first frame.

In addition to rapidity, pseudorapidity ($\eta$) is another important kinematic variable, defined as:
\begin{equation}
    \eta = \frac{1}{2} \ln{\left( \frac{ \vert \vec{p} \vert+p_z}{\vert \vec{p} \vert -p_z}\right)} = -\ln{\left( \tan \frac{\theta}{2}\right)}
\end{equation}
where $|\vec{p}| = \sqrt{p_T^2 + p_z^2}$, and $\theta$ is the angle between $\vec{p}$ and the z-axis, with $p_z = |\vec{p}| \cos(\theta)$. 
For high-momentum or low-mass particles, where $|\vec{p}| \gg m$, $E \approx |\vec{p}|$, and thus $y \approx \eta$.

In high-energy collisions, the observation of a plateau in charged particle production at mid-rapidity indicates approximate longitudinal boost invariance in this central rapidity region. This symmetry implies that the system’s dynamics are invariant under boosts along the beam axis. To analyze the kinematics under such conditions, it is convenient to express the space-time coordinates $x^{\mu} = (t, x, y, z)$ in the Milne coordinate system, where the time ($t$) and longitudinal position ($z$) are replaced by the longitudinal proper time ($\tau$) and space-time rapidity ($\eta_s$). These are defined as: \beq \eta_s = \frac{1}{2} \ln{ \left( \frac{t+z}{t-z} \right) }, \ \ \tau = \sqrt{t^2 - z^2} \eeq Boost invariance holds if the system expands with a longitudinal velocity $v_z = z/t$, meaning in Milne coordinates the corresponding velocity $v_{\eta_s} = 0$. The advantage of using Milne coordinates is that it transforms a longitudinally expanding fluid in the laboratory frame into a static fluid in a coordinate system that naturally expands along the beam direction. This simplifies the description of the dynamics in scenarios where boost invariance approximately holds.

\subsection{Stages of relativistic heavy-ion collision}
\begin{figure}
  \centering
  \includegraphics[width=1\textwidth]{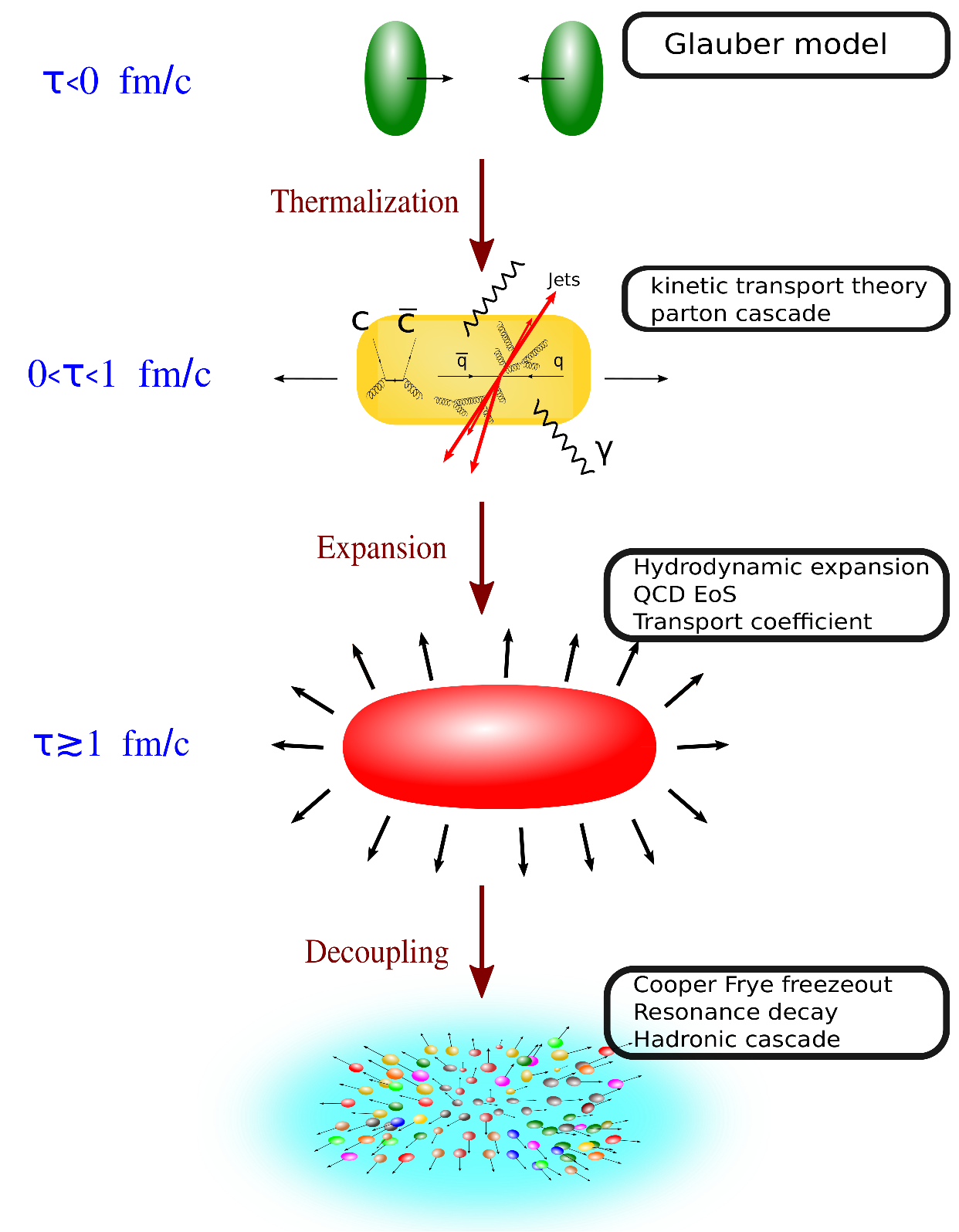} 
  \caption{Diagram depicting the stages of a heavy-ion collision. The outline of the diagram is adapted from ref. \cite{Heinz:2004qz}.  }
 \label{fig:stagesHIC}
\end{figure}

A typical relativistic heavy ion collision event at RHIC or LHC progresses through several distinct stages: initial interaction, thermalization, expansion, hadronization, and decoupling (or Freezeout) \cite{Heinz:2004qz,Busza:2018rrf}. These stages are illustrated in Fig. \ref{fig:stagesHIC}. Before the collision, as the nuclei move at relativistic speeds, they undergo Lorentz contraction along the z-axis, causing the nuclei to appear flattened in the laboratory frame. As the nuclei penetrate each other, the passage time is extremely brief, typically less than 0.2 fm/c for an Au+Au collision at $\sqrt{s_{NN}}=200$ GeV \cite{Shen:2017bsr}. During this short passage, the incident partons strongly interact, losing a portion of their kinetic energy in the interaction region. Most of these interactions involve small momentum transfers, resulting in the production of low-momentum (soft) particles. However, a small fraction of the incident partons experience hard scattering, characterized by a momentum transfer $\mathcal{Q}^2 \approx p_{T}^2 \gg 1 $ GeV$^2$, leading to the formation of high-energy jets or heavy-flavored quark-antiquark pairs \cite{CDF:2005fny,D0:2011jpq,H1:2003asu,ZEUS:1998tig,Salam:2010nqg,Dong:2019byy,Rapp:2009my,Heinz:2004qz}.

Unlike elementary particle collisions, the particles produced in a heavy ion collision do not simply escape into the surrounding vacuum. Instead, they undergo multiple rescatterings with each other. Immediately after the initial impact, the medium is highly out of equilibrium. However, if the density of the constituents within the medium is sufficiently high—as is typically the case in central to mid-central heavy ion collisions—these interactions can lead to the formation of a locally equilibrated, thermalized medium with a very high energy density \cite{Hwa:1985tv,Zhu:2017oei,Baier:2000sb,Berges:2014yta,Kurkela:2014tea}.

From mid-rapidity transverse energy measurements, the energy density of the system produced after about $\tau \sim 1$ fm/c in a central Au+Au collision at $\sNN=200$ GeV is estimated to be around 5 GeV/fm$^3$ \cite{STAR:2004moz}. This is about five times greater than the energy density predicted by lattice QCD for the phase transition from hadronic matter to the quark-gluon plasma (QGP). This indicates the formation of the QGP medium at RHIC. At even higher energies, such as those achieved at the LHC, the initial energy density is predicted to reach around 10 GeV/fm$^3$ \cite{Busza:2018rrf}, with lattice QCD calculations suggesting that matter in equilibrium at this energy density has a temperature of approximately 300 MeV \cite{HotQCD:2014kol} (around $3 \times 10^{12}$ Kelvin).

The produced thermalized medium exerts thermal pressure, and the pressure gradient relative to the surrounding vacuum drives the expansion of the QGP fireball. This tiny fireball expands into the surrounding medium, displaying collective behavior that can be modeled as the evolution of a hydrodynamic fluid in a vacuum \cite{Teaney:2001av,Heinz:2004ar,Jeon:2015dfa,Romatschke:2009im,Heinz:2013th,Kolb:2003dz,Leupold:2011zz,Bhalerao:2015iya}. As the fireball expands, it cools down, and its energy density decreases. When the energy density of any region of the fireball falls below a critical energy density of approximately 1 GeV/fm$^3$, the partons within the fireball combine to form hadrons in a process known as hadronization \cite{Heinz:2004qz,Busza:2018rrf,Florkowski:2010zz,Vogt:2007zz}.

After hadronization, the produced hadrons continue to interact via both inelastic and elastic collisions. The stage, where all scattering ceases and the system fully decouples, is known as freezeout, which occurs in two phases: chemical and kinetic freezeout. The point at which all inelastic collisions stop is called the chemical freezeout. After chemical freezeout, the hadrons in the system no longer change identity, and no new hadrons are produced. However, elastic collisions continue to change the momenta of the hadrons. In heavy ion collisions, the large number of produced hadrons ensures that these elastic collisions are efficient in maintaining thermal equilibrium. The signature of this thermal equilibrium before kinetic freezeout is reflected in the momentum spectra of the final hadrons. The point at which all elastic collisions cease is known as the kinetic freezeout.

\subsection{Basic observables}
The final outcome of a typical heavy ion collision event is a large number of hadrons. All the information, from the initial impact to the final freezeout, is usually inferred from the measured momentum spectra of these hadrons. Numerous distinct observables have been proposed, measured, and interpreted based on the momentum distribution of the produced hadrons. These observables have proven invaluable in extracting relevant information about the system at different stages of its evolution and in probing the properties of the produced QGP. We will focus on a few fundamental observables that are particularly relevant to the results discussed in this thesis.

The Lorentz-invariant single-particle distribution of a specific hadron species $i$ in momentum space is expressed as:
\beq
E \frac{dN_i}{d^3p} = \frac{dN_i}{p_T dp_T dy d\phi}
\eeq
To calculate the total yield of that hadron type, we integrate over the phase space:
\beq
N_i = \int p_T dp_T dy d\phi \frac{dN_i}{p_T dp_T dy d\phi}
\eeq
Experimental measurements are often conducted within specific kinematic ranges, hence the limits of integration in theoretical or model calculations are adjusted accordingly to facilitate direct comparison. Additionally, observables in experiments are typically averaged over multiple events within a particular centrality class, and model calculations account for this as well.

To obtain the rapidity distribution of a given particle species $i$, we integrate the triple differential spectrum over transverse momentum ($p_T$) and azimuthal angle ($\phi$): \begin{equation} \frac{dN_i}{dy} = \int p_T  dp_T  d\phi  \frac{dN_i}{p_T  dp_T  dy  d\phi}. \end{equation}

To convert this to the pseudorapidity distribution $dN_i/d\eta$, one can use the Jacobian transformation: \begin{equation} \frac{dy}{d\eta} = \frac{|\vec{p}|}{E}, \end{equation} which allows for calculating $dN_i/d\eta$ for each particle species. The total charged particle pseudorapidity distribution $dN_{\text{ch}}/d\eta$ is then obtained by summing the contributions from all charged species. The mid-rapidity charged hadron yield is directly proportional to the initial total entropy deposited in the nuclear overlap region, providing an estimate of the centrality or impact parameter of the collision event \cite{Song:2008si,Shen:2014lye}.

Additionally, the $p_T$ differential spectrum and mean $p_T$ ($\la p_T \ra$) of hadron species $i$ within a rapidity window $dy$ can be calculated as follows:
\beq
\frac{dN_i}{d y p_T dp_T} =  \int d\phi \frac{dN_i}{p_T dp_T dy d\phi}
\eeq   

\beq
\langle p_T \rangle = \frac{ \int p_T \left( \frac{dN_i}{p_T dp_T dy d\phi} \right)  p_T dp_T dy d \phi }{ \int  \left( \frac{dN_i}{p_T dp_T dy d\phi} \right) p_T dp_T dy d \phi}
\label{eq:menpt}
\eeq
The measurements of the $p_T$ differential spectra and $\la p_T \ra$ provide essential insights into the bulk evolution of the QGP fluid \cite{Ollitrault:2007du,Borghini:2005kd,Parida:2024ckk,Samanta:2023kfk,Bozek:2009dw,STAR:2017sal}.

Anisotropic flow coefficients are key observables in heavy ion collisions, providing vital information about the initial state and the medium properties of the fluid \cite{Yan:2014nsa,Shen:2014lye,Romatschke:2007mq,Song:2009gc,Giacalone:2020ymy,Bernhard:2018hnz,Bozek:2011ua,Bozek:2009dw}. In these collisions, the initial spatial anisotropy is transformed into final-state momentum space anisotropy due to the strong collective behavior of the medium, which is reflected in the azimuthal angle distribution of the produced particles \cite{Ollitrault:1992bk}. This anisotropy in the transverse plane can be quantified by expanding the invariant particle distribution in a Fourier series \cite{Voloshin:1994mz,Poskanzer:1998yz}:
\beq
\frac{dN_i}{p_T dp_T dy d\phi} =  \frac{dN_i}{p_T dp_T dy} \frac{1}{2 \pi}  \left(  1 + 2 \sum_{n=1}^{\infty} \left( v_n(p_T,y) \cos{ n (\phi-\Psi_n) }  \right) \right)
\label{eq:fourier}
\eeq
Here, $\Psi_n$ is the $n$th-order event plane angle, and $v_n$ is the $n$th-order flow coefficient \cite{Voloshin:1994mz,Poskanzer:1998yz,Bilandzic:2008nx}. These flow coefficients can be calculated from particle tracks as:
\beq
v_n(p_T,y) = \langle \langle \cos{ n (\phi-\Psi_n) } \rangle \rangle
\eeq 
where the first angular bracket represents an average over all tracks in a given event, and the second bracket represents an average over all events within a centrality class. The first flow harmonic, $v_1$, is referred to as directed flow; the second-order flow coefficient, $v_2$, is known as elliptic flow; and the third-order coefficient, $v_3$, is termed triangular flow, and so on. This thesis primarily focuses on directed flow, so we will elaborate further on this observable.

Directed flow is characterized by the first-order harmonic coefficient in the Fourier series expansion of the azimuthal distribution of produced particles. It is typically measured with respect to a symmetry plane angle, denoted as $\Psi$, of the collision: 
\beq
v_1(p_T,y) = \langle \langle  \cos(\phi-\Psi) \rangle \rangle
\eeq
The directed flow is classified into two types: rapidity-even ($v_1^{\text{even}}$) and rapidity-odd ($v_1^{\text{odd}}$) directed flow. The The rapidity-even component, $v_1^{\text{even}}$, is measured relative to the first-order event plane angle ($\Psi^{\text{EP}}_{1}$), while the rapidity-odd component is measured with respect to the first order spectator plane angle ($\Psi^{\text{SP}}_{1}$). The selection of $\Psi$ for measuring a specific type of $v_1$ is guided by it's correlation with different symmetry plane angles.

Rapidity-odd directed flow arises from the breaking of forward-backward symmetry or boost invariance in the collision geometry. It develops either parallel or anti-parallel to the impact parameter direction at different rapidities. Consequently, determining the direction of $\vec{b}$ in each collision event is essential. As noted in previous sections, in the context of heavy ion collisions, the impact parameter direction is taken as the positive x-axis of the collision event.
Thus, identifying the direction of $\vec{b}$ is equivalent to determining the positive x-side of the collision event. The importance of establishing the direction of $\vec{b}$ (or the positive x-axis) for measuring $v_1^{\text{odd}}$ is further discussed in \nameref{ch:appendix1}. The angle $\Psi^{\text{SP}}_{1}$ provides an estimate of the angle between the x-axis of the laboratory coordinate system and the impact parameter direction. The experimental techniques for determining $\Psi^{\text{SP}}_{1}$ are elaborated in refs. \cite{STAR:2008jgm,ALICE:2013xri}.

\begin{figure}
  \centering
  \includegraphics[width=0.7\textwidth]{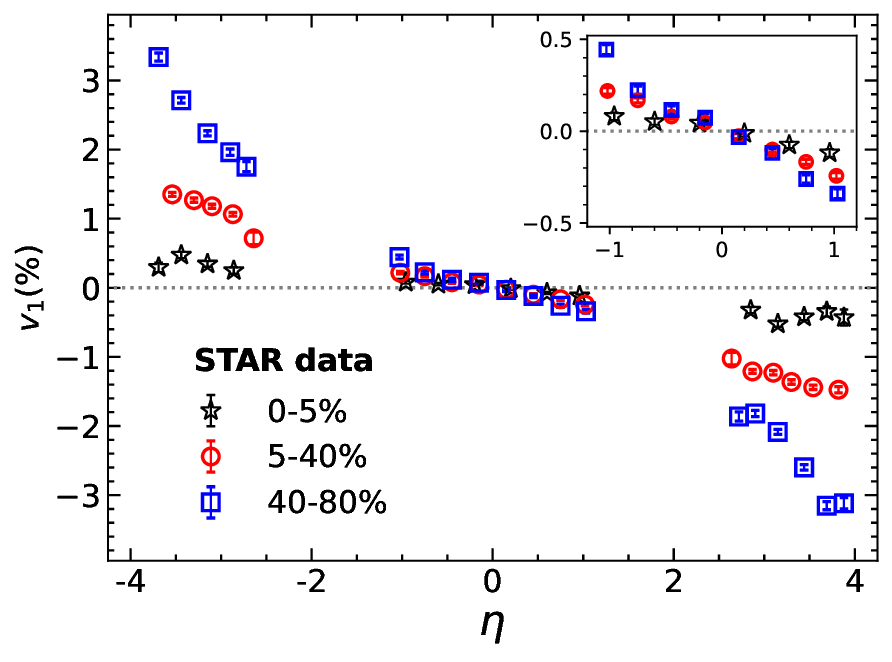} 
  \caption{ The pseudorapidity ($\eta$) dependence of the rapidity-odd directed flow of charged hadrons measured by the STAR collaboration for different centralities in Au+Au collisions at $\sNN = 200$ GeV. The inset provides a detailed view of the mid-$\eta$ region. The experimental data is taken from \cite{STAR:2008jgm}.  }
  \label{fig:v1oddSTAR}
\end{figure}

The $v_1^{\text{odd}}$ is calculated by rotating the azimuthal angle of each track by $\Psi^{\text{SP}}_{1}$ of the respective event and then averaging over all tracks and ultimately across all events:
\beq
v_1^{\text{odd}} = \langle \langle  \cos(\phi-\Psi^{\text{SP}}_{1}) \rangle \rangle.
\eeq 
Considering an event where $\Psi^{\text{SP}}_{1}=0$, we find that $v_1^{\text{odd}}=\la cos \phi \ra = \la p_x / p_T \ra$. While $p_T$ is always positive, $p_x$ can be both positive and negative for the produced particles. This indicates that $v_1^{\text{odd}}$ will be non-zero if the left ($p_x<0$) and right ($p_x>0$) symmetry is broken or if there is asymmetric particle production along $\phi=0$ and $\phi=\pi$. Fig. \ref{fig:v1oddSTAR} shows the $v_1^{\text{odd}}$ of charged hadrons as a function of pseudorapidity ($\eta$) for 0-5\%, 5-40\%, and 40-80\% centrality, measured by the STAR collaboration in Au+Au collisions at $\sNN=200$ GeV. The $v_1$ is observed to be an odd function of $\eta$, with a zero crossing at $\eta=0$. The $v_1^{\text{odd}}$ is often quantified by measuring it's slope of at mid-rapidity, given by $\frac{dv_1}{dy}$ or $\frac{dv_1}{d\eta}$. This slope increases with centrality, reflecting greater longitudinal symmetry breaking from central to peripheral collisions.

In symmetric nucleus collisions, such as Au+Au or Pb+Pb, the rapidity-even $v_1$ arises from event-by-event fluctuations in the energy deposition by participants. It is measured relative to the first order event plane angle, $\Psi^{\text{EP}}_{1}$ \cite{Retinskaya:2012ky,Jia:2012hx}. This angle, $\Psi^{\text{EP}}_{1}$, is determined from the produced particles and is random due to fluctuations, having no correlation with the direction of $\vec{b}$ :
\beq
v_1^{\text{even}} = \langle \langle  \cos(\phi-\Psi^{\text{EP}}_{1}) \rangle \rangle.
\eeq 
In symmetric collisions, the source of rapidity-even $v_1$ is fluctuations. However, in asymmetric collisions, such as Cu+Au, this can be non-zero not only due to fluctuations but also due to the collision geometry \cite{Bozek:2012hy,STAR:2016cio}. As discussed earlier, $v_1^{\text{odd}}$ is zero at mid-rapidity($y=0$) due to geometrical symmetry, while the even component can be non-zero there as a result of fluctuations. In ultra-central collisions, $v_1^{\text{odd}}$ remains zero, but its even counterpart persists due to finite fluctuations. The magnitude of $v_1^{\text{odd}}$ is strongly dependent on collision energy. As the energy increases, boost invariance is better approximated, leading to a reduction in $v_1^{\text{odd}}$, whereas $v_1^{\text{even}}$ exhibits very little energy dependence since it is primarily driven by fluctuations \cite{STAR:2018gji}. Studying both types of $v_1$ and comparing them with model predictions is crucial for understanding the initial conditions, transport properties of the QCD medium, and the nature of the transition from the quark-gluon plasma (QGP) to the hadronic phase, among other aspects \cite{Bozek:2010bi,Du:2022yok,Stoecker:2004qu,Nara:2016phs,Singha:2016mna,Bozek:2022svy}. 

It is important to note that in this thesis, whenever directed flow is mentioned, it refers specifically to rapidity-odd directed flow. The rapidity-even directed flow will be explicitly mentioned when discussed or presented in the thesis.

\section{Motivation of this thesis}

\begin{figure}
  \centering
  \includegraphics[width=0.8\textwidth]{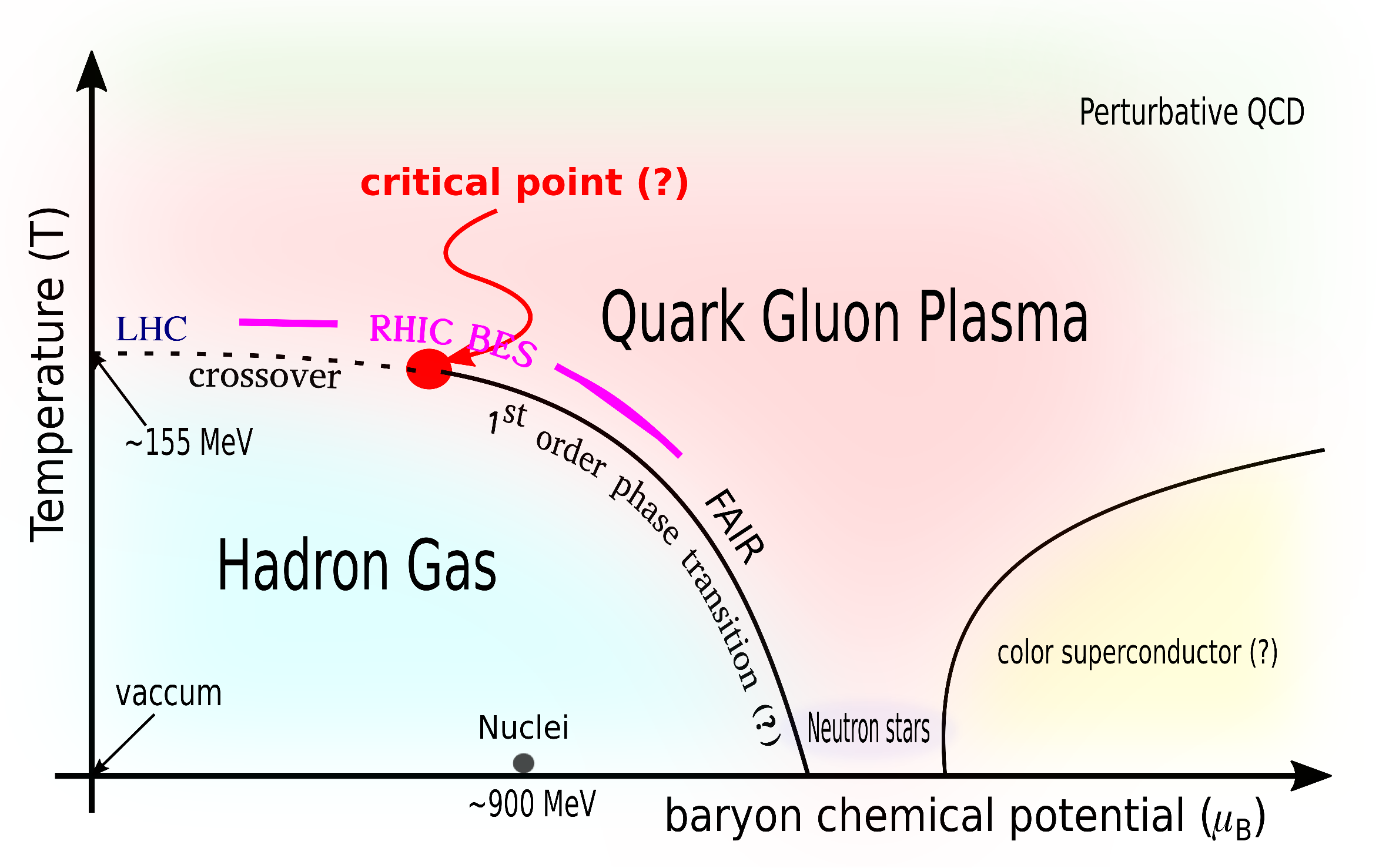} 
  \caption{ A sketch of the QCD phase diagram in $(T,\mu_B)$ plane. }
  \label{fig:phases}
\end{figure}

In addition to probing the signatures and studying the properties of the QGP, a major goal of heavy-ion collision experiments is to investigate the QCD phase diagram in detail \cite{Akiba:2015jwa,Stephanov:2004wx,Guenther:2022wcr}. This research aims to provide insight into the nature of the phase transition between the hadron gas phase and the QGP phase at different conserved charge densities. A sketch of the conjectured QCD phase diagram is shown in Fig. \ref{fig:phases}, illustrating the phase structure in the plane of temperature ($T$) and baryon chemical potential ($\mu_B$). The baryon chemical potential, $\mu_B$, is related to the net baryon density ($n_B$), with an increase in $n_B$ corresponding to a higher value of $\mu_B$. At low temperatures and $\mu_B$, the energy scale is below the QCD confinement scale, $\Lambda_{\text{QCD}}$, leading to the existence of QCD matter in the form of color singlets or bound states. In this region, QCD matter remains as a hadron gas. Conversely, at high temperatures and $\mu_B$, QCD matter transitions to the QGP phase. It is believed that QCD matter with very low temperature but high $\mu_B$ might exist inside neutron stars \cite{Annala:2019puf,Alford:2000sx}. At even smaller $\mu_B$ and very low temperatures, matter exists in the form of atomic nuclei, which is the most common form of matter we experience. Additionally, in the region of low temperature and very high $\mu_B$, there is speculation about the existence of a color superconducting phase, where quarks are expected to form Cooper pairs near the Fermi surface, similar to conventional superconductors \cite{Alford:2007xm}.

First-principle lattice QCD calculations indicate that the transition from hadron gas to QGP at zero $\mu_B$ is a smooth crossover occurring around $T \approx 155$ MeV \cite{HotQCD:2018pds,Aoki:2006we,Borsanyi:2020fev,Guenther:2022wcr}. At larger $\mu_B$, effective theory models, such as the Nambu-Jona-Lasinio (NJL) model and the non-linear sigma model, predict a first-order phase transition between the hadronic phase and the QGP phase \cite{Asakawa:1989bq,Scavenius:2000qd,Antoniou:2002xq}. Consequently, it is expected that a critical point (CP) exists at finite $\mu_B$, where the nature of the transition changes from first-order to crossover. While there are many theoretical predictions regarding the precise location of the CP in the QCD phase diagram, these predictions vary widely across a large range of $\mu_B$ \cite{Luo:2017faz}. Moreover, lattice QCD calculations at finite $\mu_B$ face significant challenges due to the fermion sign problem \cite{Nagata:2021ugx,Goy:2016egl}. Additionally, perturbative QCD is not applicable at this energy scale. Given these challenges, it is currently difficult to get the exact position of the CP in the QCD phase diagram using first-principle calculations. Instead, researchers rely on phenomenological approaches and heavy-ion collision experiments. In this methodology, expected criticality signals are estimated using phenomenological models that assume the existence of the CP. Systematic heavy-ion collision experiments are then conducted at different collision energies, indirectly scanning the phase diagram to detect where such criticality signals might occur. Several experimental programs are actively searching for the CP, including those at the Relativistic Heavy Ion Collider (RHIC) at Brookhaven National Laboratory (BNL) \cite{Tlusty:2018rif,Luo:2017faz}, the High-Acceptance Di-Electron Spectrometer (HADES) at GSI \cite{HADES:2009aat,HADES:2022gdr}, and the NICA experiment in Dubna \cite{Kekelidze:2014dta}.

\begin{figure}[htbp]
    \centering
    \begin{minipage}{0.45\textwidth} 
        \centering
        \includegraphics[width=\textwidth]{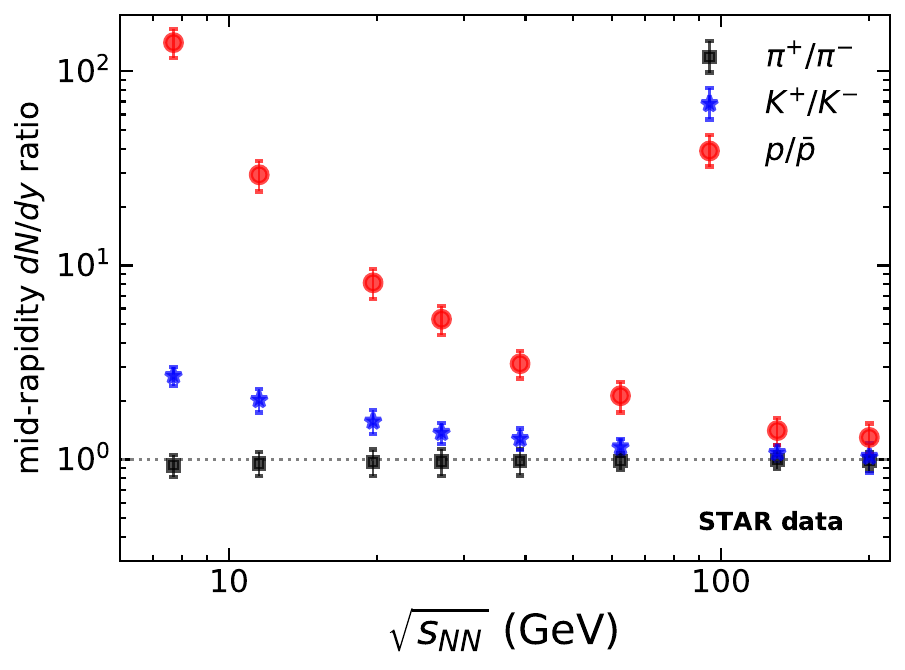}
        \caption{Experimental measurements of mid-rapidity particle yield ratios ($\pi^{+}/\pi^{-}$, $K^{+}/K^{-}$, and $p/\bar{p}$) as a function of collision energy $\sqrt{s_{\text{NN}}}$. The data is from the STAR collaboration, reported in Refs. \cite{STAR:2017sal,STAR:2008med}. }
        \label{fig:part_ratio_data}
    \end{minipage}
    \hfill 
    \begin{minipage}{0.45\textwidth} 
        \centering
        \includegraphics[width=\textwidth]{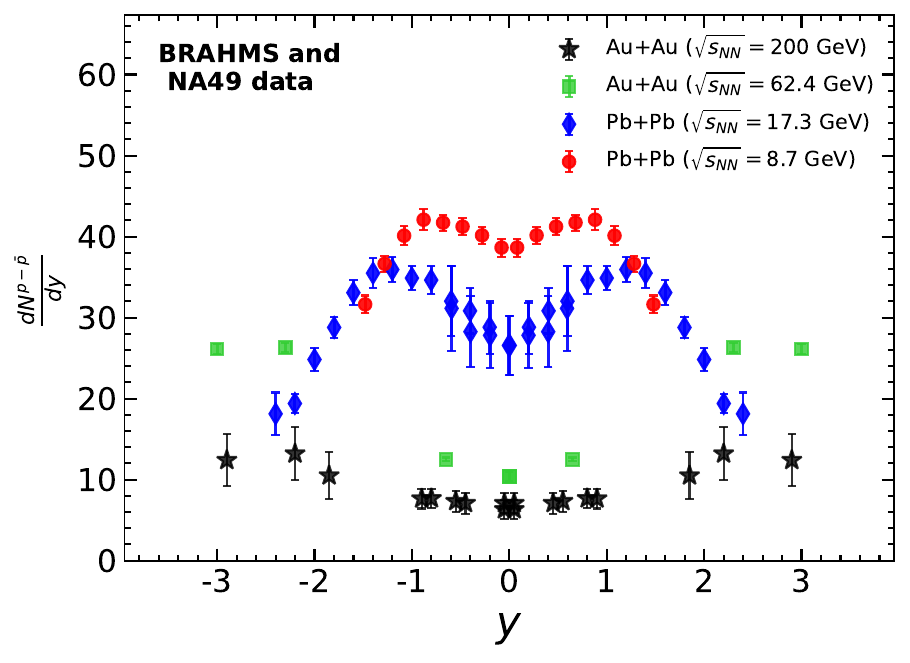}
        \caption{Experimental data of the rapidity-differential net-proton yield at various collision energies. The data is from the BRAHMS collaboration \cite{BRAHMS:2003wwg,BRAHMS:2009wlg} and NA49 collaboration \cite{NA49:1998gaz,NA49:2010lhg,NA50:2002edr}.   }
        \label{fig:rap_netp_data}
    \end{minipage}
    \label{fig:combined_figures}
\end{figure}

The experimental measurement of the mid-rapidity yield ratios between $\pi^{+}/\pi^{-}$, $K^{+}/K^{-}$, and $p/\bar{p}$ as a function of collision energy is shown in Fig. \ref{fig:part_ratio_data}. In the framework of a grand canonical ensemble for particle production, these yield ratios reflect the average chemical potentials of conserved charges in the produced system. Specifically, the ratios follow: $\pi^{+}/\pi^{-} \propto \exp{(\mu_Q)}$, $p/\bar{p} \propto \exp{(2\mu_Q + 2\mu_B)}$, and $K^{+}/K^{-} \propto \exp{(2\mu_Q + 2\mu_S)}$, where $\mu_Q$, $\mu_S$, and $\mu_B$ are the chemical potentials corresponding to electric charge, net strangeness, and net baryon number, respectively. Experimental data show that the $\pi^{+}/\pi^{-}$ ratio remains close to unity across the collision energy range of $\sNN = 7.7$ to 200 GeV, indicating that the system has a small net charge chemical potential. However, for $p/\bar{p}$, a significant increase is observed at lower $\sNN$, suggesting a larger net baryon chemical potential ($\mu_B$) in the produced medium. This is further supported by the rapidity distribution of the net-baryon yield across different $\sNN$ values, as shown in Fig. \ref{fig:rap_netp_data}, which demonstrates increased baryon stopping in the mid-rapidity region as the collision energy decreases. These observations imply that, at lower collision energies, a medium with a larger $\mu_B$ is produced in the mid-rapidity region, granting access to the high $\mu_B$ region of the QCD phase diagram. Therefore, it is believed that relevant observables sensitive to the QCD critical point can be measured at various collision energies, potentially providing signatures of the critical point.

It has been proposed that the cumulants of the net baryon number ($\kappa_j$) exhibit strong sensitivity to the QCD critical point, which are related to the baryon susceptibilities ($\chi_j^B$) of the QCD equation of state through the following relation \cite{Bzdak:2019pkr,Luo:2017faz,Gupta:2011wh}:
\beq
\chi_{j}^{B} \equiv \frac{\partial^j (p/T^4) }{\partial (\mu/T)^j} = \frac{1}{VT^3} \kappa_j
\eeq
Model calculations suggest that, near the QCD critical point, higher-order cumulants of the baryon number are expected to diverge, as they are proportional to higher powers of the correlation length ($\xi$) \cite{Stephanov:2008qz,Stephanov:2011pb}: 
\beq
\kappa_2 = \la (\delta N)^2 \ra \sim \xi^2, \ \ \kappa_3 = \la (\delta N)^3 \ra \sim \xi^{9/2}, \ \ \kappa_4 = \la (\delta N)^4 \ra \sim \xi^{7}
\eeq
Here, $N$ represents the baryon number, and $\delta N = N - \langle N \rangle$ is the event-by-event fluctuation of the baryon number from its event-averaged value, $\langle N \rangle$.
In this context, the cumulants of event-by-event fluctuations in net proton or proton yields are being measured at various collision energies in search of the QCD critical point \cite{STAR:2020tga,STAR:2021iop}. 

Notably, these measurements are conducted in the mid-rapidity region. However, event-by-event fluctuations in the initial baryon stopping and the subsequent transport of baryons to mid-rapidity could introduce an additional source of fluctuation in the net proton yield. This could interfere with the observation of critical fluctuations, which are the key signals of interest. To distinguish critical signals, it is essential to establish baseline predictions for the behavior of observables in the absence of a critical point, where fluctuations are entirely non-critical. Models that accurately incorporate initial baryon stopping and its subsequent dynamics, while also successfully describing bulk observables, can provide a reliable baseline. This underscores the need for a thorough understanding of both initial baryon stopping and it's dynamics in the produced system.

In addition to the search for the QCD critical point, several other interesting physical phenomena are being investigated at RHIC, where baryon stopping and dynamics are expected to play a crucial role. One notable example is the search for the signals of the electromagnetic (EM) field. In this regard, the STAR collaboration has proposed the centrality dependence of the splitting of rapidity-odd directed flow slope ($\Delta (dv_1/dy)$) between oppositely charged hadrons, such as $\pi^{+}-\pi^{-}$, $K^{+}-K^{-}$, and $p-\bar{p}$, as a promising observable \cite{STAR:2023jdd}. A key observation is the sign change of $\Delta (dv_1/dy)$ in peripheral collisions, which is believed to result from the presence of a strong EM field \cite{STAR:2023jdd,Gursoy:2018yai}. Importantly, in these measurements, the splitting between $p$ and $\bar{p}$ shows the strongest centrality dependence and is interpreted as a robust signal of the EM field. However, the $v_1$ splitting between $p$ and $\bar{p}$ can not be attributed solely to the EM field, which affects the flow of oppositely charged protons and antiprotons differently. The difference in baryon number between $p$ and $\bar{p}$ could also be responsible for their different flow. The inhomogeneous distribution of net baryons within the medium and its evolution could lead to a splitting in their flow \cite{Bozek:2022svy}. Similarly, the difference in strangeness quantum number between $K^{+}$ and $K^{-}$ can cause a splitting in $v_1$ due to the anisotropic evolution of net strangeness in the medium. Therefore, understanding the dynamics of conserved charges in the system formed in low-energy collisions is essential for interpreting EM field signals accurately.

Moreover, in non-central collisions of relativistic heavy ion nuclei, the system carries a large amount of angular momentum. After the collision, part of this angular momentum is transferred to the locally thermalized fireball. The hydrodynamic response generates fluid vorticity and potentially spin polarization, which ultimately  reflected in the phase space distribution and polarization states of the emitted particles \cite{Liang:2004ph,Liang:2004xn,Becattini:2007sr,Betz:2007kg,Karpenko:2016jyx}. In this context, polarization is being measured for $\Lambda$ and $\bar{\Lambda}$ particles, as they are well-suited for such studies due to their weak decay \cite{STAR:2017ckg,STAR:2023nvo,Niida:2024cmt}. Their decay products, $p$ and $\bar{p}$, tend to be emitted along the spin direction of the parent $\Lambda$ and $\bar{\Lambda}$, respectively. Observations have shown a difference in the polarization between $\Lambda$ and $\bar{\Lambda}$ \cite{STAR:2017ckg}. Since $\Lambda$ and $\bar{\Lambda}$ carry opposite baryon numbers, it is expected that baryon stopping and dynamics play a role in this polarization behavior as well \cite{Karpenko:2024xla,Ryu:2021lnx,Wu:2022mkr,Sahoo:2024yud}.

In addition to the observables discussed above, other observables which are being investigated in lower energy collisions for their potential to reveal important physics is likely to be influenced by baryon stopping and dynamics. As we explore lower collision energies, it becomes increasingly crucial to understand the system's dynamics in the presence of conserved charges. In particular, developing a clear understanding of initial baryon stopping and its subsequent evolution is essential, as net baryon number is observed to be the dominant conserved charge \cite{STAR:2017sal,STAR:2008med,NA49:1998gaz,NA49:2010lhg,NA50:2002edr,BRAHMS:2003wwg,BRAHMS:2009wlg}.

For over two decades, research in heavy-ion collisions has predominantly concentrated on the dynamics of baryon-free quark-gluon plasma (QGP) produced in ultra-relativistic high-energy heavy-ion collisions, where the mid-rapidity region of the resulting system exhibits nearly zero conserved charge density \cite{Akiba:2015jwa,Arslandok:2023utm}. In this context, hydrodynamic models have proven particularly successful in explaining experimental data \cite{Ollitrault:2007du,Hirano:2008hy,Heinz:2013th,Bhalerao:2015iya}. Phenomenological studies conducted within the relativistic viscous hydrodynamic framework, which primarily solve the conservation equations of the energy-momentum tensor, have yielded a deeper understanding of the dynamics of baryon-free QGP. However, the study of QGP dynamics in the presence of finite baryon density is gaining increased attention, particularly at lower collision energies, where experimental data indicates that a significant amount of baryon stopping occurs in the mid-rapidity region \cite{Denicol:2018wdp,Pihan:2024lxw,Shen:2023awv,Shen:2022oyg,Ryu:2021lnx,Shen:2020jwv,Shen:2017bsr,Li:2018fow,Du:2022yok,Bozek:2022svy}. Furthermore, at lower collision energies, the breaking of boost invariance necessitates full 3+1D simulations of the evolving system \cite{STAR:2014clz,STAR:2019vcp,STAR:2017okv,Denicol:2018wdp,Pang:2015zrq}. This approach requires solving the conservation equations for net baryon current alongside the energy-momentum tensor within a comprehensive 3+1D hydrodynamic framework. Currently, hydrodynamic models that incorporate finite baryon density evolution have been developed \cite{Denicol:2018wdp,Wu:2021fjf,Plumberg:2024leb,Du:2019obx}, and corresponding phenomenological studies are ongoing \cite{Denicol:2018wdp,Pihan:2024lxw,Shen:2023awv,Shen:2022oyg,Ryu:2021lnx,Shen:2020jwv,Shen:2017bsr,Li:2018fow,Du:2022yok,Bozek:2022svy}.

Three crucial ingredients are required as inputs to study the dynamics of the baryonic quark-gluon plasma (QGP) medium within the hydrodynamic model:
\begin{itemize}
\item Initial conditions,
\item Equation of state (EoS), and
\item Transport coefficients.
\end{itemize}
In the baryon-free scenario, we have already acquired insight regarding the initial energy-momentum deposition, the EoS at zero baryon chemical potential, and transport coefficients such as shear and bulk viscosity. However, in the presence of baryons, additional inputs are necessary: the initial net baryon distribution, the EoS at finite baryon chemical potential, and transport coefficients associated with baryon diffusion.

From these three components:
\begin{itemize}
\item The EoS at finite baryon density can be derived by extending the lattice QCD EoS to non-zero $\mu_B$ using the Taylor expansion method \cite{Monnai:2024pvy,Monnai:2019hkn,Noronha-Hostler:2019ayj}.
\item The baryon diffusion coefficient are obtained from both kinetic theory \cite{Denicol:2018wdp} and strongly interacting theories \cite{Rougemont:2015ona}, and it can also be constrained through model-to-data comparisons.
\item Most importantly, the initial baryon deposition, which is related to the initial baryon stopping within the system, remains largely unknown. Therefore, it is crucial to model the initial baryon deposition accurately and to constrain the models through comparisons with experimental data by identifying relevant observables sensitive to initial baryon stopping.
\end{itemize}

In this context, it has been observed that the directed flow of hadrons is sensitive to the initial three-dimensional distribution of matter \cite{Bozek:2010bi,Bozek:2011ua,Hirano:2002ds,Hirano:2001yi,Shen:2020jwv,Du:2022yok,Ryu:2021lnx,Parida:2022lmt,Jiang:2021foj,Jiang:2021ajc}. Notably, the splitting of 
$v_1$ between baryons and anti-baryons is particularly responsive to the initial baryon deposition profile \cite{Bozek:2022svy,Shen:2020jwv,Du:2022yok}. Therefore, it is believed that the 
$v_1$ of identified hadrons can provide valuable insights into the initial baryon distribution within the medium. Additionally, comparing the 
$v_1$ values of baryons and anti-baryons can help distinguish between different models of initial baryon stopping \cite{Bozek:2022svy,Shen:2020jwv,Du:2022yok}.

\begin{figure}
  \centering
  \includegraphics[width=0.7\textwidth]{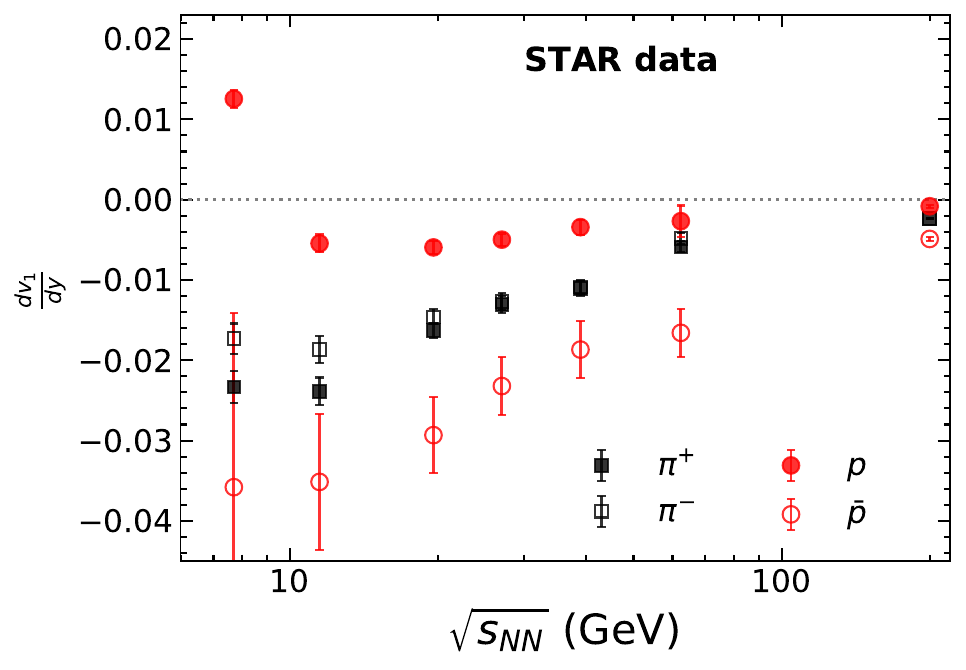} 
  \caption{ Experimental measurements of the mid-rapidity slope of rapidity-odd directed flow ($dv_1/dy$) for $\pi^{\pm}$, $p$, and $\bar{p}$ as a function of collision energy $\sNN$ for 10-40\% centrality. Data is from the STAR collaboration \cite{STAR:2014clz}.   }
  \label{fig:v1_pippbar_STAR}
\end{figure}

In Fig. \ref{fig:v1_pippbar_STAR}, the mid-rapidity slope of rapidity odd directed flow ($dv_1/dy$) for $\pi^{\pm},p$ and $\bar{p}$ is plotted as a function of collision energy. It is observed that the splitting between $p$ and $\bar{p}$ increases with decreasing collision energy. Analyzing this observation alongside the experimental measurements of the proton to anti-proton yield ratio (see Fig. \ref{fig:part_ratio_data}) indicates a close relationship between the $v_1$ of baryons, anti-baryons, and the degree of net baryon stopping. This relationship emphasizes the importance of studying the observable $v_1$ to better understand initial baryon stopping.

The observed splitting of directed flow between protons and anti-protons, along with the non-monotonic behavior of $dv_1/dy$ for protons, has posed significant challenges for models to capture \cite{Singha:2016mna,Shen:2020jwv}. Within a hydrodynamic framework, these challenges primarily arise from lack of accurate modeling of initial baryon stopping. Thus, it is crucial to develop a model of initial baryon deposition that can explain the experimental data on $v_1$ of identified hadrons and be utilized as input in hydrodynamic models to understand baryon dynamics in heavy ion collisions.

\begin{itemize}
\item To address this issue, this thesis proposes an initial net-baryon profile that, upon further hydrodynamic evolution, produces the directed flow of $p$, $\Lambda$ , and their antiparticles in agreement with data from the STAR Beam Energy Scan program \cite{STAR:2014clz,STAR:2019vcp,STAR:2017okv}.

\item Additionally, using our initial baryon profile, we have extracted the baryon diffusion coefficient through model-to-data comparisons for the first time. Our calculations demonstrate that the transverse momentum ($p_T$) differential splitting of $v_1$ between baryons and anti-baryons is highly sensitive to the baryon diffusion coefficient. If measured and compared with model calculation, this splitting can provide an additional constraint on the baryon diffusion coefficient.

\item Furthermore, this thesis emphasizes the importance of the rapidity-even $v_1$ of baryons and anti-baryons. We show that the rapidity differential measurement of the splitting of even component of $v_1$ between protons and anti-protons can simultaneously constrain both the initial baryon stopping model and the magnitude of the baryon diffusion coefficient.

\item We have also investigated the observable measured by STAR in the search for electromagnetic (EM) field signals—the centrality-dependent splitting of the $v_1$ slope, $\Delta (dv_1/dy)$, between positively and negatively charged hadrons \cite{STAR:2023jdd}. Our model calculations, which do not account for any electromagnetic field effects but focus solely on baryon diffusion, effectively capture the centrality trend of the $v_1$ slope splitting between protons and anti-protons. In this context, we will discuss how initial baryon stopping provides a significant background for such electromagnetic field signals.
\end{itemize}

Our baryon stopping model, combined with the extracted baryon diffusion coefficient that aligns well with experimental data, serves as a reliable input for hydrodynamic simulations at finite baryon density. This model could provide baseline predictions in the search for both the QCD critical point and electromagnetic field signals.

\section{Outline of the thesis}
The thesis is structured as follows:

In the next chapter, we will provide a brief overview of the simulation framework utilized in this thesis work. This chapter will cover the numerical codes employed in our work, along with a succinct explanation of the modeling details and the equations that are numerically solved within the code.

\textbf{Chapter 3} begins with a discussion on the initial energy deposition and the generation of directed flow in a baryon-free fluid before delving into the complexities of baryonic fluids. We will start by demonstrating, through a 1+1D toy model, how asymmetric energy deposition leads to asymmetric flow in the medium, ultimately resulting in directed flow. Following this, we will revisit the previously established, phenomenologically successful tilted energy deposition model proposed by Bozek and Wyskiel \cite{Bozek:2010bi}. We will compare this model with a more recent tilted source model introduced by the CCNU group \cite{Jiang:2021foj}, showing that the CCNU model is essentially a generalization of the Bozek-Wyskiel model. This will be substantiated through mathematical proofs and numerical simulations presented in this chapter. At the conclusion of this chapter, we will highlight that in baryon-free fluid simulations, particles of similar mass, such as protons and anti-protons  exhibit similar directed flow. However, the experimentally observed splitting of $v_1$ between protons and anti-protons serves as a clear indication of the presence of non-zero baryon density in the medium. This finding underscores the necessity of transitioning to the study of baryon inhomogeneity in the medium and the directed flow of identified hadrons, which will be addressed in the subsequent chapters of the thesis.

In \textbf{Chapter 4}, we will explore how finite baryon density influences the dynamics of the medium. Using a simple 1+1D toy model, we will illustrate how the interplay between baryon advection along the fluid velocity direction and baryon diffusion determines the net baryon current. This example provides a preliminary understanding of the crucial role of baryon diffusion in shaping baryon flow within the medium. We will then introduce a novel two-component initial net-baryon deposition model, where baryon deposition occurs from both participant and binary collision sources \cite{Parida:2022ppj,Parida:2022zse}. We will discuss how this model can be interpreted within the framework of a baryon junction picture. Using our model as input to the hydrodynamic code, we will examine the detailed mechanism through which initial baryon stopping, and its subsequent evolution with diffusion, affects the directed flow of mesons, baryons, and anti-baryons. By performing comprehensive simulations across an energy range of $\sqrt{s_{\text{NN}}} = 7.7$ to $200$ GeV, we will show our model's ability to capture experimental data for rapidity-odd directed flow of all identified hadrons \cite{STAR:2014clz,STAR:2017okv}. Specifically, we will highlight the model’s success in capturing the elusive directed flow splitting between baryons and anti-baryons across a wide range of collision energies, which is not yet captured by any existing model. In the same chapter, we will also investigate two additional phenomena related to directed flow. The first is the effect of the hadronic stage on the directed flow of $K^{*0}$ resonances \cite{Parida:2023tdx}. The second one is the splitting of elliptic flow between hadrons produced in different regions of phase space \cite{Parida:2022lmt}.

In \textbf{Chapter 5}, we will present our model study of an observable interpreted as a signal of the electromagnetic field by the STAR collaboration \cite{STAR:2023jdd}. This observable is the centrality and system size dependence of the splitting of the mid-rapidity slope of directed flow ($\Delta \frac{dv_1}{dy}$) between oppositely charged hadrons. In this context, we will discuss how initial baryon stopping and its subsequent diffusion provide a significant background for electromagnetic field signals observed in these quantities \cite{Parida:2023ldu}.

In \textbf{Chapter 6}, we will examine observables that are sensitive to the baryon diffusion coefficient, one of the fundamental transport coefficients of the QCD medium. We will explore how these observables can place tight constraints on the baryon diffusion coefficient. Additionally, we will present our method for extracting the baryon diffusion coefficient through model-to-data comparisons \cite{Parida:2023rux}.

In \textbf{Chapter 7}, we will present our model predictions of rapidity-even $v_1$ for identified hadrons, an observable that has not yet been measured in experiments. This chapter will highlight the importance of measuring the rapidity differential $v_1^{\text{even}}$ splitting between protons and anti-protons to gain phenomenological insight into the baryon junction picture, as well as to enable a precise extraction of the baryon diffusion coefficient.

\def \la{\langle}
\def \ra{\rangle}
\chapter{The simulation framework}
\label{ch:framework}
The system created in relativistic heavy ion collisions undergoes multiple stages during its space-time evolution, from the initial interaction to the final freeze-out. The underlying physics of each stage is different, and various models is usually implemented to simulate the system at different stages. Hybrid models, which integrate hydrodynamic fluid evolution during the dense stages with hadronic transport in the later stages, have proven particularly effective in reproducing experimental observations from heavy ion collisions \cite{Shen:2014vra,Shen:2014lye,Denicol:2018wdp,Bernhard:2018hnz}. The phenomenological studies presented in this thesis are conducted using such a multi-stage hybrid model. This chapter outlines the framework we used for numerical simulation.

The hybrid model employed in this study consists of multiple components:
\begin{itemize}
\item Initial Conditions (Glauber Model): The Glauber model \cite{Glauber:2006gd,Bialas:1976ed,Miller:2007ri,Yagi:2005yb} is used to set up the initial conditions for the hydrodynamic evolution. It provides the initial spatial distribution of energy and net-baryon density at a constant proper time, $\tau$.
\item Hydrodynamic Evolution (MUSIC): The hydrodynamic evolution is carried out using the publicly available code, MUSIC \cite{Schenke:2010nt, Paquet:2015lta, Schenke:2011bn,Denicol:2018wdp}. It performs the 3+1D relativistic viscous hydrodynamic evolution with finite net-baryon density, accounting for the baryon dissipative current. 
\item Particlization (iSS): Particlization is handled by the iSS code \cite{https://doi.org/10.48550/arxiv.1409.8164,https://github.com/chunshen1987/iSS,Shen:2014vra,Shen:2014lye}, which converts the hydrodynamic fluid to an ensemble of particles from a constant energy density hypersurface using the Cooper-Frye prescription.
\item Hadronic Transport (UrQMD): The UrQMD code \cite{Bass:1998ca, Bleicher:1999xi} is employed for simulating the late-stage hadronic interactions and resonance decays until freeze-out, where the dynamics are governed by the Boltzmann equation.
\end{itemize}

\section{Glauber model}
The Glauber model assumes the nucleus-nucleus collision as a series of multiple nucleon-nucleon interactions \cite{Glauber:2006gd,Bialas:1976ed,Miller:2007ri,Yagi:2005yb}. Nucleons are assumed to follow straight line trajectory and are not deflected after the collision which is a reasonable approximation in the context of ultra-relativistic nuclear collisions. The model assumes the size of the nucleus to be significantly larger than the range of the nucleon-nucleon interaction. Moreover, during the collsion, the nucleon-nucleon inelastic cross section ($\sigma_{\text{NN}}^{\text{in}}$) is assumed to be same as in vaccum.

The Glauber model calculations require two essential inputs. The first input is the nucleon density distribution of a nucleus, while the second is the energy dependence of the inelastic nucleon-nucleon cross-section, where the latter is obtained from experimental measurements of proton-proton or proton-neutron collision cross-sections \cite{ParticleDataGroup:2006fqo,ParticleDataGroup:2022pth}. For heavy nuclei such as gold (Au) or lead (Pb), the nucleon density distribution is typically assumed to be same to the nuclear charge density distribution \cite{Miller:2007ri,Yagi:2005yb}. This can be parameterized using a Fermi or Woods-Saxon distribution, expressed as follows:
\begin{equation}
\rho_{\text{WS}} (r,\theta,\phi) = \frac{\rho_{0}}{e^{\frac{r-R(\theta)}{a}}+1}
\label{eq:Wdsxn}
\end{equation}
where,
\begin{equation}
R(\theta) = R_0 \left[ 1+ f(\theta) \right]
\end{equation}
In this expression, $R_0$ and $a$ represent the radius and skin depth parameters, respectively. The normalization factor $\rho_0$ is determined by the following equation:
\begin{equation}
\int_{r=0}^{\infty} \int_{\theta=0}^{\pi} \int_{\phi=0}^{2\pi} \rho_{\text{WS}}(r,\theta,\phi) \  r^{2}  \sin \theta \ dr d\theta d\phi = 1
\end{equation}
Deformation in the nucleus can be incorporated by suitably taking the form of $f(\theta)$ with appropriate deformation parameters. For example, the prolate-shaped Uranium nucleus is parameterized by taking $f(\theta) = \beta_{2} Y_{20} (\theta) + \beta_{4} Y_{40}(\theta)$ , where the $\beta$ are deformation parameters and $Y_{20}$ and $Y_{40}$ are the spherical harmonics. For a spherical nucleus, $f(\theta)=0$ . The parameters ($R_0,a,\beta_2, \beta_4$) of the nucleon density distribution are constrained from low-energy electron-nucleus scattering experiments \cite{Anni:1994ey,Friar:1973wy,DeVries:1987atn}. The Woods-Saxon parameters for a few nuclei are mentioned in Table~\ref{tab:wdsxn_params}, and we have used these parameterizations for nucleus construction in the Glauber model of our study.

\begin{table}
  \centering
  \begin{tabular}{lccccc}
    \toprule
    Nucleus & Mass number($A$) & $R_0$ & $a$ & $f(\theta)$  \\
    \midrule
    Pb & 208 & 6.66 & 0.45 & 0  \\
    Au & 197 & 6.38 & 0.53 & 0  \\
    Cu & 63 & 4.2 & 0.59 & 0  \\
    \bottomrule
  \end{tabular}
  \caption{Woods-Saxon parameters for Cu, Au, and Pb nuclei utilized in the Glauber model calculations\cite{Shou:2014eya,DeVries:1987atn}.}
  \label{tab:wdsxn_params}
\end{table}

Given these inputs, Glauber modeling of high-energy heavy ion collisions can be carried out using two formalisms. One is known as the optical Glauber model, and the other is the Monte Carlo Glauber model \cite{Wong:1995jf,Miller:2007ri}. The primary distinction between these approaches lies in the representation of nucleons within the nucleus. The optical Glauber model envisions the nucleus as filled with a smooth Woods-Saxon type distribution of nucleonic matter, whereas the Monte Carlo Glauber model assumes localized nucleons, with nucleonic positions sampled using the Monte Carlo method. In essence, the Monte Carlo Glauber model introduces fluctuations in nucleonic positions within the nucleus, which is absent in the optical Glauber model.

In the optical Glauber model, the thickness function of the colliding nuclei at any transverse point $(x,y)$ is given by \cite{Miller:2007ri}:
\beq
T(x,y) = \int dz \rho_{\text{WS}}(x,y,z) 
\label{eq:thickness_glau1}
\eeq
Using this thickness function, the total number of binary collisions for a collision event with impact parameter $b$ can be calculated as follows \cite{Bialas:1976ed,Kharzeev:1996yx,Miller:2007ri}:
\beq
N_{coll} = \int dx dy N_{coll}(x,y) = AB \int dx dy \sigma_{\text{NN}}^{\text{in}} T_{+}(x-b/2,y) T_{-}(x+b/2,y) 
\eeq 
Here, $T_{+}$ and $T_{-}$ represent the thickness functions of the projectile and target nuclei, respectively, centered at the transverse plane coordinates 
$(x,y)=(b/2,0)$ and $(x,y)=(-b/2,0)$. $A$ and $B$ denote the mass numbers of the projectile and target nuclei, respectively.
Furthermore, the number of participants from the projectile ($N_{+}$) and the target ($N_{-}$) can be calculated separately using the following equations \cite{Bialas:1976ed,Kharzeev:1996yx,Miller:2007ri}:
\beq
N_{+} = \int dx dy N_{+}(x,y) = A \int dx dy T_{+}(x - b/2,y) \left[  1 - \left(  1 - T_{-}(x + b/2,y) \sigma_{\text{NN}}^{\text{in}}  \right)^{B}  \right] 
\eeq
\beq
N_{-} = \int dx dy N_{-}(x,y) = B \int dx dy T_{-}(x + b/2,y) \left[  1 - \left(  1 - T_{+}(x - b/2,y) \sigma_{\text{NN}}^{\text{in}}  \right)^{A}  \right] 
\eeq

On the other hand, in the MC Glauber model, nucleon positions are sampled within the nucleus using the Woods-Saxon distribution described in Eq. \ref{eq:Wdsxn}.  In spherical polar coordinates, the probability of sampling a nucleon at position ($r,\theta, \phi$) is given by:
\begin{equation}
    P(r,\theta,\phi) = \rho(r,\theta,\phi) r^2 \sin{\theta} dr d \theta d \phi.
    \label{wdsxn_samp_prob}
\end{equation}
In our calculations, we employed the acceptance-rejection method to sample the nucleons.\footnote{While sampling, one could impose a minimum distance threshold between nucleons within the nucleus; however, we did not implement such a threshold in our calculations.} After sampling the nucleon positions in both the projectile and target nuclei in spherical polar coordinates, we convert their coordinates to Cartesian form:
\begin{center}
$x=r \sin \theta \cos \phi$\\
$y=r \sin \theta \sin \phi$\\
$z=r \cos \theta$
\end{center}
Subsequently, we set the center of the nucleon distributions for both the projectile and target nuclei to 
$(x=0,y=0,z=0)$ before randomly orienting both nuclei.

In collisions involving spherically symmetric nuclei, the initial orientation of the nuclei does not influence the outcome due to the inherent rotational symmetry of the Woods-Saxon density profile ($\rho_{WS}$) along all three spatial axes (x, y, and z). As a result, event-averaged observables remain unchanged regardless of the nuclei’s orientation in such cases. However, the situation is different for collisions involving deformed nuclei, such as uranium-uranium (U+U) collisions. The nuclear deformation preserves rotational symmetry about one axis which we call the major axis or the z-axis. As a result, rotations about the z-axis do not lead to any physically distinct configurations. The orientation of this major axis is designated by two angles. Thus, only two Euler angles are needed to describe the relevant orientations. In our model, the initial orientation of each nucleus is assigned using the following rotational procedure.

To rotate a nucleus in a random direction defined by ($\theta_{*},\phi_{*}$), we generate a random number $\theta_{*}$ distributed between $0$ and $\pi$ using a $\sin(\theta)$ probability, while $\phi_*$ is generated uniformly between $0$ and $2\pi$. The positions of the nucleons in the rotated nucleus A can then be calculated as:
\begin{equation}
\begin{pmatrix}
x^{\prime}\\
y^{\prime}\\
z^{\prime}
\end{pmatrix}
=
\begin{pmatrix}
\cos\phi_{*} & -\sin\phi_{*} & 0 \\
\sin\phi_{*} & \cos\phi_{*} & 0 \\
0 & 0 & 1
\end{pmatrix}
\begin{pmatrix}
\cos\theta_{*} & 0 & -\sin\theta_{*}\\
0 & 1 & 0 \\
\sin \theta_{*} & 0 & \cos \theta_{*}
\end{pmatrix}
\begin{pmatrix}
x\\
y\\
z
\end{pmatrix}
\end{equation}
Afterward, an impact parameter is sampled with a probability density $P(b)=2\pi b db$. Then we shift the x-coordinate of the center of the projectile to $x=b/2$ and the center of the target to $x=-b/2$.

After establishing the nucleus configuration, we determine which nucleons participate in the collision. It is determined based on the condition that a nucleon from the projectile will interact with a nucleon from the target nucleus if the distance $d$ between them satisfies \cite{Miller:2007ri}:
\beq
d \le \sqrt{\frac{\sigma_{\text{NN}}^{\text{in}}}{\pi}}
\eeq
By counting the total number of binary interactions and how many nucleons have interacted at least once, we can calculate the number of binary collisions $N_{coll}$ and the number of participants $N_{part}=N_{+}+N_{-}$.

The positions of the participants and binary collisions \footnote{The midpoint of two interacting nucleons is considered as the position for the binary collision source.} are treated as point sources of energy deposition due to the collision. Consequently, this creates an energy deposition profile in the transverse plane that consists of multiple delta distributions located at the participant and binary collision points. However, such a delta-like profile is not suitable for further dynamical evolution because the large gradients can lead to numerical instabilities in hydrodynamic simulations \cite{RihanHaque:2012wp}. Therefore, a smooth profile of energy distribution is required as input, which can be constructed from the smoothed profiles of participants and binary collisions. In our MC Glauber model calculations, we generated smooth profiles for both participants and binary collisions in the transverse plane by employing Gaussian smearing:
\beq
N_{\pm}(x,y) =  \sum_{i}^{N_{\pm}} \frac{1}{2 \pi \sigma_w^2 }  \exp{ \left(  \frac{ (x-x_i)^2 + (y-y_i)^2   }{2 \sigma_w^2} \right) } 
\eeq
and similarly,
\beq
N_{coll}(x,y) =  \sum_{i}^{N_{coll}} \frac{1}{2 \pi \sigma_w^2 }  \exp{ \left(  \frac{ (x-x_i)^2 + (y-y_i)^2   }{2 \sigma_w^2} \right) } 
\eeq
where ($x_i,y_i$) represents the transverse positions of the delta-type sources. In our model calculations, we set 
$\sigma_w = 0.4$ fm. 

Finally, using these participant and binary collision profiles, we constructed the initial distributions for both energy density and net-baryon density, which serve as inputs to the hydrodynamic models. The explicit expressions used to generate these initial profiles will be discussed in detail in the upcoming chapters.

\section{(3+1)D relativistic viscous hydrodynamics with non-zero baryon density}
We perform the fluid evolution using the publicly available MUSIC code \cite{Denicol:2018wdp}, which numerically solves the conservation equations for the energy-momentum tensor ($T^{\mu \nu}$) and the net-baryon current ($N^{\mu}_B$):
\begin{equation}
 \partial_{\mu} T^{\mu \nu} = 0 \\
 \label{eq:hydro1_tmunu}
\end{equation}
\begin{equation}
 \partial_{\mu} J^{\mu}_{B} = 0 
 \label{eq:hydro1_nmu}
\end{equation}
where the $T^{\mu \nu}$ and the $J^{\mu}_B$ are given by:
\beq
T^{\mu \nu} = ( \epsilon + p ) u^{\mu} u^{\nu} - p g^{\mu \nu} - \Pi \Delta^{\mu \nu} + \pi^{\mu \nu}
\eeq
and
\beq
J^{\mu}_{B} = n_B u^{\mu} + q^{\mu}_{B}
\eeq
In these equations:
\begin{itemize}
\item $\partial_{\mu} \ (\mu = 0,1,2,3)$ represents the covariant derivative.
\item $\epsilon$ and $n_B$ denote the energy density and net-baryon density in the fluid rest frame, while $p$ represents the local equilibrium pressure.
\item $g^{\mu \nu}$ is the metric tensor.
\item $u^{\mu}$ is the four-velocity of the fluid, and $\Delta^{\mu \nu} = g^{\mu \nu} - u^{\mu} u^{\nu}$ is the projection operator that satisfies the relation $u_{\mu} u_{\nu} \Delta^{\mu \nu} = 0$.
\item $\pi^{\mu \nu}$ and $\Pi$ represent the shear stress tensor and bulk viscous pressure, respectively.
\item $q^{\mu}_{B}$ is the diffusion current associated with the net-baryon density.
\end{itemize}

In MUSIC, the time evolution of the dissipative currents $\pi^{\mu \nu}$, $\Pi$, and $q^{\mu}_{B}$ is governed by the following relaxation-type equations derived from kinetic theory \cite{Denicol:2014vaa,Molnar:2013lta,Denicol:2012cn,Denicol:2018wdp,Denicol:2014vaa} :
\begin{equation}
    \tau_{\pi} \Dot{\pi}^{\langle  \mu \nu \rangle} + \pi^{\mu \nu} = 2 \eta \sigma^{\mu \nu} - \delta_{\pi \pi} \pi^{\mu \nu} \theta - \tau_{\pi \pi} \pi^{\lambda \langle \mu } \sigma^{\nu \rangle}_{\lambda}  + \phi_7 \pi^{\langle \mu}_{\alpha} \pi^{ \nu \rangle \alpha} + \lambda_{\pi \Pi} \Pi \sigma^{\mu \nu} 
\end{equation}

\begin{equation}
\tau_\Pi \Dot{\Pi} + \Pi =  -\zeta \theta - \delta_{\Pi \Pi} \Pi \theta + \lambda_{\Pi \pi} \pi^{\mu \nu} \sigma_{\mu \nu} 
\end{equation}

\begin{equation}
\tau_B \Dot{q}^{\la \mu \ra}_{B} + q^{\mu}_{B} =  \kappa_B  \nabla^{\mu} \left( \frac{\mu_B}{T} \right)   - \delta_{qq} q^\mu_{B} \theta - \lambda_{qq} (q_{B})_{\nu} \sigma^{\mu \nu}  
\end{equation}
In these equations, $\dot{A} \equiv u^{\mu} \partial_{\mu} A, \ A^{\la \mu \ra} \equiv \Delta^{\mu}_{\alpha} A^{\alpha}, \  A^{\la  \mu \nu \ra} \equiv \frac{1}{2} \left( \Delta^{\mu}_{\alpha} \Delta^{\nu}_{\beta} + \Delta^{\mu}_{\beta} \Delta^{\nu}_{\alpha} - \frac{2}{3} \Delta^{\mu \nu} \Delta_{\alpha \beta} \right) A^{\alpha \beta} $, $\sigma^{\mu \nu} \equiv \frac{1}{2}\left( \Delta_{\mu \rho} \partial^{\rho} u^\nu +\Delta_{\nu \rho} \partial^{\rho} u^{\mu} \right) - \frac{1}{3}\theta \Delta^{\mu \nu}$ and $\theta \equiv \partial_{\mu} u^{\mu}$.

The coefficients $\eta$, $\zeta$, and $\kappa_B$ are first-order transport coefficients, which are used as inputs in the hydrodynamic model and are constrained by model-to-data comparison. Additionally, the second-order transport coefficients ($\tau_\Pi, \tau_{\pi}, \delta_{\pi \pi}, \tau_{\pi \pi}, \phi_7, \lambda_{\pi \Pi}, \delta_{\Pi \Pi}, \lambda_{\Pi \pi}, \delta_{qq}, \lambda_{qq}$) are determined from kinetic theory calculations in the massless limit and are kept fixed in the code \cite{Denicol:2010xn,Denicol:2012cn,Molnar:2013lta,Denicol:2014vaa,Denicol:2018wdp}. The specific values of these second-order transport coefficients used in the calculations are listed in Table \ref{tab:2nd_trans_coeff}.
\begin{table}
  \centering
  \begin{tabular}{lcccccccccc}
    \toprule
   $\tau_\Pi$ &  $\tau_{\pi}$ & $\delta_{\pi \pi}$ & $\tau_{\pi \pi}$ & $\phi_{7}$ & $\lambda_{\pi \Pi}$ & $\delta_{\Pi \Pi}$ &  $\lambda_{\Pi \pi}$ &  $\delta_{qq}$ &  $\lambda_{qq}$ \\
    \midrule
    $\frac{\zeta}{15 (1/3-c_s^2)^2 (\epsilon+p) }$ & $\frac{5 \eta}{\epsilon+p}$ &   $\frac{4}{3}\tau_{\pi}$ & $\frac{10}{7}\tau_{\pi}$ & $\frac{9}{70p}$ & $\frac{6}{5}$ & $\frac{2}{3} \tau_\Pi$ &  $\frac{8}{5}(1/3-c_s^2) \tau_\Pi$ & $\tau_B $ & $\frac{3}{5}\tau_B$ \\
    \bottomrule
  \end{tabular}
  \caption{The value of the second order transport coefficients taken in the evolution equation of the dissipative currents $\pi^{\mu \nu}, \Pi$ and $q^{\mu}_{B}$ \cite{Denicol:2010xn,Denicol:2012cn,Molnar:2013lta,Denicol:2014vaa,Denicol:2018wdp}. }
  \label{tab:2nd_trans_coeff}
\end{table}

In the hydrodynamic model, the impact of shear and bulk viscosities, $\eta$ and $\zeta$, on final-state observables has been extensively studied \cite{Romatschke:2007mq,Song:2007fn,Dusling:2007gi,Bozek:2011ua,Teaney:2003kp,Shen:2014lye,Song:2009gc,Heinz:2005bw,Niemi:2011ix,Niemi:2012ry,Dusling:2011fd,Ryu:2017qzn,Roy:2012np,Bozek:2017kxo,Ryu:2015vwa}. The values of $\eta/s$ and $\zeta/s$ (where $s$ is the entropy density), are quantitatively extracted through model-to-data comparisons \cite{JETSCAPE:2020mzn,Shen:2023awv,Jahan:2024wpj,Parkkila:2021tqq,Bernhard:2018hnz,Yang:2022ixy}. However, the baryon diffusion coefficient, $\kappa_B$, which is another essential property of the QCD medium, has not been explored much \cite{Rougemont:2015ona,Denicol:2018wdp,Li:2018yvx,Greif:2017byw,Jaiswal:2015mxa,Albright:2015fpa}. At lower collision energies, where baryon stopping becomes more significant in the mid-rapidity region, the dynamics of baryons within the QCD medium can be better examined \cite{Denicol:2018wdp,Du:2024wjm}. This creates an opportunity to estimate the baryon diffusion coefficient in the produced medium \cite{Denicol:2018wdp}. Recent progress has been made in this direction, and the effect of $\kappa_B$ on various observables is now being actively studied \cite{Denicol:2018wdp,Li:2018yvx,Wu:2022mkr,Parida:2022ppj,Du:2022yok,Wu:2021fjf}. 
In the MUSIC code, a temperature and baryon chemical potential-dependent form of $\kappa_B$ has been implemented, derived by solving the Boltzmann equation under the relaxation time approximation \cite{Denicol:2018wdp}: 
\begin{equation}
\kappa_B = \tau_{B} n_B \left( \frac{1}{3} \coth{\left(\frac{\mu_B}{T}\right)} - \frac{n_B T}{\epsilon + p } \right) 
\label{eq:kappaB_form}
\end{equation}
Here, the relaxation time $\tau_B$ is an unknown parameter, and it is assumed to be inversely proportional to temperature, as expected for a massless system. The proportionality constant, $C_B$, is introduced as: 
\beq \tau_B = \frac{C_B}{T} \label{eq:tauB_CB}\eeq
Thus, $C_B$ remains an undetermined parameter in the expression for $\kappa_B$ and can be adjusted to vary the magnitude of the diffusion coefficient. Studying the effect of $\kappa_B$ effectively becomes an investigation of $C_B$. This parameter can be constrained by comparing model results to experimental data after identifying observables that show notable sensitivity to the baryon diffusion coefficient.

Apart form initial condition and transport coefficient another import input to hydrodynamic model is the Equation of state. During the hydrodynamic evolution we have used a state-of-art EoS developed by Monnai et. al \cite{Monnai:2019hkn, HotQCD:2012fhj, Ding:2015fca, Bazavov:2017dus}.  This EoS takes the Lattice QCD EoS at vanishing chemical potential for the high temperature region. Then, following the taylor expansion method the pressure at finite chemical potential has been calculated. Moreover, for the low temperature region the pressure calculated using the Hadron resonance gas model has been taken. The pressure obtained from Lattice QCD and HRG are then matched near the cross-over region.  This equation of state has three variants dubbed as NEoS-B, NEoS-BS and NEOS-BQS \cite{Monnai:2019hkn}. In NEoS-B, the strangeness and charge chemical potential both are taken to be zero $\mu_Q = \mu_S=0$ where the baryon density is considered as the only conserved quantity. In the NEoS-BS case, the strangeness nutrality condition $n_S=0$ has been implemented and $\mu_Q$ is taken as zero.  This EoS provides non-zero $\mu_S$ in the presence of net-baryon density in the medium, since for $n_B \ne 0$ the $\mu_S$ has to be non-zero to satisfy the condition $n_S=0$.  Additionaly, In the NEoS-BQS, along with strangeness neutrality another constraint of net-baryon to charge density to be 0.4 has been introduced. Hence, in this case, $\mu_S$ and $\mu_Q$ both are non-zero for a system with non zero net baryon density. In our hydrodynamic simulation in the presence of net-baryon density we take NEoS-BQS EoS and for the simulation without any conserved charge we have utilized the $\mu_B=0$ slice of it.

\section{Particlisation}
Hydrodynamics modelling is based on the assumption that the evolving system maintains local thermal equlibrium. This assumption is reasonable for the dense and high temperature QGP phase. But as the system expands and cools down, it will reach a point where the system becomes so dilute that the local thermal equlibrium might not be a good assumption to hold there. In this context, microscopic transport models are well-suited to simulate the dynamics of this less dense, low-temperature phase. Moreover, in transport models, the chemical and kinetic freeze-out emerge naturally from the underlying microscopic dynamics. To transit from a continuous hydrodynamic description to a microscopic transport model, it is necessary to convert the fluid representation into an ensemble of particles. This process is referred to as "particlisation". It is important to note that particlisation is not linked to physical processes like freeze-out or hadronization; rather, it serves as a numerical method for transitioning between different model representations of the system. This shift from macroscopic hydrodynamic modeling to microscopic transport modeling is valid in regions of energy density or temperature near the QCD crossover. In this temperature or energy density window, the mean free path of the medium constituents is such that both macroscopic and microscopic models can describe the system dynamics.

In the framework, we have used the iSS code \cite{https://doi.org/10.48550/arxiv.1409.8164,https://github.com/chunshen1987/iSS,Shen:2014vra,Shen:2014vra,Shen:2014lye} to perform the particlisation at a constant energy density hypersurface. The particles are sampled from the surface with the momentum distribution given by the Cooper-Frye formula.
\beq
E \frac{dN_i}{d^3p} = \frac{g_i}{(2 \pi)^3}\int_{\Sigma} d^3 \Sigma_{\mu}(x^{\mu}) p^{\mu} f_i( x^{\mu}, p^{\mu})
\eeq
The left-hand side of the equation represents the invariant yield of the $i$-th type of hadron species. On the right-hand side, the integration is performed over the entire hypersurface $\Sigma$, where  $d^3 \Sigma_{\mu}(x^{\mu})$  denotes the normal vector to the surface element at a specific spacetime position $x^{\mu}$. The term $g_i$ represents the degeneracy factor, while $p^\mu$ is the four-momentum of the produced particles. The function $f_i(x^{\mu}, p^{\mu})$ denotes the single-particle distribution function for the $i$-th hadron species and depends on all the hydrodynamic field variables—namely, 
$u^{\mu}, \epsilon, p, T, \mu_{i}, \pi^{\mu \nu}, \Pi, q^{\mu}_{B}$—evaluated at the freeze-out hypersurface. Generally, 
$f_i(x^{\mu}, p^{\mu})$ comprises both an equilibrium part and out-of-equilibrium  corrections.
\beq
f_i( x^{\mu}, p^{\mu}) =  f_{i}^{\text{eq}} + \delta f_{i}^{\text{shear}}  + \delta f_{i}^{\text{bulk}} + \delta f_{i}^{\text{diffusion}}
\eeq
The equilibrium distribution $f_i(x^{\mu}, p^{\mu})$ is defined as the Maxwell-Jüttner distribution:
\beq
f_{i}^{\text{eq}} = \left[ \exp{ \left( {\frac{p^{\mu} u_{\mu} - ( \mu_B B_i + \mu_S S_i + \mu_Q Q_i )   }{T} } \right) } + \Theta_i \right]^{-1} 
\eeq
In this expression, $u^{\mu}, T$ and $(\mu_B, \mu_Q, \mu_S)$ represent the fluid four-velocity, temperature, and chemical potentials, respectively, evaluated at a cell of the freeze-out hypersurface. These quantities can vary from one freeze-out cell to another. The variables $B_i, S_i, Q_i$  denote the baryon number, strangeness, and electric charge of the $i$-th hadron species, respectively. The $\Theta_i=+1$  for fermions and -1 for bosons, accounting for the effects of quantum statistics.

The forms of the non-equilibrium corrections are typically derived from kinetic theory calculations. These are usually different in different methods of calculations, such as the 14-moment method or the Chapman-Enskog method \cite{McNelis:2021acu,Monnai:2009ad,McNelis:2019auj,Jaiswal:2014isa,Bhalerao:2013pza}. In our analysis, we have adopted the following expressions for $\delta f$ of the shear and bulk.
\beq
\delta f_{i}^{\text{shear}} = f_{i}^{\text{eq}} (1 + \Theta_i f_{i}^{eq}) \frac{p^{\mu} p^{\nu} \pi_{\mu \nu} }{ 2 T^2 (\epsilon+P) }
\eeq
\beq
\delta f_{i}^{\text{bulk}} = f_{i}^{\text{eq}} (1 + \Theta_i f_{i}^{\text{eq}}) \frac{\tau_\Pi}{\zeta} \left[ \frac{p^\mu u_\mu}{T} (\frac{1}{3} - c_s^2) - \frac{1}{3} \left( \frac{m}{T}\right)^2 \frac{T}{p^\mu u_\mu}  \right] \Pi
\eeq
For the non-equilibrium correction related to baryon diffusion, we utilized the expression derived in Ref. \cite{Denicol:2018wdp}. It is given by:
\beq
\delta f_{i}^{\text{diffusion}} = f_{i}^{\text{eq}} (1 + \Theta_i f_{i}^{\text{eq}}) \left(  \frac{n_B}{\epsilon + p} - \frac{B_i}{p^\mu u_\mu}  \right) \frac{p^{\la \mu \ra} (q_B)_{\mu}}{\hat{\kappa}_B}
\label{eq:deltaf_diff_CF}
\eeq
Here, $\hat{\kappa}_B = \frac{\kappa_B}{\tau_B}$, and $B_i$ denotes the baryon number of the sampled particle species. In hydrodynamic simulations with a non-zero $\kappa_B$ it is crucial to include $\delta f_{i}^{\text{diffusion}}$ in the Cooper-Frye formula during the particlisation process to ensure the conservation of net baryon number. Furthermore, the $\delta f_{i}^{\text{diffusion}}$ correction significantly influences final state observables, such as $p_T$ spectra and elliptic flow coefficients, as discussed in Ref. \cite{Denicol:2018wdp}. In the upcoming chapters we will also discuss about the effect of $\delta f_{i}^{\text{diffusion}}$ on the directed flow of baryons and anti-baryons as observed in our study.  

\section{Hadronic transport}

The hadrons sampled from the freezeout hypersurface are subsequently evolved through the UrQMD code \cite{Bass:1998ca, Bleicher:1999xi}, which simulates further hadronic interactions, resonance formation, and decay until the kinetic freezeout occurs. In this framework, the dynamics of the hadronic system are governed by the relativistic Boltzmann equation:
\beq
p^{\mu} \partial_{\mu} f_i(x^{\mu},p^{\mu}) = C_{i}^{coll}(x^{\mu},p^{\mu})
\eeq
In this equation, $f_i(x^{\mu},p^{\mu})$ represents the single-particle distribution function for species $i$, while the collision kernel $C_{i}^{coll}(x^{\mu},p^{\mu})$ accounts for both binary collisions and $2 \rightarrow n $ processes. During the transport simulation, the hadrons experience multiple instances of elastic and inelastic scattering within the medium, ultimately resulting in the production of stable hadrons. By incorporating these late-stage microscopic dynamics, the framework yields a more realistic particle production and freezeout scenario. 

In our model calculations, we generate multiple Cooper-Frye sampling events from a single freezeout hypersurface obtained from a specific hydrodynamic evolution. Each set of hadrons from the sampling events is evolved independently through the transport simulation, a method commonly referred to as oversampling in the literature. The particles produced from all oversampling events are then combined to compute the observables corresponding to that particular hydrodynamic evolution stemming from a single initial condition event. This approach is computationally efficient and provides robust statistics. But averaging over multiple particlisation samples certainly suppresses some event-by-event thermal fluctuations. Therefore, it is crucial to ensure that the observables of interest are not overly sensitive to these fluctuations.

\def \JYP{\text{\tiny{J}}}
\def \BW{\text{\tiny{B}}}
\def \la{\langle}
\def \ra{\rangle}

\chapter{Initial energy deposition and directed flow in a baryon less fluid}
\label{ch:tilt}
At the highest RHIC and LHC energies, the density of conserved charges in the mid-rapidity region is very low \cite{Ciacco:2023ekv,BRAHMS:2003wwg}. Hence, the evolution of the conserved charges is often neglected in hydrodynamic models \cite{Bozek:2010bi,Bozek:2011ua,Hirano:2002ds,Hirano:2001yi,Parida:2022lmt,Jiang:2021foj,Jiang:2021ajc}.  In such cases, the directed flow ($v_1$) primarily probes the initial three-dimensional distribution of energy density \cite{Bozek:2010bi,Jiang:2021foj,Jiang:2021ajc,Parida:2022lmt,Chatterjee:2017ahy}. Measurements at top RHIC and LHC energies show that the $v_1$ of charged hadrons exhibits a negative slope near mid-rapidity \cite{STAR:2005btp,STAR:2008jgm,PHOBOS:2005ylx,STAR:2011gzz}. This feature has been successfully reproduced in hydrodynamic models using a tilted initial energy deposition profile, proposed by Bozek and Wyskiel (BW) \cite{Bozek:2010bi}. In this chapter, we will review this phenomenologically successful tilted initial condition (TIC) model. This review will provide insight into the generation of directed flow in a baryon less fluid. This foundational understanding will be helpful for comprehending the dynamics of the system discussed in later chapters, particularly when the complexity of non-zero baryon density will be introduced into the hydrodynamic models. Additionally, in this chapter we will discuss about another initial condition model recently proposed by the CCNU group, which also generates a tilted profile in the reaction plane \cite{Jiang:2021foj, Jiang:2021ajc}, capturing the negative slope of charged hadron $v_1$. We will explore the relationship between the tilted profiles of the Bozek-Wyskiel model and the CCNU group model in this chapter.

In the hydrodynamic model, directed flow is understood to arise from the presence of dipolar asymmetry in the initial energy distribution profile in the transverse plane at non-zero $\eta_s$ \cite{Bozek:2010bi,Jiang:2021foj,Jiang:2021ajc}. To comprehend the generation of $v_1$, it is essential to understand how this asymmetry in the initial energy distribution leads to the development of asymmetric flow. Therefore, we will now illustrate through a simple example how an initial asymmetric deposition of matter or energy density in the initial state can lead to the development of asymmetric flow during the fluid evolution.

\begin{figure}[htbp]
  \centering
  \includegraphics[width=0.9\textwidth]{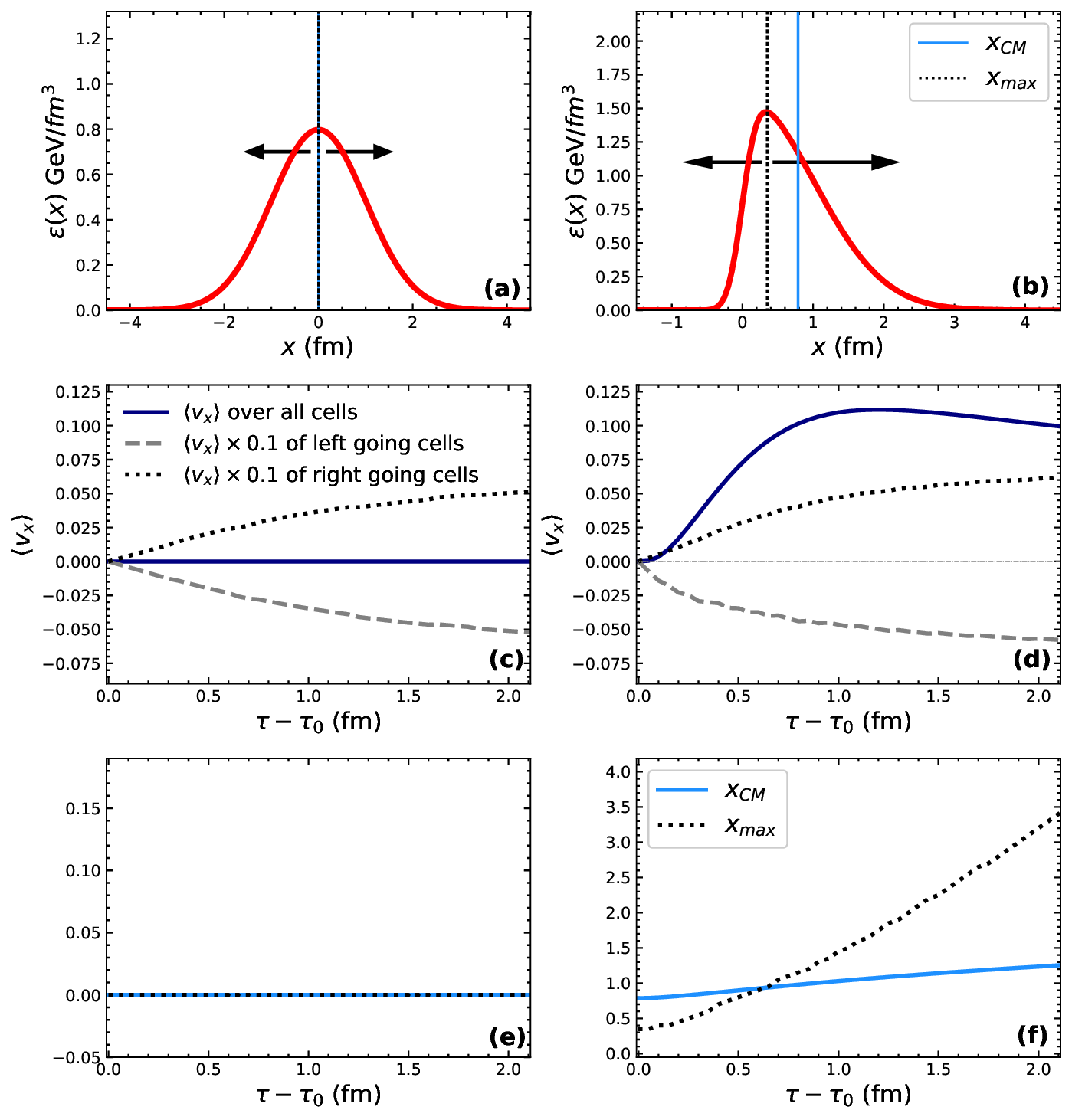} 
  \caption{This figure demonstrates how asymmetry in the initial energy deposition leads to the generation of asymmetric flow in one dimension. Panels (a) and (b) depict two configurations of initial energy density distribution: (a) shows a symmetric energy distribution, while (b) shows an asymmetric energy distribution. In both plots, $x_{\text{CM}}$ represents the mean of the energy density distributions, and $x_{\text{max}}$ denotes the point where the energy densities are maximum. Panels (c) and (d) present the time evolution of the average velocity $\langle v_x \rangle$ (defined in Eq. \ref{eq:avgvx1D})  for the symmetric and asymmetric profiles, respectively, obtained from 1+1D ideal hydrodynamic simulations with a conformal equation of state. The plots also separately show $\langle v_x \rangle$ for matter moving to the left ($v_x < 0$) and to the right ($v_x > 0$), alongside the net $\langle v_x \rangle$. The $\langle v_x \rangle$ values for the matter moving to left and right are scaled down by a factor of 10. Panels (e) and (f) show the time evolution of $x_{\text{CM}}$ and $x_{\text{max}}$ for the symmetric and asymmetric profiles, respectively. }
  \label{fig:two_1D_dist}
\end{figure}

Consider a simple one-dimensional scenario with two configurations of fluid distribution, as illustrated in Fig. \ref{fig:two_1D_dist}. In Figs. \ref{fig:two_1D_dist}(a) and (b), the energy density profile of the fluid is plotted against $x$. The mean of the energy density distributions in both configurations is labeled as $x_{\text{CM}}$, while the point where the energy densities are maximum is labeled as $x_{\text{max}}$. Assuming the fluid is relativistic and massless, the pressure of the fluid is related to its energy density by the equation of state, $p = \frac{\epsilon}{3}$. Thus, at the point where the energy density is maximum, the pressure is also maximum, and the pressure gradient ($\partial_x p$) is zero at $x_{\text{max}}$. We performed a 1+1D ideal hydrodynamic evolution of the two configurations, starting from the initial time $\tau = \tau_0 = 1$ fm, and the results are presented in Figs. \ref{fig:two_1D_dist}(c), (d), (e), and (f).

In fluid dynamics, the time evolution of matter and the subsequent development of the fluid flow velocity are driven by the pressure gradient \cite{Bozek:2010bi,Bozek:2011ua,Ollitrault:2007du,Schenke:2010nt,Shen:2014vra}:
\beq
\partial_\tau u_x = -\frac{1}{\epsilon+p} \partial_x p 
\label{eq:1D_hydro_evo_eq}
\eeq
Here, $u_x$ is the $x$-component of the fluid four-velocity. In 1+1D, the four-velocity is given by $u_{\mu} = (u_{\tau}, u_x)$ with $u_{\tau}^2 = 1 + u_x^2$. As a result, there will be no flow at the point of maximum pressure, where the pressure gradient is zero ($\partial_x p = 0$). The matter to the right of the maximum pressure position, $x_{\text{max}}$, will flow to the right, while the matter to the left of $x_{\text{max}}$ will flow to the left. In the first scenario, where $x_{\text{CM}}$ coincides with $x_{\text{max}}$, the left-right symmetry of the pressure gradient results in an equal amount of energy flowing to both the left and right, leading to a net fluid velocity ($v_x$) of zero, as shown in Fig. \ref{fig:two_1D_dist}(c). However, in the second configuration, depicted in Fig. \ref{fig:two_1D_dist}(b), the left-right symmetry is broken because $x_{\text{max}}$ and $x_{\text{CM}}$ are not aligned. Due to the pressure gradient, the flow will separate at $x_{\text{max}}$, but since $x_{\text{CM}}$ is located to the right of $x_{\text{max}}$, more matter is concentrated on the right side. This causes more matter to flow to the right than to the left, ultimately resulting in asymmetric flow.

\beq
\langle v_x \rangle = \frac{\int dx \epsilon(x) v_x(x) }{ \int dx \epsilon(x) }
\label{eq:avgvx1D}
\eeq

In the case of asymmetric energy distribution, the pressure gradient is steeper on the left side of $x_{\text{max}}$ compared to the right. As a result, the average velocity ($\langle v_x \rangle$), calculated using Eq. \ref{eq:avgvx1D}, of the matter moving leftward is greater in magnitude than that of the matter moving rightward. In Fig. \ref{fig:two_1D_dist}(d), the $\langle v_x \rangle$ of the matter moving to left and to right is presented for the asymmetric matter deposition scenario. It is observed that the $\langle v_x \rangle$ of the matter moving to left has a larger magnitude than that moving towards the right. However, since $x_{\text{CM}}$ is initially located to the right of $x_{\text{max}}$, there are more matter moving to the right\footnote{In our calculations, we observed that the ratio between the amount of matter moving to right and left is always less than 1, starting from approximately 0.2 at $\tau - \tau_0 = 0.1$ fm and increasing to approximately 0.75 at $\tau - \tau_0 = 5$ fm.}. Consequently, when averaging over all cells, the net flow is in the rightward direction, leading to $\langle v_x \rangle > 0$ for all $\tau$. The $\langle v_x \rangle$ calculated by averaging over all cells is plotted as solid lines in Fig. \ref{fig:two_1D_dist}(d).

Another key aspect of this asymmetric energy distribution case is that the magnitude of flow velocity is greater in the direction of the steeper pressure gradient. Consequently, particles produced from cells with $v_x<0$ will have relatively higher $p_T$ compared to those produced from the cells with $v_x>0$, even though the majority of particles are produced in the right region with smaller $p_T$. This indicates that, in a dipolar asymmetric scenario, particles with relatively higher $p_T$ will exhibit flow in the opposite direction to those with lower $p_T$. This behavior is precisely observed in the $p_T$ differential $v_1$, where higher $p_T$ particles display a different sign of $v_1$ compared to lower $p_T$ particles \cite{STAR:2008jgm,Singha:2016mna}.

In Fig. \ref{fig:two_1D_dist}(e) and (f), we illustrate how the positions of $x_{\text{CM}}$ and $x_{\text{max}}$ evolve with time. In the first scenario, due to symmetry, both $x_{\text{CM}}$ and $x_{\text{max}}$ remain unchanged throughout the evolution, maintaining their initial positions. However, in the second scenario, early fluid evolution drives the system towards symmetry, causing $x_{\text{max}}$ to shift to the right and align with $x_{\text{CM}}$ after approximately 0.6 fm. By that time, however, significant left-right asymmetry in the flow has already developed. While the fluid tends toward symmetric flow after $x_{\text{CM}}$ and $x_{\text{max}}$ align, the inertia of the fluid propels further flow, pushing $x_{\text{max}}$ past $x_{\text{CM}}$ and continuing to grow. This interplay between fluid inertia and the tendency toward symmetric flow results in a decrease in $\langle v_x \rangle$ after $x_{\text{max}}$ surpasses $x_{\text{CM}}$.

In summary, this simple example demonstrates that the misalignment between the mean position ($x_{\text{CM}}$) and the maximum energy density (or pressure\footnote{It’s important to note that in the presence of conserved charge density, the pressure depends on both energy density and the conserved charge density, meaning the point of maximum pressure may not coincide with the point of maximum energy density.}) position ($x_{\text{max}}$) in the initial matter deposition profile leads to asymmetric pressure gradient ($\partial_x p$) on both sides of $x_{\text{max}}$ which ultimately generates an asymmetric flow. In the final state, this left-right asymmetry in the flow will be reflected through the directed flow coefficient of measured hadrons.

\section{Tilted initial condition (TIC) }
\label{Sec:TIC}
In asymmetric systems, the observed difference in particle production between positive and negative rapidities indicate an asymmetric energy deposition by the participant nucleons along the rapidity \cite{Bzdak:2009xq,PHOBOS:2004fzb,Bzdak:2009dr}. Nucleons moving toward positive rapidity tend to deposit more energy in the forward region ($\eta_s > 0$), while those moving toward negative rapidity deposit more energy in the backward region ($\eta_s < 0$). Incorporating this type of energy deposition asymmetry, Bozek and Wyskiel proposed an initial condition model in 2010, where the energy deposition leads to an initial profile that appears tilted in the reaction plane. In this model, the initial energy density at a constant proper time $\tau_0$ is expressed as \cite{Bozek:2010bi,Bozek:2011ua}:
\beq
  \epsilon(x,y,\eta_{s}; \tau_0) =\epsilon_{0} \left[ \left( N_{+}(x,y) f_{+}(\eta_{s}) + N_{-}(x,y) f_{-}(\eta_{s})  \right) \times \left( 1- \alpha \right) + N_{coll} (x,y)  \epsilon_{\eta_s} \left(\eta_{s}\right) \alpha \right] 
 \label{eq.tilt}
\eeq
Here, $N_+(x,y)$ and $N_-(x,y)$ represent the density of participants from the projectile and target nuclei, respectively, while $N_{coll}(x,y)$ accounts for the contribution of binary collisions to the energy deposition at a specific transverse position. Both $N_{\pm}(x,y)$ and $N_{coll}(x,y)$ are calculated using the Glauber model. The parameter $\epsilon_0$ serves as a normalization factor for the energy density profile and can be adjusted within the model to accurately reflect the charged particle multiplicity for a given centrality. The energy deposition scheme used to determine the transverse energy density profile is the two-component model, where the free parameter $\alpha$ regulates the relative contributions from participant and binary collision sources. In the model calculations, $\alpha$ is chosen to best capture the centrality dependence of the mid-rapidity charged particle yield\footnote{Glauber model calculations indicate that $\alpha = 0.14$ best explains the centrality dependence of charged particle yield in Au+Au collisions at $\sqrt{s_{NN}} = 200$ GeV.}. The function $\epsilon_{\eta_s}(\eta_s)$ defines a symmetric space-time rapidity ($\eta_s$) distribution at a given transverse position $(x,y)$.
\begin{equation}
\epsilon_{\eta_s}(\eta_s) = \exp \left( -\frac{ \left( \vert \eta_{s} \vert - \eta_{0} \right)^2}{2 \sigma_{\eta}^2}
\Theta (\vert \eta_{s} \vert - \eta_{0} ) \right)
\label{eq:plateau_fall}
\end{equation}

where $\Theta(...)$ represents the Heaviside theta function, defined as:
\[
\Theta(\eta_s) = \begin{cases}
0, & \text{if } \eta_s < 0 \\
1, & \text{if } \eta_s \geq 0
\end{cases}
\]
Thus, $\epsilon_{\eta_s}(\eta_s) = 1$ for $\vert \eta_s \vert < \eta_0$, and beyond that, it follows a Gaussian distribution with variance $\sigma_{\eta}^2$. This parameterization of the rapidity distribution involves two parameters ($\eta_0$ and $\sigma_\eta$), which are typically tuned to match the experimental measurements of the pseudorapidity distribution of the charged particle yield. In Au+Au collisions at $\sqrt{s_{\text{NN}}}=200$ GeV, the pseudorapidity distribution of the charged particle yield is well reproduced with $\eta_0=1.3$ and $\sigma_\eta=1.5$.

\begin{figure}
  \centering
  \includegraphics[width=0.5\textwidth]{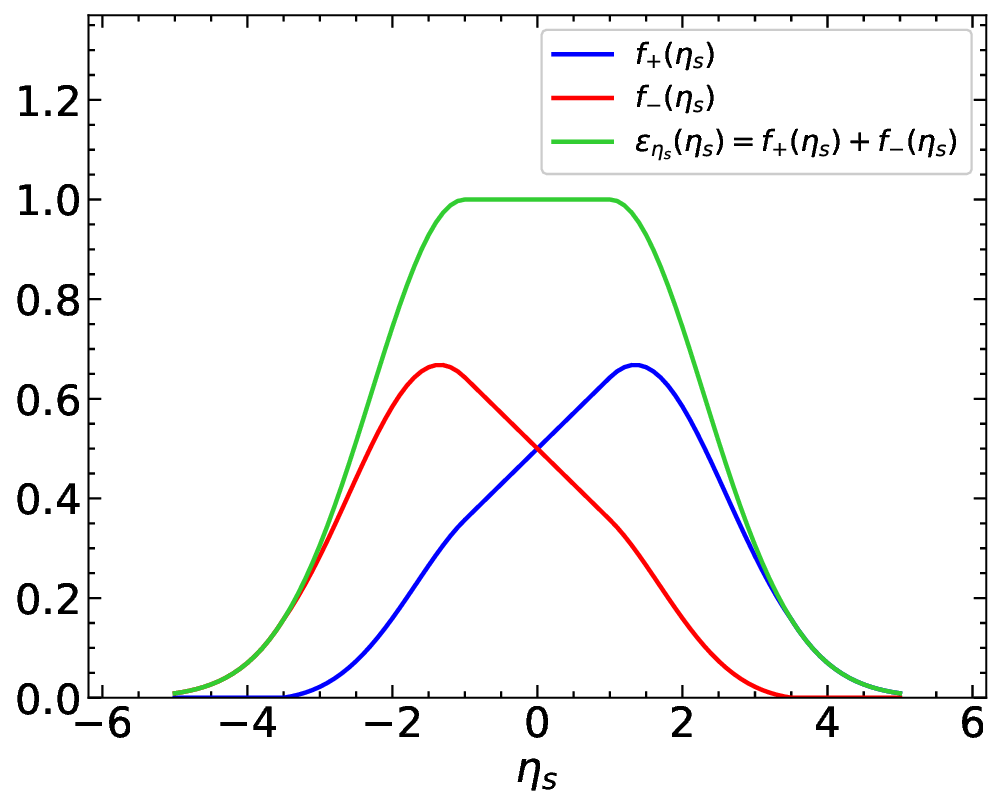} 
  \caption{ Space-time rapidity distribution profiles of left ($f_{-}(\eta_s)$) and right ($f_{+}(\eta_s)$) going participants for the initial energy density in a tilted initial condition. The figure also shows $\epsilon_{\eta_s}(\eta_s)$, which is an even function of $\eta_s$ and fulfills the relation $\epsilon_{\eta_s}(\eta_s) = f_{+}(\eta_s) + f_{-}(\eta_s)$.  }
  \label{fig:tilt_func}
\end{figure}

The functions $f_{+}(\eta_s)$ and $f_{-}(\eta_s)$ account for the asymmetric deposition of matter in the forward and backward rapidity regions and are defined as follows:
\begin{equation}
    f_{+,-}(\eta_s) = \epsilon_{\eta_s}(\eta_s) \epsilon_{F,B}(\eta_s)
\end{equation}
where
\begin{equation}
    \epsilon_{F}(\eta_s) = 
    \begin{cases}
    0, & \text{if } \eta_{s} < -\eta_{m}\\
    \frac{\eta_{s} + \eta_{m }}{2 \eta_{m}},  & \text{if }  -\eta_{m} \le \eta_{s} \le \eta_{m} \\
    1,& \text{if }  \eta_{m} < \eta_{s}
\end{cases}
\label{tilt_prof}
\end{equation}
and 
\begin{equation}
    \epsilon_{B} (\eta_s) = \epsilon_F(-\eta_s)
\label{eq:forward_eps_backward_envelop_relation}
\end{equation}
Figure \ref{fig:tilt_func} illustrates $f_{+}(\eta_s)$ and $f_{-}(\eta_s)$, which represent the emission functions for forward and backward-going participants as a function of $\eta_s$. Adding $f_{+}(\eta_s)$ and $f_{-}(\eta_s)$ yields $\epsilon_{\eta_s}(\eta_s)$, a forward-backward symmetric function in $\eta_s$. This model introduces a free parameter, $\eta_m$, which is used to control the amount of tilt in the energy density profile in the reaction plane \cite{Bozek:2010bi}.

\begin{figure}[htbp]
    \centering
    \begin{minipage}{0.5\textwidth} 
        \centering
        \includegraphics[width=\textwidth]{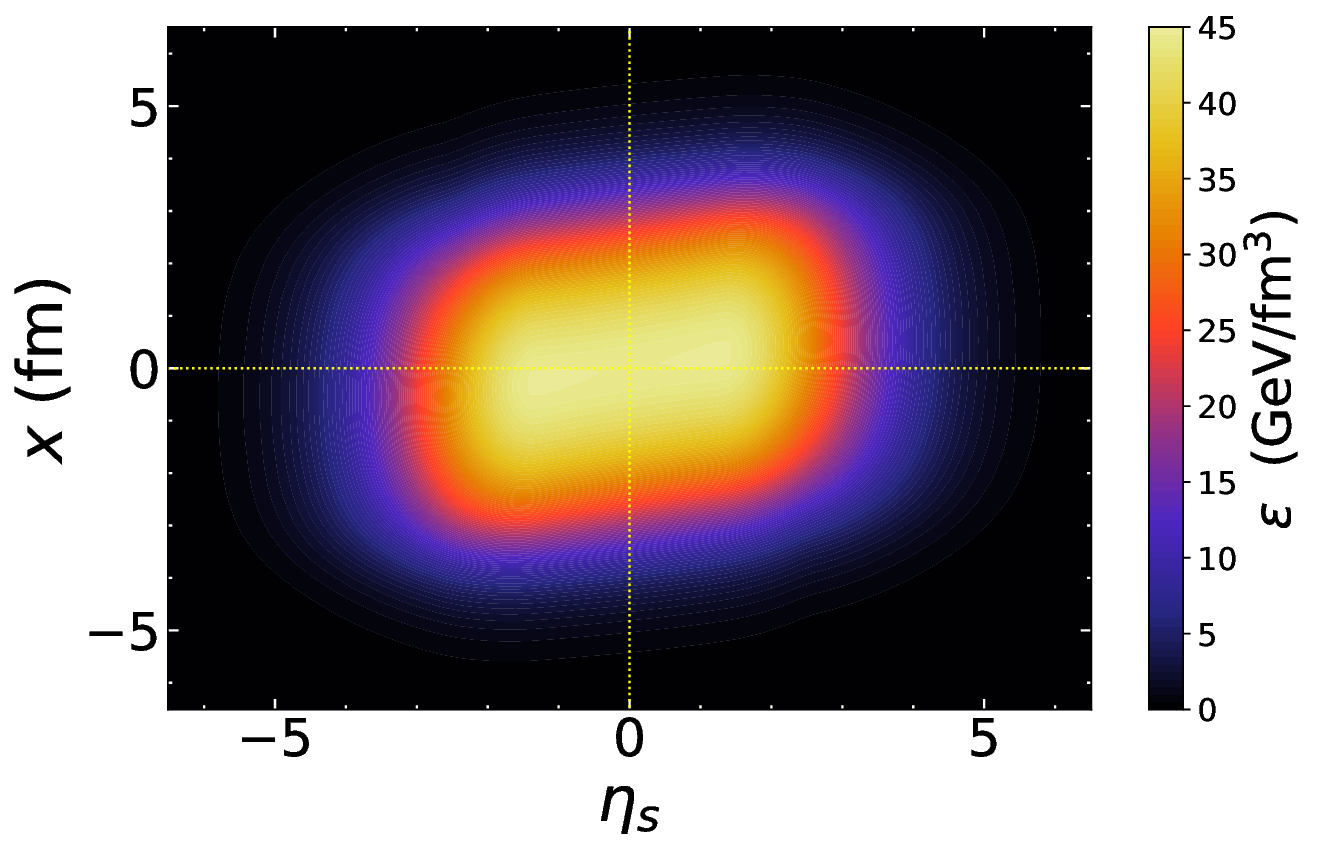}
        \caption{Color contour of the tilted initial energy density distribution in the $x-\eta_s$ plane (at $y=0$) for Au+Au collisions at $\sqrt{s_{NN}} = 200$ GeV. The initial energy density is obtained from a Glauber model calculation with an impact parameter of $b=7.2$ fm, which represents the average impact parameter for the 20-30\% centrality class. The tilt parameter ($\eta_m$) for this profile is $2.5$. }
        \label{fig:tilt_dist_xetas_conta}
    \end{minipage}
    \hfill 
    \begin{minipage}{0.45\textwidth} 
        \centering
        \includegraphics[width=\textwidth]{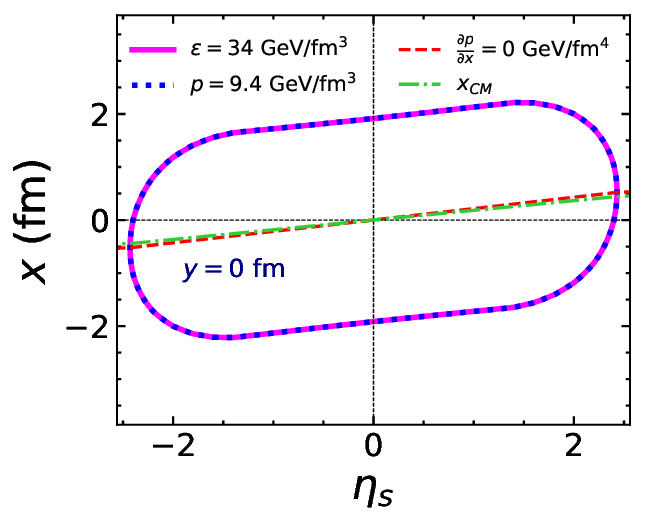}
        \caption{Contour of constant energy density ($\epsilon = 34$ GeV/fm$^3$) in the $x-\eta_s$ plane (at $y=0$) for the energy density profile shown in Fig. \ref{fig:tilt_dist_xetas_conta}. This figure also displays the $\eta_s$ dependence of the $x$ coordinate where $\partial_x p = 0$ (the location of maximum pressure points) along with the $x$ coordinate of the center of mass of the transverse energy density distribution ($x_{\text{CM}}$) at each $\eta_s$.   }
        \label{fig:tilt_dist_xetas_contb}
    \end{minipage}
    \label{fig:tilt_dist_xetascont}
\end{figure}

\begin{figure}
  \centering
  \includegraphics[width=1.0\textwidth]{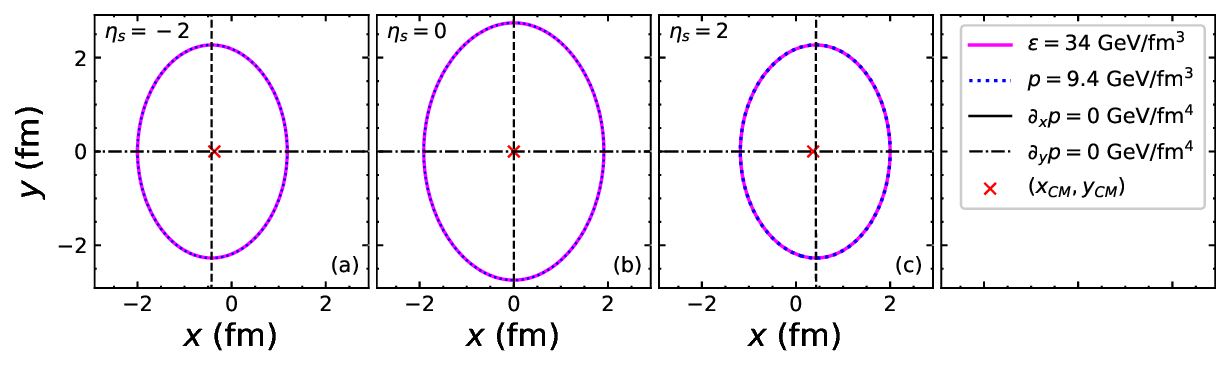} 
  \caption{Contours of constant energy density for the tilted energy density profile shown in Fig. \ref{fig:tilt_dist_xetas_contb} in the transverse plane at (a) $\eta_s = -2$, (b) $\eta_s = 0$, and (c) $\eta_s = 2$. The lines where $\partial_x p = 0$ and $\partial_y p = 0$ (indicating the locations of maximum pressure points) are shown, along with the coordinates of the center of mass of the transverse energy density distribution.}
  \label{fig:tilt_xy_dist}
\end{figure}

For illustrative purposes, we have presented color contour plots of the initial three-dimensional distribution of the tilted energy density profile in the reaction plane, obtained using the optical Glauber model for Au+Au collisions with an impact parameter $b=7.2$ fm, as shown in Fig. \ref{fig:tilt_dist_xetas_conta}. The tilt parameter $\eta_m$ set to 2.5. According to our Glauber model calculations, $b=7.2$ fm corresponds to the average impact parameter for the 20-30\% centrality class. In our calculation, we have aligned the impact parameter direction along the $x$-axis, thereby setting the reaction plane angle to zero from the outset. The resulting fireball produced in the model is deformed in the reaction plane, and a clear breaking of forward-backward symmetry can be observed along the longitudinal direction. Additionally, Fig. \ref{fig:tilt_dist_xetas_contb} shows the contour of constant energy density $\epsilon = 34$ GeV/fm$^3$, corresponding to $p = 9.4$ GeV/fm$^3$, in the $x-\eta_s$ plane at $y=0$. In the same figure, we have also plotted the $\eta_s$ dependence of the $x$ coordinate where $\partial_x p=0$, which indicates the $x$ coordinate of the maximum pressure points in the transverse energy density distribution. Furthermore, we have depicted the $x$ coordinate of the center of mass of the transverse energy density distribution ($x_{\text{CM}}$) at each $\eta_s$. The center of mass of the energy density distribution in the transverse plane at each $\eta_s$ is calculated using the following formula:
\beq
(x_\text{CM}(\eta_s),y_{\text{CM}}(\eta_s)) = \left( \frac{ \int dx dy \  x \ \epsilon(x,y,\eta_s)  }{\int dx dy \  \epsilon(x,y,\eta_s) }, \frac{ \int dx dy \ y \  \epsilon(x,y,\eta_s)  }{\int dx dy \ \epsilon(x,y,\eta_s) } \right)
\eeq

In Fig. \ref{fig:tilt_dist_xetas_contb}, it is evident at non-zero $\eta_s$, there is a very small difference between the $x_{\text{CM}}$ and the $x$ coordinate where $\partial_x p = 0$. As previously discussed, this misalignment between the center of mass position and the point of maximum pressure in the transverse plane indicates a dipolar asymmetric deposition, which ultimately leads to asymmetric flow\footnote{It is worth noting that due to the symmetry of the collision, the $y$ coordinate of the energy density distribution's center of mass ($y_{\text{CM}}$) is consistently zero across all $\eta_s$, and the $y$ coordinates of the points where $\partial_y p = 0$ align with $y_{\text{CM}} = 0$. Therefore, as discussed earlier, since the center of mass and the maximum pressure point coincide, this will result in symmetric flow along the $y$-axis in opposite directions.}. This misalignment is further illustrated in Fig. \ref{fig:tilt_xy_dist}, where the contours of energy density in the transverse plane are plotted at (a) $\eta_s = -2$, (b) $\eta_s = 0$, and (c) $\eta_s = 2$. At $\eta_s = 0$, the center of mass of the transverse energy density lies on the line where $\partial_x p = 0$ and $\partial_y p = 0$, indicating a symmetric deposition. However, at $\eta_s = -2$ and $\eta_s = 2$, the center of mass of the transverse energy density distribution lies on the $\partial_y p = 0$ line but deviates from the $\partial_x p = 0$ line, signaling a clear dipolar asymmetry along the $x$-axis. Additionally, the center of mass of the energy density is positioned to the right of the $\partial_x p = 0$ line at $\eta_s = -2$, while it is located to the left of the $\partial_x p = 0$ line at $\eta_s = 2$, showing a reflection symmetry along the longitudinal direction.

The asymmetry in energy deposition along the longitudinal direction, $\eta_s$, can be quantified by calculating the first-order eccentricity as a function of $\eta_s$, defined as \cite{Teaney:2010vd,Shen:2020jwv}:
\begin{equation}
    \Vec{\mathcal{E}_1} (\eta_s) = \mathcal{E}_1 (\eta_s) e^{i \Psi_{1}(\eta_s)} =  - \frac{\int d^2r \Tilde{r}^3  e^{i \Tilde{\phi} } \epsilon(r,\phi,\eta_s) }{\int d^2r \Tilde{r}^3 \epsilon(r,\phi,\eta_s)}
\label{eq:Eps1}
\end{equation}
where,
\begin{equation}
\Tilde{r}(x,y,\eta_s) = \left[ \left( x - x_{\text{CM}}(\eta_s) \right)^2 + \left( y - y_{\text{CM}}(\eta_s) \right)^2 \right]^{\frac{1}{2}}  
\end{equation}
and
\begin{equation}
 \Tilde{\phi}(x,y,\eta_s) = \arctan\left[ \left( x - x_{\text{CM}}(\eta_s) \right) /  \left( y - y_{\text{CM}}(\eta_s) \right) \right]   
\end{equation}

\begin{figure}[htbp]
    \centering
    \begin{minipage}{0.49\textwidth} 
        \centering
        \includegraphics[width=\textwidth]{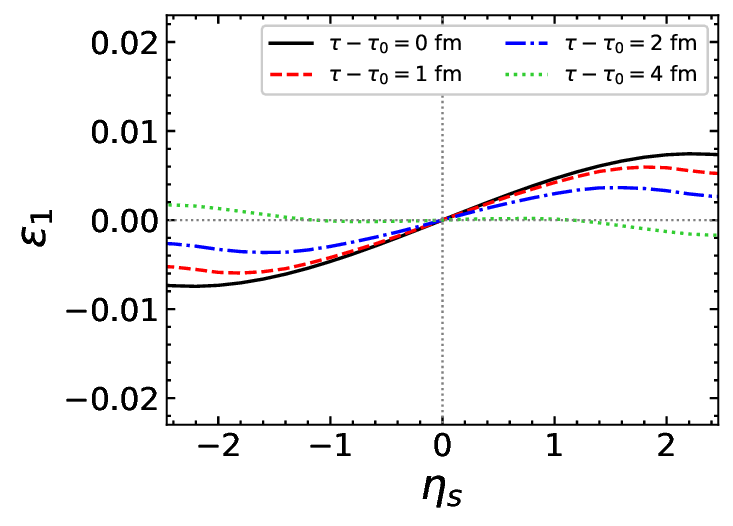}
        \caption{Time evolution of the first-order eccentricity, $\mathcal{E}_1(\eta_s)$ (defined in Eq. \ref{eq:Eps1}), obtained from the hydrodynamic evolution of the tilted energy deposition profile shown in Fig. \ref{fig:tilt_dist_xetas_conta}.}
        \label{fig:tilt_e1_vx_avg_a}
    \end{minipage}
    \hfill 
    \begin{minipage}{0.45\textwidth} 
        \centering
        \includegraphics[width=\textwidth]{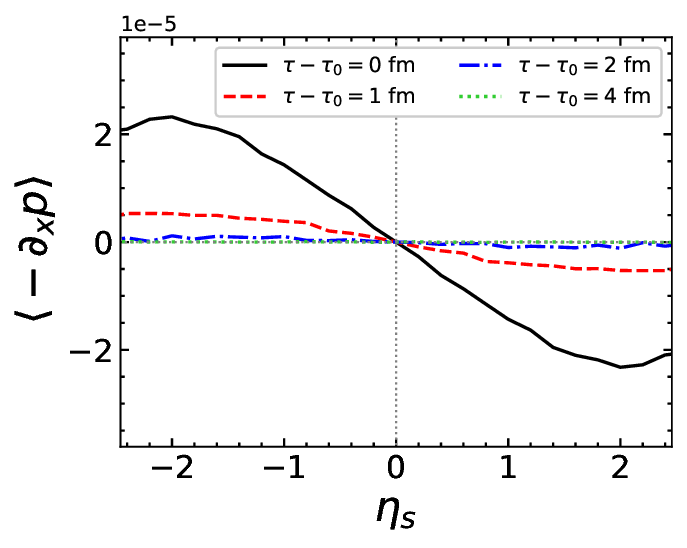}
        \caption{Time evolution of the $\la -\partial_x p  \ra (\eta_s)$ (defined in Eq. \ref{eq:dhoxpavg}), obtained from the hydrodynamic evolution of the tilted energy deposition profile shown in Fig. \ref{fig:tilt_dist_xetas_contb}.}
        \label{fig:tilt_e1_vx_avg_c}
    \end{minipage}
    \hfill 
    \begin{minipage}{0.45\textwidth} 
        \centering
        \includegraphics[width=\textwidth]{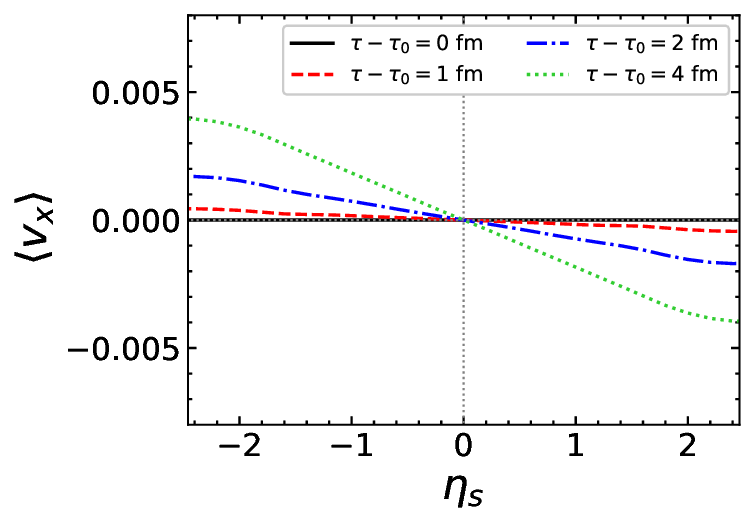}
        \caption{Time evolution of the $\la v_x \ra(\eta_s)$ (defined in Eq. \ref{eq:vxavgetasshift}), obtained from the hydrodynamic evolution of the tilted energy deposition profile shown in Fig. \ref{fig:tilt_dist_xetas_contb}.}
        \label{fig:tilt_e1_vx_avg_b}
    \end{minipage}
    \label{fig:tilt_e1_vx_avg}
\end{figure}

In Fig. \ref{fig:tilt_e1_vx_avg_a}, we present the plot of $\mathcal{E}_1$ as a function of $\eta_s$ for the initial energy distribution depicted in Fig. \ref{fig:tilt_dist_xetas_contb}, shown as a black solid line. It is evident that $\mathcal{E}_1$ is an odd function of $\eta_s$, with its magnitude being smaller in the mid-$\eta_s$ region and increasing significantly at larger $\eta_s$, indicating an enhanced dipole asymmetry at higher $\eta_s$ values.

Apart from the conventional quantification of the dipolar asymmetry through the $\mathcal{E}_1$, the dipolar asymmetry can also be quantified by calculating the average pressure gradient along x-direction ($ \la -\partial_x p \ra$), which can be defined as follows :
\beq
\la -\partial_x p  \ra (\eta_s) = \frac{\int d^2r (-\partial_x p ) \epsilon(x,y,\eta_s) }{\int d^2r \epsilon(x,y,\eta_s)}
\label{eq:dhoxpavg}
\eeq
The $\la -\partial_x p (\eta_s) \ra$ of the energy distribution at initial time $\tau_0$ has been plotted in black solid lines in Fig. \ref{fig:tilt_e1_vx_avg_c}. It can be observed that $\la -\partial_x p (\eta_s) \ra$ has a negative mid-rapidity slope. This is also a odd function of $\eta_s$ and shows larger asymmetry at large $\eta_s$.

To explore how this dipole asymmetry evolves into asymmetric flow over time through fluid dynamical evolution, we conducted a 3+1D viscous hydrodynamic simulation of this energy density profile, starting from an initial time of $\tau_0=0.4$ fm, using the publicly available MUSIC code \cite{Schenke:2010nt, Paquet:2015lta, Schenke:2011bn}. At the onset of the hydrodynamic evolution, we assumed a Bjorken flow ansatz for the fluid, with the fluid velocity given by $u^{\mu}(\tau_0;x,y,\eta_s) = (\cosh{\eta_s}, 0, 0, \sinh{\eta_s})$. The hydrodynamic evolution utilized the NEoSB equation of state (EoS) derived from Lattice QCD and the HRG EoS \cite{Monnai:2019hkn}. A constant shear viscosity to entropy density ratio ($\eta/s = 0.08$) was applied during the evolution, while bulk viscosity was set to zero.

The left-right deformation of the initial matter profile in the transverse plane generates asymmetric flow at different $\eta_s$ values. This asymmetric flow can be quantified by measuring the average $v_x$ of the fluid at various $\eta_s$, which can be calculated using the following formula:
\beq
\la v_x \ra (\eta_s) = \frac{\int d^2r v_x \epsilon(x,y,\eta_s)}{\int d^2r \epsilon(x,y,\eta_s)}
\label{eq:vxavgetasshift}
\eeq

The time evolution of the generated asymmetric flow due to the deformed distribution in the TIC is shown by plotting the $\eta_s$ dependency of $\langle v_x \rangle$ in Fig. \ref{fig:tilt_e1_vx_avg_b}. As the fluid evolves, the magnitude of $\mathcal{E}_1(\eta_s)$ decreases due to the isotropization of the fluid, which reduces the $\la -\partial_x p (\eta_s) \ra$ as well \cite{Bozek:2010bi,Jiang:2021ajc}. Conversely, the flow asymmetry increases with evolution, as shown by the time evolution of $\langle v_x \rangle$. Notably, $\langle v_x \rangle$ exhibits a negative slope at $\eta_s = 0$. This negative slope in $\langle v_x \rangle(\eta_s)$ at $\eta_s = 0$ will be reflected in the final-state mid-rapidity $v_1(\eta)$ slope of charged hadrons.

\begin{figure}
  \centering
  \includegraphics[width=0.5\textwidth]{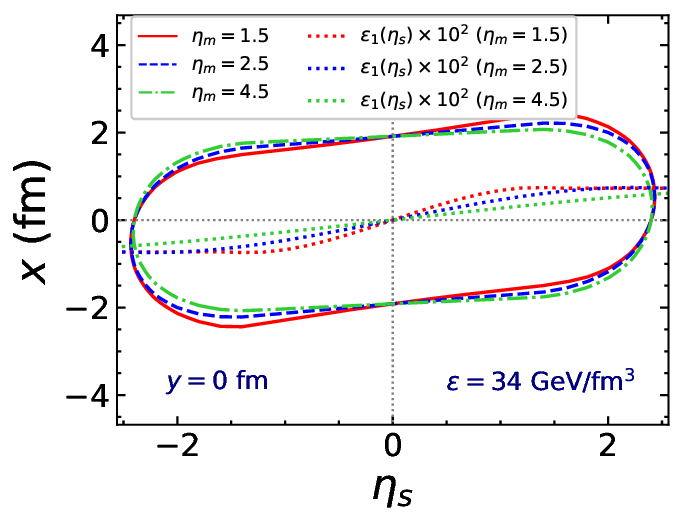} 
  \caption{Contours of constant energy density ($\epsilon = 34$ GeV/fm$^3$) in the $x-\eta_s$ plane (at $y=0$) for tilted energy density profiles with different tilt parameters ($\eta_m$), demonstrating effect of tilt parameters on the distribution. Additionally, the first-order eccentricities, $\mathcal{E}_1(\eta_s)$ (defined in Eq. \ref{eq:Eps1}), obtained from the initial tilted energy deposition profiles for different $\eta_m$ values, are also shown.}
  \label{fig:etas_x_eps_cont_diff_tilt}
\end{figure}

In the TIC model depicted in Fig. \ref{fig:tilt_dist_xetas_conta}, we have used $\eta_m = 2.5$, but we also examine two additional $\eta_m$ values for comparison: $\eta_m = 1.5$ and $\eta_m = 4.5$. It can be observed in Fig. \ref{fig:etas_x_eps_cont_diff_tilt} that, as $\eta_m$ increases, the dipole asymmetry $\mathcal{E}_1(\eta_s)$ decreases. This dependence of $\mathcal{E}_1$ on $\eta_m$ is evident from Eq. \ref{tilt_prof}. As $\eta_m \rightarrow \infty$, the values of $\epsilon_F(\eta_s)$ and $\epsilon_B(\eta_s)$ approach $\frac{1}{2}$, leading to symmetric energy deposition.

\begin{figure}
  \centering
  \includegraphics[width=1.0\textwidth]{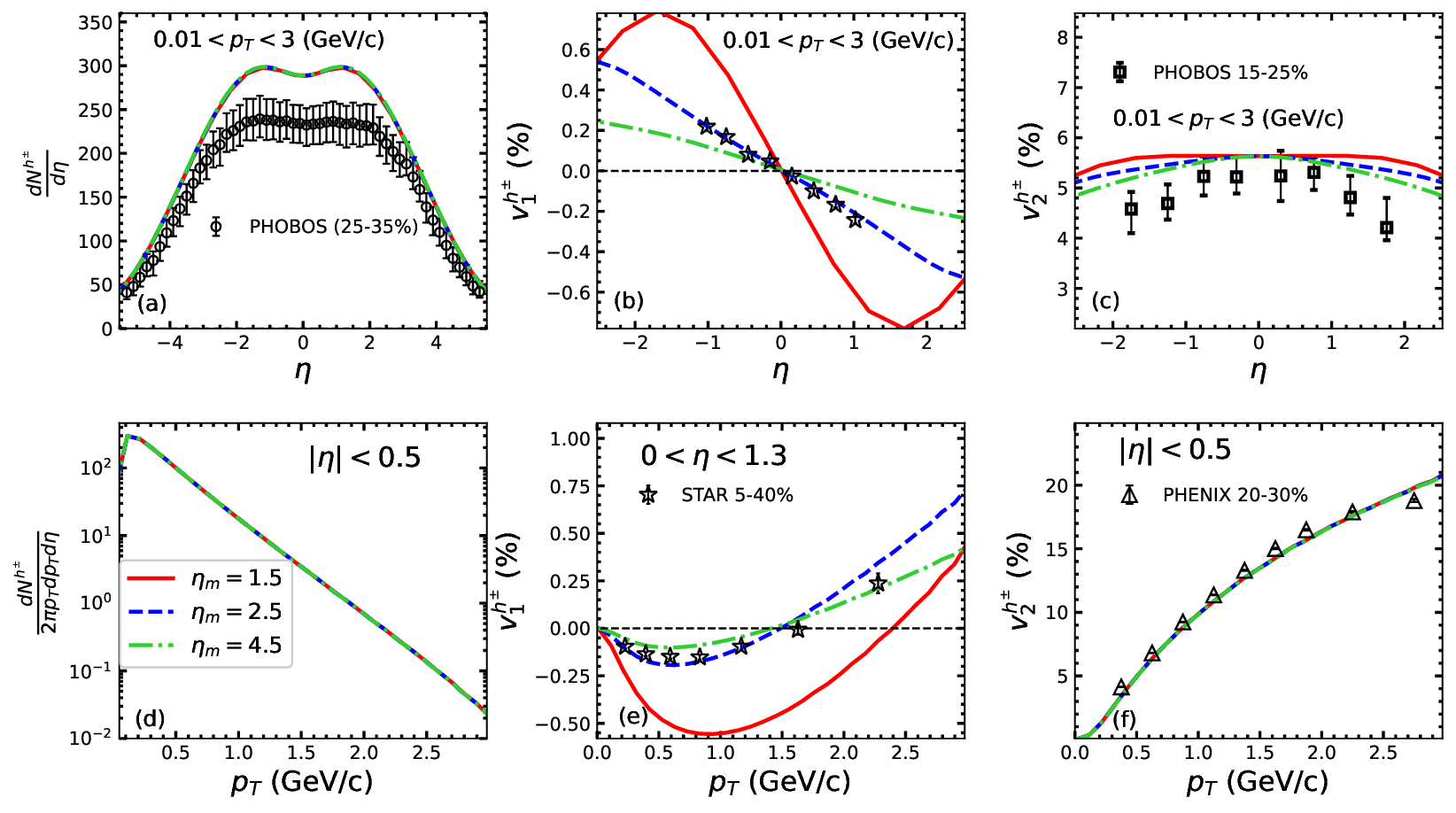} 
  \caption{ pseudorapidity ($\eta$) dependence of (a) yield (b) $v_1$ and (c) $v_2$ of charged hadrons along with $p_T$-differential (d) Invariant yield (e) $v_1$ and (f) $v_2$ of charged hadrons are plotted. The results from model calculations using tilted initial conditions with various tilt parameters are compared. The model calculations are done for 20-30\% centrality. Experimental data are also included for comparison \cite{Back:2002wb,STAR:2008jgm,PHOBOS:2004vcu,PHENIX:2011yyh} though they are not precisely of the 20-30\% centrality class for all observables, except for the $p_T$-differential $v_2$.}
\label{fig:v1_v2_eta_shift_tilt}
\end{figure}

The results of the hydrodynamic simulations with the tilted energy density profile (shown in Fig. \ref{fig:tilt_dist_xetas_contb}) for Au+Au collisions at $\sqrt{s_{NN}} = 200$ GeV are shown in Fig. \ref{fig:v1_v2_eta_shift_tilt}. These figures also include results from different tilt parameters in the tilted initial condition (TIC) for comparison. Experimental data from various collaborations are also plotted for reference \cite{Back:2002wb,STAR:2008jgm,PHOBOS:2004vcu,PHENIX:2011yyh}. Note that the model calculations are performed for an impact parameter of $b = 7.2$ fm, corresponding to the 20-30\% centrality range. Since experimental data for all observables in this exact centrality range are not available, data from centrality bins closest to 20-30\% are used. For example, experimental results for $v_1$ as a function of $y$ and $p_T$ are taken from the 5-40\% centrality range. The Glauber model indicates that for this centrality, $b = 6.8$ fm, which is close to $b = 7.2$ fm.

Figure \ref{fig:v1_v2_eta_shift_tilt} compares the effects of different tilt parameter ($\eta_m$) on final state observables. In panels (a), (b), and (c) of Fig. \ref{fig:v1_v2_eta_shift_tilt}, the pseudorapidity ($\eta$) dependence of the charged particle yield, $v_1$, and $v_2$ are plotted whereas in panels (d), (e), and (f), the $p_T$-differential invariant spectra, $v_1$, and $v_2$ at mid-rapidity ($|\eta| < 0.5$) for charged hadrons are shown. The results indicate that the tilt parameter has no effect on the $\frac{dN_{ch}}{d\eta}$ distribution. Furthermore, there is alsoe no difference in the mid-rapidity $p_T$-differential invariant yield of charged hadrons. In Fig. \ref{fig:v1_v2_eta_shift_tilt}(f), the $p_T$-differential elliptic flow ($v_2$) at mid-rapidity is presented. It is observed that the the $\eta_m$  do not affect $v_2(p_T)$, which is also reflected in the $\eta$-differential $v_2$ shown in Fig. \ref{fig:v1_v2_eta_shift_tilt}(c), where all initial configurations yield similar $p_T$-integrated $v_2$ values around $\eta = 0$. However, deviations are observed at larger $\eta$, likely due to different correlations between longitudinal and transverse flows during the evolution of the various initial configurations.

The most pronounced effect of the initial configuration is seen in both the $p_T$ and $\eta$-differential $v_1$. The mid-rapidity slope of $v_1(\eta)$, $\frac{dv_1}{d\eta}$, shows a negative sign. However, as the tilt increases, the magnitude of $\frac{dv_1}{d\eta}$ also increases, which is consistent with expectations. Noticably, an interesting feature emerges in the $p_T$-differential $v_1$. In the forward rapidity region ($0 < \eta < 1.3$), $v_1$ shifts from negative at lower $p_T$ to positive at higher $p_T$, indicating opposite flow directions for particles with low and relatively higher $p_T$. This behavior is expected, as discussed previously. The $\la v_x \ra < 0$ in the forward rapidity region in TIC implies more of the fluid flows along negative x-direction as compared to positive $x-$direction as the $x_{\text{CM}}$ at each $\eta_s$ is below the maximum pressure points. However, the steeper pressure gradient toward positive $x$ generates higher transverse velocities for fluid moving in that direction. This results in an enhanced production of higher $p_T$ particles from these fluid cells, leading to a positive $v_1$ for the higher $p_T$ particles.

\section{An alternative ansatz of tilted profile}
Similar to the concept introduced by Bozek and Wyskiel (BW) in 2010 \cite{Bozek:2010bi}, which proposed that nucleons tend to deposit more energy or entropy along their direction of motion, an alternative approach of the initial energy distribution has been presented in Ref. \cite{Jiang:2021foj} by Z.-F. Jiang, C. B. Yang, and Q. Peng (JYP). The JYP proposal also suggests an asymmetric  matter deposition along space-time rapidity by the forward and backward moving participants. Consequently, this parameterization results in a tilted profile within the reaction plane. In the phenomenological model of JYP, the energy density at a constant proper time ($\tau_{0}$) is expressed as :
\begin{equation}
\epsilon\left(x,y,\eta_{s}; \tau_{0}\right) = \epsilon_{0}^{\JYP} \mathcal{W}\left( x, y, \eta_{s} \right) f(\eta_{s})
\end{equation}
Here, $\epsilon_{0}^{\JYP}$ serves as a normalization parameter. The weight function $ \mathcal{W} \left( x, y, \eta_{s} \right)$ is defined as:
\begin{equation}
 \mathcal{W} \left( x, y, \eta_{s} \right) = (1-\alpha^{\JYP}) \mathcal{W}_{N}( x, y, \eta_{s} ) + \alpha^{\JYP} N_{coll}(x,y)   
\end{equation}
with
\beq
\begin{aligned}
\mathcal{W}_{N} \left( x, y, \eta_{s} \right) & =  \left[ N_{+}(x,y) + N_{-}(x,y) \right] +  \left[ N_{+}(x,y) - N_{-}(x,y) \right]  H_{t}  \tan{ \left( \frac{\eta_{s}}{\eta_{t}}\right)} 
\end{aligned}
\label{jyp_tilt_wn}
\eeq
Here, $H_{t} \tan{ \left( \frac{\eta_s}{\eta_t} \right) }$ parameterizes the imbalance between energy deposition in forward and backward space-time rapidities ($\eta_s$) at any transverse point (x,y).

Upon comparing the expressions for initial deposited energy density in both models, it is observed that the JYP model is essentially a generalization of the BW model, and under certain parameter conditions of the JYP model, it reduces to the BW model. The mapping between the parameters in the BW and JYP models is derived as follows.

The expression for the initial energy density in the JYP model is \cite{Jiang:2021foj}:
\begin{equation}
\begin{aligned}
\epsilon & \left(x,y,\eta_{s}\right)  = \epsilon_{0}^{\JYP} W\left( x, y, \eta_{s} \right) f(\eta_{s}) \\
& =  \epsilon_{0}^{\JYP} \left[   (1-\alpha^{\JYP})   \Big( \left( N_{+}(x,y) + N_{-}(x,y) \right) + \left( N_{+}(x,y) - N_{-}(x,y) \right)  H_{t}  \tan{ \left( \frac{\eta_{s}}{\eta_{t}}\right)} \Big) + \alpha^{\JYP}  N_{coll}(x,y)   \right]f(\eta_{s}) \\
& =  \epsilon_{0}^{\JYP}  \alpha^{\JYP} \left[  \frac{ (1-\alpha^{\JYP})}{  \alpha^{\JYP}}  \Big( \left( N_{+}(x,y) + N_{-}(x,y) \right) + \left( N_{+}(x,y) - N_{-}(x,y) \right)  H_{t}  \tan{ \left( \frac{\eta_{s}}{\eta_{t}}\right)} \Big) +   N_{coll}(x,y)   \right]f(\eta_{s}) \\
\label{eq.JYP_tilt_anz}
\end{aligned}
\end{equation}
The three-dimensional profile of the initial energy density around mid-rapidity ($-\eta_m < \eta_s < \eta_m$) in the BW model is given by:
\begin{equation}
\begin{aligned}
\epsilon & \left(x,y,\eta_{s}\right) \\ & = \epsilon_{0}^{\BW} \Big[ \Big( N_{+}(x,y) f_{+}(\eta_s) + N_{-}(x,y) f_{-}(\eta_s) \Big) (1-\alpha^{\BW}) + \alpha^{\BW} N_{coll}(x,y) \Big] f(\eta_{s}) \\
& = \epsilon_{0}^{\BW} \Big[ \Big( N_{+}(x,y) \left( \frac{\eta_s + \eta_m }{2 \eta_m } \right) +  N_{-}(x,y)  \left( \frac{-\eta_s + \eta_m }{2 \eta_m } \right) \Big) (1-\alpha^{\BW}) + \alpha^{\BW} N_{coll}(x,y) \Big] f(\eta_{s}) \\
& = \epsilon_{0}^{\BW} \alpha^{\BW} \left[  \left( \frac{1-\alpha^{\BW}}{2 \alpha^{\BW}} \right) \left(  \left( N_{+}(x,y) + N_{-}(x,y) \right)  + \left( N_{+}(x,y) -  N_{-}(x,y) \right) \frac{\eta_s}{\eta_m}   \right) +  N_{coll}(x,y) \right] f(\eta_{s}) \\
\label{eq.BW_tilt_anz}
\end{aligned}
\end{equation}
By comparing Eq. \ref{eq.JYP_tilt_anz} with Eq. \ref{eq.BW_tilt_anz} and matching the coefficients of $\left[ N_{+}(x,y) \pm N_{-}(x,y) \right]$, the following relationships are obtained:
\begin{equation}
\frac{1-\alpha^{\JYP}}{\alpha^{\JYP}} = \frac{1- \alpha^{\BW} }{ 2 \alpha^{\BW} }
\label{eq.alpha_relation_anz}
\end{equation}

\begin{equation}
\epsilon_{0}^{\JYP} \alpha^{\JYP} = \epsilon_{0}^{\BW} \alpha^{\BW}
\label{eq.e0_relation_app}
\end{equation}
and
\begin{equation}
\begin{aligned}
 & H_t \tan{\left( \frac{\eta_s}{\eta_t}\right)} = \frac{\eta_s}{\eta_m} \\
 & \implies H_{t} \Big(\frac{\eta_s}{\eta_t} + \frac{1}{3} \Big(\frac{\eta_s}{\eta_t}\Big)^3 + ... \Big) = \frac{\eta_s}{\eta_m} \\
 & \implies \frac{H_t}{\eta_t} \left( 1 + \mathcal{O} \left( \frac{\eta_s}{\eta_t} \right)^2 \right)   = \frac{1}{\eta_m}
 \label{eq.JYP_BW_tilt_rltn}
\end{aligned}
\end{equation}

For a large value of $\eta_t$, as $\mathcal{O} \left( \frac{\eta_s}{\eta_t} \right)^2$ tends towards zero, the tilt parameter of both models becomes related by the following equation:

\begin{equation}
\frac{H_t}{\eta_t} \approx \frac{1}{\eta_m}
\label{eq.JYPBW.large_etat_Ht_etam_reltn}
\end{equation}

Based on this insight, we conclude that for large $\eta_t$, the expression of the initial energy profile in the JYP model reduces to that of the BW model, with the parameters of both models related by Eqs. \ref{eq.JYPBW.large_etat_Ht_etam_reltn}, \ref{eq.alpha_relation_anz}, and \ref{eq.e0_relation_app}. However, for small $\eta_t$, the initial longitudinal gradient of deposited matter differs between the two models. The BW model assumes a linear $\eta_s$ dependency in the longitudinal profile ($ \pm \frac{\eta_s}{2 \eta_m} + \frac{1}{2}$), while the JYP model assumes a $\tan{\left( \eta_s \right)}$ function, for the energy deposition by the participants along $\eta_s$ near mid-rapidity. In general, for an arbitrary $\eta_t$, the two models can be considered as representing two distinct initial conditions. Hence, our goal is to verify through numerical simulation whether at large $\eta_t$ JYP is equivalent to the BW model and at small $\eta_t$ how JYP model yield different results for $v_1$, and how the two scenarios of energy deposition could be distinguished.

\begin{figure}
  \centering
  \includegraphics[width=0.7\textwidth]{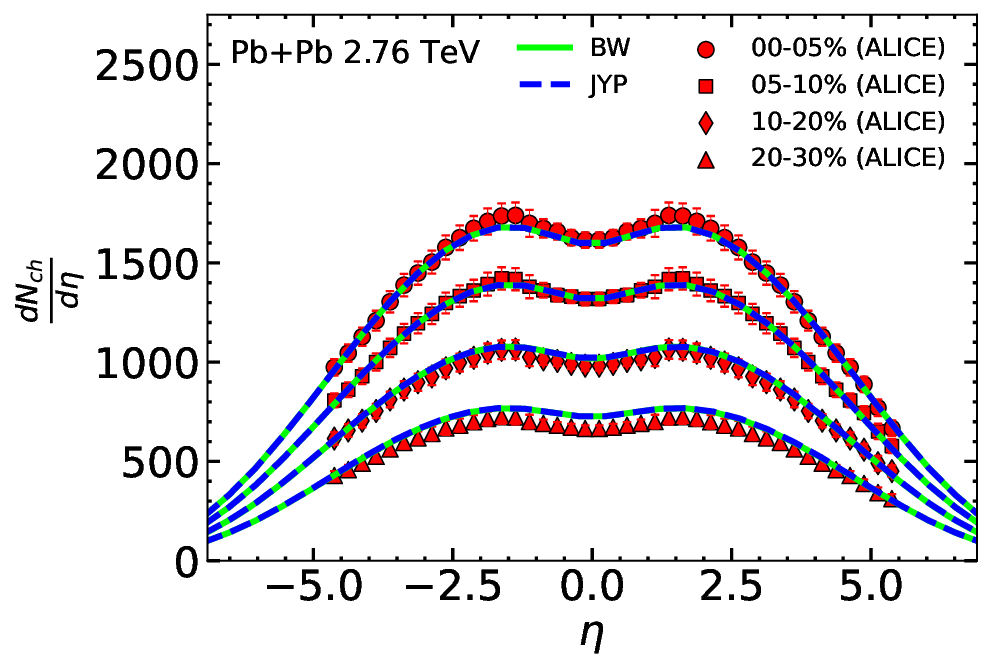} 
  \caption{Charged particle yield as a function of pseudo-rapidity for 0-5\%, 5-10\%, 10-20\% and 20-30\% centrality class in Pb+Pb collisions at $\sqrt{s_{NN}} = 2.76$ TeV. Experimental data from the ALICE collaboration \cite{ALICE:2013jfw} is compared with hydrodynamic model calculations using two different initial condition models of JYP~\cite{Jiang:2021foj} and BW~\cite{Bozek:2010bi}.}
  \label{fig:JYPBW:dnchdeta_eta}
\end{figure}

\begin{center}
\begin{table}[h!]
\centering
\begin{tabular}{ |p{2.5cm}|p{2.5cm}|p{2.5cm}|  }
\hline
\multicolumn{3}{|c|}{Pb+Pb at $\sNN =$ 2.76 TeV} \\
\hline
Parameter(unit) & BW & JYP  \\
\hline
$b$(fm) & 5.6 & 5.6 \\
\hline
$\epsilon_{0}^{\BW}$(GeV/fm$^3$) & 31.0 & - \\
\hline
$\epsilon_{0}^{\JYP}$(GeV/fm$^3$) & -   & 17.6 \\
\hline
$\alpha^{\BW}$ & 0.14 & - \\
\hline
$\alpha^{\JYP}$ & - & 0.245 \\
\hline
$H_{t}$ & - & 7.0 \\
\hline
$\eta_{t}$ & - & 8.0 \\
\hline
$\eta_{m}$ & 2.9 & - \\
\hline
$\eta_{0}$ & 2.0 & 2.0 \\
\hline
$\sigma_{\eta}$ & 1.8 & 1.8 \\
\hline
$\tau_{0}$(fm) & 0.4 & 0.4 \\
\hline
$\eta / s $ & 0.08 & 0.08 \\
\hline
$\zeta / s$ & 0 & 0 \\
\hline
$T_{f}$(MeV) & 150 & 150 \\
\hline
\end{tabular}
\caption{Parameters used during simulation of 10-20\% Pb+Pb collisions at $\sNN = 2760$ GeV 
for different initial conditions.}
\label{table:JYPBW:param2.76TeV}
\end{table}
\end{center}
For this purpose, we performed hydrodynamic simulations of Pb+Pb collisions at $\sqrt{s_{NN}} = 2.76$ TeV using both the initial condition models, JYP and BW. For the JYP model, we set $\eta_t=8.0$ as done in the original work by JYP in Ref. \cite{Jiang:2021foj}. Initially, we tuned $\epsilon_{0}$, $\alpha$, $\eta_0$, and $\sigma_{\eta_s}$ in the BW model to capture the centrality and rapidity dependence of charged particle yield. Then, to achieve the same energy deposition in the transverse plane at $\eta_s=0$ for the JYP model, we followed the relations given by Eqs. \ref{eq.alpha_relation_anz} and \ref{eq.e0_relation_app} to set the energy normalization $\epsilon_{0}$ and hardness factor $\alpha$ of the JYP model. Subsequently, we conducted the hydrodynamic simulation, maintaining all other parameter values as detailed in Table \ref{table:JYPBW:param2.76TeV}. With this setup, in Fig \ref{fig:JYPBW:dnchdeta_eta}, we observed that after hydro evolution, the rapidity and centrality dependence of charged particle yield are identical for both models.

\begin{figure}
  \centering
  \includegraphics[width=0.9\textwidth]{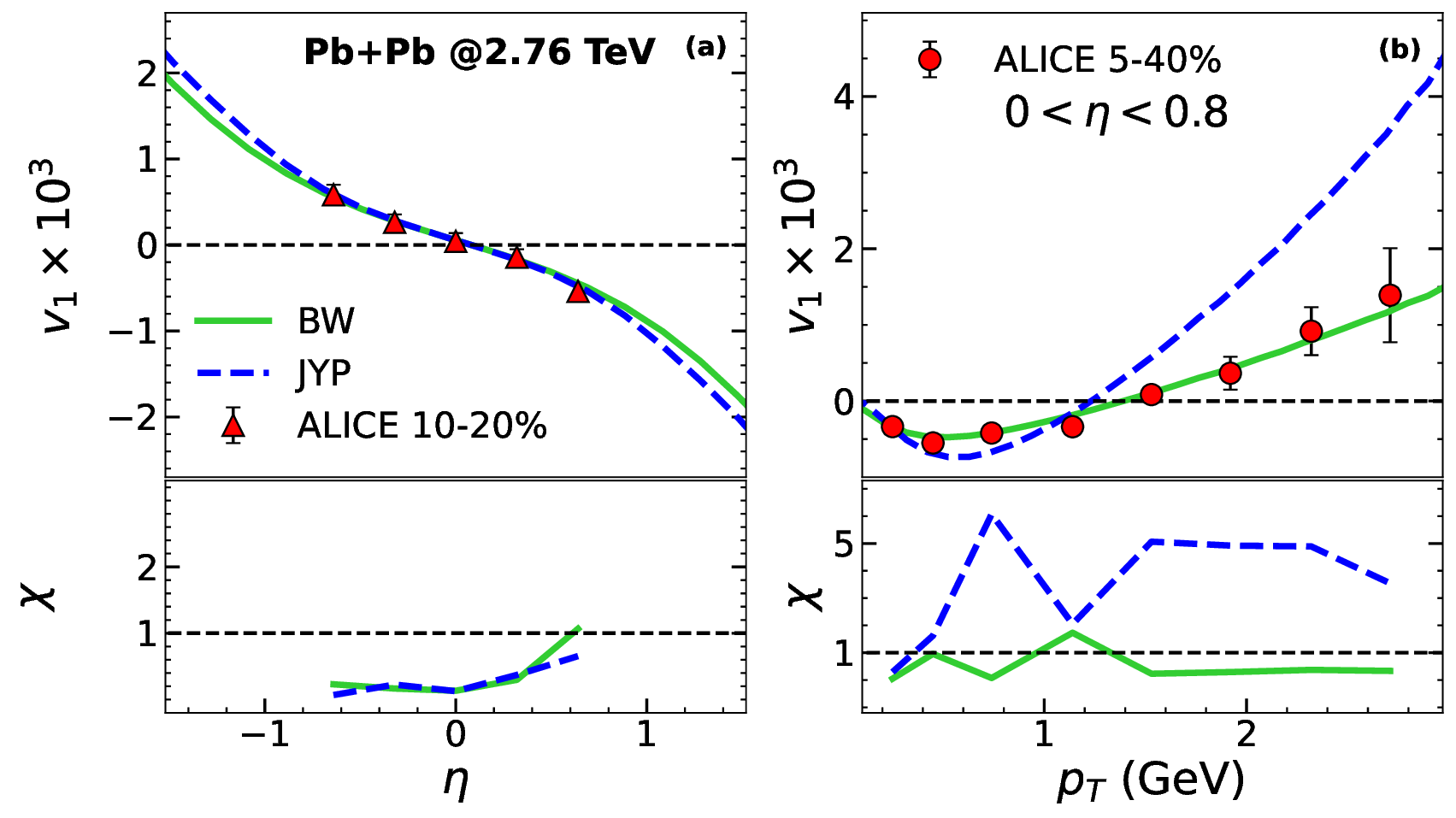} 
  \caption{Charged particle $v_1$ as a function of pseudorapidity (a) and transverse momentum (b) in Pb+Pb collisions at $\sqrt{s_{NN}} = 2.76$ TeV. Experimental data from the ALICE collaboration~\cite{ALICE:2013xri} is compared with hydrodynamic calculations using two different initial condition models by JYP~\cite{Jiang:2021foj} and BW~\cite{Bozek:2010bi}. The difference between the hydrodynamic model calculations and experimental data points is plotted in sub-panels of panels (a) and (b) in units of experimental error (see Eq.~\ref{eq:chi}). Solid lines represent calculations based on Bozek-Wyskiel initial conditions, while dashed lines depict calculations using the JYP initial matter deposition profile.}
  \label{fig:BWJYP:PbPb2.76:v1_eta_pt}
\end{figure}

\begin{figure}
  \centering
  \includegraphics[width=0.9\textwidth]{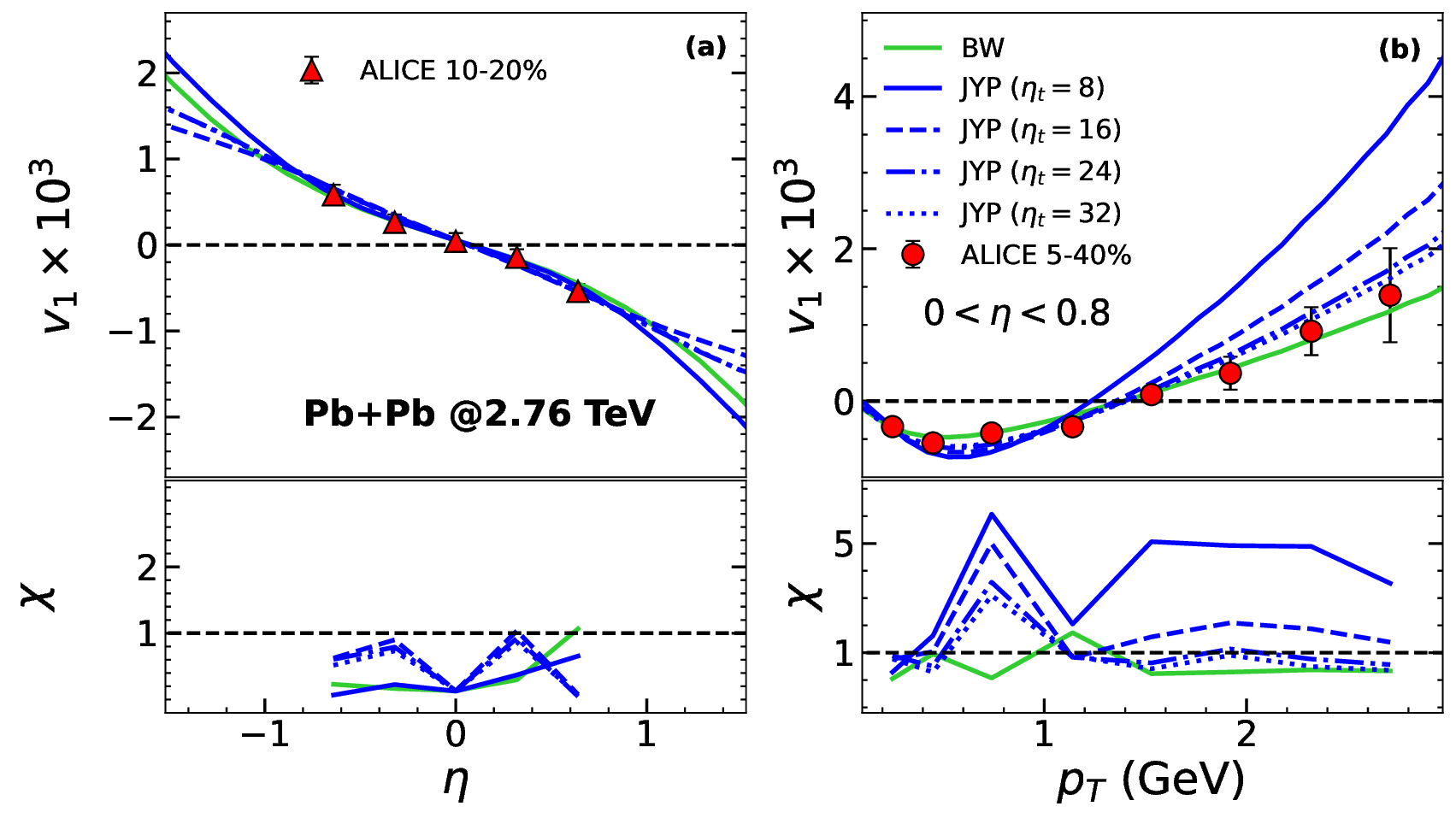} 
  \caption{Charged particle $v_1$ as a function of (a) pseudorapidity and (b) transverse momentum are depicted for various JYP model parameters, $\eta_t$, in Pb+Pb collisions at $\sqrt{s_{NN}} = 2.76$ TeV. Results from BW model calculations are represented by solid green lines for reference. Experimental data from Ref.~\cite{ALICE:2013xri} is compared with hydrodynamic model calculations.}
  \label{fig:BWJYP:PbPb2.76:v1_eta_pt_diff_etat}
\end{figure}

We proceeded by comparing the $v_1$ results obtained from both models. In Fig.~\ref{fig:BWJYP:PbPb2.76:v1_eta_pt}, a model-to-data comparison of $v_1$ of charged hadrons as a function of $\eta$ and $p_T$ is presented in panels (a) and (b) respectively. A sub-panel is placed at the bottom of each main panel, displaying the $\chi$ value, which quantifies the deviation of the model values from the experimental data. This $\chi$ value is calculated as follows:
\begin{equation}
\chi = \frac{ \vert \text{Model calculations} - \text{Mean of experimental data} \vert }{\text{Experimental Error}}
\label{eq:chi}
\end{equation}
Upon observation, we noted that for a reasonable choice of tilt parameters $H_t$ and $\eta_m$, both models adequately capture the $\eta$ dependency of $v_1$. However, a noticeable discrepancy in $p_T$-differential $v_1$ between the two models emerges above $p_T=1$ GeV.

Delving deeper into the observed discrepancy, we systematically adjusted the $\eta_t$ parameter within the JYP model, while keeping all other parameters same as listed in Table~\ref{table:JYPBW:param2.76TeV}. The resulting $v_1$ versus $\eta$ and $p_T$ plots for varying $\eta_t$ are shown in Fig.~\ref{fig:BWJYP:PbPb2.76:v1_eta_pt_diff_etat}(a) and (b) respectively. For each $\eta_t$, we fine-tuned the $H_t$ parameter to align the $v_1(\eta)$ curve with the experimental data within the range $-1 < \eta < 1$. Notably, we observed a significant alteration in the $v_1(p_T)$ behavior with varying $\eta_t$ within the JYP model calculations. As we increased the $\eta_t$ from 8 to 32 in steps of 8 units, the $v_1(p_T)$ values in the JYP model progressively approached to those obtained from the BW model calculations. This hydrodynamic simulation revealed that the $p_T$-differential $v_1$ is highly sensitive to the initial energy deposition, even more so than its $\eta$ dependence. Furthermore, the numerical findings from this simulation supported the statement that the JYP model serves as a generalization of the BW model and converges to it when the parameter $\eta_t$ takes a large value.

\section{Directed flow of identified particles in a baryonless fluid}

\begin{figure}
  \centering
  \includegraphics[width=0.7\textwidth]{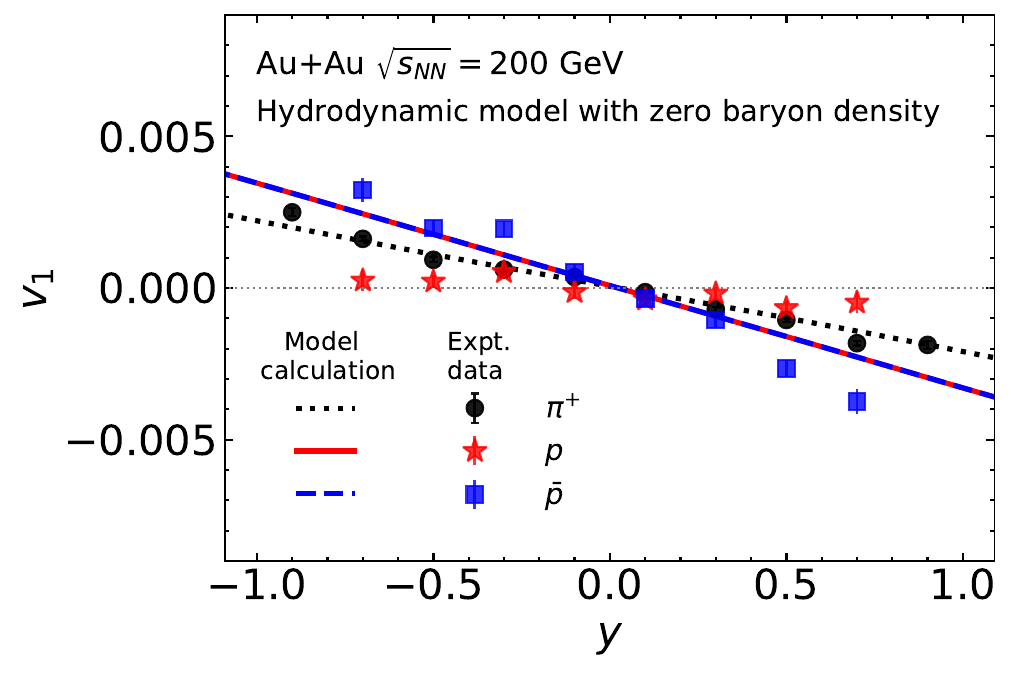} 
  \caption{Rapidity differential directed flow ($v_1$) of $\pi^{+}$, $p$, and $\bar{p}$ for Au+Au collisions at $\sqrt{s_{NN}} = 200$ GeV in 10-40\% centrality. The hydrodynamic model calculations, performed with zero baryon density, are shown as lines and compared with experimental data from the STAR collaboration \cite{STAR:2014clz}. }
  \label{fig:v1_pippbar_baryonless}
\end{figure}

It has been observed that the negative slope of the directed flow ($v_1$) for charged hadrons can be well described by hydrodynamic model simulations, especially when a tilted initial energy density profile is used \cite{Bozek:2010bi,Jiang:2021foj}. Extending this to the directed flow of identified particles, it is expected that in simulations of a baryonless fluid, a mass ordering in $v_1$ will emerge—similar to what has been observed in the case of elliptic flow for various hadron species \cite{Bozek:2010bi}. Furthermore, hadrons with similar masses are anticipated to exhibit similar flow behavior \cite{Ollitrault:2007du}. This expectation has been confirmed by our model calculations. The results of our model are presented in Fig. \ref{fig:v1_pippbar_baryonless}, where the rapidity differential $v_1$ for $\pi^+$, $p$, and $\bar{p}$ is shown for Au+Au collisions at $\sqrt{s_{NN}} = 200$ GeV, along with experimental data from the STAR collaboration \cite{STAR:2014clz}. Our results show that protons and antiprotons, which are more massive than pions, exhibit a larger magnitude of $v_1$ compared to $\pi^+$. Importantly, $p$ and $\bar{p}$ display same $v_1(y)$ behavior, which is expected due to their similar masses. However, the experimental data reveal a significant splitting between the $v_1$ of protons (baryons) and antiprotons (anti-baryons). This baryon-antibaryon splitting observed in experiments can be explained by introducing non-zero baryon density in fluid dynamic simulations \cite{Bozek:2022svy}. In particular, accurate modeling of the initial net-baryon profile, alongside the initial energy density profile, is crucial for reproducing the experimental $v_1$ splitting between baryons and antibaryons, as well as the ordering in the mid-rapidity slope of $v_1(y)$ of $\pi^+$, $p$, and $\bar{p}$ \cite{Bozek:2022svy}. Therefore, this study of $v_1$ for identified hadrons, alongside comparisons with experimental data, would offer valuable insights into both the initial conditions and the evolution of baryon-rich matter in heavy-ion collisions.

\section{Chapter summary}
In this chapter, rapidity-odd directed flow has been studied in a baryon less hydrodynamic model. From the investigation, we observed that the initial left-right asymmetry of the pressure gradient along the impact parameter direction in the transverse plane at non-zero $\eta_s$ leads to the generation of asymmetric flow which ultimately reflects in the directed flow measurement of observed hadrons. We review the previously proposed phenomenologically successful tilted initial condition (TIC) model, which explains the experimentally observed negative sign of the mid-rapidity slope of rapidity differential $v_1$ of charged hadrons. This model is proposed by Bozek and Wyskiel \cite{Bozek:2010bi} which introduces asymmetric matter deposition by participants moving along forward and backward rapidities. Additionally, the chapter examines another recently proposed initial condition model by the CCNU group \cite{Jiang:2021foj}, which generates a tilted profile in the reaction plane similar to the Bozek-Wyskiel (BW) TIC model. It is shown that the CCNU model generalizes the BW TIC model, and under certain parameter limits, it reduces to the BW model. This is demonstrated both mathematically and through hydrodynamic simulation results of rapidity and $p_T$-differential $v_1$.

In the end, we have demonstrated that in hydrodynamic simulations with zero baryon density, hadron species such as $p$ and $\bar{p}$ exhibit identical $v_1(y)$ due to their similar masses. However, the observed splitting of $v_1$ between $p$ and $\bar{p}$ in experimental data highlights the need for hydrodynamic evolution with non-zero baryon density to accurately capture this feature. In this context, we conclude that comparing the $v_1$ of identified hadrons between model predictions and experimental data could offer valuable insights into both the initial energy distribution and the initial net-baryon stopping.

\def \la{\langle}
\def \ra{\rangle}
\chapter{Initial baryon stopping and splitting of baryon-antibaryon directed flow}
\label{ch:baryon_tilt}

In relativistic heavy ion collisions, the colliding ions deposit significant energy, baryon, and electric charges into a dense, hot fireball that then expands and evolves under extreme conditions \cite{STAR:2017sal,BRAHMS:2009wlg,BRAHMS:2003wwg,NA49:2005cix,Andronic:2017pug,Shen:2022oyg}. Understanding this evolution relies on the framework of relativistic hydrodynamics, which has been remarkably effective in describing RHIC observables with a minimal set of parameters characterizing the initial state and the dynamics with dissipative currents \cite{Arslandok:2023utm,Schenke:2010nt,Bozek:2011ua,Shen:2023awv,Shen:2020mgh,Shen:2020jwv,De:2022yxq,Karpenko:2015xea,Du:2019obx,Werner:2024ntd,Schafer:2021csj}. This framework has been particularly successful at high energies, where the fireball is nearly baryon-free \cite{
Hirano:2002ds,Hirano:2005xf,Niemi:2011ix,Song:2010aq,Schenke:2010nt,Bozek:2011ua}, but expanding the model to include baryon evolution poses additional challenges that are only beginning to be addressed \cite{Denicol:2018wdp,Pihan:2024lxw, Du:2022yok, Bozek:2022svy,Shen:2020jwv, Shen:2023awv,Shen:2022oyg,Shen:2017bsr,Du:2018mpf}.

At lower beam energies, RHIC experiments reveal substantial baryon stopping at mid-rapidity, evidenced by rising baryon chemical potentials ($\mu_B$) at freeze-out in thermal models as collision energies decrease \cite{Andronic:2017pug,Chatterjee:2017yhp,STAR:2017sal,BRAHMS:2009wlg,BRAHMS:2003wwg,NA49:2005cix,Andronic:2017pug,Shen:2022oyg}. The phenomenon of baryon stopping, driven by nucleon collisions in the initial stages, has been explored through various phenomenological models, as no first-principle description yet exists \cite{Li:2018fow,Li:2018ini,Li:2016wzh,Du:2022yok,Bozek:2022svy,Shen:2020jwv,Pihan:2024lxw,Shen:2023awv,Shen:2022oyg,Shen:2017bsr,Du:2018mpf}. In studies involving hydrodynamic evolution, the initial baryon profile which is related to the baryon stopping have been either obtained from parametric models where the thermalised distribution of energy-momentum and baryon is obtained at a constant proper time hypersurface \cite{Du:2022yok,
Bozek:2022svy,Shen:2020jwv}, or from more microscopic models where the transition from pre-equilibrium dynamics to hydrodynamics is implemented locally through dynamical rules \cite{Shen:2023awv,Shen:2022oyg,Shen:2017bsr,Du:2018mpf}. Such strategies attempt to approximate the initial three dimensional distribution of energy and net-baryon density in the fireball, which significantly influence final state observables.

One crucial observable influenced by the initial deposition scheme is the directed flow ($v_1$) of identified hadrons, especially the baryon and anti-baryon flow is very sensitive to the initial net-baryon distribution \cite{Du:2022yok, Bozek:2022svy, Shen:2020jwv}. There have been several efforts to construct phenomenologically successful baryon stopping models to capture the elusive beam energy dependence of the $v_1$ splitting between baryon and anti-baryons in a hydrodynamic framework \cite{Du:2022yok, Bozek:2022svy, Shen:2020jwv,Jiang:2023fad}. A natural ansatz for the transverse baryon deposition is to consider it to be proportional to the distribution of the participant nucleons, carriers of the baryon quantum number \cite{Du:2022yok, Bozek:2022svy, Shen:2020jwv,Denicol:2018wdp}. This simple participant scaling of the deposited baryon charge, however, substantially overestimates the rapidity odd $v_1$ of proton and anti-proton \cite{Shen:2020jwv}. It has been shown that such a participant scaling ansatz along with a tilted energy distribution of the fireball can capture the proton-anti-proton $v_1$ splitting at $\sNN$ = 200 GeV \cite{Bozek:2022svy}. In addition to the above participant deposition, a rapidity even deposition motivated by baryon junction model has been shown to successfully explain the pion and proton rapidity $v_1$ across $\sNN = 7.7 - 200$ GeV \cite{Du:2022yok}. However, no model so far is able to capture simultaneously the $v_1$ of proton and anti-proton across collsion energy.

In this chapter we will describe our proposal of a new ansatz for the initial thermalised baryon charge distribution that manages to describe the identified hadrons $v_1$ across $\sNN = 7.7-200$ GeV including the elusive double sign change in the mid-rapidity slope of net-baryon $v_1$ between $\sNN = 7.7 - 39$ GeV.

Before delving into the specifics of our model and the results for $v_1$, we first illustrate how the presence of baryons influences the medium dynamics using a simple 1+1D toy model. This example helps elucidate how asymmetries or inhomogeneities in the initial energy and baryon distributions, and their subsequent evolution, interplay to produce asymmetric flow patterns in both energy and net-baryon density. In particular, this example highlights the role of baryon diffusion in shaping the baryon flow within the medium.

Following this simple toy model, we provide a detailed discussion of the initial energy and baryon density profiles adopted in our hydrodynamic simulations for BES energies. Special attention is given to the initial baryon distribution, as we propose this new deposition scheme that has been demonstrated to be phenomenologically successful. We then examine the impact of this baryon profile on the directed flow of identified hadrons, focusing on baryons and anti-baryons.

Next, we present our simulation results across a wide range of collision energies ($\sNN=7.7-200$ GeV), utilizing carefully chosen model parameters. These parameters are calibrated to capture several experimental data of bulk observables, including rapidity-differential charged particle yields, net-proton yields, mid-rapidity $p_T$ spectra and mean $p_T$ of identified hadrons, and the elliptic flow ($v_2$) of charged hadrons. Most importantly, we show our model predictions for the directed flow ($v_1$) of all identified hadrons over this energy range, showing good agreement with experimental data. We also explore how $v_1$ is influenced by factors such as deviations from Bjorken flow and the effects of hadronic transport.

In the latter part of this chapter, we address two additional interesting phenomena related to $v_1$. The first involves the significant impact of hadronic transport on the directed flow of the $K^{*0}$ resonance, highlighting it's sensitivity to this late stage dynamics of the system \cite{Parida:2023tdx}. The second is the splitting of elliptic flow among hadrons originating from distinct regions of phase space, specifically characterized by $\Delta v_2 = v_2(p_x > 0) - v_2(p_x < 0)$. This splitting is shown to be driven by rapidity-odd $v_1$ and $v_3$ \cite{Parida:2022lmt}.

\section{1+1D toy model demonstrating dynamics in the presence of baryons}
\begin{figure}[htbp]
  \centering
  \includegraphics[width=1.0\textwidth]{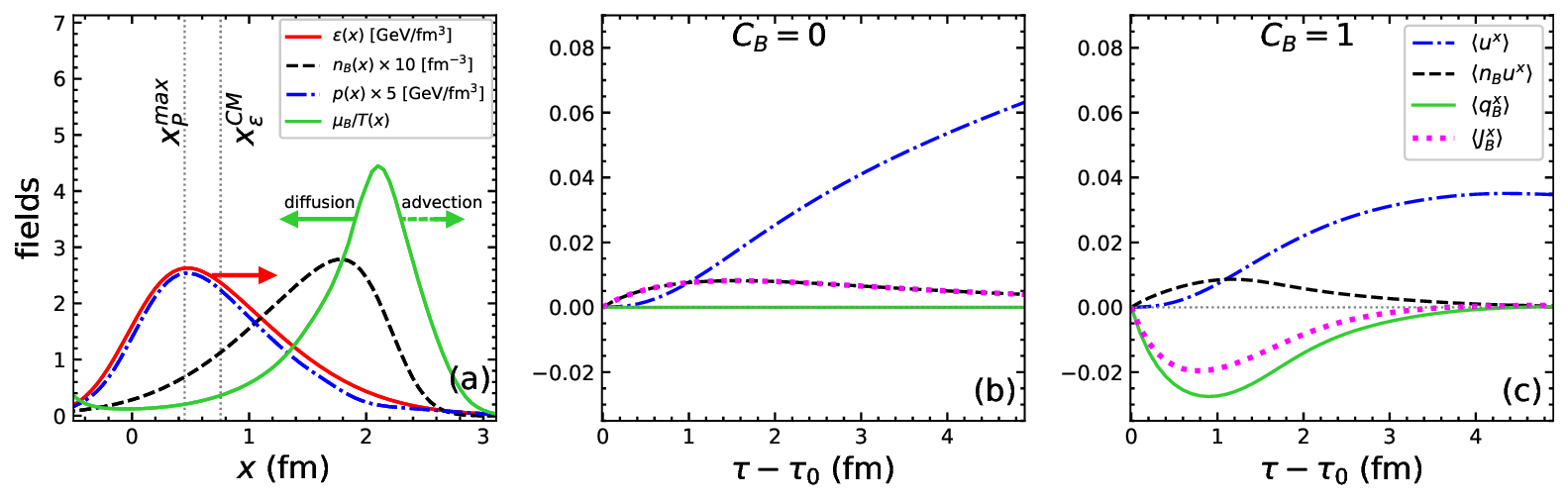} 
  \caption{Dynamics of baryon flow in the presence and absence of baryon diffusion.
Panel (a) shows the initial one-dimensional energy and baryon density distributions. Panels (b) and (c) display the time evolution of flow variables in the cases without baryon diffusion ($C_B=0$) and with baryon diffusion ($C_B=1$), respectively, highlighting the effects of baryon diffusion on the flow dynamics. }
  \label{fig:1D_dist_demo}
\end{figure}

In Fig. \ref{fig:1D_dist_demo}(a), we take a one-dimensional energy and baryon density distribution for demonstration purpose. We perform a 1+1D hydrodynamic evolution of this profile using MUSIC code \cite{Denicol:2018wdp} under two scenarios: without baryon diffusion ($C_B = 0$) and with baryon diffusion ($C_B = 1$). The corresponding time evolution of flow variables is illustrated in Fig. \ref{fig:1D_dist_demo}(b) and Fig. \ref{fig:1D_dist_demo}(c), respectively. The goal is to highlight the contrasting nature of the dynamics with and without baryon diffusion for the chosen initial configuration of energy and net-baryon density profiles.
\newline \\ 
\textbf{Case 1: Evolution without baryon diffusion}

First, we consider the scenario where baryon diffusion is absent ($C_B = 0$). As shown in Fig. \ref{fig:1D_dist_demo}(a), the initial baryon distribution differs significantly from the energy density distribution. Using the NEoSB equation of state \cite{Monnai:2019hkn}, the pressure is determined, which is a function of both energy density and net-baryon density, $p = p(\epsilon, n_B)$. The resulting pressure distribution closely resembles the energy density distribution, with minor deviations visible, particularly at locations where the baryon density peaks. The pressure exhibits a noticeable suppression in the baryon dense region.

As it has been discussed in the previous chapter, in a hydrodynamic model the flow is generated by the pressure gradient (see Eq. \ref{eq:1D_hydro_evo_eq}). Hence, matter located to the right of the position of maximum pressure (where $-\partial_x p = 0$) acquire a positive velocity ($v_x > 0$), while those on the left move with negative velocity ($v_x < 0$). Notably, the center of mass of the energy density distribution ($x^{CM}_{\epsilon}$) lies to the right of the peak position of the pressure distribution ($x_{p}^{max}$). Since the left-right flow separation occurs at $x_{p}^{max}$, and a larger amount of the matter is located to its right-as evident from the fact that $x^{CM}_{\epsilon}$ lies right to $x_{p}^{max}$-this configuration leads to most of the energy flowing with $v_x > 0$. As a result, the system exhibits a net positive average flow velocity, $\langle u_x \rangle$, which is defined as :
\beq
\la u^{x} \ra = \frac{\int dx u^{x} \epsilon}{\int dx \epsilon}
\label{eq:avguxepsw}
\eeq
The time evolution of $\langle u^x \rangle$, obtained from numerical simulations, is shown as dashed-dotted line in Fig. \ref{fig:1D_dist_demo}(b), indicating a positive value.

In the case where $C_B = 0$, there is no response to the baryon gradient, meaning that the baryon profile does not experience any thermodynamic force to diffuse. However, since the flow velocity varies across different cells, the baryon density will be advected differently in each cell. Since most of the baryon density lies to the right of the maximum pressure point, or in cells where $u^x > 0$, it is expected that the baryons on an average will also flow with a positive $u^x$. To quantify the baryon flow in this scenario, we compute the quantity $\langle n_B u^x \rangle$, defined as:

\beq
\langle n_B u^{x} \rangle = \frac{\int dx (n_B u^{x}) \epsilon}{\int dx \epsilon }
\eeq
In Fig. \ref{fig:1D_dist_demo}(b), the time evolution of $\langle n_B u^x \rangle$ is shown as a dashed line, also indicating a positive value. Since there is no baryon diffusion in the $C_B = 0$ case, the baryon diffusion coefficient $\kappa_B = 0$ (as seen in \ref{eq:kappaB_form} and \ref{eq:tauB_CB}). As a result, the baryon diffusion current, $q^{x}_B \sim \kappa_B \partial_x (\frac{\mu_B}{T})$, generated by the spatial gradient of $\frac{\mu_B}{T}$, becomes zero. Therefore, the net-baryon current ($J_{B}^{x} = n_B u^x$) leads to the baryon flow primarily towards the right.
\newline \\
\textbf{Case 2: Evolution with baryon diffusion}

In contrast, the dynamics in the presence of baryon diffusion ($C_B = 1$) is very different. In this case, the spatial gradient of $\frac{\mu_B}{T}$ generates baryon diffusion. Since the gradient of $\frac{\mu_B}{T}$ is steeper to the left of its peak position, baryons diffuse more towards the left than towards the right. Meanwhile, the energy density, influenced by the pressure gradient, preferentially flows towards the right along the $x$-direction. This creates a competition between the energy and baryon flows, significantly altering the dynamics. Interestingly, baryon diffusion also affects the energy flow through the equation of state (EoS), as the pressure depends on both the energy density ($\epsilon$) and baryon density ($n_B$). In Fig. \ref{fig:1D_dist_demo}(c), the time evolution of $\langle u^x \rangle$ is shown as dashed-dotted line, and it is clear that this evolution differs considerably from the $C_B = 0$ case.

In our hydrodynamic model, we work in the Landau frame, where the flow velocity $u^x$ is defined as the flow direction of the energy density. In the presence of baryon diffusion, there is advection of baryons along $u^x$, which corresponds to the equilibrium component of the baryon flow. However, the non-equilibrium component is the baryon flow perpendicular to $u^x$, known as the baryon diffusion current $q_B^x$. It is important to note that the relationship $u_\mu q_B^\mu = 0$ holds, indicating that the baryon diffusion current is orthogonal to the flow velocity \cite{Denicol:2018wdp}.

Both the equilibrium component ($n_B u^x$), generated by the advection of baryons, and the non-equilibrium component ($q_B^x$), generated by baryon diffusion, determine the net-baryon current ($J_B^x$) or the net-baryon flow \cite{Denicol:2018wdp}. At any given time, the baryon diffusion current can be quantified by calculating $\langle q_B^x \rangle$, which is defined as:
\beq
\la q^{x}_{B} \ra = \frac{\int dx q_B^{x} \epsilon}{\int dx \epsilon }
\eeq
In Fig. \ref{fig:1D_dist_demo}(c), we plot the time evolution of $\langle q_B^x \rangle$ as solid line. Initially, the gradient of $\frac{\mu_B}{T}$ is steep to the left of its peak, which leads to a large negative value of $\langle q_B^x \rangle$. Over time, this magnitude decreases as the homogeneity emerges. Meanwhile, as $u^x > 0$, the equilibrium component $\langle n_B u^x \rangle$ remains positive. Ultimately, the combined effect of $\langle q_B^x \rangle$ and $\langle n_B u^x \rangle$ generates a negative $\langle J_B^x \rangle$, defined as:
\beq
\la J_{B}^{x} \ra = \frac{\int dx J_B^{x} \epsilon}{\int dx \epsilon }.
\eeq
This is in stark contrast to the $C_B = 0$ case, where $\langle J_B^x \rangle > 0$.

In conclusion, the dynamics of the fluid in the presence of non-zero baryon density depend on both the initial configuration of the energy and baryon density distributions, as well as on baryon diffusion. When baryon diffusion is neglected, the flow of baryons is solely driven by the pressure gradient, meaning that the relative initial distribution of energy and baryons plays a crucial role in determining the flow dynamics. However, when baryon diffusion is included, an additional thermodynamic force emerges due to the gradient of $\frac{\mu_B}{T}$, which can significantly influence the flow of baryons.

In the hydrodynamic model, the gradients of field variables generate the flow. Therefore, quantifying these relevant gradients in the initial state can serve as a valuable tool for understanding the development of flow and predicting final flow coefficients. In particular, the average gradients along the impact parameter direction in the transverse plane, at any space-time rapidity, provide crucial insight about the directed flow. Later in this chapter, we will introduce some initial state estimators of directed flow constructed from the gradient of field variables and demonstrate their ability to predict the final directed flow of mesons, baryons, and anti-baryons.

\section{Initial profile of energy and net-baryon density}
\label{sec2}
In this section, we describe the initial energy and net-baryon density profiles used as inputs for the hydrodynamic evolution. First, the transverse distributions of participant and binary collision sources are constructed following the methodology outlined in Ref. \cite{Shen:2020jwv,Denicol:2018wdp,Du:2022yok}. In this approach, Monte Carlo (MC) Glauber simulations are performed for a specified centrality class. For each MC Glauber event, the participant and binary collision sources are rotated by the second-order participant plane angle and smeared in the transverse plane using a Gaussian profile with a parametric width $\sigma_{w} = 0.4$ fm. By averaging over 25,000 initial MC Glauber configurations, smooth transverse profiles of participants and binary collision sources are obtained. These profiles are then used to construct the event averged three-dimensional distributions of energy and net-baryon density for a centrality.

For the initial energy distribution, we adopt the tilted initial condition model proposed by Bozek and Wyskiel in Ref. \cite{Bozek:2010bi}. The form of the energy density profile $\epsilon(x, y, \eta_s; \tau_0)$ at the initial proper time $\tau_0$, as implemented in our simulations, is detailed in Eqs. \ref{eq.tilt}–\ref{eq:forward_eps_backward_envelop_relation} in the previous chapter. This model generates a tilted energy density profile in the reaction plane with the tilt controlled by the parameter $\eta_m$. Such a tilted initial profile induces transverse expansion with a subtle sideward motion, leading to a negative slope in the rapidity-odd directed flow, $v_1(y)$, for mesons at mid-rapidity. This model has been shown to successfully reproduce the rapidity-dependent $v_1$ of charged hadrons observed at the LHC and the highest RHIC energies \cite{Bozek:2010bi,Jiang:2021ajc,Jiang:2021foj,Parida:2022lmt}.

The initial net-baryon density profile, $n_{B} \left( x, y, \eta_s ; \tau_{0} \right)$ , is parameterized as follows:
\begin{equation}
  n_{B} \left( x, y, \eta_s ; \tau_{0} \right) = N_{B} \left[ \left( N_{+}(x,y)  f_{+}^{n_B}(\eta_{s}) + N_{-}(x,y)f_{-}^{n_B}(\eta_{s})  \right) \times \left( 1- \omega \right) + N_{coll} (x,y)  f^{n_B}_{\text{coll}}\left(\eta_{s}\right) \omega \right] 
 \label{eq:two_component_baryon_profile}    
\end{equation}
where $N_{B}$ is a normalization factor, and $\omega$ is a free parameter representing the relative contribution of participant and binary collision sources.

\begin{figure}
  \centering
  \includegraphics[width=0.5\textwidth]{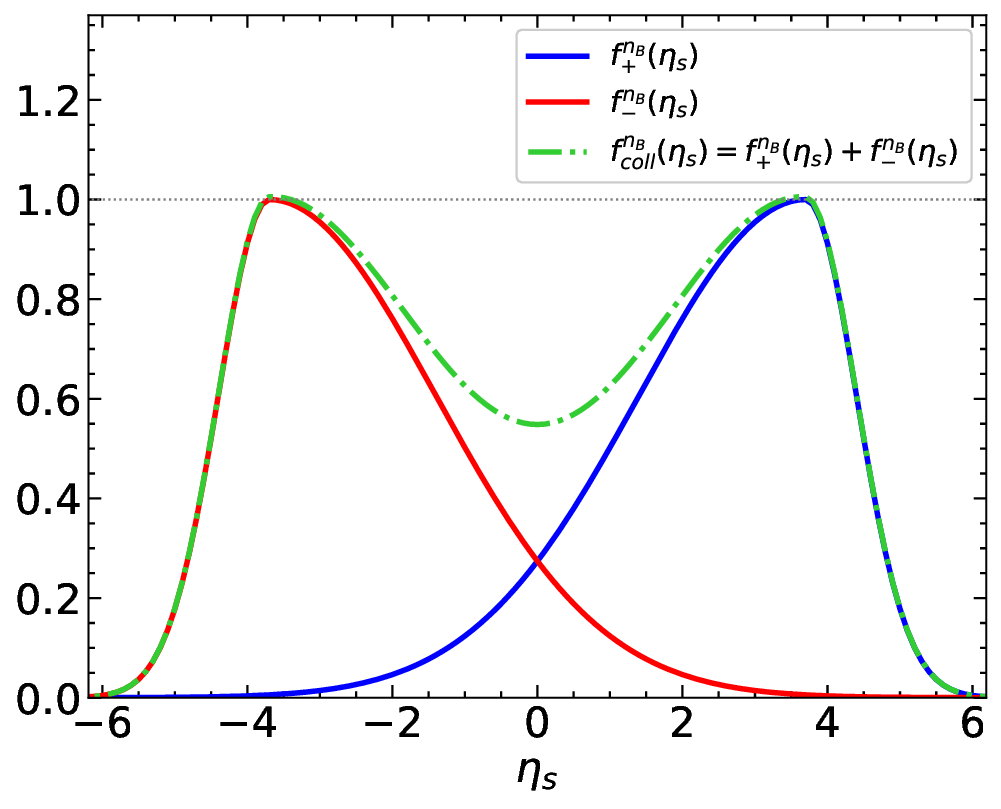} 
  \caption{The rapidity envelop profiles $f_{+}^{n_B}(\eta_{s}), f_{-}^{n_B}(\eta_{s})$ and $f_{coll}^{n_B}(\eta_s)$ are plotted by taking $\eta_{0}^{n_{B}}=3.7,$ $\sigma_{B,-}= 2.3$ and $\sigma_{B, + } = 0.7$. }
  \label{fig:rap_envelop_baryon}
\end{figure}

The rapidity envelope profiles for the forward and backward net-baryon distributions are defined as \cite{Denicol:2018wdp}:
\begin{equation}
    f_{+}^{n_{B}}   \left( \eta_s \right) =  \left[  \theta\left( \eta_s - \eta_{0}^{n_{B} } \right)   \exp{- \frac{\left( \eta_s - \eta_{0}^{n_{B} }  \right)^2}{2 \sigma_{B, + }^2}}   +  \theta\left(  \eta_{0}^{n_{B} } - \eta_s \right)   \exp{- \frac{\left( \eta_s - \eta_{0}^{n_{B} }  \right)^2}{2 \sigma_{B, - }^2}}   \right]
\label{eq:forward_baryon_envelop}  
\end{equation}
\begin{equation}
    f_{-}^{n_{B}}  \left( \eta_s \right) =  \left[   \theta\left( \eta_s + \eta_{0}^{n_{B} } \right)   \exp{- \frac{\left( \eta_s + \eta_{0}^{n_{B} }  \right)^2}{2 \sigma_{B, - }^2}}   + \theta\left( -\eta_s -  \eta_{0}^{n_{B} }  \right)   \exp{- \frac{\left( \eta_s + \eta_{0}^{n_{B} }  \right)^2}{2 \sigma_{B, + }^2}}   \right]
\label{eq:backward_baryon_envelop}    
\end{equation}
These rapidity envelopes are constrained by comparison to experimental data on the rapidity dependence of net-proton yields. Additionally, we have taken a forward-backward symmetric rapidity envelop profile which is multiplied with $N_{coll}$ sources. We consider a simple form for the rapidity symmetric profile.
\beq
f_{coll}^{n_B}(\eta_s)  = f_{+}^{n_B}(\eta_{s}) + f_{-}^{n_B}(\eta_{s})
\label{eq:fcollnb_default}
\eeq
For visualization, the profiles $f_{+}^{n_B}(\eta_{s}), f_{-}^{n_B}(\eta_{s})$ and $f_{coll}^{n_B}(\eta_s)$  are shown in Fig. \ref{fig:rap_envelop_baryon}, using parameters $\eta_{0}^{n_{B}}=3.7,$ $\sigma_{B,-}= 2.3$ and $\sigma_{B, + } = 0.7$.
The normalization factor $N_B$ in Eq.~\ref{eq:two_component_baryon_profile} is not a free parameter, rather 
it is constrained by the initially deposited net baryon carried by the participants \cite{Denicol:2018wdp}. 
\begin{equation}
      \int  \tau_{0}  n_{B} \left( x, y, \eta, \tau_{0} \right) dx  dy  d\eta  = N_{\text{part}}
      \label{eq:constraint_net_baryon_is_npart}
\end{equation}
This parameterization differs from the conventional approach, where the baryon transverse profile is typically taken as proportional to $N_{\pm}(x,y)$ only \cite{Du:2022yok,Shen:2020jwv,Bozek:2022svy,Denicol:2018wdp}. The inclusion of binary collision sources is inspired from microscopic models like LEXUS \cite{Jeon:1997bp,De:2022yxq}, which suggest that baryon stopping depends on the number of binary collisions. 
The $N_{coll}$ term can also be interpreted within the baryon junction picture, where baryon number is carried by the non-perturbative gluon vertex rather than the valence quarks in a nucleon \cite{Kharzeev:1996sq,Rossi:1977cy,Artru:1974zn}. This perspective introduces two distinct sources of baryon deposition: single junctions and double junctions \cite{Kharzeev:1996sq,Sjostrand:2002ip,Lewis:2022arg}. Single junction stopping includes one unit of baryon number and exhibit a forward-backward asymmetry in their rapidity deposition profile, depending on whether the junction originates from the forward- or backward-going nucleus. Double junctions, on the other hand, are formed by pairs of junctions, belonging to nucleons from the two colliding nuclei. These sources carry two units of baryon number and are expected to produce a deposition profile that is flat in rapidity \cite{Kharzeev:1996sq,Sjostrand:2002ip,Lewis:2022arg,Du:2022yok}. Thus, the single junction sources are expected to scale as the participant nucleons while the double junction sources as the pair of participating nucleons which are known as binary collision sources within the Glauber model. Consequently, our baryon deposition ansatz in Eq. \ref{eq:two_component_baryon_profile} incorporates contributions from both participant and binary collision sources. The participant sources have a rapidity-dependent asymmetric profile, consistent with the single junction stopping picture. The binary collision sources are extended by rapidity even profile, generalising the double junction stopping picture of a rapidity flat deposition. We will show in next chapters that the rapidity odd directed flow is not sensitive to the choice of the rapidity even profile of the binary collision term and a plateau profile also works equally well.

\begin{figure}
  \centering
  \includegraphics[width=0.6\textwidth]{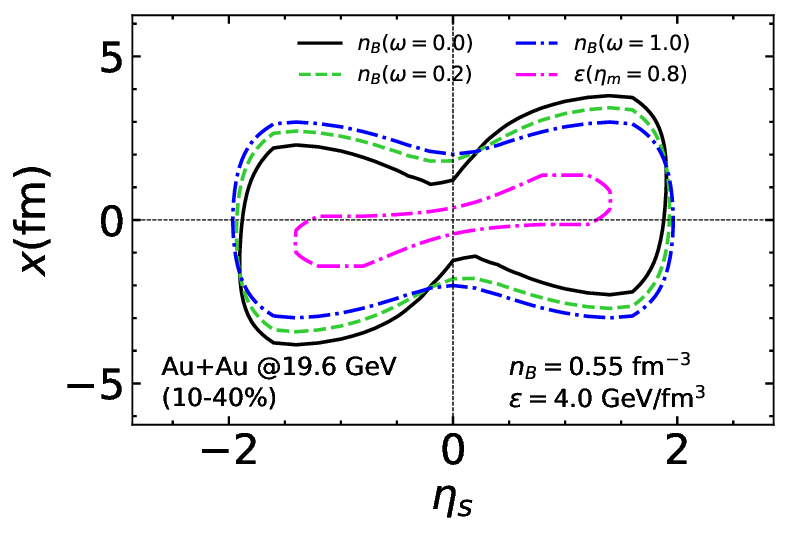} 
  \caption{Contours of constant baryon density for different $\omega$ values are plotted in the reaction plane for Au+Au Collisions of 10-40\% Centrality, $\sqrt{s_{NN}}=19.6$ GeV. baryon density: $n_B=0.55$ fm$^{-3}$. Additionally, a contour of constant energy density ($\epsilon = 4.0$ GeV/fm$^{3}$) is  plotted with $\eta_m = 0.8$.}
  \label{fig:baryon_prof_contour}
\end{figure}

The two-component net-baryon deposition profile that we have proposed in Eq.~\ref{eq:two_component_baryon_profile} results in a tilted baryon configuration in the reaction plane of the collision. The free parameter $\omega$ in the expression of initial baryon distribution governs the degree of tilt in the baryon profile. Conversely, the parameter $\eta_m$ governs the tilt observed in the energy distribution \cite{Bozek:2010bi}. Consequently, the relative tilt between the energy and net baryon density can be adjusted by simultaneously varying the $\eta_m$ and $\omega$ parameters of the model. The tilted energy and baryon distribution is illustrated in Fig. \ref{fig:baryon_prof_contour}, where contours of constant baryon density are plotted for different $\omega$ values, with the contour of constant energy density (with $\eta_m=0.8$). It is observed that the maximum tilt in the baryon distribution occurs when $\omega$ is zero. However, increasing the value of $\omega$ from 0 to 1, indicating a greater contribution from the $N_{coll}$ sources in the initial baryon distribution, results in a decrease in the tilt of the baryon profile. The dependence of the baryon profile tilt on $\omega$ is evident from the rapidity profile of baryon deposition due to the $N_{coll}$ term. Equation \ref{eq:two_component_baryon_profile} indicate that the rapidity profile of baryon deposition from the $N_{coll}$ term is forward-backward symmetric, while baryons deposited by participant sources are asymmetric in rapidity, as characterized by $f_{+}(n_B)$ and $f_{-}(n_B)$. The parameter $\omega$ controls the relative weight between participant and binary collision sources, thereby altering the initial baryon tilt independently of the matter tilt.

\section{Hydrodynamic evolution and hadronic transport}
\label{sec3}
For the hydrodynamic evolution of the initial energy and baryon density profiles, we utilize the publicly available MUSIC code \cite{Denicol:2018wdp,Schenke:2010nt, Paquet:2015lta, Schenke:2011bn}. The hydrodynamic equations governing the conserved quantities and dissipative currents, including those for finite baryon density, are solved within the framework of MUSIC. The detailed formulation of these equations is provided in Chapter \ref{ch:framework}. For the model calculation of a given centrality, we perform a single-shot hydrodynamic evolution of the event averaged smooth initial condition constructed by the procedure detailed in previous section. 

In this study, other conserved charges—net strangeness ($n_S$) and net electric charge density ($n_{Q}$)—are not independently evolved. Instead, they are constrained locally to satisfy the following conditions \cite{Monnai:2019hkn}:
\beqa
n_S&=&0\label{eq.ns}\\ 
n_Q&=&0.4n_B\label{eq.nq}
\eeqa
In the initial conditions, we have assumed the Bjorken flow ansatz \cite{Bjorken:1982qr} :
\beq
u^{\mu}(\tau_0,x,y,\eta_s) = (\cosh{\eta_s},0,0, \sinh{\eta_s})
\eeq
The simulation incorporates a temperature ($T$) and baryon chemical potential ($\mu_B$) dependent baryon diffusion coefficient ($\kappa_B$). The functional form of $\kappa_B$ is given in Eq.~\ref{eq:kappaB_form}, where its magnitude is regulated by the parameter $C_B$, which serves as a model parameter controlling baryon diffusion within the medium. During the hydrodynamic evolution, we employ the lattice QCD-based equation of state (EoS) NEoS-BQS \cite{Monnai:2019hkn}, which accounts for finite baryon density and enforces the constraints specified in Eqs.~\ref{eq.ns} and \ref{eq.nq}. The transition from fluid to particles is carried out using the Cooper-Frye prescription on a hypersurface defined by a constant energy density $\epsilon_{f} = 0.26$ GeV/fm$^{3}$. This process is performed using the iSS code \cite{https://doi.org/10.48550/arxiv.1409.8164,https://github.com/chunshen1987/iSS,Shen:2014vra,Shen:2014vra,Shen:2014lye}. The resulting primary hadrons are subsequently passed through UrQMD to simulate the late-stage hadronic evolution \cite{Bass:1998ca, Bleicher:1999xi}.

\section{Effect of baryon tilt on net-proton yield and $v_1$ of identified hadrons}
Using our proposed model for initial baryon deposition, we first investigate its impact on baryonic observables, specifically the net-proton yield and the directed flow ($v_1$) of identified hadrons, with a focus on protons and anti-protons. To gain a preliminary understanding of how the initial baryon profile (or its geometry) influences final-state observables, we perform simulations with $C_B=0$, effectively neglecting the role of baryon diffusion during the hydrodynamic evolution. While this approach provides insight into the influence of initial conditions, but still it is to be kept in mind that the baryon diffusion is expected to have a significant impact on these observables \cite{Denicol:2018wdp}. The aspect of baryon diffsuion will be explored in the next section.

The spatial distribution of participant and binary collision sources are different in the transverse plane, thereby changing the value of $\omega$ in our model alters the transverse distribution of the baryon profile. However, it is interesting to investigate how variations in $\omega$ influence the $\eta_s$ distribution of the net baryon profile in the initial state. This investigation is crucial because the $\eta_s$ distribution of $n_B$ directly impacts the rapidity-differential net-proton yield in the final state. The space-time rapidity dependence of the net baryon profile ($n_B(\eta_s)$) can be obtained by integrating the distribution over the transverse plane. The expression for $n_B(\eta_s)$ can be calculated as:
\begin{equation}
    n_B(\eta_s; \tau_0) = \int dx \ dy \ n_B(x,y,\eta_s; \tau_0)
\label{eq:n_B(etas)}
\end{equation}

Substituting Eq. \ref{eq:two_component_baryon_profile} into Eq. \ref{eq:n_B(etas)} and integrating over $x$ and $y$, we obtain:
\begin{equation}
   \begin{aligned}
       n_B(\eta_s; \tau_0) = N_B \left[  (1-\omega)  \left( N_{+} f_{+}(\eta_s) + N_{-} f_{-}(\eta_s) \right) + \omega N_{coll} \left( 
 f_{+}(\eta_s) + f_{-}(\eta_s) \right)  \right]
 \label{eq:etas_nb_trans_integ}
   \end{aligned} 
\end{equation}
where $N_{+}, N_{-}$, and $N_{coll}$ are the transverse plane-integrated participant and binary collision numbers. 

\begin{figure}
  \centering
  \includegraphics[width=1.0\textwidth]{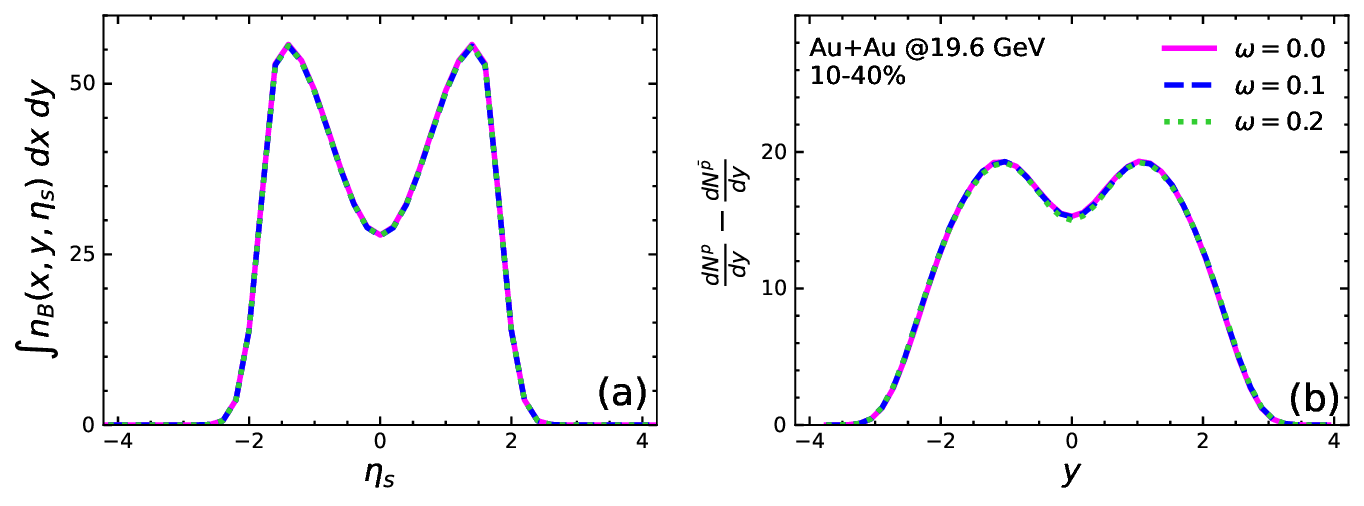} 
  \caption{(a) The effect of $\omega$ on the initial space-time rapidity ($\eta_s$) distribution of the net baryon profile and (b) the effect of $\omega$ on the final-state rapidity-differential net proton yield for Au+Au collisions of 10-40\% centrality at $\sqrt{s_{NN}}=19.6$ GeV.}
  \label{fig:etas_vs_netp_diff_omega}
\end{figure}

Furthermore, using the constraints from Eq. \ref{eq:constraint_net_baryon_is_npart} and utilizing the facts $ \int_{-\infty}^{\infty} d \eta_s f_{+}(\eta_s) = \int_{-\infty}^{\infty}  d \eta_s f_{-}(\eta_s)$, $f_{-}(\eta_s) = f_{+}(-\eta_s)$ , and also exploiting  $n_B(\eta_s) = n_B(-\eta_s)$ for symmetric collisions, Eq. \ref{eq:etas_nb_trans_integ} simplifies to:

\begin{equation}
 n_B(\eta_s; \tau_0) =  \frac{(N_{+} + N_{-})/\tau_0}{ 2 \int_{-\infty}^{\infty} d \eta_s f_{+}(\eta_s)  } \times 
   \left( 
 f_{+}(\eta_s) + f_{-}(\eta_s) \right)   
\end{equation}
This shows that the transverse-integrated and $\eta_s$-differential net baryon distribution is independent of $\omega$. To validate this numerically, we plotted the $\eta_s$ distribution of $n_B$ at 19.6 GeV in Fig. \ref{fig:etas_vs_netp_diff_omega}(a) for various $\omega$ values. We find that the $\eta_s$ distribution remains unaffected by $\omega$. Additionally, we conducted a hydrodynamic simulation with these initial conditions which have different $\omega$ values and calculated the rapidity-differential net proton yield. The resulting rapidity-differential net proton yield for different $\omega$ is depicted in Fig. \ref{fig:etas_vs_netp_diff_omega}(b). It is observed that the net proton yield is independent of the $\omega$ parameter. 

When conducting hydrodynamic simulations without accounting for the net-baryon density within the medium ($n_B=0$), the directed flow of identified particles originates primarily from the initially tilted distribution of energy or entropy density. However, in the case of a baryonic fireball, it becomes feasible to generate the directed flow of hadrons with a forward-backward symmetric energy density profile but with a tilted baryon profile. 
This happens due to the underlying dynamics of the fluid. In hydrodynamic evolution, fluid dynamics is primarily governed by the pressure distribution within the medium. In a baryon-free fireball scenario, the pressure is solely determined by the energy density, resulting in a symmetric pressure distribution for a symmetric energy density profile, which does not induce any asymmetry in the flow. However, when considering a non-zero baryon density in the fluid, the pressure $p$ becomes dependent on both energy $\epsilon$ and net-baryon density $n_B$, $p=p(\epsilon,n_B)$, thereby influencing the dynamics. Consequently, with a tilted baryonic profile, the distribution of baryons could impact the momentum space distribution of both mesons and baryons in the final state through the equation of state (EoS).

\begin{figure}
  \centering
  \includegraphics[width=0.6\textwidth]{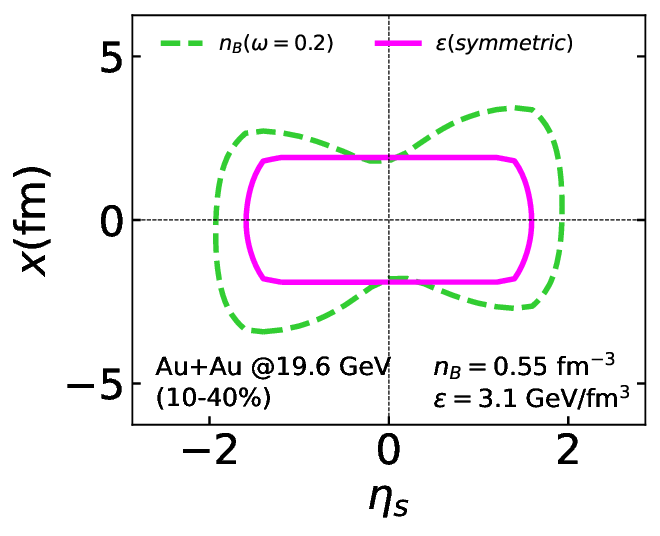} 
  \caption{Contour of constant baryon density for  $\omega=0.2$ is plotted in the reaction plane for Au+Au Collisions of 10-40\% Centrality, $\sqrt{s_{NN}}=19.6$ GeV. baryon density: $n_B=0.55$ fm$^{-3}$. Additionally, a contour of constant energy density ($\epsilon = 3.1$ GeV/fm$^{3}$) is plotted for a profile which is forward-backward symmetric.}
  \label{fig:contour_matter_symmetric}
\end{figure}

\begin{figure}
  \centering
  \includegraphics[width=1.0\textwidth]{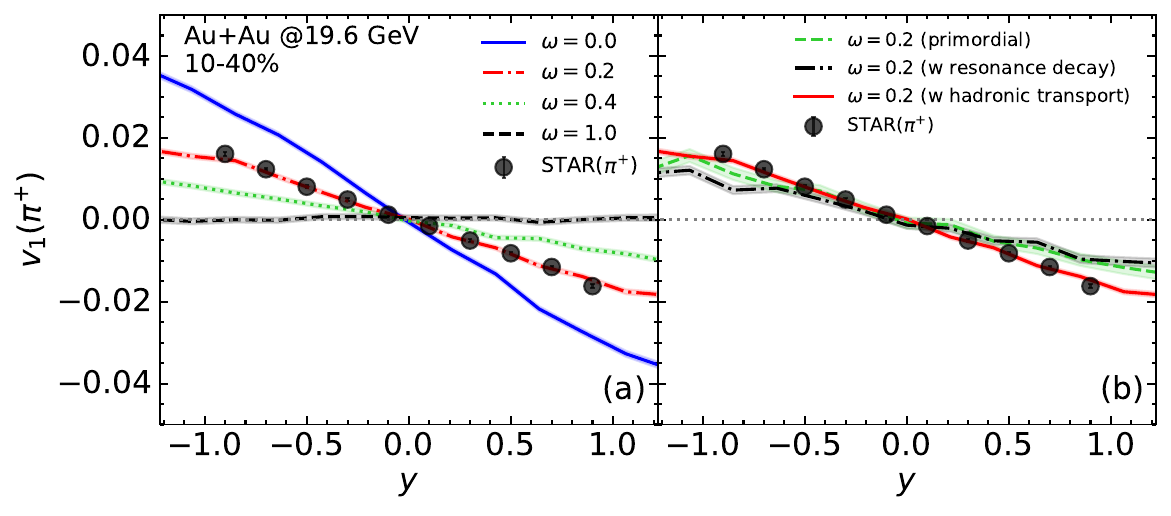} 
  \caption{(a) Rapidity differential directed flow ($v_1$) of $\pi^{+}$ for Au+Au collisions at 10-40\% centrality and $\sqrt{s_{NN}} = 19.6$ GeV. The figure shows the effect of baryon tilt on meson $v_1$. The results correspond to symmetric initial energy density profiles with varying degrees of baryon tilt parameterized by $\omega = 0.0, 0.2, 0.4$, and $1.0$. For $\omega = 1.0$, the pressure profile achieves forward-backward symmetry, leading to zero $v_1$ of $\pi^{+}$. Conversely, increasing baryon tilt with lower $\omega$ values induces significant $v_1$ due to pressure asymmetry. (b) The effect of resonance decays and hadronic transport on $v_1(\pi^{+})$ is also shown, indicating minimal contribution compared to the equation of state (EoS) effects. The experimental data of $v_1(y)$ of $\pi^{+}$ is from STAR collaboration \cite{STAR:2014clz}. }
  \label{fig:v1_pi_symm_epsprof}
\end{figure}

To illustrate this concept, we conducted a hydrodynamic simulation with a symmetric energy density but a tilted baryon density profile, varying the parameter $\omega$. The resulting distributions are depicted in Fig. \ref{fig:contour_matter_symmetric} as contour plots. While the energy density profile remains forward-backward symmetric, we intentionally introduce a tilt in the baryon distribution. For visualization purposes, only the baryon density contour for $\omega = 0.2$ is shown. Subsequently, we analyzed the rapidity ($y$) differential directed flow of $\pi^{+}$ obtained from the hydrodynamic model calculations, considering symmetric initial energy density profiles but with tilted baryon profiles corresponding to $\omega = 0.0, 0.2, 0.4$, and $1.0$. The results are plotted in Fig. \ref{fig:v1_pi_symm_epsprof}. It is noteworthy that for $\omega = 1.0$, the baryon profile has no tilt, resulting in a symmetric pressure distribution within the fireball and consequently yielding zero $v_1$ for $\pi^{+}$. Conversely, when $\omega = 0.0$, the maximum tilt in the baryon profile induces maximum pressure asymmetry at any non-zero rapidity, leading to a significant $v_1$ for $\pi^{+}$. As $\omega$ increases from 0 to 1, the baryon tilt diminishes, resulting in a smaller magnitude of $v_1$ for $\pi^{+}$. Additionally, the impact of resonance decay and hadronic transport on $v_1(\pi^{+})$ is depicted in Fig. \ref{fig:v1_pi_symm_epsprof}(b). Notably, the primary contribution to $v_1(\pi^{+})$ arises from the effect of the equation of state (EoS), whereas the contribution from the resonance decay of higher baryons or the effects of hadronic interactions in the late stage of evolution is minimal.

\begin{figure}
  \centering
  \includegraphics[width=1.0\textwidth]{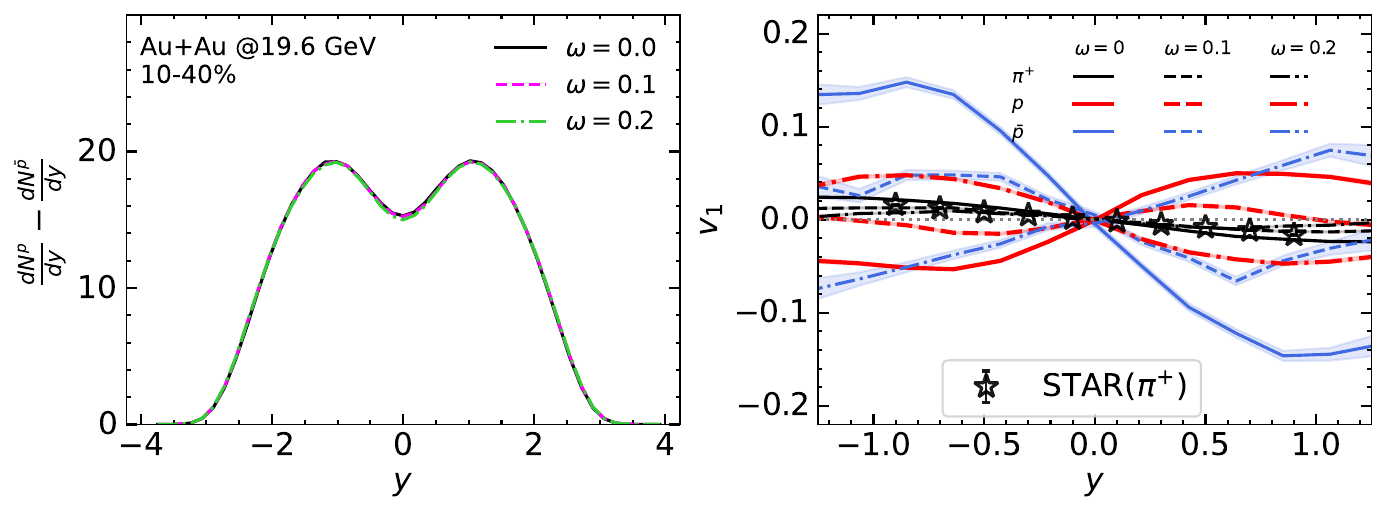} 
  \caption{The rapidity differentail net-proton yield (left) and the directed flow ($v_1$) of $\pi^{+}$, protons, and anti-protons as a function of rapidity (right) for Au+Au collisions at 10–40\% centrality and $\sqrt{s_{NN}}=19.6$ GeV. Results are shown for three baryon tilt parameter values: $\omega = 0.0$ (solid lines), $0.1$ (dashed lines), and $0.2$ (dashed-dotted lines). The tilt parameter ($\eta_m$) was adjusted independently for each $\omega$ to match with the experimental data of $v_1(y)$ of $\pi^{+}$. Significant impact of $\omega$ on $v_1(y)$ of both $p$ and $\bar{p}$ has been observed. Especially, the mid-rapidity slope of $v_1(y)$ of protons changes sign from positive to negative as $\omega$ increases, while no significant dependence of $\omega$ is observed on the net-proton rapidity distribution. The experimental data of $v_1(y)$ of $\pi^{+}$ is from STAR collaboration \cite{STAR:2014clz}.}
  \label{fig:v1_pi_p_pbar_diff_omega}
\end{figure}

With the above insight regarding the impact of baryon tilt on $v_1$ of mesons, we proceed to investigate the impact of the $\omega$ parameter on the directed flow ($v_1$) of protons and anti-protons in Au+Au collisions of 10-40\% centrality at $\sqrt{s_{NN}}=19.6$ GeV. Three baryon distribution profiles were considered with $\omega = 0.0, 0.1$, and $0.2$. Building upon previous observations indicating a significant influence of $\omega$ on the $v_1(y)$ of $\pi^{+}$, we adjusted the tilt parameter ($\eta_m$) independently for each $\omega$ value to ensure that the resulting model calculations capture the experimental data of $v_1(y)$ of $\pi^{+}$. Subsequently, we investigated the impact of the baryon tilt on both the directed flow ($v_1$) of protons and anti-protons, as well as the rapidity distribution of net-protons. The results obtained after performing the simulation are presented in Fig. \ref{fig:v1_pi_p_pbar_diff_omega}. Our observation indicate that there is no discernible dependence of $\omega$ on the rapidity differential net-proton distribution. However, our model calculations revealed a strong dependence of the $v_1(y)$ of protons and anti-protons on the baryon distribution, as anticipated. Notably, we observed a sign change in the mid-rapidity slope of the $v_1(y)$ of protons, transitioning from positive to negative with increasing $\omega$ from 0 to 0.2.

\begin{figure}[htbp]
  \makebox[\textwidth]{
    \begin{subfigure}{0.4\textwidth}
      \centering
      \includegraphics[height=.2\paperwidth]{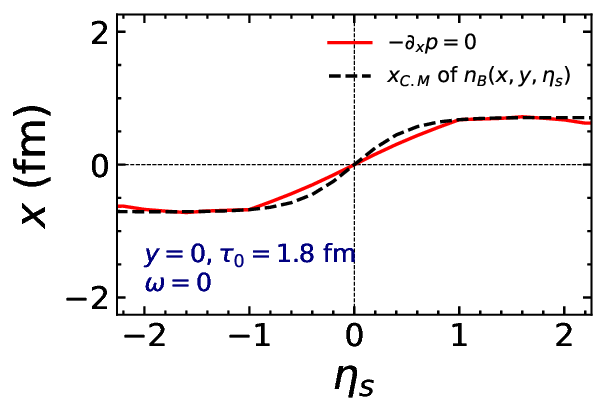}
      \caption{$\omega=0$}
    \end{subfigure}
    \hspace{0.03\textwidth} 
    \begin{subfigure}{0.4\textwidth}
      \centering
      \includegraphics[height=.2\paperwidth]{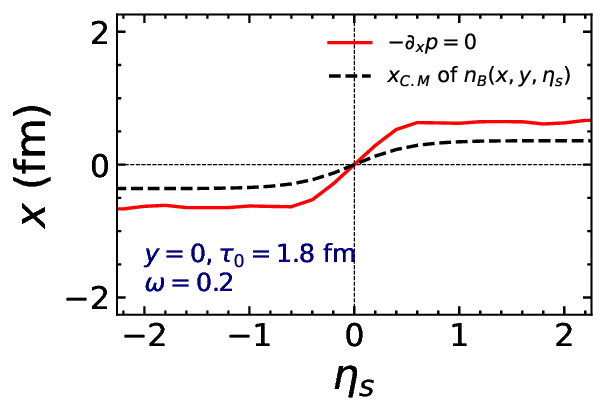}
      \caption{$\omega=0.2$}
    \end{subfigure}
    \hspace{0.03\textwidth} 
  }
  \centering
  \caption{Contour plots of $(-\partial_x p) = 0$ in the $\eta_s - x$ plane for (a) $\omega = 0.0$ and (b) $\omega = 0.2$, depicting the flow separation lines. Dashed lines represent the $x$-coordinate of the center of mass of the baryon distribution in the reaction plane. The transition in the $v_1$ sign of proton in Fig. \ref{fig:v1_pi_p_pbar_diff_omega} is attributed to the relative positioning of the baryon center of mass with respect to the flow separation line at different $\eta_s$.}
\label{fig:dhoxp_baryon_CM_tilt_effect}
\end{figure}

This sign change can be understood by analyzing the relative distribution of baryons with respect to the energy density distribution at the initial stage. Fig. \ref{fig:dhoxp_baryon_CM_tilt_effect}(a), and (b) illustrate the contours of $(-\partial_x p) = 0$ in the $\eta_s - x$ plane for $\omega = 0$, and $0.2$, respectively. Additionally, the $x$-coordinate of the center of mass of the baryon distribution in the reaction plane is overlaid. For $\omega = 0.0$, the center of mass of the baryon distribution is above the $(-\partial_x p) = 0$ line at $\eta_s > 0$. At the location where $(-\partial_x p) = 0$, the flow $u^x$ is zero, marking the separation line for flow direction. Matter above this line begin to flow with positive $u^x$, while matter below begin to flow with negative $u^x$. Consequently, at $\omega = 0.0$, most baryons exhibit positive $u^x$, resulting in a positive $v_1$ for protons at forward rapidity. Conversely, for $\omega = 0.2$, the center of mass of the baryon distribution is below the $(-\partial_x p) = 0$ line at $\eta_s > 0$, leading to a negative $v_1$ for protons at forward rapidity.

\section{Effect of baryon diffusion on net-proton yield and $v_1$ of identified hadrons}

As demonstrated in Ref. \cite{Denicol:2018wdp}, baryon diffusion significantly influences the rapidity differential net-proton yield. For a given initial baryon density distribution, non-zero baryon diffusion causes baryons to flow rapidly toward the mid-rapidity region, leading to a higher net-proton yield at mid-rapidity compared to the zero-diffusion case. However, this effect can be compensated by appropriately adjusting the model parameters $\eta_0^{n_B}$ (the position of the baryon peak) and $\sigma_{\pm}^{n_B}$ (the width of the baryon distribution) for different values of $C_B$. For instance, setting the baryon peak at a larger $\eta_s$ for the non-zero baryon diffusion case and at a smaller $\eta_s$ for the zero diffusion case allows both scenarios to reproduce the experimental mid-rapidity net-proton yield. This demonstrates that the impact of baryon diffusion on the net-proton distribution can be effectively traded off by carefully tuning the initial distribution parameters.
We observed this trade-off in our analysis. By selecting $\eta_0^{n_B} = 1.5$ and $\sigma_{-}^{n_B} = 0.9$ for $C_B = 0$, and $\eta_0^{n_B} = 1.8$ and $\sigma_{-}^{n_B} = 0.8$ for $C_B = 1$, we successfully reproduced the mid-rapidity net-proton yield as well as its rapidity dependence in Au+Au collisions at $\sqrt{s_{NN}}=19.6$ GeV.

However, the effect of baryon diffusion on $v_1$ is particularly interesting. To explore this, we studied Au+Au collisions at 10–40\% centrality and $\sqrt{s_{NN}} = 19.6$ GeV. In the initial state, we set the baryon tilt parameter $\omega = 0.15$ and the energy tilt parameter $\eta_m = 0.8$. The asymmetries along the impact parameter direction ($x$-axis) in the initial state could be quantified using gradient-based estimators, which serve as predictors for the development of directed flow in the energy and baryon density distributions. These initial state estimators are defined as follows:
\beq
\la (-\partial_x p) \ra_{\epsilon} (\eta_s) = \frac{ \int  d^2 r_{\perp} (-\partial_x p)   \epsilon(x,y,\eta_s) }{\int  d^2 r_{\perp}    \epsilon(x,y,\eta_s)}
\label{eq:estimator_1}
\eeq
\beq
\la (-\partial_x p) \ra_{n_B} (\eta_s) = \frac{ \int  d^2 r_{\perp} (-\partial_x p)   n_B(x,y,\eta_s) }{\int  d^2 r_{\perp}    n_B(x,y,\eta_s)}
\label{eq:estimator_2}
\eeq
\beq
\la -\partial_x \frac{\mu_B}{T} \ra_{n_B} (\eta_s) = \frac{ \int  d^2 r_{\perp} (-\partial_x \frac{\mu_B}{T} )   n_B(x,y,\eta_s) }{\int  d^2 r_{\perp} n_B(x,y,\eta_s)}
\label{eq:estimator_3}
\eeq
where $d^2r_{\perp}=dx dy$. These quantities could provide qualitative insights into the initial state asymmetries and their role in shaping the directed flow of mesons and baryons during the system evolution.

\begin{figure}
  \centering
  \includegraphics[width=1.0\textwidth]{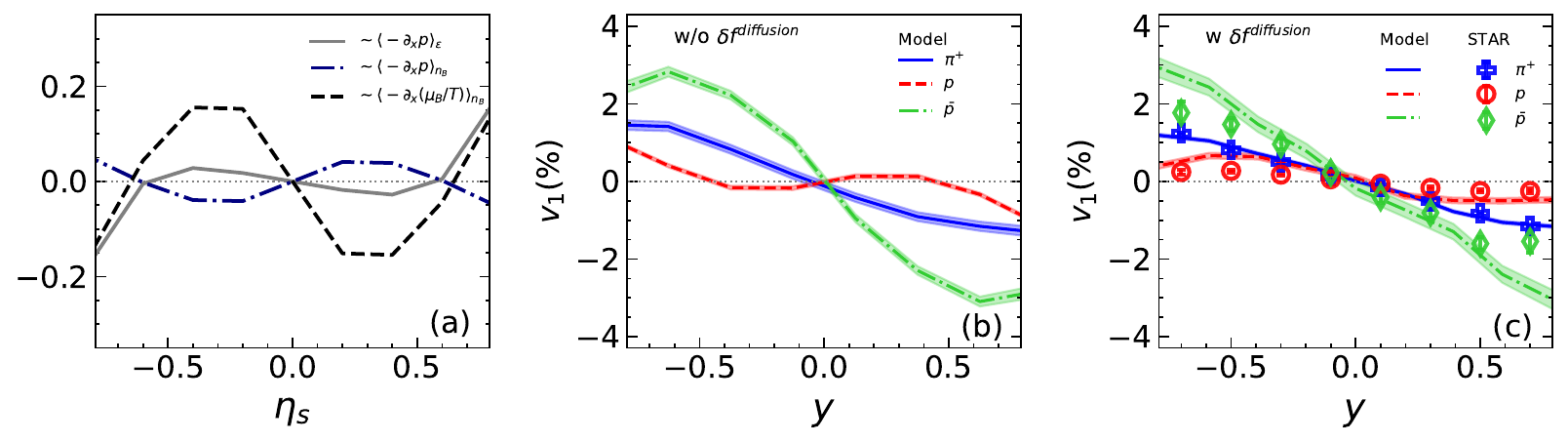} 
  \caption{Panel (a): Initial state predictors for baryon and energy flow, $\langle (-\partial_x p) \rangle_{\epsilon}$, $\langle (-\partial_x p) \rangle_{n_B}$, and $\langle -\partial_x \frac{\mu_B}{T} \rangle_{n_B}$, as a function of $\eta_s$. Panels (b) and (c): Rapidity differential directed flow of $\pi^{+}$, protons, and anti-protons in Au+Au collisions at $10-40\%$ centrality and $\sqrt{s_{NN}}=19.6$ GeV, comparing results with and without the inclusion of baryon diffusion ($\delta f^{\text{diffusion}}$). Panel (b) shows the $v_1$ without baryon diffusion ($\delta f^{\text{diffusion}}=0$), while panel (c) includes the baryon diffusion current. The experimental data of $v_1(y)$ of $\pi^{+},p$ and $\bar{p}$ is from STAR collaboration \cite{STAR:2014clz}.}
  \label{fig:v1_w_wo_diff}
\end{figure}

In Fig. \ref{fig:v1_w_wo_diff}(a), we present the predictors as functions of $\eta_s$. To facilitate comparison among different estimators and make them dimensionless, each predictor is normalized by the average magnitude of its respective gradient. For instance, $\langle (-\partial_x p) \rangle_{\epsilon} (\eta_s)$ is divided by $\langle |-\partial_x p| \rangle_{\epsilon} (\eta_s)$. It is important to note that while these initial state estimators effectively indicate the sign of the developed flow, their relative magnitudes may not reliably predict the relative magnitudes of the final flow coefficients for mesons and baryons.

From earlier discussions, it is evident that the flow of energy density is driven by pressure gradient. Consequently, at each $\eta_s$ slice, $\langle (-\partial_x p) \rangle_{\epsilon} (\eta_s)$ provides insight into the development of $\la u^x \ra$ of the energy density. In the $\eta_s>0$ region near mid-rapidity, $\langle (-\partial_x p) \rangle_{\epsilon} < 0$, indicating that $\la u^x \ra$ will be negative in the forward rapidity region. This, in turn, is expected to result in a negative $v_1$ for mesons in this region.

The other two predictors provide insights into the baryonic flow. The predictor $\langle (-\partial_x p) \rangle_{n_B}$ captures the average flow of baryon induced by the pressure gradient along $x-$ direction, which represents the equilibrium component of the baryon flow. In other words, it reflects the flow of baryons along the $u^x$ direction. The sign of $\langle (-\partial_x p) \rangle_{n_B}$ at any given $\eta_s$ is determined by the relative initial distribution of baryons in the transverse plane, with respect to the pressure distribution. The relative tilt between the baryon and energy density profiles decides the sign of $\langle (-\partial_x p) \rangle_{n_B}$ across different $\eta_s$ in our model. 

On the other hand, since the baryon diffusion current arises in response to the gradient of $\mu_B/T$, the estimator $\langle -\partial_x \frac{\mu_B}{T} \rangle_{n_B}$ determines the flow of baryons along $x-$direction driven by baryon diffusion. As shown in Fig. \ref{fig:v1_w_wo_diff}(a), at non-zero $\eta_s$, the equilibrium component of the baryon current and the baryon diffusion have opposite signs. This implies that there is a competition between the advection of baryons along $u^x$ and the diffusion current $q_B^x$. The combined effect of these two will ultimately determine the sign of the net baryon current.

To evaluate the predictability of these estimators, we perform a simulation with $C_B=1$. The final rapidity differential directed flow of $\pi^{+}$, $p$, and $\bar{p}$ are shown in Figs. \ref{fig:v1_w_wo_diff}(b) and \ref{fig:v1_w_wo_diff}(c). In the case of Fig. \ref{fig:v1_w_wo_diff}(b), when sampling particles from the freezeout hypersurface, we set $\delta f^{\text{diffusion}}=0$, whereas in Fig. \ref{fig:v1_w_wo_diff}(c), this term is included. The inclusion of $\delta f^{\text{diffusion}}$ in the Cooper-Frye formula accounts for contributions from the baryon diffusion current, $q_{B}^{\mu}$ \cite{Denicol:2018wdp}. When we neglect the effect of baryon diffusion ($\delta f^{\text{diffusion}}=0$), the flow of baryons is governed solely by the equilibrium component of the baryon current, as estimated by $\langle (-\partial_x p) \rangle_{n_B}$. In this case, the baryon flow is positive for $\eta_s > 0$, resulting in a positive mid-rapidity slope for the proton $v_1$. However, when baryon diffusion is included, there is a competition between the equilibrium and diffusion currents, causing the proton $v_1$ slope to change sign, as observed in Fig. \ref{fig:v1_w_wo_diff}(c). It is important to note that $\delta f^{\text{diffusion}}$ is also non-zero for $\pi^{+}$, which have zero baryon number. This is because variations in the baryon chemical potential can alter the thermal pressure, thereby affecting the momentum distributions of mesons \cite{Denicol:2018wdp}. Consequently, the $v_1$ of $\pi^{+}$ is also influenced by baryon diffusion.

\section{Parameter selection for the simulation at BES energies}
To investigate our model predictions for the directed flow ($v_1$) of identified hadrons at various collision energies, we performed a comprehensive simulation of Au+Au collisions at seven different values of $\sqrt{s_{NN}}$. The specific energies considered in our study include $\sqrt{s_{NN}} = 7.7, 11.5, 19.6, 27, 39, 62.4$, and 200 GeV.

In the model, different free parameters determine the initial distribution of energy and net-baryon density, as well as influence the fluid dynamics during hydrodynamic evolution. The parameters governing the initial energy distribution include $\epsilon_0$, $\alpha$, $\eta_0$, $\sigma_\eta$, and $\eta_m$. On the other hand, the baryon distribution is determined by $\eta_{0}^{n_B}$, $\sigma_{B,+}$, $\sigma_{B,-}$, and $\omega$. We systematically adjust these parameters to capture different experimental data in our model calculations. The parameters present in the expression of initial energy and net-baryon distribution and the corresponding experimental data used for their calibration is provided in Table \ref{table:BES_simulation_param_fix}.

\begin{center}
\begin{table}[h!]
\centering
\begin{tabular}{ |p{4.5cm}|p{8.5cm}|  }
\hline
\textbf{Model parameter(unit)} & \textbf{Experimental data used for the calibration}   \\
\hline
$\epsilon_{0}$ (GeV/fm$^3$) &  Mid-rapidity yield of charged particles or the mid-rapidity yield of $\pi^{+}$. \\
\hline
$\alpha$ & Centrality dependence of the mid-rapidity yield of charged hadrons. \\ 
\hline
$\eta_0, \sigma_\eta$ & Pseudo-rapidity distribution of charged particle yield. \\
\hline
$\eta_{0}^{n_B}, \sigma_{B,-}, \sigma_{B,+}$ & Rapidity differential net proton yield. \\
\hline
$\eta_m, \omega$ & Rapidity dependence of the $v_1$ of $\pi^{+}, p$ and $\bar{p}$ \\
\hline
\end{tabular}
\caption{The parameters and the corresponding experimental data used for their calibration.}
\label{table:BES_simulation_param_fix}
\end{table}
\end{center}

Additionally, hydrodynamic evolution is mainly influenced by the initial starting time of hydrodynamics ($\tau_0$), the shear viscosity to entropy density ratio ($\eta/s$), the baryon diffusion coefficient ($C_B$), and the particlization energy density ($\epsilon_f$). As collision energies decrease, the colliding nuclei take longer durations to traverse through each other. At energies as low as 7.7 GeV, this traversal time exceeds 3 fm/c \cite{Shen:2017bsr}. Hence, in our model calculations, we set the initial starting time of hydrodynamics to be greater than this passage time at each respective energy. Furthermore, for simplicity, we maintain a fixed value of $\eta/s=0.08$ across all collision energies. A temperature($T$) and baryon chemical potential ($\mu_B$)-dependent $\eta/s$ could offer a more quantitative agreement with experimental data \cite{Shen:2020jwv,Shen:2023awv}. Since this is a qualitative study, we have opted to use a constant $\eta/s$. In future studies, when aiming for more quantitative predictions or paramter extraction through Baysian analysis, it would be more appropriate to incorporate a $T$ and $\mu_B$-dependent $\eta/s$. Our model observations reveal that particle ratios in central collisions are highly sensitive to the particlization energy density ($\epsilon_f$). Through analysis at $\sNN = $ 19.6 and 200 GeV, we found that selecting $\epsilon_f =0.26$ GeV/fm$^3$ allows us to capture the proton-to-anti-proton yield ratio at mid-rapidity. Previous studies also support this choice, as it yields satisfactory particle ratios across various collision energies \cite{Monnai:2019hkn}. This motivated to adopt $\epsilon_f =0.26$ GeV/fm$^3$ in our simulations. Additionally, the $C_B$ is not well constrained for the medium produced in heavy-ion collsions. Hence, to initiate a detailed simulation, we incorporate non-zero baryon diffusion within the fluid. For collisions at $\sNN \geq 19.6$ GeV, we set $C_B=1.0$. However, for $\sNN = 11.5$ and $7.7$ GeV, we opt for $C_B=0.5$, as we did not find a suitable parameter set for $C_B=1$ to concurrently explain the $v_1$ of protons and anti-protons at these lower energies. The parameter values taken in our simulation is detailed in Table \ref{param_for_model}.

\begin{center}
\begin{table}[h!]
\centering
\begin{tabular}{|p{0.9cm}|p{0.4cm}|p{0.7cm}|p{1.1cm}|p{0.55cm}|p{0.4cm}|p{0.4cm}|p{0.5cm}|p{0.6cm}|p{0.6cm}|p{0.45cm}|p{0.55cm}|}
\hline 
$\sqrt{S_{NN}}$ \tiny{(GeV)} & $C_B$ & $\tau_0$\tiny{(fm)} &$\epsilon_{0}$ \tiny{(GeV/fm$^{3}$)} & $\alpha$ &  $\eta_{0}$ & $\sigma_{\eta}$ & $\eta_{0}^{n_{B}}$ & $\sigma_{B,-}$ & $\sigma_{B,+}$ & $\eta_m$ & $\omega$ \\ \hline
200  & 1.0 & 0.6  &  8.0 & 0.14  &  1.3  &  1.5  &  4.6  &  1.6   &  0.1  & 2.2 & 0.25  \\ 
\hline
62.4 & 1.0 & 0.6  &  5.4 & 0.14  &  1.4  &  1.0  &  3.0  &  1.0   &  0.1  & 1.4 & 0.25  \\
\hline 
39   & 1.0 & 1.0  &  3.0 & 0.12   &  1.0  &  1.0  &  2.5  &  1.0   &  0.1  & 1.1 & 0.20  \\ 
\hline
27   & 1.0 & 1.2  &  2.4 & 0.11  &  1.3  &  0.7  &  2.3  &  1.1   &  0.2  & 1.1 & 0.11  \\ 
\hline
19.6 & 1.0 & 1.8  &  1.55 & 0.1  &  1.3  &  0.4  &  1.8  &  0.8   &  0.3 & 0.8 & 0.15  \\ 
\hline
11.5 & 0.5 & 2.6  &  0.9 & 0.1  &  0.9  &  0.4  &  1.2  &  0.55  &  0.2  & 0.4 & 0.22   \\
\hline 
7.7  & 0.5 & 3.6  &  0.55 & 0.1  &  0.9  &  0.4  &  0.9  &  0.35  &  0.2  & 0.3 & 0.35  \\ 
\hline
\end{tabular}
\caption{ Model parameters used in the simulations at different $\sqrt{s_{NN}}$ . }
\label{param_for_model}
\end{table}
\end{center}

\section{Results}
\subsection{Yield, spectra, $\la p_T \ra$ and $v_2$}
\begin{figure}
  \centering
  \includegraphics[width=1.0\textwidth]{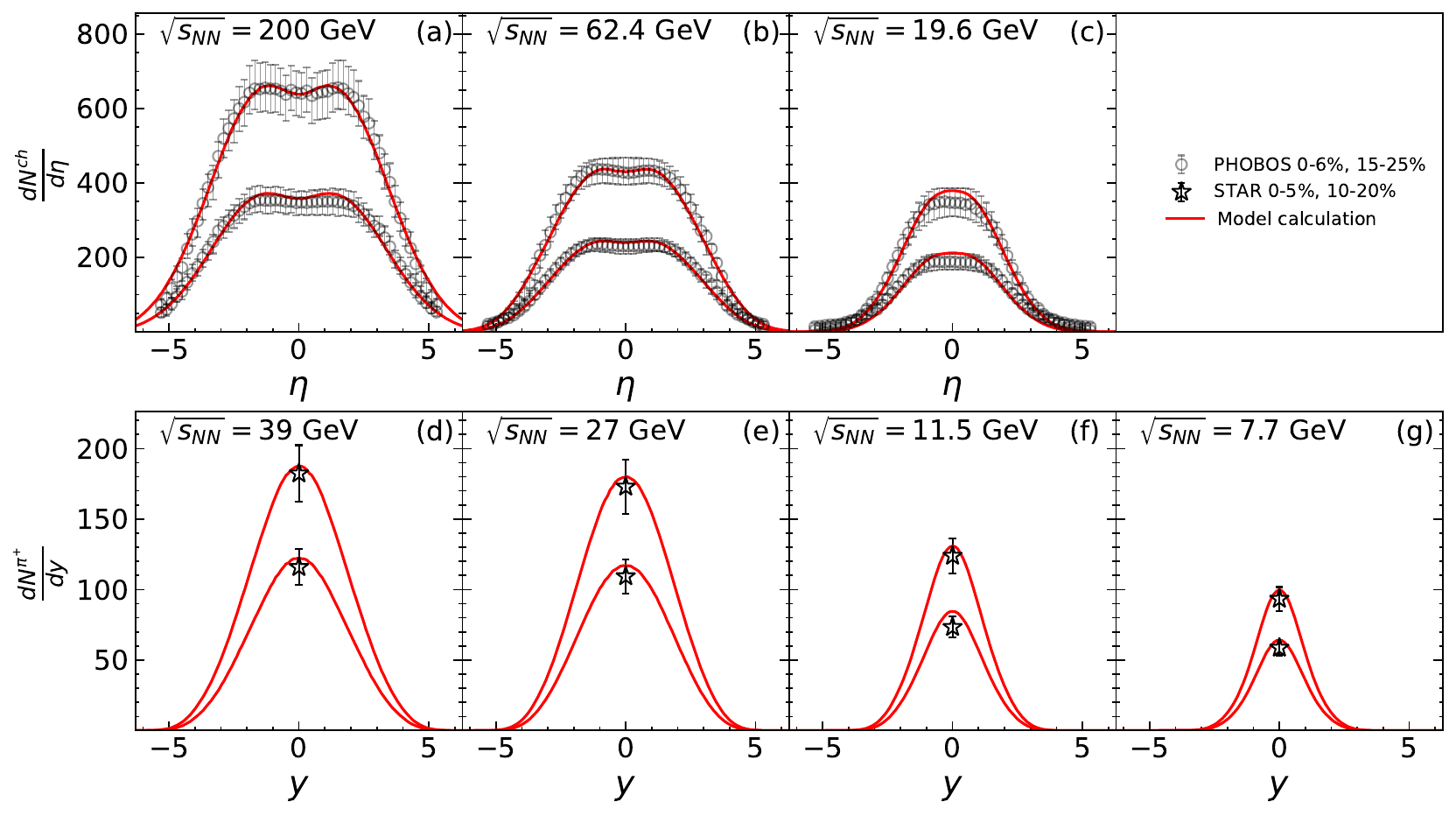} 
  \caption{Pseudorapidity ($\eta$) dependence of charged hadron yield for 0-6\% and 15-25\% centrality Au+Au collisions at $\sqrt{s_{NN}} = 200$, 62.4, and 19.6 GeV (Panels a-c). Model calculations (lines) are compared with PHOBOS data \cite{Back:2002wb}. Panels (d-g) show the rapidity ($y$) dependence of $\pi^{+}$ yield for 0-5\% and 10-20\% centrality at $\sqrt{s_{NN}} = 39$, 27, 11.5, and 7.7 GeV, respectively, with STAR collaboration measurements for mid-rapidity included for comparison \cite{STAR:2017sal}.}
\label{fig:dnchdeta_eta_BES}
\end{figure}

The pseudorapidity ($\eta$) dependence of charged hadron yield is illustrated in Fig. \ref{fig:dnchdeta_eta_BES}(a), \ref{fig:dnchdeta_eta_BES}(b), and \ref{fig:dnchdeta_eta_BES}(c) for 0-6\% and 15-25\% Au+Au collisions at $\sqrt{s_{NN}} = $ 200, 62.4, and 19.6 GeV, respectively. Model calculations are compared with measurements from the PHOBOS collaboration \cite{Back:2002wb}. The $\eta$ distribution is well captured by appropriately selecting parameters $\epsilon_0$, $\eta_0$, and $\sigma_\eta$ governing the initial space-time rapidity distribution of energy density. Additionally, by adjusting parameter $\alpha$, centrality dependence is adequately reproduced. However, neither the $\eta$-differential charged particle yield nor the rapidity-differential identified particle measurement is available at other considered energies, where we calibrated $\epsilon_0$ using the mid-rapidity yield of $\pi^{+}$. Observations from existing experimental data show that the rapidity-dependent charged hadron yields at different energies follow a similar distribution when the pseudo-rapidity is scaled by the respective beam rapidity \cite{Du:2023efk}. Therefore, we adjust $\eta_0$ and $\sigma_{\eta}$ to match that scaled distribution at energies where neither the charged particle nor any identified particle's rapidity distribution has been experimentally measured. In Fig. \ref{fig:dnchdeta_eta_BES}(d), \ref{fig:dnchdeta_eta_BES}(e), \ref{fig:dnchdeta_eta_BES}(f), and \ref{fig:dnchdeta_eta_BES}(g), the rapidity ($y$) dependence of the $\pi^{+}$ yield is plotted for 0-5\% and 10-20\% centrality at $\sqrt{s_{NN}} = $ 39, 27, 11.5, and 7.7 GeV, respectively. Mid-rapidity measurements from the STAR collaboration are also included for comparison \cite{STAR:2017sal}.

\begin{figure}
  \centering
  \includegraphics[width=1.0\textwidth]{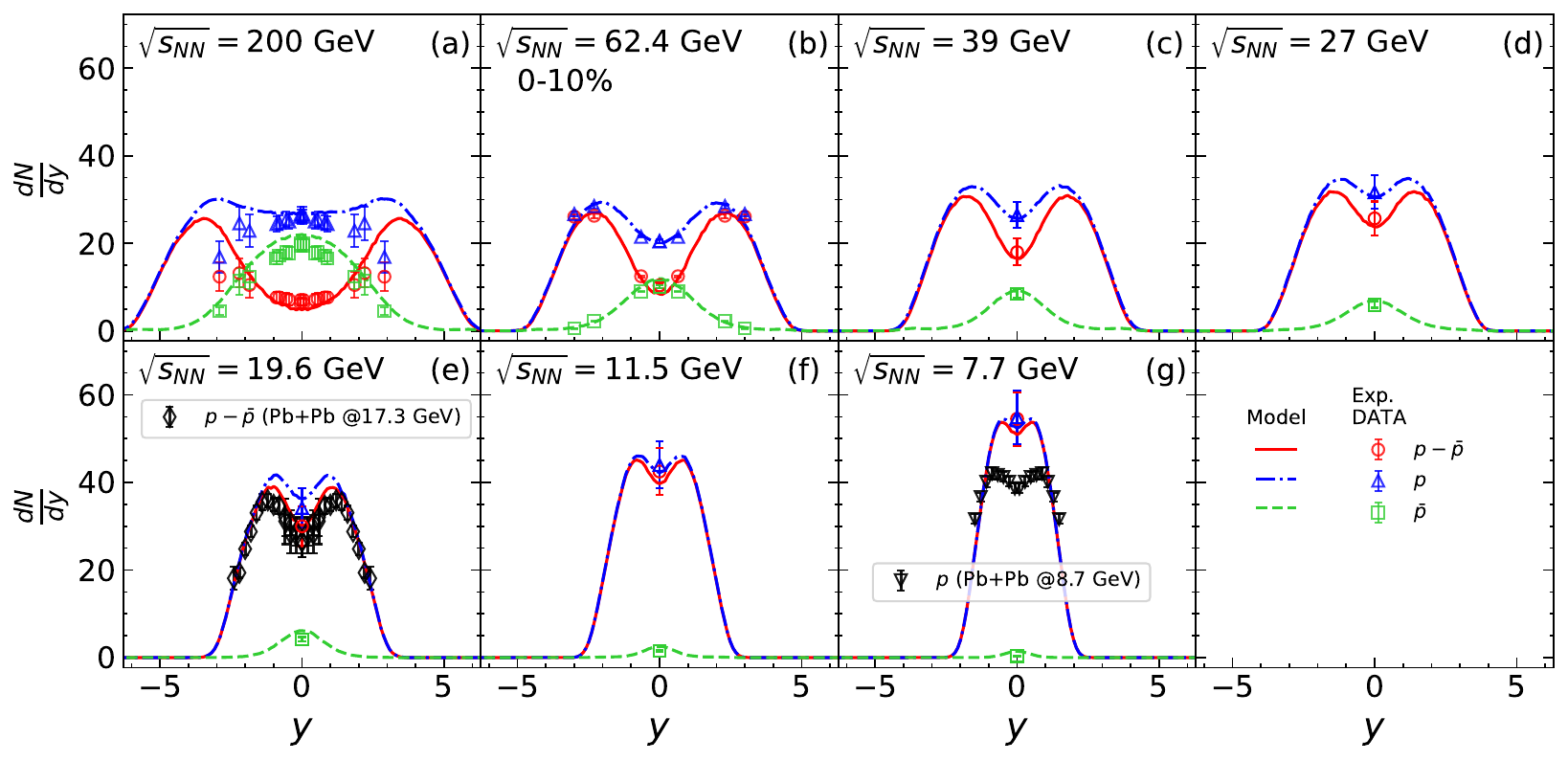} 
  \caption{Rapidity distributions of protons, anti-protons, and net-protons in central Au+Au collisions at various $\sqrt{s_{NN}}$ energies (200, 62.4, 39, 27, 19.6, 11.5, and 7.7 GeV). Model calculations (represented by lines), which include contributions from weak decays for proton and anti-proton yields, are compared with experimental data from the BRAHMS and STAR collaborations \cite{BRAHMS:2003wwg,BRAHMS:2009wlg,STAR:2017sal}. The rapidity-differential net-proton measurements at 200 and 62.4 GeV are taken from the BRAHMS collaboration\cite{BRAHMS:2003wwg,BRAHMS:2009wlg}, while only mid-rapidity data are available from STAR at other energies \cite{STAR:2017sal}. The net-proton distribution data from NA49 at 17.3 GeV is used to constrain the model for Au+Au collisions at 19.6 GeV \cite{NA49:2010lhg}. The proton distribution measurement in Pb+Pb collsions at $\sNN = 8.7$ GeV has been plotted in panel (g) for reference \cite{NA49:2010lhg}.  }
  \label{fig:netp_y_BES_central}
\end{figure}

The initial distribution of net-baryon along $\eta_s$ is crucial to be constrained in the model using experimental data, as the longitudinal gradient of baryon chemical potential could significantly influence the flow development of baryons in the transverse plane \cite{Bozek:2022svy,Du:2022yok}. As the rapidity-differential net-proton distribution is sensitive to the initial baryon configuration \cite{Denicol:2018wdp}, we carefully chose our model parameters to capture this distribution. Fig. \ref{fig:netp_y_BES_central} depict the rapidity distributions of protons, anti-protons, and net-protons for central Au+Au collisions at $\sqrt{s_{NN}}$ = 200, 62.4, 39, 27, 19.6, 11.5 and 7.7 GeV. Our model calculations show good agreement with experimental measurements. To compare with experimental data, we included contributions from weak decays in the calculation of proton and antiproton yields.  The rapidity distribution of net-proton is available for $\sqrt{s_{NN}}$ = 200 and 62.4 GeV by the BRAHMS collaboration \cite{BRAHMS:2003wwg,BRAHMS:2009wlg}, while at other considered energies, only mid-rapidity measurements are available from the STAR collaboration \cite{STAR:2017sal}. The rapidity-differential net-proton measurements for Pb+Pb collisions at $\sqrt{s_{NN}}$ = 17.3 GeV and 8.7 GeV have been conducted by the NA49 collaboration \cite{NA49:2010lhg}. We utilized the experimental data of net-proton distribution at 17.3 GeV to constrain the model parameters of initial net-baryon distribution for Au+Au collisions at 19.6 GeV, while the net-proton data at 8.7 GeV has been plotted with the model calculation of Au+Au at $\sqrt{s_{NN}}$ = 7.7 GeV for reference. Furthermore, we successfully captured the rapidity distribution of protons and anti-protons separately, indicating that the chosen freeze-out energy density represents a suitable combination of temperature and baryon chemical potential at particlisation. 

\begin{figure}
  \centering
  \includegraphics[width=1.0\textwidth]{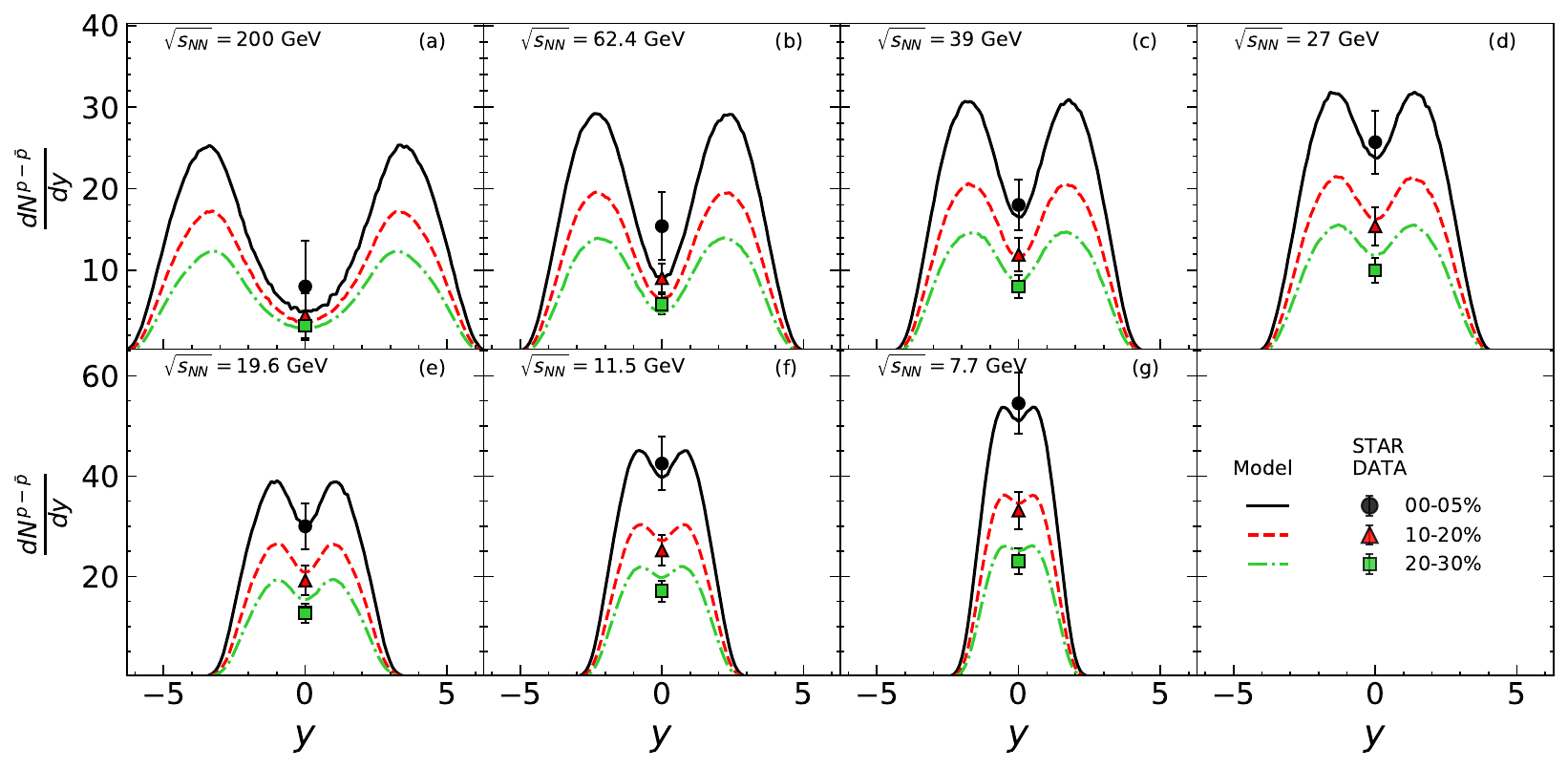} 
  \caption{Rapidity-dependent net-proton yield in Au+Au collisions for centrality classes 0-5\%, 10-20\%, and 20-30\% across energies from $\sqrt{s_{NN}}$ = 200 to 7.7 GeV. The model predictions (represnted by lines) are compared with STAR collaboration data for mid-rapidity net-proton yields \cite{STAR:2017sal}. }
  \label{fig:netp_y_cent_depen}
\end{figure}

Figure \ref{fig:netp_y_cent_depen} shows the rapidity-dependent net-proton yield in Au+Au collisions for three different centrality classes: 0-5\%, 10-20\%, and 20-30\%, across a range of energies from $\sqrt{s_{NN}}$ = 200 to 7.7 GeV. Although experimental data of the rapidity-differential net-proton is not available at all these centralities, the STAR collaboration provides mid-rapidity net-proton yield data \cite{STAR:2017sal}, which are shown alongside our model calculations. The model parameters were calibrated to match the net-proton yield for the 0-5\% centrality class, while the results for the other centrality classes are model predictions. Notably, our model successfully captures the centrality-dependent baryon stopping at mid-rapidity, from central to mid-central collisions.

\begin{figure}
  \centering
  \includegraphics[width=1.0\textwidth]{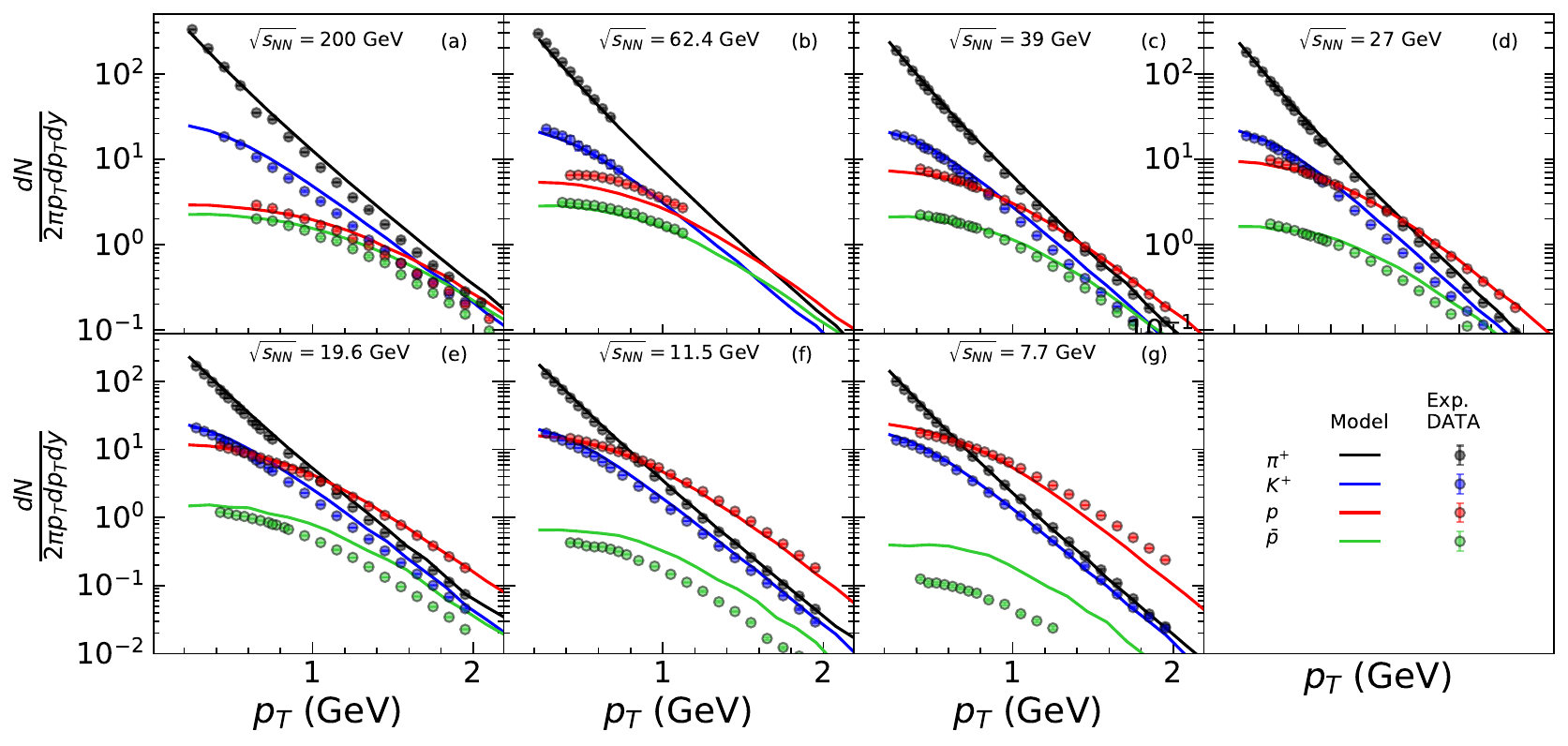} 
  \caption{The $p_T$ spectra of $\pi^{+}, K^{+}, p$ and $\bar{p}$ in 0-5\% centrality class of Au+Au collisions at $\sqrt{s_{NN}} = 200,\ 62.4,\ 39, \ 27, \ 19.6, \ 11.5$ and $7.7$ GeV. The model claculations (lines) are compared with the experimental measurements \cite{PHENIX:2003iij,STAR:2017sal,STAR:2008med}. }
  \label{fig:pT_spectra_BES}
\end{figure}

We present the transverse momentum ($p_T$) spectra of identified particles $\pi^{+}, K^{+}, p$, and $\bar{p}$ in Au+Au collisions at BES energies for 0-5\% centrality in Fig. \ref{fig:pT_spectra_BES}.  Experimental measurements from the PHENIX collaboration \cite{PHENIX:2003iij} are compared with numerical calculations at $\sqrt{s_{NN}}=200$ GeV. Notably, the proton spectra in the PHENIX measurement were corrected for weak decays. Consequently, in our model calculations, we disabled weak decays to facilitate comparison with experimental data. However, at other energies, we included weak decay corrections in the proton and anti-proton spectra to compare our model calculations with data from the STAR collaboration \cite{STAR:2017sal,STAR:2008med}. Our numerical results exhibit good agreement with experimental measurements for the mid-rapidity $p_T$ spectra of $\pi^{+}, K^{+}$, and $p$ across all considered energies. However, discrepancies between the model and data are evident for the $\bar{p}$ spectra, particularly as collision energy decreases. At lower collision energies, the $p_T$-integrated yield of $\bar{p}$ is notably suppressed compared to that of $p$, resulting in the net proton yield predominantly reflecting the $p_T$-integrated proton yield. Our initial baryon profile, calibrated based on the net-proton rapidity distribution, captures the $p$ spectra effectively but the subtle effect on the highly suppressed $\bar{p}$ yield is largely overlooked. The anti-proton to proton yield ratio at mid-rapidity is very small at lower collision energies which can be better captured by fine-tuning the freezeout energy density. Moreover, in our model calculations, we employ the grand canonical ensemble framework for particle production at the freezeout hypersurface. At lower collision energies, effects such as canonical suppression and the local conservation of baryon number play a crucial role in shaping the hadronic chemistry \cite{Gorenstein:2000yt,Braun-Munzinger:2020jbk,Becattini:2004rq,Cleymans:1990mn,Cleymans:1997ib,Hamieh:2000tk,Vovchenko:2021kxx,Shen:2022oyg}. Further refinements in the particle production mechanism in our model could lead to a better agreement between the model predictions and experimental data of the $p_T$ spectra at lower $\sNN$.

\begin{figure}[htbp]
    \centering
    \begin{minipage}{0.47\textwidth} 
        \centering
        \includegraphics[width=\textwidth]{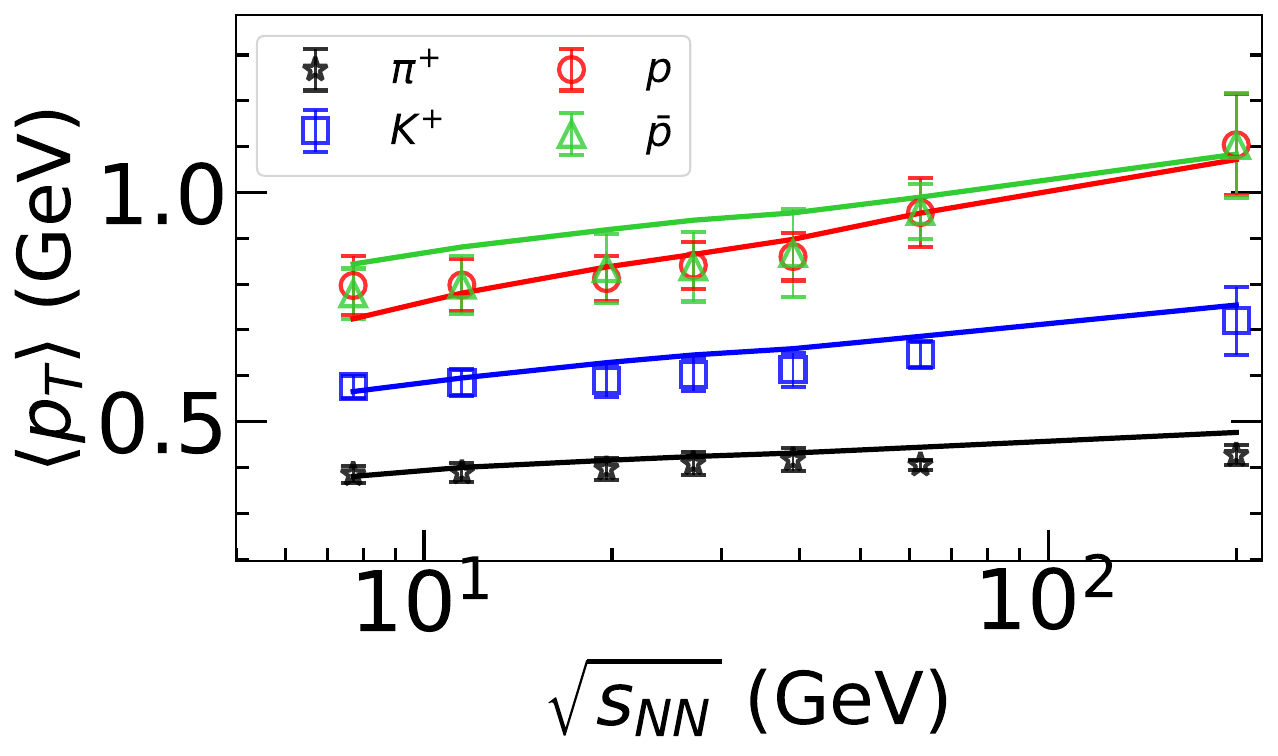}
        \caption{ Average transverse momentum $\langle p_T \rangle$ of $\pi^+, K^{+},p$ and $\bar{p}$ as a function of collision energy ($\sqrt{s_{NN}}$) for Au+Au collisions in the 0-5\% centrality class. The model predictions (represnted by lines) are compared with experimental data \cite{STAR:2017sal}.}
        \label{fig:meanpT_BES}
    \end{minipage}
    \hfill 
    \begin{minipage}{0.47\textwidth} 
        \centering
        \includegraphics[width=\textwidth]{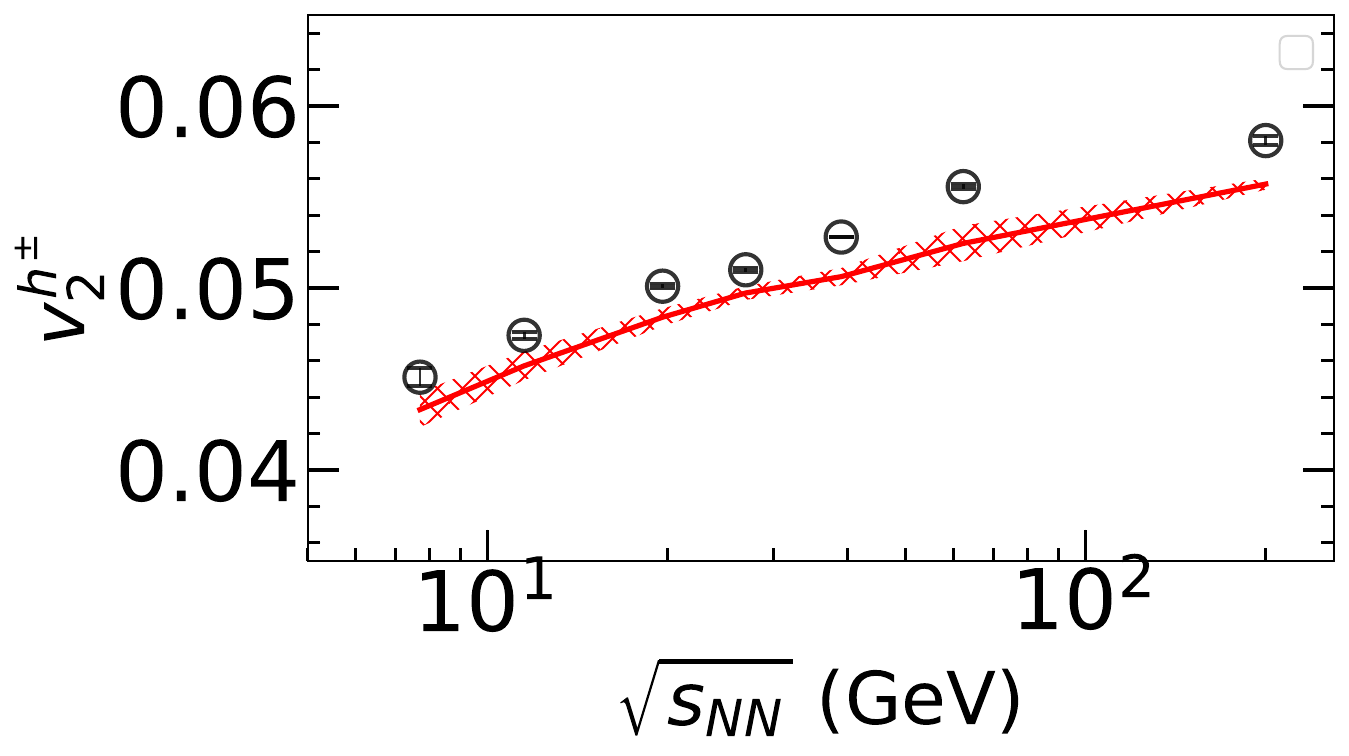}
        \caption{ Mid-rapidity elliptic flow coefficient $v_2$ of charged hadrons as a function of collision energy ($\sqrt{s_{NN}}$) for Au+Au collisions in the 10-40\% centrality class. Model calculation (line with band) is compared with experimental data \cite{STAR:2012och,STAR:2008ftz}. }
        \label{fig:v2_sNN_BES}
    \end{minipage}
    \label{fig:combined_figures}
\end{figure}

The study of the mean transverse momentum ($\langle p_T \rangle$) is important as it reflects the radial flow effects within the created medium, influenced by the slope of the $p_T$ spectra. Figure \ref{fig:meanpT_BES} presents the $\langle p_T \rangle$ for identified particles $\pi^{+}$, $K^{+}$, $p$, and $\bar{p}$ as a function of collision energy for Au+Au collisions with 0-5\% centrality. Our model successfully captures the experimental data from STAR \cite{STAR:2017sal}. We observe distinct radial flow effects for $\pi^{+}$, $K^{+}$, and $p$, owing to their different masses, resulting in a more pronounced blue-shift effect for more massive particles. Regarding the collision energy dependence, we find that the mean momenta of $\pi^{+}$, $K^{+}$, and $p$ exhibit a mild increase with collision energy due to enhanced radial flow. Particularly interesting is the observed splitting in $\langle p_T \rangle$ between $p$ and $\bar{p}$, becoming more prominent with decreasing collision energy, attributed to baryon stopping effects \cite{Shen:2020jwv}. In our model, we start the hydrodynamic evolution from a constant proper time ($\tau_0$) with zero transverse velocity but it's worth mentioning that before $\tau_0$, a pre-equilibrium stage exists, potentially contributing additional radial flow \cite{Liu:2015nwa,Dore:2023qxr,Kurkela:2018vqr,Chattopadhyay:2017bjs}. This is a factor not accounted in our model \cite{Kurkela:2018vqr,Shen:2017bsr,Du:2018mpf,Shen:2022oyg,Bozek:2022cjj}. Incorporating dynamical initial conditions and pre-equilibrium evolution could enhance the accuracy of the $\langle p_T \rangle$ prediction \cite{Liu:2015nwa}.

In Figure \ref{fig:v2_sNN_BES}, the elliptic flow $v_2$ of charged hadrons is plotted as a function of collision energy for Au+Au collisions with 10-40\% centrality. Our model qualitatively captures the measured trend of the beam energy dependence of $v_2$ \cite{STAR:2012och,STAR:2008ftz}. The elliptic flow coefficient $v_2$ stem from initial asymmetry in the transverse plane. As the collision energy decreases, the deposited energy or entropy diminishes, resulting in a shorter evolution time of the medium before freeze-out. Consequently, there is a reduced conversion of initial spatial anisotropy into final momentum anisotropy, leading to a decrease in $v_2$ with decreasing collision energy in the model.

\subsection{$v_1$ of identified hadrons}
\begin{figure}
  \centering
  \includegraphics[width=1.0\textwidth]{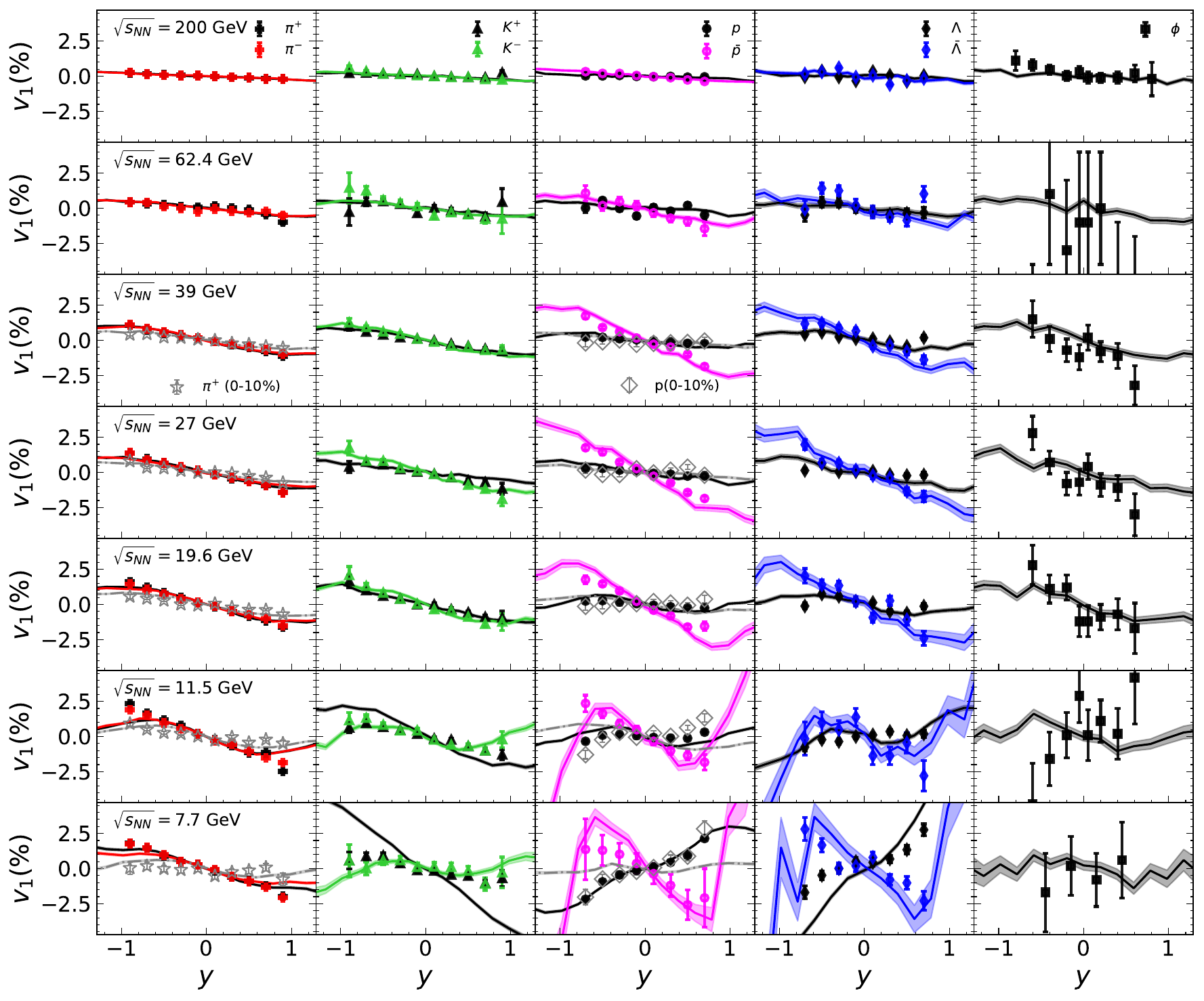} 
  \caption{Rapidity dependence of directed flow ($v_1$) for various particle species in Au+Au collisions at $\sqrt{s_{NN}} = 200$ to 7.7 GeV. Rows correspond to different collision energies, while columns represent $v_1$ for $\pi^{\pm}$, $K^{\pm}$, $p-\bar{p}$, $\Lambda-\bar{\Lambda}$, and $\phi$. Results are shown for 10-40\% centrality, with additional comparisons for $\pi^{+}$ and $p$ in the 0-10\% centrality class at lower collision energies. Model calculations (lines with bands) are compared with the experimental measurements (different symbols) of STAR collaboration~\cite{STAR:2014clz,STAR:2017okv}. The available measurements and model calculations of the $v_1$ of $\pi^+$ and $p$ for $0-10\%$ centrality are plotted in grey colored symbols and lines respectively. The bands represent the statistical uncertainty in the model calculations. }
  \label{fig:v1_y_BES}
\end{figure}

After calibrating the model parameters for the initial energy and baryon density profiles to reproduce basic observables such as yield, spectra, $\langle p_T \rangle$, and $v_2$, we now present the directed flow ($v_1$) of identified particles. The relative tilt between the matter and baryon profiles as well as the baryon diffusion determines the sign and the magnitude of the splitting between proton and anti-proton directed flow. Therefore, by suitably choosing ($\eta_{m}$, $\omega$) at each $\sqrt{s_{NN}}$, we are able to describe the rapidity dependence of $v_1$ for $\pi^{+}, p$, and $\bar{p}$ simultaneously in the 10-40\% centrality range. However, the $v_1$ of other hadrons are our model predictions. In Fig. \ref{fig:v1_y_BES}, our model calculations of the rapidity dependence of directed flow ($v_1$) of identified hadrons are plotted for Au+Au collisions at $\sqrt{s_{NN}} = 200$ to 7.7 GeV alongwith experimental data from STAR collaboration \cite{STAR:2014clz,STAR:2017okv}. Each row in the figure corresponds to results from a particular collision energy, while the directed flow of various particle species is displayed in different columns. The top row showcases the rapidity dependence of directed flow coefficients in Au+Au collisions at $\sqrt{s_{NN}}=200$ GeV, with subsequent rows presenting results for lower energies in descending order. On the other hand, columns 1, 2, 3, 4, and 5 depict $v_1$ for $\pi^{\pm}$, $K^{\pm}$, $p-\bar{p}$, $\Lambda-\bar{\Lambda}$, and $\phi$, respectively. Additionally, the $v_{1}$ of $\pi^{+}$ and $p$ in the 0-10\% centrality class is also plotted at energies below $\sqrt{s_{NN}} = 39$ GeV for comparison. The model calculations shows a quite good agreement with experimental data \cite{STAR:2014clz,STAR:2017okv}.

We have observed a distinct splitting between the directed flows ($v_1$) of protons and anti-protons, which increases with decreasing collision energy. Additionally, similar to protons and anti-protons, there exists a discernible splitting between $\Lambda$ and $\bar{\Lambda}$, owing to their opposite baryon numbers. The phenomenon of baryon stopping prominently contributes to the generation of this splitting as collision energy decreases.

The inhomogeneous deposition and further evolution of net-baryon density in the medium give rise to inhomogeneities for other conserved charges like strangeness and electric charge via the constraints in Eqs.~\ref{eq.ns} and ~\ref{eq.nq}, leading to correlations between the corresponding chemical potentials: $\mu_B$, $\mu_Q$, and $\mu_S$. This results in a splitting in directed flow of hadrons with different quantum numbers but the same mass, similar to the $v_1$ splitting of $p$ and $\bar{p}$. The difference in directed flow ($v_1$) between $\pi^{+}$ and $\pi^{-}$ arises solely from the inhomogeneity of $\mu_{Q}$ within the medium. However, both $\mu_S$ and $\mu_Q$ contribute to the difference in the directed flow coefficient of $K^{+}$ and $K^{-}$. In our current model calculations, we did not observe a significant splitting in $v_1$ between $\pi^{+}$ and $\pi^{-}$, consistent with experimental findings. However, noticeable splitting emerge in the $v_1$ between $K^{+}$ and $K^{-}$. While the split between $K^{+}$ and $K^{-}$ in the model aligns with data for $\sqrt{s_{NN}} > 11.5$ GeV, it overestimates the data at $\sqrt{s_{NN}} = 11.5$ and $7.7$ GeV. Furthermore, our model fails to capture the $v_1$ of $\phi$ particles at these collsion energies. These disparities emphasize the importance of evolving all conserved charges independently in fluid dynamical simulations. 

The directed flow measurements of $\pi^{+}$ and $p$ at 0-10\% centrality are also displayed in Fig.~\ref{fig:v1_y_BES} alongside 10-40\% centrality. Our model calculations successfully capture the centrality dependence of $v_1$ for $\pi^{+}$ at all considered collision energies but fail to do so for $p$ below $\sqrt{s_{NN}} = 19.6$ GeV. This discrepancy suggests that the mechanism of baryon stopping from central to mid-central collisions differs at lower energies, and our model is unable to provide a proper gradient of initial net-baryon density at different centralities. 
Additionally, for simplicity, our model calculations assume the same initial proper time ($\tau_0$) and the same $\omega$ parameter value for both 0-10\% and 10-40\% centralities, whereas these parameters are expected to vary with centrality. Moreover, the model does not incorporate pre-equilibrium dynamics, which could influence the system evolution differently in the initial phase for different centralities. Incorporating pre-equilibrium dynamics or a centrality-dependent $\omega$ and $\tau_0$, could improve agreement with the data. Further exploration into the dependence of baryon deposition in transverse plane with centrality and the corresponding measurements of proton directed flow at different centralities could potentially provide additional insights into baryon stopping mechanism.

\begin{figure}
  \centering
  \includegraphics[width=1.0\textwidth]{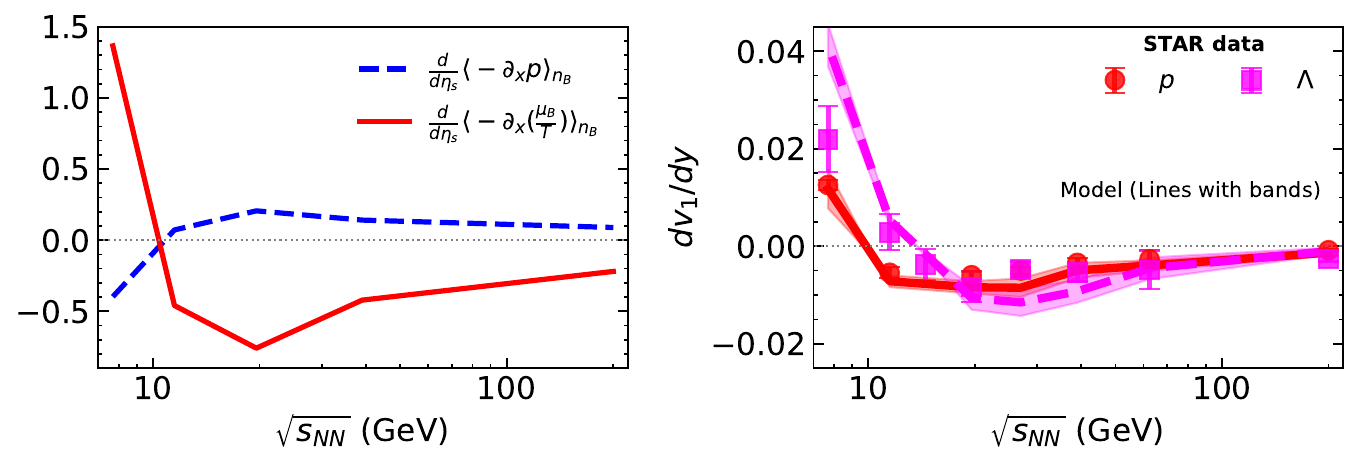} 
  \caption{ (Left) Mid-rapidity slopes of the initial estimators $\langle -\partial_x p \rangle_{n_B}$ and $\langle -\partial_x (\mu_B / T) \rangle_{n_B}$ as functions of $\sqrt{s_{NN}}$, illustrating the roles of baryon advection and diffusion in determining the sign and magnitude of the proton and $\Lambda$ directed flow shown in right panel. The interplay between these components explains the observed non-monotonic variation and sign change in $v_1$ slopes with collision energy, particularly the dominance of diffusion effects at $\sqrt{s_{NN}} = 7.7$ GeV. The experimental data is from STAR collaboration \cite{STAR:2014clz,STAR:2017okv}.}
  \label{fig:sign_change_proton_w_estimators}
\end{figure}

We now focus on the directed flow ($v_1(y)$) of protons ($p$) and lambdas ($\Lambda$), whose mid-rapidity slope exhibit a striking non-monotonic dependence on collision energy ($\sqrt{s_{NN}}$), including a sign change between $\sqrt{s_{NN}} = 11.5$ GeV and $\sqrt{s_{NN}} = 7.7$ GeV. This behavior is observed in both experimental data and our model. To investigate this feature, we plotted the mid-rapidity slopes of the initial estimators $\langle -\partial_x p \rangle_{n_B}$ and $\langle -\partial_x (\mu_B / T) \rangle_{n_B}$, as defined in Eqs. \ref{eq:estimator_2} and \ref{eq:estimator_3}, as functions of collision energy in left side panel of Fig. \ref{fig:sign_change_proton_w_estimators}. The estimator $\langle -\partial_x p \rangle_{n_B}$ represents the equilibrium component of the baryon current, predicting the part of baryon flow that advects with the fluid velocity ($u^x$). This contribution depends on the initial geometry, specifically the relative tilt between the energy and baryon-density profiles. We observed that, without the diffusion term (i.e., $\delta f^{\text{diffusion}} = 0$), this component leads to a positive $v_1$ slope for protons at mid-rapidity. However, the inclusion of the diffusion, reverses the sign of the proton $v_1$ slope, making it negative and consistent with experimental data at $\sNN \ge 11.5$ as shown in Fig. \ref{fig:sign_change_proton_w_estimators}.

At $\sqrt{s_{NN}} = 7.7$ GeV, the behavior changes as the slope of $\langle -\partial_x (\mu_B / T) \rangle_{n_B}$ itself switches sign and strongly dominates. This reversal is due to the baryon peaks, typically located far from mid-rapidity at higher collision energies, moving closer together at lower energies, resulting in a baryon-rich mid-rapidity region and very different distribution of $\frac{\mu_B}{T}$ in the medium. Consequently, the baryon gradient driving the diffusion becomes significant and plays a crucial role in determining the sign of the $v_1$ slopes for protons and $\Lambda$ baryons. This change in slope of proton and lambda directed flows at $\sqrt{s_{NN}} = 7.7$ GeV shown in right panel of Fig. \ref{fig:sign_change_proton_w_estimators} was previously associated with a signature of first-order phase transition \cite{Stoecker:2004qu,Brachmann:1999xt,Csernai:1999nf,Rischke:1995pe,Steinheimer:2014pfa,Nara:2016phs,Nara:2016hbg,Steinheimer:2022gqb}. Now, it's essential to find out whether such a sign change primarily arises from initial baryon stopping and it's diffusion or reflects the equation of state (EoS) effect. Therefore, a thorough investigation is necessary to disentangle the respective contributions of the equation of state and baryon dynamics to the $v_1$ behavior of baryons across $\sNN$.

\begin{figure}[htbp]
    \centering
    \begin{minipage}{0.47\textwidth} 
        \centering
        \includegraphics[width=\textwidth]{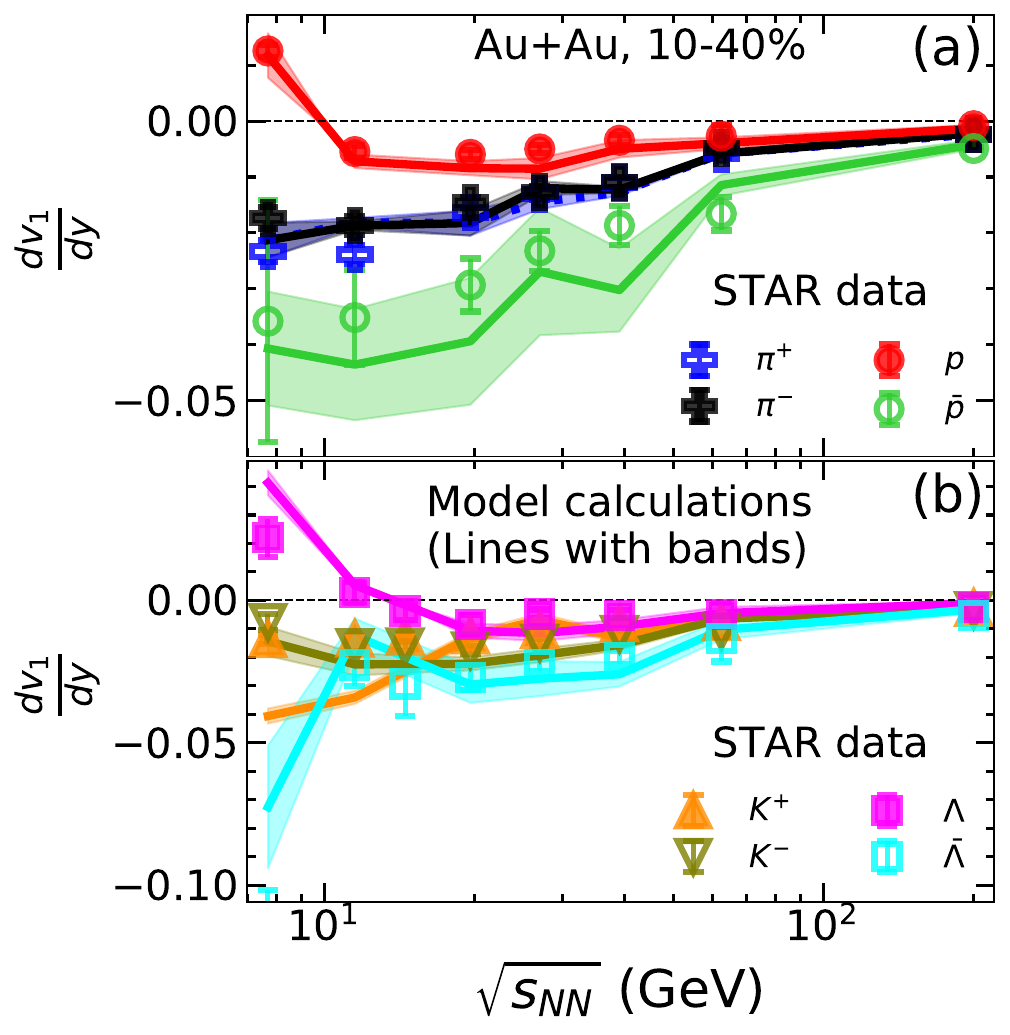}
        \caption{ Beam energy dependence of directed flow slope $(dv_1/dy)$ of identified hadrons at mid-rapidity for 10-40\% Au+Au collisions. Model calculations (lines with bands) are compared with experimental measurements. The model calculation for a particular particle species
is plotted as a line having the same color as the symbol of
experimental data. The experimental meaurements are from
STAR collaboration \cite{STAR:2014clz,STAR:2017okv}.    }
        \label{fig:dv1_dy_BES}
    \end{minipage}
    \hfill 
    \begin{minipage}{0.47\textwidth} 
        \centering
        \includegraphics[width=\textwidth]{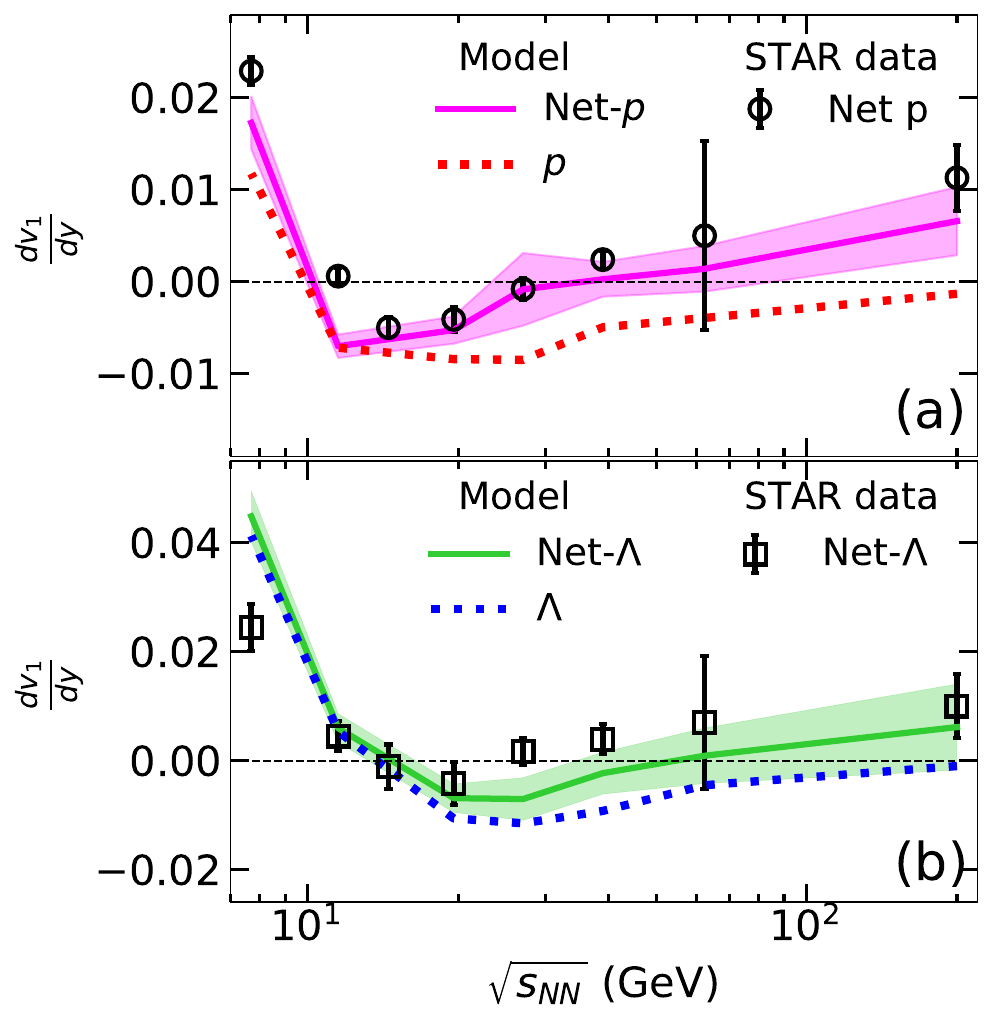}
        \caption{Beam energy dependence of the directed flow slope $(dv_1/dy)$ of (a) net-proton and (b) net-lambda for 10-40\% Au+Au collisions.
   The model calculations (lines with bands) are compared with experimental measurements of STAR collaboration~\cite{STAR:2014clz,STAR:2017okv}. The $(dv_1/dy)$ of $p$ and $\Lambda$ are also plotted for comparison. At lower $\sNN$, the $dv_1/dy$ of proton and net-proton as well as $\Lambda$ and nat-$\Lambda$ shows same magnitude and sign. }
        \label{fig:dv1_dy_nep_netl}
    \end{minipage}
    \label{fig:combined_figures}
\end{figure}

The mid-rapidity slope of the directed flow for identified particles is plotted as a function of collision energy for 10-40\% Au+Au collisions in Fig. \ref{fig:dv1_dy_BES}. Panel (a) displays the slopes of $\pi^{\pm}$, $p$, and $\bar{p}$, while panel (b) presents the slopes of $K^{\pm}$, $\Lambda$, and $\bar{\Lambda}$. In our model calculations of the directed flow slope, we adopted the same fitting function and fitting range in rapidity as specified in the experimental papers \cite{STAR:2014clz,STAR:2017okv}. A notable observation is the change in sign of the $v_1$-slope for $p$ and $\Lambda$ at lower collsion energies, whereas the slope of their corresponding anti-particles remains negative across all collsion energy. Notably, the $\frac{dv_1}{dy}$ of baryons and anti-baryons consistently align on opposite sides of the $v_1$ slope for $\pi^+$. The magnitudes of these slopes from model calculation describes the experimental data well. However, the model calculations overestimate the split between the $\frac{dv_1}{dy}$ of $K^{+}-K^{-}$ and $\Lambda-\bar{\Lambda}$ at $\sqrt{s_{NN}} = 7.7$ GeV.

STAR collaboration has also measured the $v_1(y)$ of net-proton and net-$\Lambda$ in Ref. \cite{STAR:2014clz,STAR:2017okv}. 
These are defined as :
\beq
 [v_1(y)]_{net-p} =  \frac{\frac{dN^{p}}{dy}[v_1(y)]_{p} - \frac{dN^{\bar{p}}}{dy} [v_1(y)]_{\bar{p}} } {\frac{dN^{p-\bar{p}}}{dy}}
\label{eq:netppv1}
\eeq
and
\beq
 [v_1(y)]_{net-\Lambda} =  \frac{\frac{dN^{\Lambda}}{dy}[v_1(y)]_{\Lambda} - \frac{dN^{\bar{\Lambda}}}{dy} [v_1(y)]_{\bar{\Lambda}} } {\frac{dN^{\Lambda-\bar{\Lambda}}}{dy}}
\label{eq:netlv1}
\eeq
Experimental results reveal a double sign change in the mid-rapidity slope of $v_1$ for net-$p$ and net-$\Lambda$ in the collision energy range $\sqrt{s_{NN}} = 39$–7.7 GeV, even though $dv_1/dy$ for protons ($p$) and lambdas ($\Lambda$) individually exhibits only a single sign change. 

These above expressions in Eq. \ref{eq:netppv1} and \ref{eq:netlv1} highlight that capturing the mid-rapidity slope of net-$p$ (net-$\Lambda$) requires simultaneously reproducing the $v_1(y)$ of  $p$ ($\Lambda$), $\bar{p}$ ($\bar{\Lambda}$), and their respective mid-rapidity yields. Since our model successfully reproduces all these quantities across the range $\sqrt{s_{NN}} = 7.7$–200 GeV, it is well-suited to describe the experimental data of $dv_1/dy$ of net-$p$ (net-$\Lambda$). Fig. \ref{fig:dv1_dy_nep_netl} presents our model calculation of the collision energy dependence of the mid-rapidity slope of $v_1$ for net-$p$ and net-$\Lambda$ baryons, along with a comparison to experimental results. The model effectively explains the observed double sign change in $dv_1/dy$ for both net-$p$ and net-$\Lambda$.

At lower $\sqrt{s_{NN}}$, the mid-rapidity yield of anti-protons ($dN^{\bar{p}}/dy$) becomes negligible compared to that of protons ($dN^p/dy$). Consequently, the mid-rapidity $v_1$ slope of net-$p$ closely resembles that of protons at lower $\sqrt{s_{NN}}$. A similar observation applies to net-$\Lambda$. In Fig. \ref{fig:dv1_dy_nep_netl}, we also plot $dv_1/dy$ for $p$ and $\Lambda$ alongside their net counterparts. At higher $\sqrt{s_{NN}}$, the slopes for net-$p$ and $p$ differ significantly, whereas at lower $\sqrt{s_{NN}}$, they exhibit similar magnitudes and signs. It is worth emphasizing that a model capable of describing only $dv_1/dy$ for protons but not for anti-protons can still capture net-$p$ behavior at lower $\sqrt{s_{NN}}$ due to the dominance of $dN^p/dy$ \cite{Du:2022yok}. However, to describe the $dv_1/dy$ of net-$p$ across the entire range of $\sqrt{s_{NN}}$, especially to capture the positive slope at higher collsion energies, the model must also accurately capture $dN^{\bar{p}}/dy$ and $dv_1/dy$ for anti-protons.

\begin{figure}
  \centering
  \includegraphics[width=1.0\textwidth]{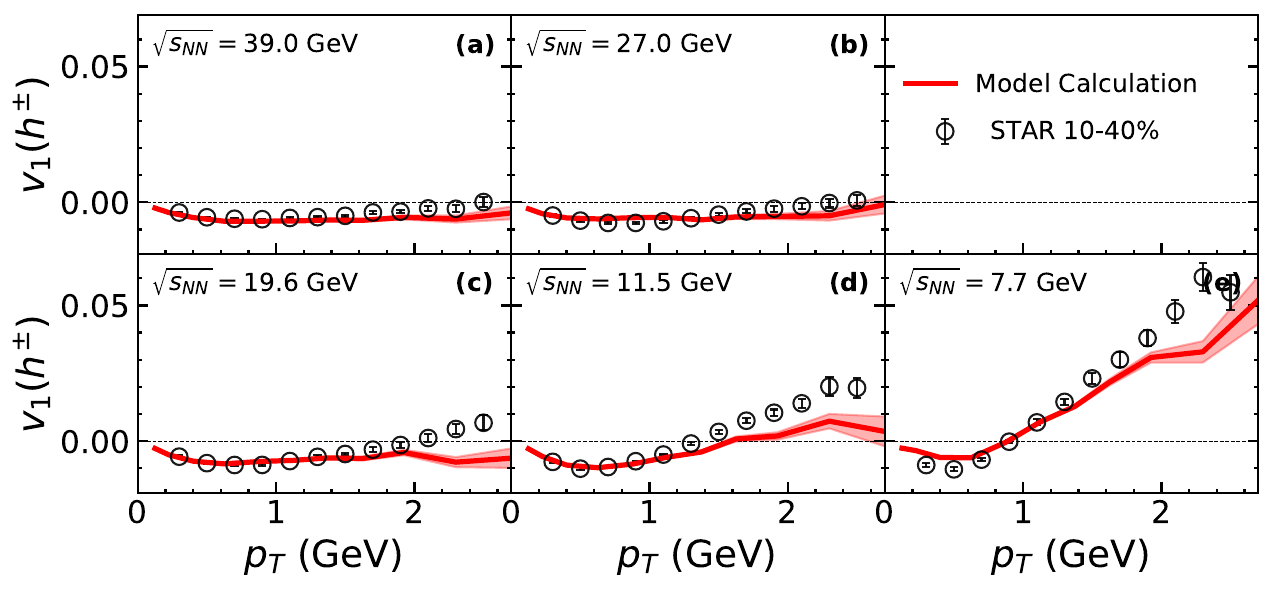}
  \caption{Transverse momentum ($p_T$) dependence of directed flow ($v_1$) for charged hadrons ($h^{\pm}$) in Au+Au collisions at $\sqrt{s_{NN}} = 39$, $27$, $19.6$, $11.5$, and $7.7$ GeV for 10–40\% centrality class. The results, obtained from model calculations, are compared with experimental data form STAR collaboration \cite{STAR:2019vcp}. }
  \label{fig:v1_ch_pT_BES}
\end{figure}

Fig. \ref{fig:v1_ch_pT_BES} presents the transverse momentum ($p_T$) dependence of directed flow ($v_1$) for charged hadrons ($h^{\pm}$), providing insights into the initial matter distribution and its evolution in the transverse plane near mid-rapidity. The results are shown for 10–40\% centrality class in Au+Au collisions at $\sqrt{s_{NN}} = 39$, $27$, $19.6$, $11.5$, and $7.7$ GeV. Our model calculations align well with experimental measurements for charged hadrons \cite{STAR:2019vcp} in $p_T<2$ GeV. Currently, experimental data on the $p_T$-differential $v_1$ of identified hadrons, particularly baryons and anti-baryons, is not available. However, such measurements, when compared with model predictions, could provide valuable insights into the initial transverse distributions of baryon density.

\section{Out-of-equilibrium correction}

\begin{figure}[htbp]
    \centering
    \begin{minipage}{0.47\textwidth} 
        \centering
        \includegraphics[width=\textwidth]{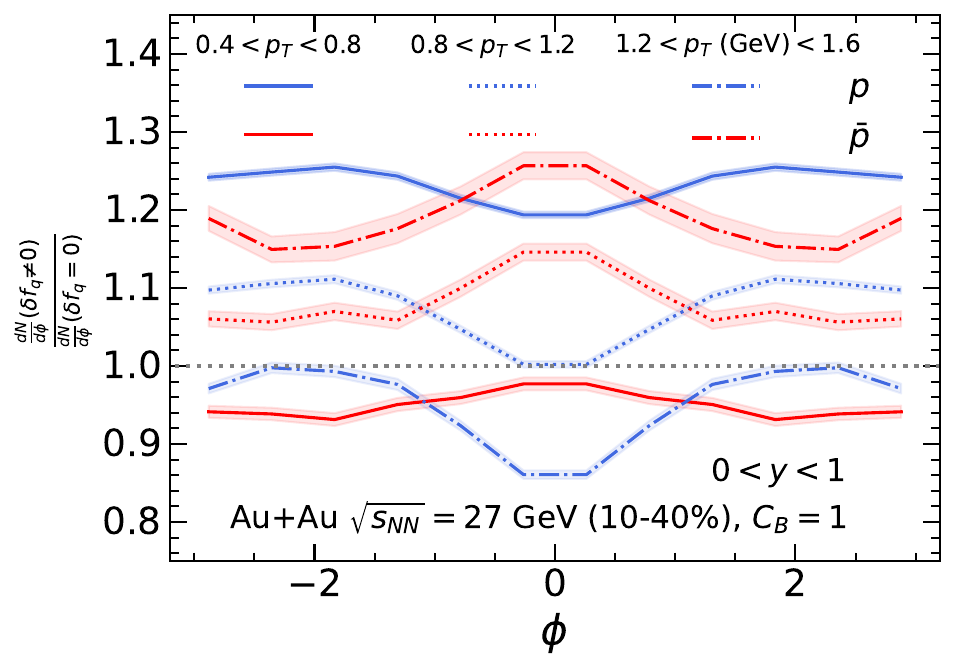}
        \caption{The impact of out-of-equilibrium corrections due to net baryon diffusion ($\delta f_q$) on the $\phi$ differential spectra of protons ($p$) and anti-protons ($\bar{p}$) in Au+Au collisions of 10-40\% centrality at $\sqrt{s_{NN}}=27$ GeV. The taken $C_B$ value at $\sNN=27$ GeV is 1.0.  The $\delta f_q$ correction in the $\phi$ spectra is presented for three different $p_T$ intervals: $0.4<p_T<0.8$ (solid lines), $0.8<p_T<1.2$ (dotted lines), and $1.2<p_T<1.6$ (dashed-dotted lines).}
        \label{fig:dndphi_27}
    \end{minipage}
    \hfill 
    \begin{minipage}{0.47\textwidth} 
        \centering
        \includegraphics[width=\textwidth]{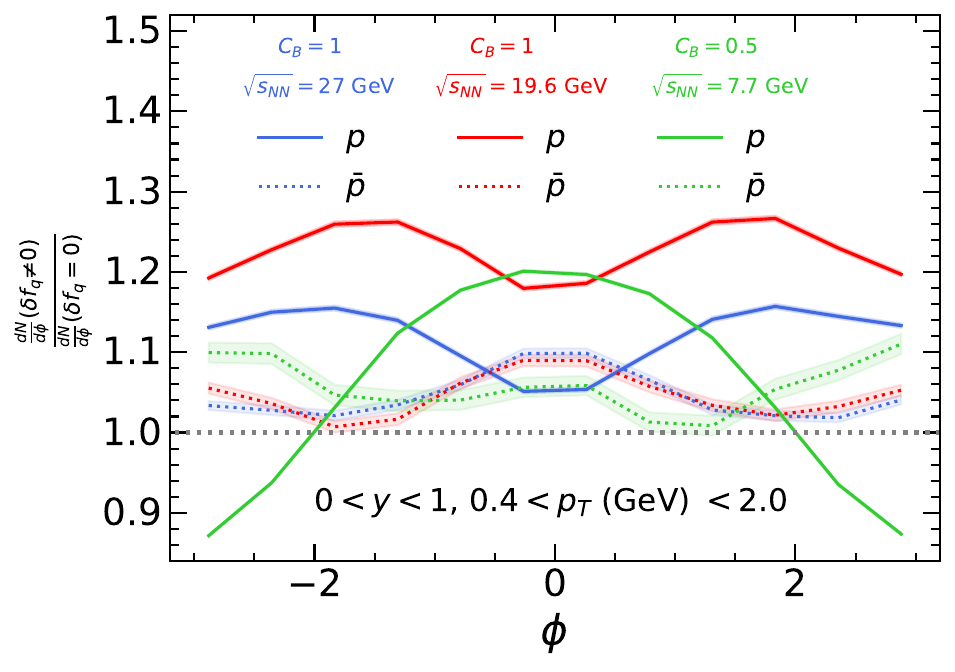}
        \caption{The effect of out-of-equilibrium corrections due to net baryon diffusion ($\delta f_q$) on the $\phi$ differential spectra of protons ($p$) and anti-protons ($\bar{p}$) in Au+Au collisions of 10-40\% centrality at $\sqrt{s_{NN}}=7.7,19.6$ and $27$ GeV. The $\delta f_q$ correction in the $\phi$ spectra is presented for $0.4 < p_T < 2.0$ and $0 < y < 1$. For $\sqrt{s_{NN}}=27$ and $19.6$ GeV, we used $C_B=1$, while for $\sqrt{s_{NN}}=7.7$ GeV, we used $C_B=0.5$.}
        \label{fig:dndphi_sNN}
    \end{minipage}
    \label{fig:combined_figures}
\end{figure}

The out-of-equilibrium correction, $\delta f^{\text{diffusion}}$ is essential in model calculations with baryon dissipation to conserve the net-baryon number during the Cooper-Frye particlisation procedure \cite{Denicol:2018wdp}. The form of $\delta f^{\text{diffusion}}$ used in our study is mentioned in Eq. \ref{eq:deltaf_diff_CF}. For the model to be valid, this correction should be much smaller than the equilibrium distribution function at any point in the phase space. In our current study, we observed that for $C_B=1.0$, the $\delta f^{\text{diffusion}}$ (in this section, we will use $\delta f_q$ as abbreviation for $\delta f^{\text{diffusion}}$) significantly affects the $v_1$ of baryons and anti-baryons. This prompted us to verify the correction due to the $\delta f_q$ on the phase space distribution of produced particles to ensure the validity of our framework at non-zero baryon diffusion. Ideally, the correction $\delta f_q$ should be examined at each point in momentum space, but for simplicity, we investigated its effect on the $\phi$ differential spectra in momentum space. Previously, the impact of $\delta f_q$ on $p_T$ spectra was studied in Ref. \cite{Denicol:2018wdp}, where the correction appeared small. In our study, we plot the ratio of the $\phi$ differential spectra of protons and anti-protons between cases where $\delta f_q \ne 0$ and $\delta f_q = 0$ at different $p_T$ intervals within the rapidity cut $0 < y < 1$ in Au+Au collisions at $\sqrt{s_{NN}} = 27$ GeV in Fig. \ref{fig:dndphi_27}. We observed that the baryon diffusion correction for protons is larger for smaller $p_T$ intervals whereas for the anti-protons, the correction is smaller for smaller $p_T$. Importantly, the correction due to $\delta f_q$ is less than 30\%, indicating that $C_B=1.0$ is reasonable to take in our model calculation.

To study the effect at different $\sqrt{s_{NN}}$, we also plotted the ratio of the $\phi$ differential spectra of protons and anti-protons between cases where $\delta f_q \ne 0$ and $\delta f_q = 0$ at different $\sqrt{s_{NN}}$ in Fig. \ref{fig:dndphi_sNN} in the $p_T$ interval $0.4 < p_T < 2$ and rapidity interval $0 < y < 1$. Here, we observed that the correction affects the $\phi$ spectra by less than 30\%. Even though $C_B$ is taken to be 1.0 for both $\sNN=27$ and $19.6$ GeV, the $\delta f_q$ correction is more significant for $\sqrt{s_{NN}} = 19.6$ GeV than for $\sqrt{s_{NN}} = 27$ GeV, due to more baryons being deposited in the mid-rapidity region. For $\sqrt{s_{NN}} = 7.7$ GeV, a smaller $C_B$ value is used ($C_B=0.5$), but the correction is still almost 20\% at mid-rapidity due to substantial baryon stopping at this collision energy.

\section{Systematic investigation of factors influencing directed flow}
This section presents a more detailed examination of various factors that influence the rapidity differential $v_1$ of identified hadrons. Our analysis encompasses several key elements, including the variation in the peak position of the initial rapidity distribution ($\eta_0^{n_B}$) of baryon density and its impact on directed flow.  Additionally, we assess our model's ability to capture the directed flow of identified hadrons by breaking the initial Bjorken flow assumption. Finally, we explore the effects of hadronic interactions. This systematic study has been performed at $\sqrt{s_{NN}}=19.6$ GeV.

\subsection{Effect of the peak position of the baryon profile}

\begin{figure}
 \begin{center}
 \includegraphics[scale=0.5]{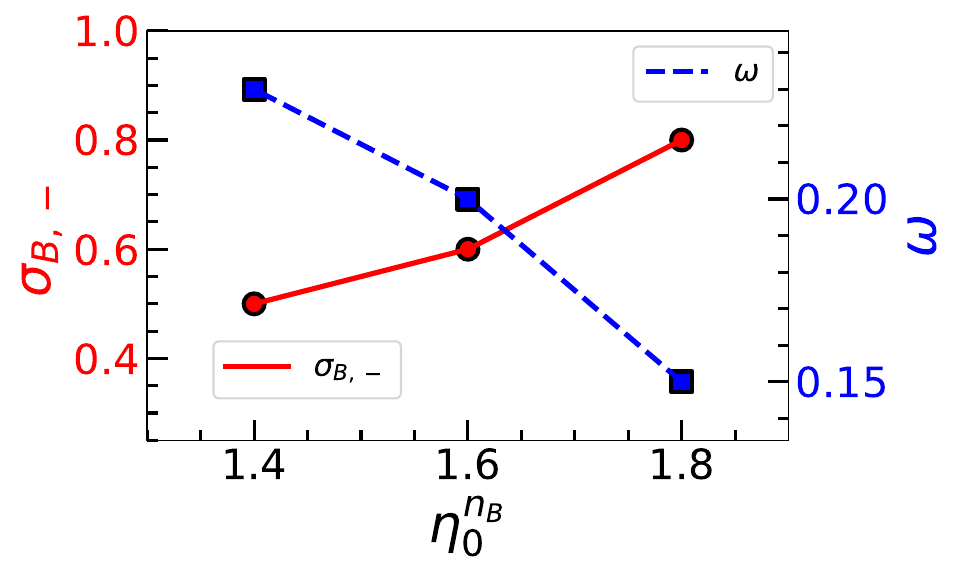}
 \caption{The variation of $\sigma_{B,-}$ and $\omega$ parameter with $\eta_{0}^{n_B}$ to describe the experimental data of rapidity distribution of net proton as well as the rapidity differntial directed flow of $\pi^{+},p$ and $\bar{p}$ in Au+Au collsions at $\sqrt{s_{NN}}=19.6$ GeV.     }
 \label{fig:eta0_peak_vary}
 \end{center}
\end{figure}

The model parameters $\eta_0^{n_B}$ and $\sigma_{B,-}$ play a crucial role in determining the longitudinal gradient of baryon density from the forward or backward $\eta_s$ regions to the mid-rapidity region during the initial stages of the collisions. By adjusting these parameters, one can effectively control the flow of baryons towards mid-rapidity \cite{Denicol:2018wdp}. Although $\eta_0^{n_B}$ and $\sigma_{B,-}$ are chosen appropriately to characterize the rapidity distribution of the net-baryon yield, it's worth noting that the selected parameters are not unique to the model. We have observed that adjusting $\eta_0^{n_B}$ towards a smaller $\eta_s$ and appropriately selecting $\sigma_{B,-}$ (which governs the Gaussian fall towards mid-rapidity) can bring the baryon peaks closer in the initial state and after the evolution we are able to capture the rapidity distribution of the net-proton yield in the final state. However, this variation in $\eta_0^{n_B}$ and $\sigma_{B,-}$ leads to significantly different initial baryon deposition in the mid-rapidity region. Remarkably, even with variations in the baryon distribution profile through these parameters, we are able to adjust our tilt parameters to effectively capture the directed flow of both baryons and anti-baryons. In Fig. \ref{fig:eta0_peak_vary}, we show the variation of $\sigma_{B,-}$ and $\omega$ with $\eta_0^{n_B}$ to capture the rapidity distribution of net-proton yield and the directed flow of identified hadrons simulataneously. We observe that a smaller value of $\sigma_{B,-}$ is required as the baryon peaks are brought closer together, which is obvious. Moreover, we note that as the peak position of the baryon density gets closer to the mid rapidity, the baryon profile in the reaction plane becomes more tilted, necessitating an increase in the $\omega$ value to maintain the relative tilt between the matter and baryon profiles. This adjustment ensures consistency in explaining the $v_1$ data of $\pi^+$, $p$, and $\bar{p}$.

\subsection{Deviation from initial bjorken flow}
\label{baryon_tilt:Bjorken_flow_break}

The rapidity odd directed flow arises when there is a breaking of boost invariance during the initial stages of relativistic heavy-ion collisions. This breaking of forward-backward symmetry can occur due to asymmetric matter deposition along the longitudinal direction or through asymmetric initial longitudinal flow. In our study, we assume a Bjorken initial flow in the longitudinal direction and no initial transverse flow, attributing the observed directed flow solely to the breaking of forward-backward symmetry in matter deposition. However, given the wide range of collision energies we consider, the breakdown of Bjorken flow is expected at lower $\sqrt{s_{NN}}$ but is not incorporated into our model. To address this, we examine our model's ability to describe the $v_1$ of identified hadrons by relaxing the initial Bjorken flow assumption. Following the approach proposed in Ref.~\cite{Ryu:2021lnx}, we introduce a modification to the Bjorken flow ansatz by setting the initial longitudinal flow at $\tau_0$ as follows:
\begin{equation}
       v_{\eta_s} = \frac{ T^{\tau \eta_s} }{ T^{\tau \tau} +P }
\end{equation}
where $P$ is the pressure. Here, the components of the energy-momentum tensor $T^{\tau \eta_s}$ and $T^{\tau \tau}$ are specified as:
\begin{equation}
    T^{\tau \tau} (x,y,\eta_s) = \epsilon(x,y,\eta_s) \cosh{y_L(x,y)}
\end{equation}
\begin{equation}
    T^{\tau \tau} (x,y,\eta_s) = \epsilon(x,y,\eta_s) \sinh{y_L(x,y)}
\end{equation}
where $y_L(x, y)$ is defined as:
\begin{equation}
    y_L(x,y) = f \eta_s^{CM}(x,y)
\end{equation}
The quantity $\eta_s^{CM}(x, y)$ can be calculated as:
\begin{equation}
    \eta_s^{CM}(x,y) = \frac{\int d\eta_s \eta_s \epsilon(x,y,\eta_s) }{ \int d\eta_s \epsilon(x,y,\eta_s) }
\end{equation}
and the parameter $f$, which lies in the range $[0, 1]$, determines the fraction of longitudinal momentum attributed to the corresponding longitudinal flow velocity. Setting $f=0$ recovers the Bjorken flow case.

\begin{figure}
 \begin{center}
 \includegraphics[scale=0.6]{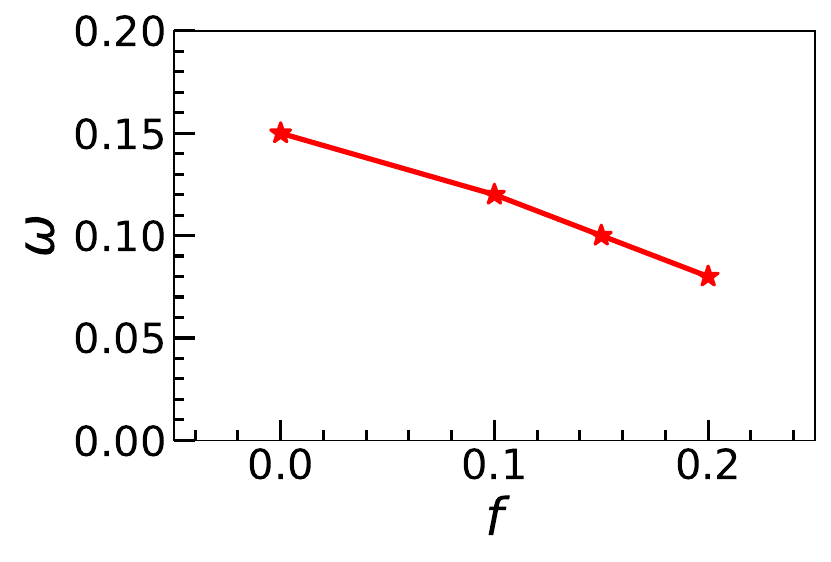}
 \caption{The variation of $\omega$ parameter with initial Bjorken flow breaking parameter $f$ to describe the experimental data of $v_1$ of $\pi^{+},p$ and $\bar{p}$ in Au+Au collsions at $\sqrt{s_{NN}}=19.6$ GeV.      }
 \label{fig:f_omega}
 \end{center}
\end{figure}

We conducted simulations at $\sqrt{s_{NN}} = 19.6$ GeV using this initial condition. Interestingly, we found that the parameter $f$ does not significantly affect the rapidity distribution of net-proton yield, but it notably impacts the directed flow ($v_1$) of hadrons. Remarkably, by varying $f$, we could select another set of tilt parameters $\eta_m$ and $\omega$ that allowed us to capture the experimental $v_1$ data for $\pi^+, p$, and $\bar{p}$. As we increased asymmetry in the longitudinal flow by raising the value of $f$, we found it necessary to decrease the tilt of the baryon profile to describe the experimental $v_1$ of identified hadrons. Fig. \ref{fig:f_omega} illustrates the dependence of $\omega$ on $f$ in our model calculations to capture the experimental $v_1$ data. Notably, we observed a decreasing trend of $\omega$ with increasing $f$. We explored whether a value of $f$ could be found for which $\omega=0$. If found, this will indicate that there is no need of two-component baryon deposition to explain the $v_1(y)$ of $\pi^{+},p$ and $\bar{p}$ and we could achive this by solely breaking the Bjorken flow. But we observed that beyond $f=0.2$, we could not find a suitable $\omega$ to explain the $v_1$ data. This underscores the necessity and advantage of our two-component baryon deposition approach in explaining the experimental data of the $v_1$ of identified hadrons.

\subsection{Influence of hadronic afterbuner}
\begin{figure}
 \begin{center}
 \includegraphics[scale=0.75]{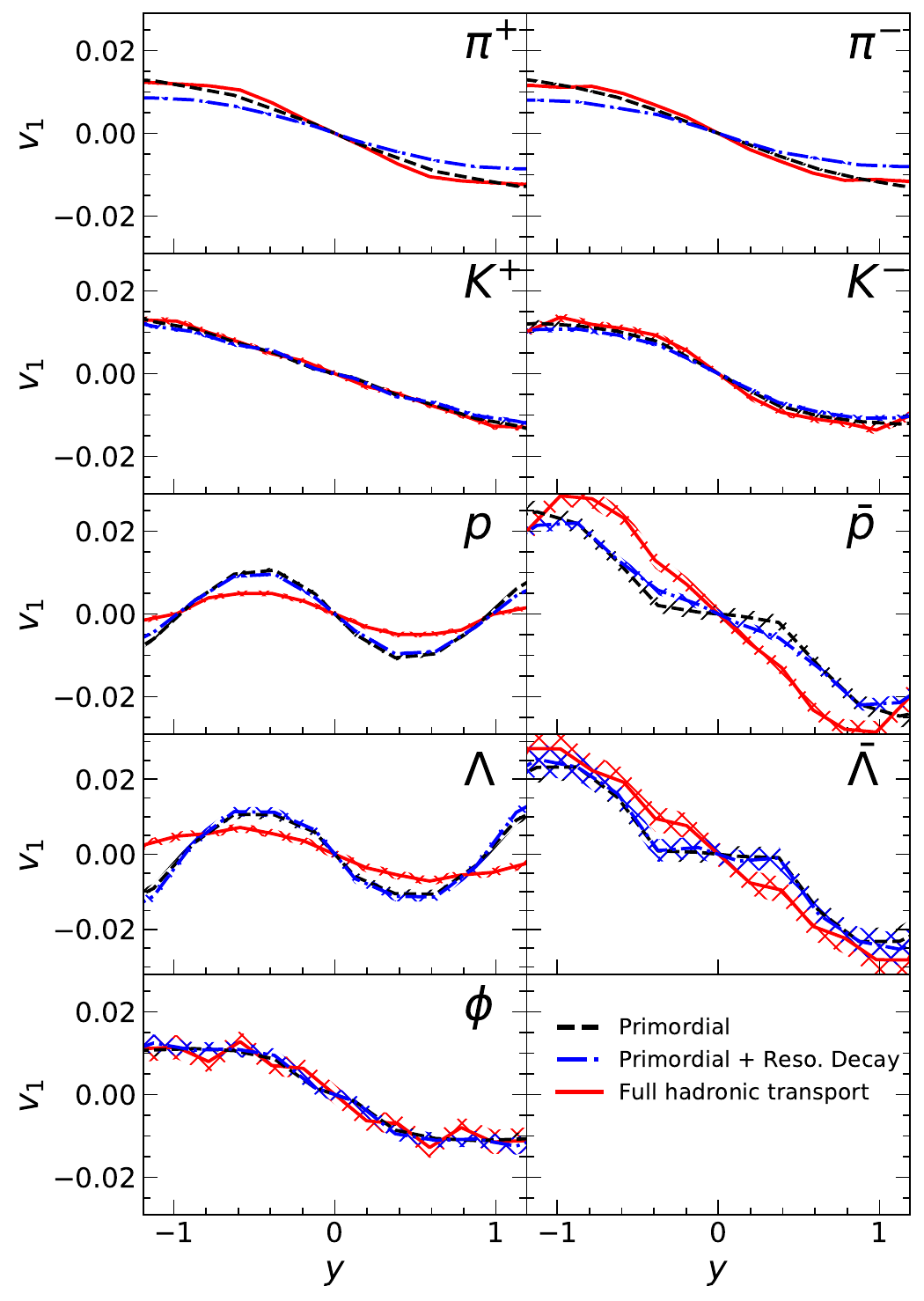}
 \caption{Effect of late stage hadronic interaction on $v_1$ of identified hadrons in Au+Au collsions of 10-40\% centrality at $\sqrt{s_{NN}}=19.6$ GeV. }
 \label{fig:afterburner}
 \end{center}
\end{figure}

In Fig. \ref{fig:afterburner}, we examine the impact of the hadronic transport stage on the $v_1$ of identified hadrons at $\sNN=19.6$ GeV. Here, we first consider the primordial hadrons sampled from the freezeout hypersurface and allow all resonances to decay without undergoing any further evolution, calculating the resulting $v_1$ after the decays. Additionally, we plot the $v_1$ after subjecting the hadrons to late-stage hadronic interactions simulated by the UrQMD code \cite{Bass:1998ca, Bleicher:1999xi}. Furthermore, we include the $v_1$ of primordially produced hadrons to isolate the effect of resonance decay alone.

We have observed that the $v_1$ of $\pi^{\pm}$ increases with hadronic interactions, as the hadronic phase allows for further system evolution, facilitating the conversion of any remaining dipole asymmetry into momentum space anisotropy \cite{Denicol:2018wdp,Sahoo:2023sgk,Pradhan:2021zbt}. However, the $v_1$ of $K^{\pm}$ remains relatively insensitive to the hadronic evolution. This might be due to their low interaction cross-section in the medium owing to their strangeness content \cite{Hirano:2007ei,Takeuchi:2015ana,Shor:1984ui}. Similarly, the $v_1$ of the $\phi$ resonance is minimally affected due to its small interaction cross-section and large lifetime, resulting in its decay predominantly occurring outside the hadronic phase, thereby keeping its phase space distribution unchanged. Conversely, it has been observed that short-lived resonances such as $K^{*0}$ exhibit significant effects during the hadronic phase \cite{Oliinychenko:2021enj,Parida:2023tdx}. The effect of hadronic phase on $v_1$ of $K^{*0}$ will be discussed in detail in the next section. Notably, the $v_1$ of baryons ($p, \Lambda$) and anti-baryons ($\bar{p}, \bar{\Lambda}$) are strongly affected by the hadronic interaction phase. It has been observed that the splitting of $v_1$ between baryons and anti-baryons increases during the late-stage evolution.  

\section{Effect of late stage hadronic interaction on the directed flow of $\phi$ and $K^{*0}$ resonances}

We have investigated the influence of late-stage hadronic interactions on directed flow of short lived resonance $K^{*0}$ and compared it with the $v_1$ of $\phi$ \cite{Parida:2023tdx}. This study has been done using three particle production formalisms within our model. In the first scenario, directed flow was calculated directly from thermally produced hadrons (primordial hadrons). The second scenario includes instantaneous decay of all primordial resonances on the hypersurface, effectively excluding any interaction effects. Finally, the third scenario incorporated the UrQMD model \cite{Bass:1998ca, Bleicher:1999xi}. Here, primordial hadrons undergo a series of binary collisions following the Boltzmann equation, simulating multiple elastic and inelastic scattering events in the hadronic phase which ultimately determines the final distribution of stable hadrons. 

\begin{figure}
  \centering
  \includegraphics[width=1.0\textwidth]{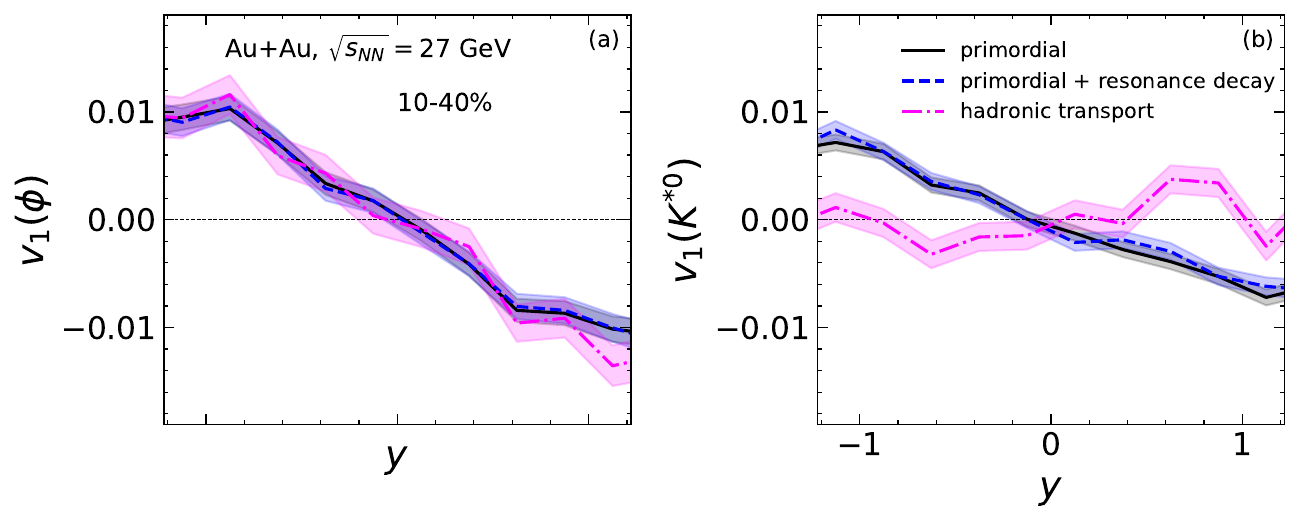} 
  \caption{The rapidity differential directed flow($v_1$) of $\phi$ and $K^{*0}$ has been plotted in panel (a) and (b) respectively for Au+Au collisons of 10-40\% centrality at $\sNN=27$ GeV . The $v_1$ of the primordial hadrons which are produced directly from the hypersurface has been represented in solid lines. The dashed lines represent the $v_1$ calculations of the hadrons after performing the decay of resonances to stable hadrons whereas the dashed-dotted lines represents the $v_1$ calculations of the hadrons after they pass through the UrQMD hadronic afterburner. The band in each line provides the statistical uncertainty in the calculation.}
  \label{fig:v1_y_kstar}
\end{figure}

First, we study the impact of hadronic interactions on the directed flow ($v_1$) of $K^{*0}$ and $\phi$ mesons in Au+Au collisions at $\sNN = 27$ GeV with 10-40\% centrality. Fig. \ref{fig:v1_y_kstar} shows the $v_1(y)$ for both the mesons. $\phi$ mesons, due to their small scattering cross-sections and long lifetimes, are minimally affected by the hadronic phase and retain their $v_1$ unchanged even after hadronic interactions \cite{Hirano:2007ei,Takeuchi:2015ana,Shor:1984ui}. In contrast, $v_1$ of $K^{*0}$ experiences a significant change during this stage. Notably, the mid-rapidity slope of $v_1(y)$ for $K^{*0}$ flips sign from negative (before interactions) to positive (after interactions).

In probing the significant change in the $v_1$ of the $K^{*0}$ meson during the hadronic phase, we analyzed variations in its yield and $v_1$ within different regions of phase space. We focused specifically on the positive rapidity region ($0 < y < 1$) and examined the behavior of hadrons flowing in both positive and negative x-directions ($p_x > 0$ and $p_x < 0$). To quantify the difference, we calculated the ratio of integrated yield ($N$) and $v_1$ between hadrons with positive and negative $p_x$. 
\begin{equation}
 \frac{N_{(p_x > 0)} }{N_{(p_x < 0)}} = \frac{ \int_{0}^{1} dy 
 \int_{-\pi/2}^{\pi/2} d\phi \int dp_T  \frac{dN}{ dp_T dy d\phi} } { \int_{0}^{1} dy 
 \int_{\pi/2}^{3 \pi/2} d\phi \int dp_T  \frac{dN}{ dp_T dy d\phi} }
 \label{Eq.int_yld}
\end{equation}
\begin{equation}
 \frac{(v_1)_{(p_x > 0)} }{(v_1)_{(p_x < 0)}} =  \frac{ \int_{0}^{1} dy 
 \int_{-\pi/2}^{\pi/2} d\phi \cos{\phi} \int dp_T  \frac{dN}{ dp_T dy d\phi} } { \int_{0}^{1} dy 
 \int_{\pi/2}^{3 \pi/2} d\phi \cos{\phi} \int dp_T  \frac{dN}{ dp_T dy d\phi} } \times  \frac{N_{(p_x < 0)} }{N_{(p_x > 0)}}
 \label{Eq.int_v1}
\end{equation}
In our model calculations, which utilize a smooth Glauber initial condition, we set the positive x-axis to align with the direction of the impact parameter vector.  Due to the well-defined direction of impact parameter vector in our initial condition, we can easily identify the $p_x$ sign of the produced hadrons in our model and can calculate the quantities mentioned in above equations.

\begin{figure}
  \centering
  \includegraphics[width=1.0\textwidth]{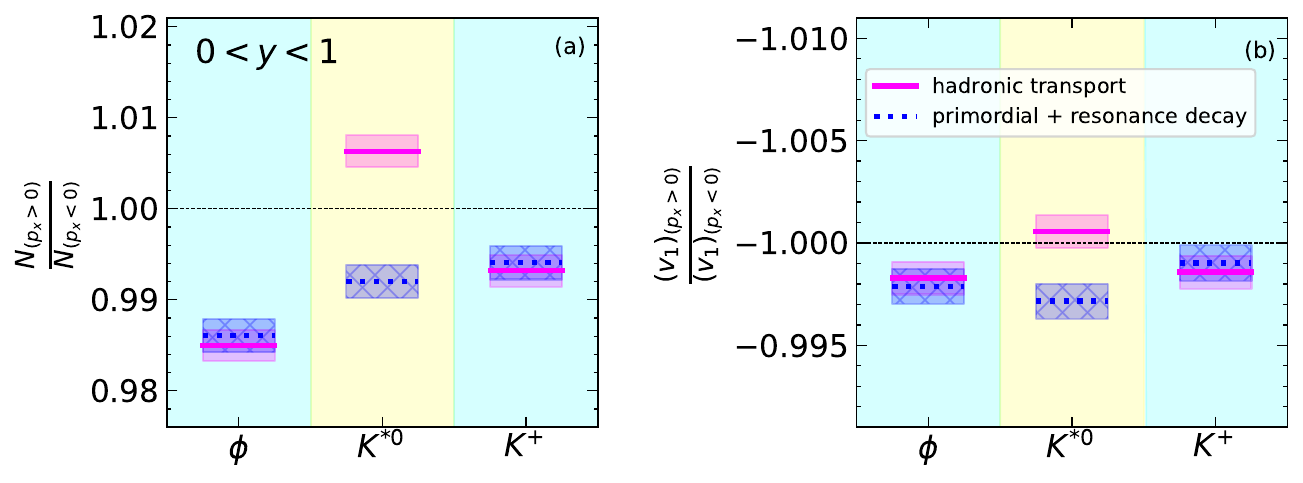} 
  \caption{(a) The rapidity integrated yield and (b) $v_1$ ratios
between hadrons of $p_x>0$ and $p_x<0$ (see Eqs. \ref{Eq.int_yld} and \ref{Eq.int_v1})
in Au+Au collisons of 10-40\% centrality at $\sNN$ = 27
GeV. The rapidity integration has been performed between
$0<y<1$. The ratios have been shown for $\phi$, $K^{*0}$
and $K^{+}$. To show the results obtained under different conditions, specifically, with or without the inclusion of hadronic transport
stages for the hadrons, we have plotted the ratios for scenarios
involving solely primordial hadrons and their resonance decay
contributions (dotted lines), as well as for cases where the
hadrons undergo interactions in the hadronic medium (solid
lines).}
  \label{fig:v1_yld_ratio_kstar}
\end{figure}

Fig. \ref{fig:v1_yld_ratio_kstar}(a) and \ref{fig:v1_yld_ratio_kstar}(b) compare the rapidity-integrated yield and directed flow ($v_1$) ratios between positive and negative $p_x$ hadrons for $\phi$, $K^{+}$ and $K^{*0}$. The results are shown with and without including a hadronic transport model (UrQMD). For $K^{+}$, the ratio of integrated yield between particles with positive and negative $p_x$ is less than one, and this value changes slightly after including hadronic interactions. Similarly, the magnitude of $v_1$ also shows a small decrease during the hadronic phase and remains less than one. These small changes indicate that the hadronic afterburner has a minimal effect on $K^{+}$. Moreover, this observation suggests that more $K^{+}$ particles are produced with negative $p_x$, leading to a negative $v_1$ of $K^{+}$ in the positive rapidity region, which persists even after the late-stage evolution. The results are similar for $\phi$ mesons because their larger lifetime leads to decay outside the fireball, minimizing the rescattering effect from the afterburner.

However, the hadronic afterburner significantly impacts the ratio of integrated yield for $K^{*0}$ mesons between positive and negative $p_x$ particles. After hadronic interactions, the ratio $N_{(p_x>0)}/N_{(p_x<0)}$ becomes greater than one, reversing the trend observed without hadronic interactions. This change in the yield ratio can be attributed to the different interactions experienced by $K^{*0}$ mesons in different momentum space regions. The initial tilted condition creates an asymmetric distribution of the hadronic fireball in both coordinate and momentum space, leading to these variations in interactions.

The short-lived $K^{*0}$ resonance decays into pions and kaons within the dense medium created in heavy ion collisions. However, if these daughter particles interact with other particles in this medium, their momenta get altered, making it impossible to reconstruct the original $K^{*0}$ from these scattered particles using invariant mass method. This is known as the "signal loss" of the $K^{*0}$ \cite{STAR:2022sir,STAR:2004bgh,STAR:2010avo,Li:2022neh,ALICE:2012pjb,ALICE:2021xyh}. Figure~\ref{fig:v1_yld_ratio_kstar}(a) shows a greater loss of signal for $K^{*0}$ particles with negative $p_x$ compared to those with positive $p_x$. This asymmetry arises from the initial uneven distribution of matter in different sides of $x$-axis in the reaction plane . This initial imbalance leads to an asymmetric evolution of the fireball, resulting in a significantly larger number of pions flowing with negative $p_x$ \cite{Bozek:2010bi,Jing:2023zrh,Jiang:2021ajc}. Consequently, $K^{*0}$ particles with negative $p_x$ encounter a denser medium, leading to a higher rate of rescattering and a more significant loss of signal. Since we can reconstruct more $K^{*0}$ with positive $p_x$, the measured $v_1$ becomes biased towards $p_x>0$ . This bias translates to an overall positive $v_1$ for $K^{*0}$ in the positive rapidity region. Due to the inherent symmetry of the collision process, the same mechanism leads to a negative $v_1$ for $K^{*0}$ in the negative rapidity region.

\begin{figure}
  \centering
  \includegraphics[width=1.0\textwidth]{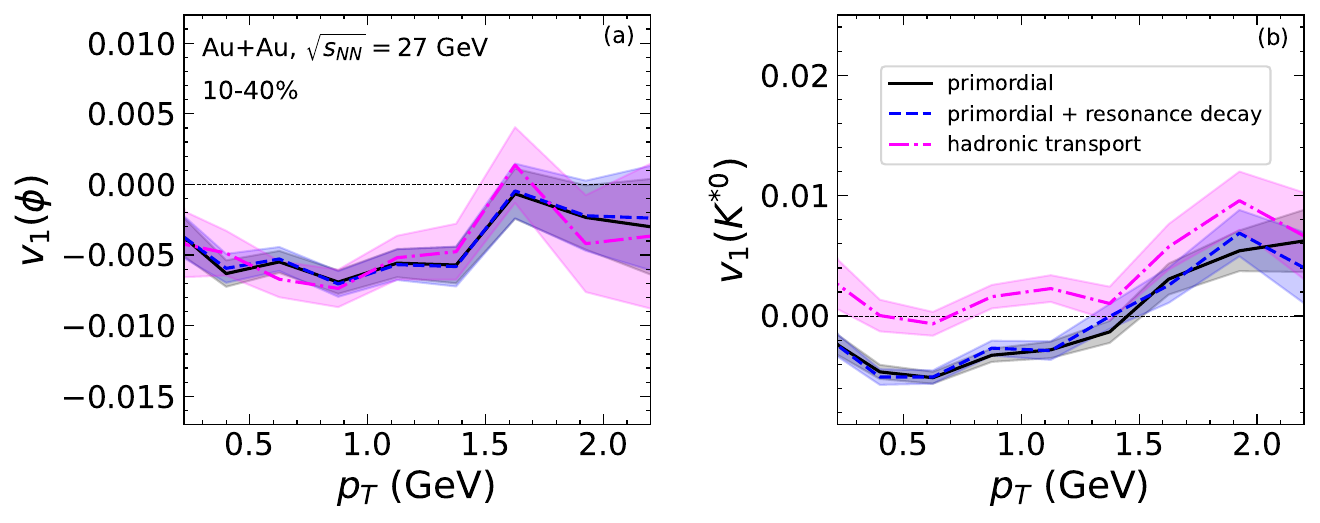} 
  \caption{The $p_T$ differential directed flow ($v_1$) has been plotted
for (a) $\phi$ and (b) $K^{*0}$ in Au+Au collisons of 10-40\% centrality at 
$\sNN$ = 27 GeV. The calculation has been done for
the particles produced within the rapidity range $0<y<1$.
The $v_1$ of the primordial hadrons which are produced directly
from the hypersurface has been represented in solid lines. The
dashed lines represent the $v_1$ calculations of the hadrons after
performing the decay of resonances to stable hadrons whereas
the dashed-dotted lines represents the $v_1$ calculations of the
hadrons after they pass through the UrQMD hadronic afterburner.}
  \label{fig:v1_pt_kstar}
\end{figure}

\begin{figure}
  \centering
  \includegraphics[width=1.0\textwidth]{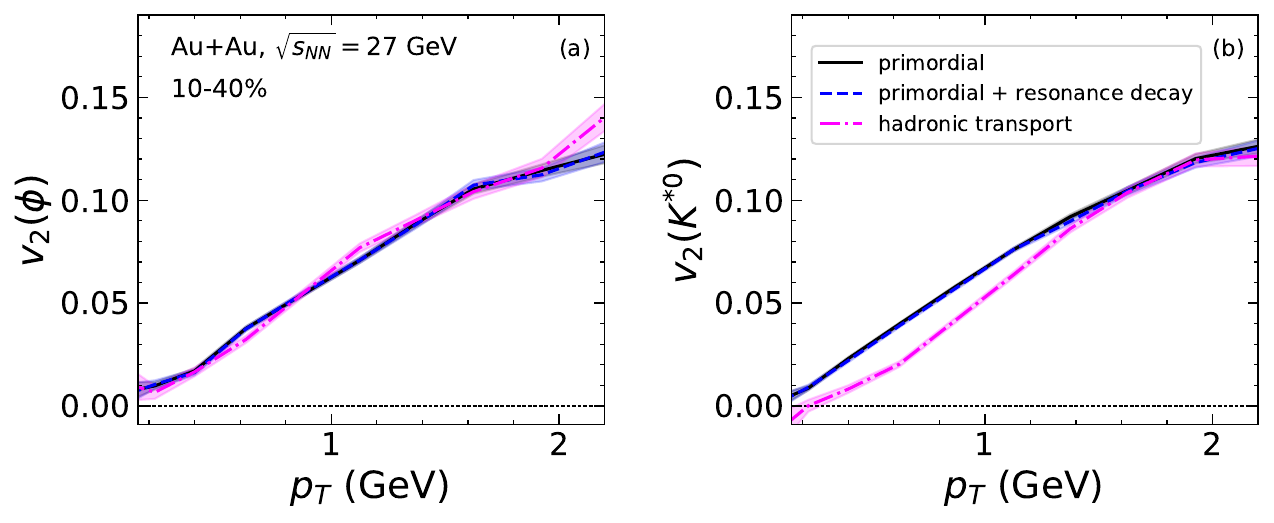} 
  \caption{Same as in Fig. \ref{fig:v1_pt_kstar} but for the $p_T$ differential elliptic flow ($v_2$) of the hadrons produced within the rapidity range $\vert y \vert < 0.5$.}
  \label{fig:v2_pt_kstar}
\end{figure}

The $p_T$ differential $v_1$ for both $\phi$ and $K^{*0}$ are depicted in Fig.\ref{fig:v1_pt_kstar}(a) and \ref{fig:v1_pt_kstar}(b) respectively. This calculation accounts for particles produced at positive rapidity within the range $0 < y < 1$. The influence of the afterburner on the $v_1$ of $\phi$ is negligible, as anticipated. However, for $K^{*0}$, a noticeable impact of hadronic interactions emerges, especially at low $p_T$. Notably, the $v_1$ values of $K^{*0}$ resonance in the higher $p_T$ region ($p_T > 1.5$ GeV/c) are already positive at the hadronization surface, suggesting they reside on the less dense side of the fireball, thereby experiencing reduced rescattering effects on their decay products. Additionally, due to their high $p_T$, their likelihood of decaying outside the fireball increases. This allows for the escape of daughters from hadronic medium without encountering rescattering, facilitating the reconstruction of $K^{*0}$.

The influence of hadronic interactions on the resonances is also evident in the elliptic flow. In Fig. \ref{fig:v2_pt_kstar}, we display the $p_T$ differential $v_2$ of $\phi$ and $K^{*0}$ resonances in the mid-rapidity region ($\mid y \mid < 0.5$). Notably, it has been observed that the magnitude of elliptic flow for $K^{*0}$ decreases for $p_T < 1.5$ GeV/c, suggesting a higher loss of $K^{*0}$ along the in-plane direction compared to the out-of-plane direction \cite{Li:2022neh,Oliinychenko:2021enj}. However, this suppression diminishes at higher $p_T$, consistent with the trends observed in the $v_1$ analysis. Conversely, the effect of hadronic transport on the $v_2$ of $\phi$ is minimal in all $p_T$ range.

\begin{figure}[htbp]
    \centering
    \begin{minipage}{0.47\textwidth} 
        \centering
        \includegraphics[width=\textwidth]{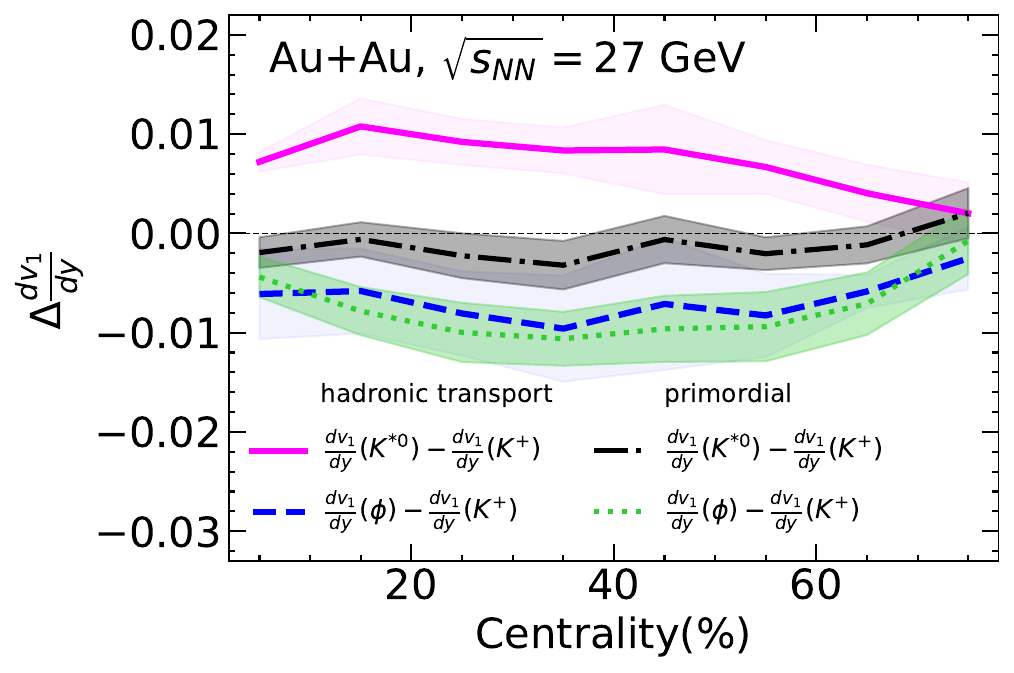}
        \caption{The splitting of directed flow slopes ( $dv_1/dy$ ) of resonances $\phi$ and $K^{*0}$ with $K^{+}$ has been plotted as a function
of centrality for Au+Au collisions at $\sNN$ = 27 GeV. The
shaded bands denote the statistical uncertainties in the model
calculations.}
        \label{fig:dv1dy_cent_kstar}
    \end{minipage}
    \hfill 
    \begin{minipage}{0.47\textwidth} 
        \centering
        \includegraphics[width=\textwidth]{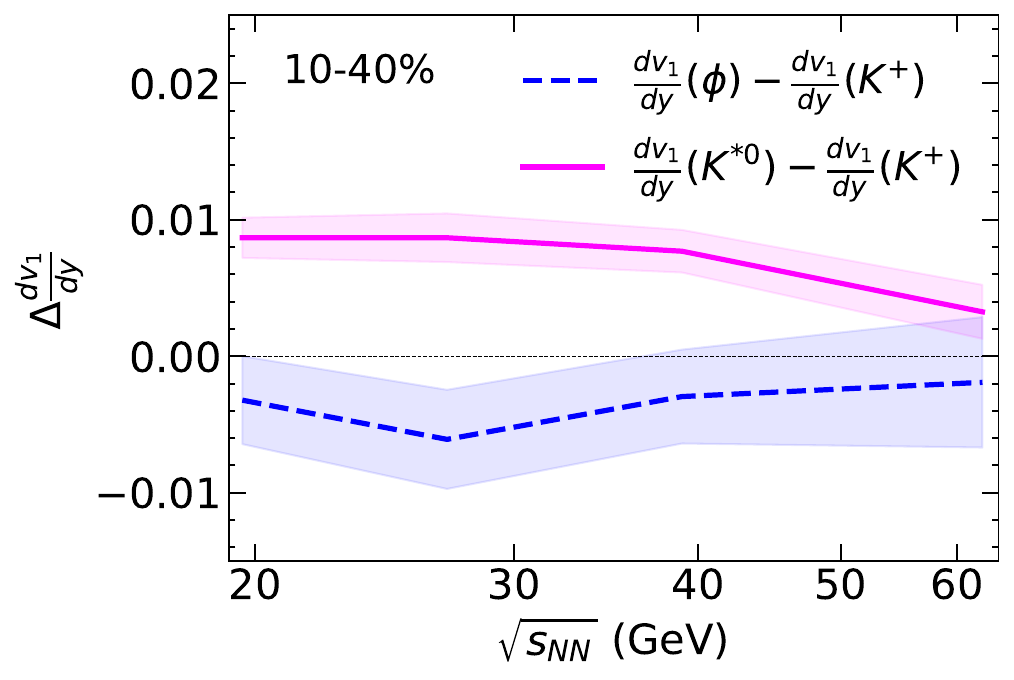}
        \caption{ The splitting of directed flow slope ($dv_1/dy$) between $K^{*0}$ and $K^{+}$, along with $\phi$ and $K^{+}$ has been plotted as a function of $\sNN$ for Au+Au collisions of
10-40\% centrality. The
shaded bands denote the statistical uncertainties in the model
calculations.}
        \label{fig:dv1dy_sNN_kstar}
    \end{minipage}
    \label{fig:combined_figures}
\end{figure}

Our investigation underscores a noteworthy contrast in how hadronic interactions affect the $v_1$ values of $\phi$ and $K^{*0}$ particles. This difference becomes especially interesting when we examine its dependence on the impact parameter (centrality) and collision energies. Such an investigation holds the potential to unlock a deeper understanding of the role played by the hadronic phase in heavy ion collisions. While the absolute magnitude and sign of the $v_1$ slope for $K^{*0}$ and $\phi$ may vary with centrality and $\sqrt{s_{NN}}$, the relative $v_1$ between $K^{*0}$ and $K^+$, or between $\phi$ and $K^+$, consistently reflects the unique influence of hadronic interactions on these resonances. To elucidate this further, we present the centrality dependence of the mid-rapidity $v_1$ slope splitting ($\Delta \frac{dv_1}{dy}$) between $K^{*0}$ and $K^{+}$ in Au+Au collisions at $\sqrt{s_{NN}} = 27$ GeV in Fig. \ref{fig:dv1dy_cent_kstar}. Remarkably, the magnitude of this splitting is minimal in peripheral collisions and gradually increases towards central collisions. In peripheral collisions, the duration of the hadronic phase lifetime is relatively shorter, coupled with a reduced production of pions, leading to a lesser afterburner effect in the hadronic phase. Conversely, in central collisions, the hadronic phase persists for a more extended period, allowing for a more pronounced influence of the hadronic afterburner.

However, the $v_1$ values of $\phi$ and $K^{+}$ appear to be minimally influenced by hadronic scatterings in all centralities. Our results reveal that the quantity $\left[ \frac{dv_1}{dy} (\phi) -\frac{dv_1}{dy} (K^{+}) \right]$ consistently maintains a negative value across all centralities. This observation suggests that $\vert \frac{dv_1}{dy} (\phi)\vert > \vert \frac{dv_1}{dy} (K^{+})\vert $ holds true for all centralities, aligning with expectations based on the mass hierarchy. Additionally, for comparison, we also present $\left[ \frac{dv_1}{dy} (K^{*0}) -\frac{dv_1}{dy} (K^{+}) \right]$ for primordial hadrons, which consistently exhibits a negative sign, following the mass hierarchy. Interestingly, the sign of $\left[ \frac{dv_1}{dy} (K^{*0}) -\frac{dv_1}{dy} (K^{+}) \right]$ shifts to positive after the hadronic transport stage at all centralities. This shift serves as a clear indication of the effect of the hadronic stage on $K^{*0}$ resonance, a phenomenon that could be measured in experiments.

Moreover, the dependence of the splitting of directed flow slope between $K^{*0}$ and $K^{+}$ on collision energy has been illustrated in Fig.~\ref{fig:dv1dy_sNN_kstar} for Au+Au collisions with 10-40\% centrality. It is evident that the splitting decreases at higher $\sqrt{s_{NN}}$ and becomes more pronounced at lower collision energies. This phenomenon is attributed to the larger tilt of the fireball at lower $\sNN$ \cite{Jiang:2023fad,Bozek:2010bi,Du:2022yok}, resulting in a more substantial asymmetric hadronic fireball that strongly affects the $v_1$ of $K^{*0}$. For comparison, the splitting between $\phi$ and $K^+$, which remains relatively constant at $\sim-0.005$, is also depicted. It's worth noting that, at lower energies, baryon stopping physics may induce non-trivial dynamics of conserved charges, potentially impacting the directed flow of these resonances, particularly that of $K^{*0}$ due to its strangeness content.

\section{Splitting of elliptic flow between hadrons with $p_x > 0$ and $p_x < 0$ }

In this section, we will elucidate how the breaking of forward-backward symmetry in non-central relativistic heavy-ion collisions not only induces the rapidity-odd component of directed flow but also creates the rapidity-odd component of higher order flow harmonics such as $v_3$ and $v_5$ \cite{STAR:2023duf,HADES:2020lob,Parida:2022lmt}. These rapidity-odd flow harmonics differ from the conventional flow harmonics measured in experiments, which are typically determined with respect to their respective event plane angles. Furthermore, we will explore how the emergence of these odd harmonics leads to the splitting of elliptic flow ($\Delta v_2$) between the final-state hadrons produced from different regions of the momentum space \cite{Chen:2021wiv,Parida:2022lmt,Zhang:2021cjt}. This splitting of $v_2$ has been proposed as a sensitive probe of the initial-state distribution of matter in the fireball.

When two non-central relativistic heavy nuclei collide, a considerable amount of angular momentum is transferred to the locally thermalized fireball \cite{Liang:2004ph,Liang:2004xn,Becattini:2007sr,Betz:2007kg,Ipp:2007ng,Becattini:2013vja,Pang:2016igs}, although the majority of this momentum is carried by the spectator nucleons. This transfer of non-zero angular momentum to the thermalized fireball initiates a rotational motion within the fluid, indicating the presence of a pronounced vorticity field \cite{Liang:2004ph,Becattini:2007sr,Becattini:2013vja,Pang:2016igs,Alzhrani:2022dpi,Ryu:2021lnx,Bhadury:2022ulr,Sahoo:2024egx}. In Ref. \cite{Chen:2021wiv}, it was suggested that the splitting of elliptic flow among particles produced on opposite sides of the impact parameter axis at non-zero rapidity is a consequence of the global vorticity. However, a subsequent study has revealed that this splitting of $v_2$ is primarily influenced by the presence of directed flow ($v_1$) within the system \cite{Zhang:2021cjt}, rather than being a direct and exclusive outcome of the global vorticity. Since $v_1$ can also be influenced by global vorticity~\cite{Csernai:2011qq,Ryu:2021lnx}, the splitting in $v_2$ ($\Delta v_2$) could be indirectly affected by vorticity.

The studies on $\Delta v_2$ in Ref.\cite{Chen:2021wiv,Zhang:2021cjt}, have utilized transport models like AMPT \cite{Lin:2004en,Jiang:2016woz} or BAMPS \cite{Xu:2004mz,Xu:2007aa}. Since $\Delta v_2$ is understood as a consequence of $v_1$ in Ref. \cite{Zhang:2021cjt}, it becomes crucial to examine this observable within a model capable of accurately reproducing the available experimental measurements of $v_1$. However, the used transport models lack in accurately describing the data of $v_1$, which raises concerns about the reliability of the predictions regarding $\Delta v_2$ in such models. To address this, we have employed our hybrid framework to investigate $\Delta v_2$, employing appropriate initial conditions consistent with explaining the data of $v_1$. Before presenting our model predictions regarding $\Delta v_2$, we have first discussed about the definition of $\Delta v_2$ and its dependence on $v_1$ below.

The azimuthal distribution of hadrons in the transverse plane perpendicular to the beam axis can be expressed using Fourier components as follows \cite{Voloshin:1994mz,Poskanzer:1998yz}:
\beq
\frac{dN}{d\phi} = \frac{1}{2\pi}\left(1 + 2 \sum_n \left(v_n \cos(n(\phi-\psi_\text{RP})) + s_n \sin(n(\phi-\psi_\text{RP}))\right) \right)
\eeq
Here, $\psi_{\text{RP}}$ denotes the angle of the reaction plane in the laboratory frame, while $v_n$ and $s_n$ are Fourier coefficients that characterize the distribution. The splitting of $v_2 =  \la \cos(2(\phi-\psi_\text{RP})) \ra $ observed in distinct regions of the final hadron momentum space is defined as \cite{Chen:2021wiv,Zhang:2021cjt} :
\beq
\Delta v_2 = {v_2}^{\text{R}} - {v_2}^{\text{L}}
\label{eq.deltav2}
\eeq
where ${v_2}^{\text{R}} = \la \cos(2(\phi^{\text{R}}-\psi_\text{RP})) \ra$ with $\phi^{\text{R}}\in((\psi_{\text{RP}}-\pi/2), (\psi_{\text{RP}}+\pi/2))$ and 
${v_2}^{\text{L}} = \la \cos(2(\phi^{\text{L}}-\psi_\text{RP})) \ra$ with $\phi^{\text{L}}\in((\psi_{\text{RP}}+\pi/2), (\psi_{\text{RP}}+3\pi/2))$. Here, $\la...\ra$ refers to averaging over the particle tracks in an event.

For experimental purposes, the second-order event plane orientation $\psi_2$ and the first-order spectator plane angle $\psi_\text{SP}$ have been suggested as suitable proxies for $\psi_\text{RP}$, which is not directly measurable. However, in the present study, $\psi_\text{SP}$ offers more advantages compared to $\psi_2$. This preference is rooted in the ambiguity associated with $\psi_2$, where $\psi_{2}=\pi$ is equivalent to $\psi_2 = 0 $, thus failing to discern between the phase spaces linked with $\phi^{\text{R}}$ and $\phi^{\text{L}}$. Conversely, the first-order spectator plane angle $\psi_\text{SP}$ provides the angle between the impact parameter direction and the positive x-direction of the event in the lab frame \cite{ALICE:2013xri,STAR:2008jgm,STAR:2011gzz,Selyuzhenkov:2011zj}, which can aid in distinguishing the phase space of the produced particles. Additionally, the recently installed Event Plane Detector at large rapidities at STAR \cite{Adams:2019fpo}, could serve as a valuable tool for determining the first-order plane.

${v_2}^{\text{R}}$ and ${v_2}^{\text{L}}$ work out to be the following~\cite{Zhang:2021cjt}:
\begin{equation}
\begin{aligned}
     v_{2}^{\text{R}} &= 
    \frac{\int_{\psi_\text{RP}-\frac{\pi}{2}}^{\psi_\text{RP}+\frac{\pi}{2}} \ \cos(2(\phi-\psi_\text{RP})) \frac{dN}{d\phi} \  d \phi } {\int_{\psi_\text{RP}-\frac{\pi}{2}}^{\psi_\text{RP}+\frac{\pi}{2}}  \frac{dN}{d\phi} d \phi} \\
    &
    \approx  \frac{v_{2}  + \frac{4 v_{1}}{3 \pi} + \frac{12 v_3}{5 \pi} - \frac{20v_5}{21 \pi}  }{ 1 + \frac{4 v_1}{\pi} - \frac{4 v_3}{3 \pi} + 
    \frac{4 v_5}{5 \pi} }
    \label{eq.v2r}
    \end{aligned}
\end{equation}

\begin{equation}
    \begin{aligned}
v_{2}^{\text{L}} &=
    \frac{\int_{\psi_\text{RP}+\frac{\pi}{2}}^{\psi_\text{RP}+\frac{3 \pi}{2}} \ \cos(2(\phi-\psi_\text{RP})) \frac{dN}{d\phi} \  d \phi } {\int_{\psi_\text{RP}+\frac{\pi}{2}}^{\psi_\text{RP}+\frac{3 \pi}{2}}  \frac{dN}{d\phi}  d \phi} \\
    &\approx
    \frac{v_{2}  - \frac{4 v_{1}}{3 \pi} - \frac{12 v_3}{5 \pi} + \frac{20v_5}{21 \pi}  }{ 1 - \frac{4 v_1}{\pi} + \frac{4 v_3}{3 \pi} 
    \frac{4 v_5}{5 \pi} }
    \label{eq.v2l}
    \end{aligned}
\end{equation}
Hence, 
\begin{equation}
\Delta v_2  \approx \frac{8v_1}{3\pi} + \frac{24v_3}{5\pi} - \frac{40v_5}{21\pi}.
\label{eq.deltav2ap}
\end{equation}

\begin{figure}[htbp]
    \centering
    \begin{minipage}{0.47\textwidth} 
        \centering
        \includegraphics[width=\textwidth]{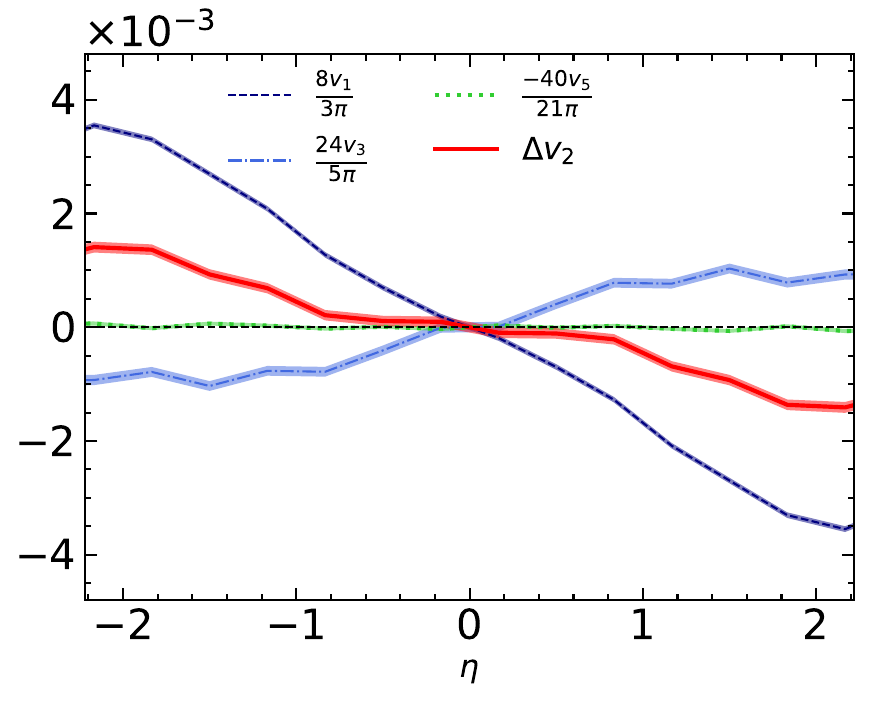}
        \caption{The prediction for $\Delta v_2$ of charged hadrons as a function of $\eta$ with a tilted initial condition in Au+Au collisions of 10-40\% centrality at $\sqrt{s_{NN}} = 200$ GeV is represented by a solid red line. Additionally, the first three dominant flow harmonics contributing to this prediction (see Eq. \ref{eq.deltav2ap}) are also presented.}
        \label{fig:Deltav2_eta}
    \end{minipage}
    \hfill 
    \begin{minipage}{0.47\textwidth} 
        \centering
        \includegraphics[width=\textwidth]{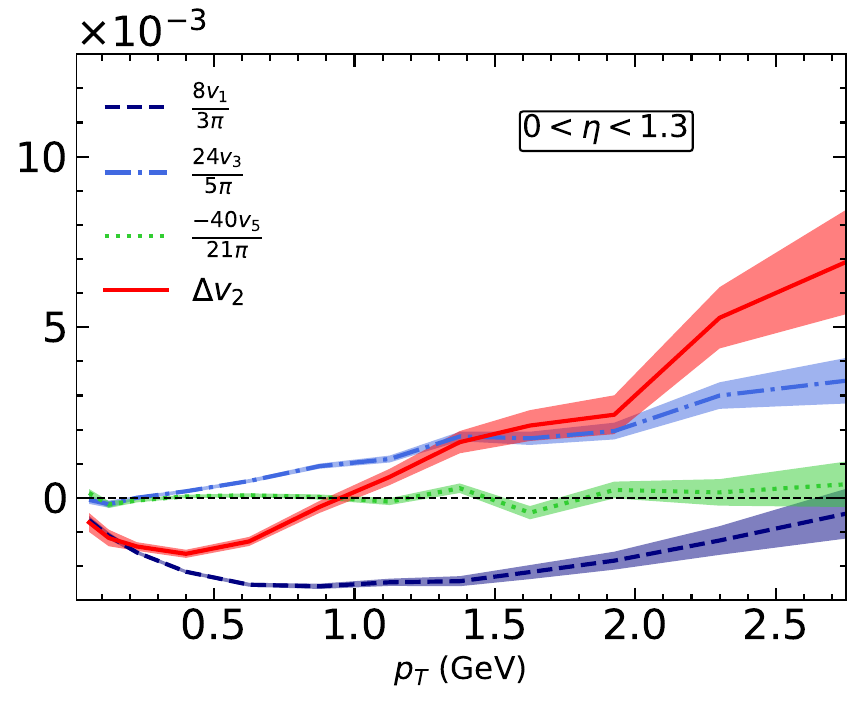}
        \caption{The prediction for $\Delta v_2$ of charged hadrons as a function of $p_T$ in Au+Au collisions of 10-40\% centrality at $\sqrt{s_{NN}} = 200$ GeV is represented by a solid red line. Additionally, the $p_T$ dependence of the first three dominant flow harmonics contributing to this prediction (see Eq. \ref{eq.deltav2ap}) are also presented. }
        \label{fig:Deltav2_pt}
    \end{minipage}
    \label{fig:combined_figures}
\end{figure}

In equations \ref{eq.v2r} and \ref{eq.v2l}, contributions beyond the seventh order odd harmonics have been omitted. Additionally, in equation \ref{eq.deltav2ap}, only terms linear in the flow harmonics have been represented, as they are sufficient for estimating $\Delta v_2$.
Furthermore, it is essential to note that all the flow harmonics on the right-hand side of Eq. \ref{eq.deltav2} are defined with respect to the reaction plane angle \cite{STAR:2023duf,HADES:2020lob} and therefore differ from the conventional flow harmonics which are measured with respect to their respective event plane angles \cite{Bhalerao:2011yg,Qiu:2012uy}. These flow harmonics in Eq. \ref{eq.deltav2} are termed as the rapidity-odd flow coefficients and stem from the geometry of the collision rather than from fluctuations, unlike the conventional odd-order flow harmonics \cite{Bhalerao:2011yg,Qiu:2012uy}. As $\Delta v_2$ primarily arises from the rapidity-odd component of the odd flow harmonics, it will also exhibit rapidity-odd behavior.

We investigate Au+Au collisions at $\sqrt{s_{NN}} = 200$ GeV. In Fig. \ref{fig:Deltav2_eta}, we present the model predictions for the $\eta$ dependence of $\Delta v_2$ of charged hadrons. Additionally, we include the odd harmonics along with their corresponding coefficients, as suggested by Eq. \ref{eq.deltav2ap}, as these play a significant role in determining $\Delta v_2$. In Fig. \ref{fig:Deltav2_eta}, we have observed that, $\Delta v_2$ emerges as a result of the interplay between the coefficients $8v_1/3\pi$ and $24v_3/5\pi$, which exhibit opposite signs. Although the $8v_1/3\pi$ term holds a marginal advantage, leading to $\Delta v_2$ adopting its sign. For $|\eta| < 1$, $\Delta v_2$ remains approximately $10^{-4}$ and experiences substantial growth for larger values of $\eta$. Notably, the contribution from $v_5$ is negligible across the entire $\eta$ range.

In Fig. \ref{fig:Deltav2_pt}, we explore the dependence of $\Delta v_2$ on transverse momentum ($p_T$). We analyze $\eta$ ranges: $0 < \eta < 1.3$.  $\Delta v_2$ displays a negative sign for smaller $p_T$, reaching a turning point at approximately $p_T \approx 0.5$ GeV, crossing zero around $p_T \approx 1$ GeV, and steadily increasing thereafter. Remarkably, for $p_T > 1.5$ GeV, $v_3$ emerges as the dominant contributor for $\Delta v_2$.

It's worth noting that the tilted profile of the initial energy and net-baryon density not only generates $v_1$ but also gives rise to the rapidity odd component of $v_3$, whose magnitude is comparable to $v_1$. The precise experimental measurements of $v_3$, $\Delta v_2$ and their comparison with the model could offer additional constraints on the initial three-dimensional distribution of matter.

\section{Chapter summary}
In this chapter, we introduced a new ansatz for the parametrized initial baryon distribution. Unlike the conventional approach, where the transverse baryon profile is assumed to be proportional to the participant distribution, our scheme incorporates contributions from both participant and binary collision sources. This two-component deposition framework is inspired by microscopic models, which suggest that baryon stopping is more in regions with higher binary collisions. Additionally, the binary collision term in our prescription can be interpreted within the baryon junction picture as a contribution from double-junction stopping. The double-junction stopping involve the stopping of two baryon units and has a forward-backward symmetric deposition profile which is similar to the deposition by binary collsions in our model. In our model, the participant sources are associated with a forward-backward asymmetric rapidity profile, while binary collisions contribute symmetrically in rapidity. A free parameter, $\omega$, is introduced to control the relative contributions of participant and binary collision sources to the initial baryon profile. We find that $\omega$ plays a crucial role in shaping the initial tilted geometry of baryon, which has significant phenomenological implications for the directed flow ($v_1$) of baryons and anti-baryons.

We implemented our baryon deposition model, coupled with a tilted fireball, as the initial condition in a hybrid framework to study Au+Au collisions across $\sqrt{s_{NN}} = 7.7-$200 GeV. This hybrid approach combines hydrodynamic evolution with a subsequent hadronic transport phase to simulate the collision dynamics. During the hydrodynamic phase, we incorporated a non-zero baryon diffusion coefficient to account for baryon transport in the medium. Using this framework, we identified regions in the parameter space where the resulting hydrodynamic evolution reproduces the observed directed flow $v_1$ of identified hadrons, including the challenging baryon-antibaryon $v_1$ splitting across the entire $\sqrt{s_{NN}}$ range. Notably, our model captures the experimentally observed double sign change in the rapidity slopes of net proton and net lambda $v_1$ between $\sqrt{s_{NN}} = 7.7$ and 39 GeV. Furthermore, we found that baryon diffusion plays a crucial role in shaping the $v_1$ of baryons and anti-baryons, particularly in driving the sign change of the proton $v_1$ observed between $\sqrt{s_{NN}} = 11.5$ and 7.7 GeV.
To ensure a comprehensive description of the system, we carefully tuned the model parameters to simultaneously describe the $v_1$ of identified hadrons with wide range of bulk observables. These include the pseudorapidity dependence of charged particle yields, the rapidity distribution of net-proton yields, the mid-rapidity $p_T$ spectra and mean $p_T$ of identified hadrons, and the elliptic flow $v_2$ of charged hadrons.

Using this model, we explored two intriguing phenomena related to $v_1$: the influence of hadronic interactions on the flow of the $K^{*0}$ resonance and the splitting of the $v_2$ of charged hadrons produced in different regions of phase space, particularly for particles with $p_x > 0$ versus $p_x < 0$.

Our analysis reveals that the $v_1$ of the $K^{*0}$ is significantly affected during the hadronic stage due to asymmetric signal loss on opposite sides of the $p_x$-axis in momentum space, a result of the tilted fireball geometry. For reference, we also studied the flow of the $\phi$ meson, which is largely unaffected by the hadronic afterburner owing to its small cross-section and longer lifetime. To investigate the energy and centrality dependence of hadronic effects on the $v_1$ of the $K^{*0}$, we analyzed the splitting between the $v_1$ of $K^{+}$ and $K^{*0}$. This splitting was found to be minimal in peripheral collisions and grew progressively larger in more central collisions, reflecting the longer duration of the hadronic phase and the increased yield of produced hadrons in central collisions. Furthermore, we observed that the impact of hadronic interactions becomes more pronounced at lower collision energies. This behavior can be attributed to the largeer tilt of the  fireball at lower energies, resulting in a more asymmetric distribution of the hadronic medium in both coordinate and momentum space. This study underscores the importance of a comprehensive, quantitative investigation of the flow coefficients of $K^{*0}$ and other resonances. In this context the comparisons between model predictions and experimental data of $v_1$ of resonances could provide valuable insights into the properties and dynamics of the hadronic phase.

The geometry of a non-central relativistic heavy-ion collision generates substantial angular momentum in the initial state, leading naturally to rapidity-odd directed flow ($v_1$). This observable provides insight into the longitudinal profile of the fireball. Recent studies have also shown that this large initial angular momentum induces a splitting in the elliptic flow, $\Delta v_2$, between hadron produced with the momentum that are parallel and anti-parallel to the impact parameter direction. The earlier works related to this used transport model, which, however, did not successfully describe the $v_1$ data. In this study, we revisited the estimation of $\Delta v_2$ at $\sqrt{s_{NN}} = 200$ GeV using our hybrid framework with a tilted initial condition, which is capable of reproducing the $v_1$ data. Consistent with prior findings, we confirmed that $v_1$ is the dominant contributor to $\Delta v_2$. However, unlike the transport model studies, our calculations revealed that the tilted initial condition also produces a significant rapidity-odd $v_3$, which contributes notably to $\Delta v_2$. These findings highlight that $\Delta v_2$ can serve as a complementary observable to $v_1$, offering additional constraints on the initial rapidity profile of the fireball in heavy-ion collisions.

\def \la{\langle}
\def \ra{\rangle}
\chapter{Baryon inhomogenity driven charge dependent directed flow}
In non-central relativistic heavy-ion collisions, fast-moving spectator nucleons, which are positively charged, generate an intense magnetic field. By applying the Biot-Savart law, this magnetic field is estimated to reach magnitudes of approximately $10^{18}$ Gauss in mid-central collision events \cite{Deng:2012pc,McLerran:2013hla, Voronyuk:2011jd, Kharzeev:2007jp, Skokov:2009qp, Tuchin:2013ie}. While spectator nucleons serve as the primary source of this electromagnetic (EM) field, participant nucleons also contribute significantly \cite{Kharzeev:2007jp, Gursoy:2018yai, Dash:2023kvr}. At the Relativistic Heavy Ion Collider (RHIC), a key objective is to detect signals associated with this EM field \cite{STAR:2023jdd,STAR:2023wjl,Taseer:2024sho}. However, this task is particularly challenging due to the rapid decay of the field \cite{Tuchin:2013apa,Gursoy:2018yai,Roy:2017yvg,Panda:2024ccj,Inghirami:2016iru}. Despite these difficulties, substantial efforts are underway to observe EM field signals and investigate the rich phenomena associated with it \cite{STAR:2009wot,ALICE:2020siw,Burnier:2011bf,Kharzeev:2010gd,Fukushima:2008xe,Zhao:2019hta,Li:2020dwr,STAR:2021mii,Hirono:2012rt,Nakamura:2022ssn,Das:2016cwd,STAR:2019clv,ALICE:2019sgg,STAR:2016cio,Bzdak:2019pkr,Kurian:2019nna,K:2021sct,Mitra:2016zdw,Singh:2023pwf,Satapathy:2022xdw,Satapathy:2021cjp}.

Charged particles within the post-collision fireball are subjected to EM fields arising from three primary effects: Coulomb, Faraday, and Lorentz forces \cite{Gursoy:2014aka,Gursoy:2018yai}. First, the spectator nucleons generate a non-zero electric field within the medium, which exerts a Coulomb force on charged particles. Second, as the magnetic field decays over time, Faraday’s law induces an electric field component in the reaction plane. Additionally, as the medium expands with strong longitudinal flow, boosting into the fluid's local rest frame causes the magnetic field of the lab frame to produce an electric field in the fluid rest frame \cite{Gursoy:2018yai}. This field acts analogously to a Lorentz force in the lab frame. Each of these electric field components aligns along the impact parameter direction ($x$-axis). Due to the geometric configuration of the collision, the Coulomb and Faraday forces align parallel to the x-axis, while the Lorentz force opposes them. The net force separates oppositely charged particles along the x-axis, an effect potentially observable in the final momentum distribution of produced particles. This separation can be quantified by measuring the directed flow splitting ($\Delta v_1$) between oppositely charged hadrons \cite{Gursoy:2014aka,Gursoy:2018yai,Zhang:2022lje,Inghirami:2019mkc,Panda:2023akn,Sun:2023adv,Coci:2019nyr}.

The STAR collaboration has measured the splitting of the mid-rapidity directed flow slope ($\Delta \frac{d v_1}{dy}$) for pairs such as $\pi^{+}-\pi^{-}$, $K^{+}-K^{-}$, and $p-\bar{p}$ across different centralities \cite{STAR:2023jdd}. These measurements reveal a change in the sign of $\Delta \frac{d v_1}{dy}$ from central collisions, where the EM field is weaker, to peripheral collisions, where the EM field is stronger. This observed sign reversal is attributed to the influence of the initial EM field, with the interplay of Coulomb, Faraday, and Lorentz forces creating a complex dependence of $\Delta v_1$ on centrality. However, this splitting in directed flow is not solely due to the EM field. Baryon inhomogeneities within the system can also produce similar splitting effects in oppositely charged hadron pairs \cite{Bozek:2022svy,Parida:2022ppj,Parida:2022zse,Jiang:2023fad}.

In relativistic heavy-ion collisions, a considerable amount of net baryon number and electric charge is deposited in the fireball, with baryon stopping being particularly prominent in the mid-rapidity region at lower collision energies \cite{STAR:2017sal,BRAHMS:2003wwg,BRAHMS:2009wlg,NA49:2010lhg}. The distribution of these conserved charges depends strongly on the collision energy and centrality \cite{Ranft:2000sf,Mehtar-Tani:2009wji,Capella:1999cz,Li:2016wzh,Li:2018ini,Bialas:2016epd}. Model calculations indicate that an initially uneven distribution of net baryons leads to a $v_1$ splitting between protons and antiprotons \cite{Bozek:2022svy,Parida:2022ppj,Parida:2022zse,Jiang:2023fad}. Similarly, inhomogeneities in the distribution and evolution of net strangeness and electric charge are expected to cause splitting between $K^{+}-K^{-}$ and $\pi^{+}-\pi^{-}$ pairs. Consequently, both conserved charge stopping and the EM field contribute to the observed $v_1$ splitting for these hadron pairs. To isolate the effect of the EM field, it is essential to disentangle it from the background contributions arising from conserved charge stopping.

In this work, we utilize the initial baryon deposition model described in the previous chapter. This model provides a consistent description of the $v_1$ of identified hadrons, effectively capturing the $v_1$ splitting between protons and antiprotons. By employing this model, we demonstrate that the initial baryon stopping and its subsequent diffusion during the evolution phase reproduces the observed sign change in $\Delta \frac{d v_1}{dy} (p-\bar{p})$ from central to peripheral collisions, as reported in experimental data \cite{STAR:2023jdd}. It is important to emphasize that the model calculations do not include any effects from the EM field. This highlights the need for further detailed investigations to accurately disentangle EM field signals from the background contributions of baryon stopping. However, the model has certain limitations, as it evolves only the net-baryon density independently, while the evolution of strangeness and electric charge is coupled to baryon density by the constraint taken in the used equation of state (EoS), NEoS-BQS \cite{Monnai:2019hkn}. Despite these constraints, our findings provide strong evidence that conserved charge stopping plays a significant role in influencing $\Delta v_1$, thereby complicating the precise extraction of EM field signals from this observable.

\section{Baryon diffusion effect on the centrality dependence of directed flow splitting}
We begin our analysis with Au+Au collisions at $\sNN = 27$ GeV, an intermediate collision energy with significant baryon stopping which provides an ideal setting to investigate the impact of baryon diffusion on the $\Delta v_1$ observable. The availability of experimental data for $\Delta v_1$ at this energy further enables direct comparisons between model predictions and observations \cite{STAR:2023jdd,STAR:2014clz,STAR:2017okv,STAR:2023wjl}. Later in this chapter, we will extend our study to include results at $\sNN = 200$ GeV.

\begin{table}[ht]
\centering
\begin{tabular}{|p{0.6cm}|p{0.6cm}|p{1.2cm}|p{0.5cm}|p{0.5cm}|p{0.5cm}|p{0.7cm}|p{0.7cm}|p{0.6cm}|p{0.6cm}|}
\hline  
$C_B$ & $\tau_0$ \tiny{(fm)} &$\epsilon_{0}$  \tiny{(GeV/fm$^{3}$)} &$\eta_{0}$ & $\sigma_{\eta}$ &  $\eta_{0}^{n_{B}}$ & $\sigma_{B,-}$ & $\sigma_{B,+}$ & $\omega$ & $\eta_m$ \\
\hline  
 0.0 & 1.2 & 2.4 & 1.3 & 0.7 & 1.9 & 1.1 & 0.2 & 0.17 & 1.0 \\ 
\hline
 1.0 & 1.2 & 2.4 & 1.3 & 0.7 & 2.3 & 1.1 & 0.2 & 0.11 & 1.1 \\ 
\hline
\end{tabular}
\caption{Parameters used in hydrodynamic simulations of Au+Au collsions at $\sNN=27$ GeV for cases with zero baryon diffusion ($C_B=0$) and non-zero baryon diffusion ($C_B=1$).}
\label{tab:param_table_27}
\end{table}

\begin{wrapfigure}{l}{0.5\textwidth} 
    \centering
    \includegraphics[width=0.45\textwidth]{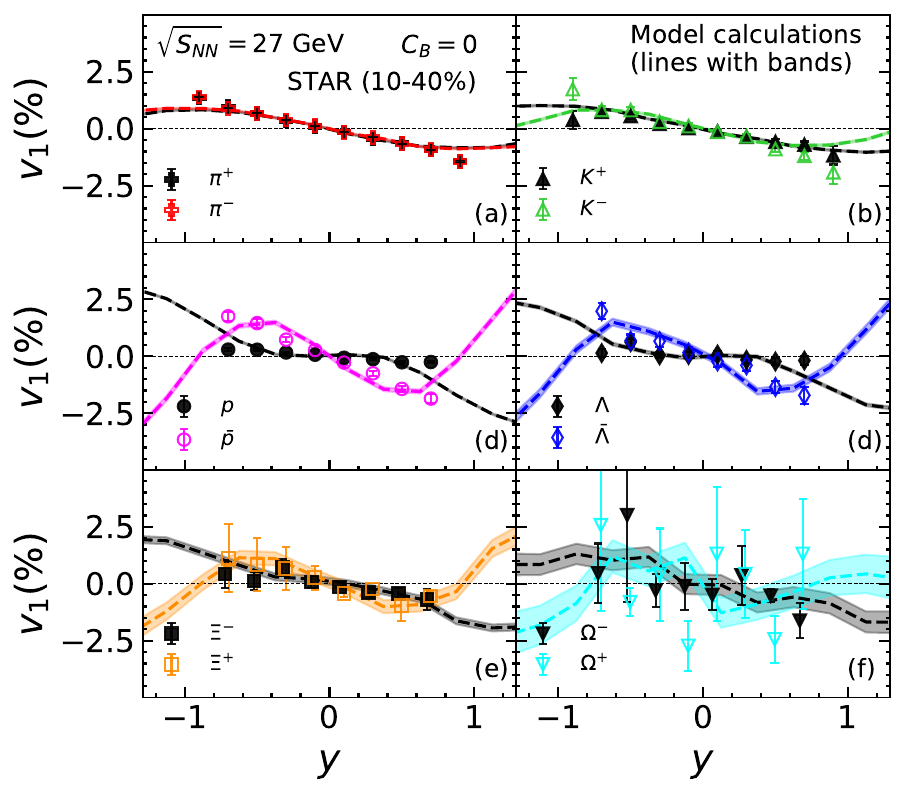}
    \caption{Rapidity-dependent directed flow ($v_1(y)$) for identified hadrons in Au+Au collisions at $\sNN = 27$ GeV, compared with STAR experimental data \cite{STAR:2014clz,STAR:2017okv,STAR:2023wjl}. Model calculations correspond to the case without baryon diffusion ($C_B=0$). Panels (a) and (b) show mesons, where black lines represent positively charged mesons and colored lines correspond to negatively charged mesons. Panels (c), (d), (e), and (f) display baryons (black lines) and anti-baryons (colored lines). Experimental data for $\pi^{+}, p, \bar{p}$ are from Ref. \cite{STAR:2014clz}; $\Lambda$ and $\bar{\Lambda}$ from Ref. \cite{STAR:2017okv}; and $\Xi^{\pm}$ and $\Omega^{\pm}$ from Ref. \cite{STAR:2023wjl}.}
    \label{fig:AuAu27idenpartv1CB0}
    \vspace{1em}
    \includegraphics[width=0.45\textwidth]{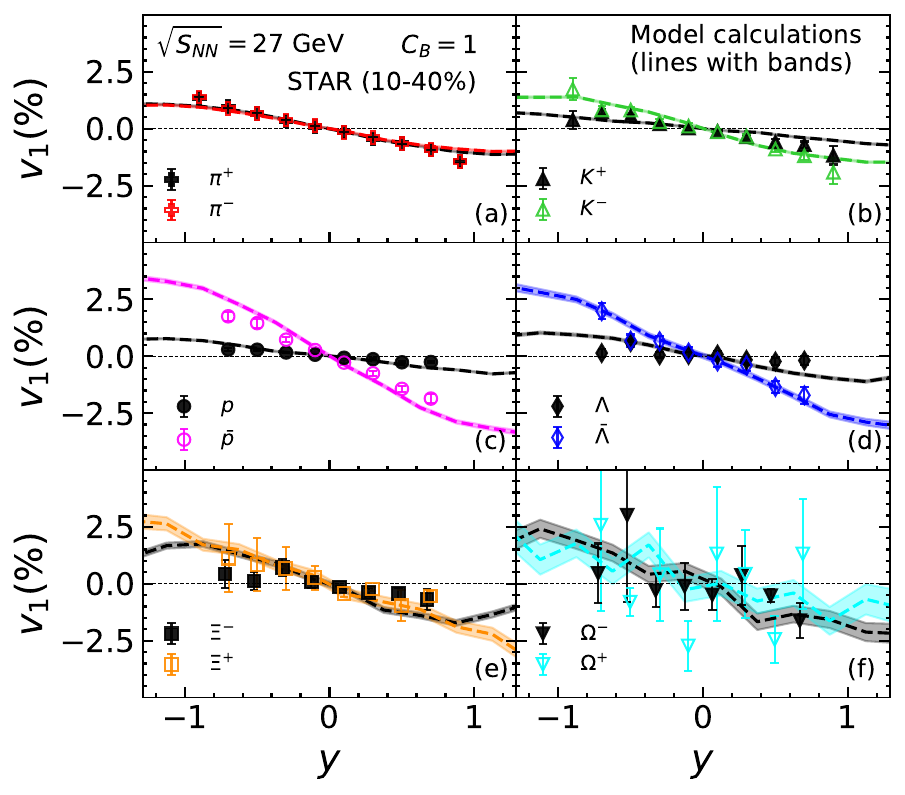}
    \caption{Same as Figure \ref{fig:AuAu27idenpartv1CB0}, but for the case with non-zero baryon diffusion ($C_B=1$).}
    \label{fig:AuAu27idenpartv1CB1}
\end{wrapfigure}

Hydrodynamic simulations were conducted for two scenarios: $C_B = 0$, representing no baryon diffusion, and $C_B = 1$, which incorporates baryon diffusion. For each case, the model parameters were independently tuned to reproduce experimental data of mid-rapidity charged particle yields, net-proton yields, and the rapidity-differential $v_1$ of identified hadrons. The parameter values used in the simulations are detailed in Table \ref{tab:param_table_27}. Figs. \ref{fig:AuAu27idenpartv1CB0} and \ref{fig:AuAu27idenpartv1CB1} display the model-to-data comparison of the rapidity-differential $v_1$ for identified hadrons in the $10\text{-}40\%$ centrality range for $C_B = 0$ and $C_B = 1$, respectively.

In these figures, black lines represent model calculations for baryons, while colored lines denote anti-baryons. For mesons, black lines correspond to positively charged hadrons, and colored lines represent negatively charged hadrons. The experimental measurements from the STAR collaboration are shown as symbols \cite{STAR:2014clz,STAR:2017okv,STAR:2023wjl}. The model calculations for both $C_B = 0$ and $C_B = 1$ successfully capture the $v_1(y)$ trends around mid-rapidity for all identified hadrons.

An important observation from our model in Figs. \ref{fig:AuAu27idenpartv1CB0} and \ref{fig:AuAu27idenpartv1CB1} is the splitting of $v_1$ between $K^+$ and $K^-$.
While the model does not explicitly initialize or independently evolve net-strangeness, the splitting arises due to the presence of non-zero and spatially inhomogeneous distribution of $\mu_S$ within the medium, induced by the strangeness neutrality condition ($n_S=0$) imposed through the equation of state (EoS) \cite{Monnai:2019hkn}.

\begin{figure}
  \centering
  \includegraphics[width=1.0\textwidth]{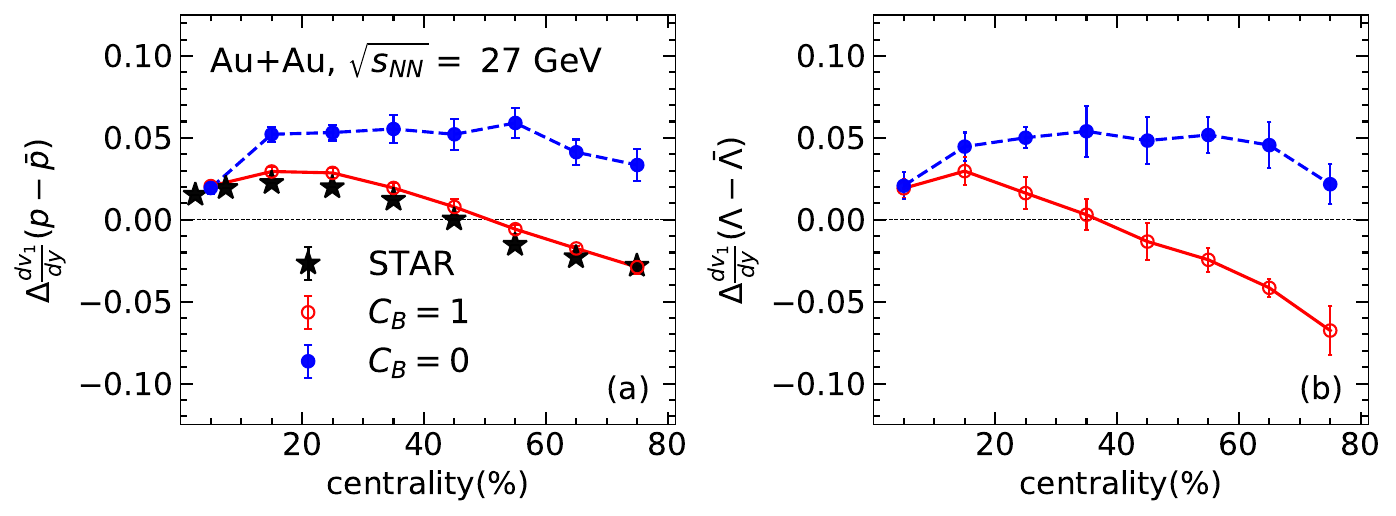} 
  \caption{Centrality dependence of the difference in mid-rapidity slope ($\Delta \frac{dv_1}{dy}$) between (a)$p$ and $\bar{p}$, and (b) $\Lambda$ and $\bar{\Lambda}$ in Au+Au collisions at $\sqrt{s_{\mathrm{NN}}} = 27$ GeV. The model calculations with baryon diffusion ($C_B = 1$) and without baryon diffusion ($C_B = 0$) are compared with the available STAR collaboration data for $p-\bar{p}$ \cite{STAR:2023jdd}.}
  \label{fig:Emback:dv1dy_cent_proton_lambda}
\end{figure}

Fig. \ref{fig:Emback:dv1dy_cent_proton_lambda} presents the model calculations of $\Delta \frac{dv_1}{dy}$ for $p-\bar{p}$ and $\Lambda-\bar{\Lambda}$ as functions of centrality in Au+Au collisions at $\sqrt{s_{\mathrm{NN}}}=27$ GeV, alongside the recent STAR collaboration measurements of $\Delta \frac{dv_1}{dy}$, which are available for $p-\bar{p}$ \cite{STAR:2023jdd}. The results for both $C_B=0$ and $C_B=1$ cases are shown for comparison. In the hydrodynamic simulations, the energy tilt parameter ($\eta_m$) and baryon tilt parameter ($\omega$) were kept constant across all centralities. 

First, we examine the splitting of $\frac{dv_1}{dy}$ between protons and anti-protons, shown in Fig. \ref{fig:Emback:dv1dy_cent_proton_lambda}(a). The distinctive sign change from positive to negative values, observed by the STAR collaboration across centrality, is also captured in the hydrodynamic simulations that include baryon diffusion, though without incorporating any electromagnetic (EM) field effects. In contrast, for the case where $C_B$=0, the splitting remains positive across all centralities. The sign change observed for $C_B$=1 at higher centralities can be attributed to the varying initial baryon inhomogeneities within the fireball and the dominance of baryon diffusion at large centralities. A detailed investigation of the reason behind this sign change is provided in the next section.

Although the sign change in $\Delta \frac{d v_1}{dy} (p-\bar{p})$ was anticipated due to the EM field effects, a similar trend is observed here as a result of baryon diffusion. This observation highlights the significant role of baryon stopping and dynamics in generating signals similar to those expected from the electromagnetic field. The model predictions for the splitting of $\frac{dv_1}{dy}$ between $\Lambda$ and $\bar{\Lambda}$ are shown in Fig. \ref{fig:Emback:dv1dy_cent_proton_lambda}(b). Baryon inhomogeneity-driven diffusion currents cause a similar splitting and sign change across centralities for $\frac{dv_1}{dy}(\Lambda-\bar{\Lambda})$, resembling the behavior observed for $\frac{dv_1}{dy}(p-\bar{p})$. Although $\Lambda$ and $\bar{\Lambda}$ are electrically neutral and therefore not expected to show a splitting due to the EM field, they carry non-zero and opposite baryon numbers. As a result, a splitting similar to that observed for $p-\bar{p}$ is naturally expected in the presence of baryon diffusion. Thus, measuring the directed flow splitting between $\Lambda$ and $\bar{\Lambda}$ in experiments would provide valuable insights into the dynamics of baryon stopping and diffusion.

\begin{figure}
  \centering
  \includegraphics[width=1.0\textwidth]{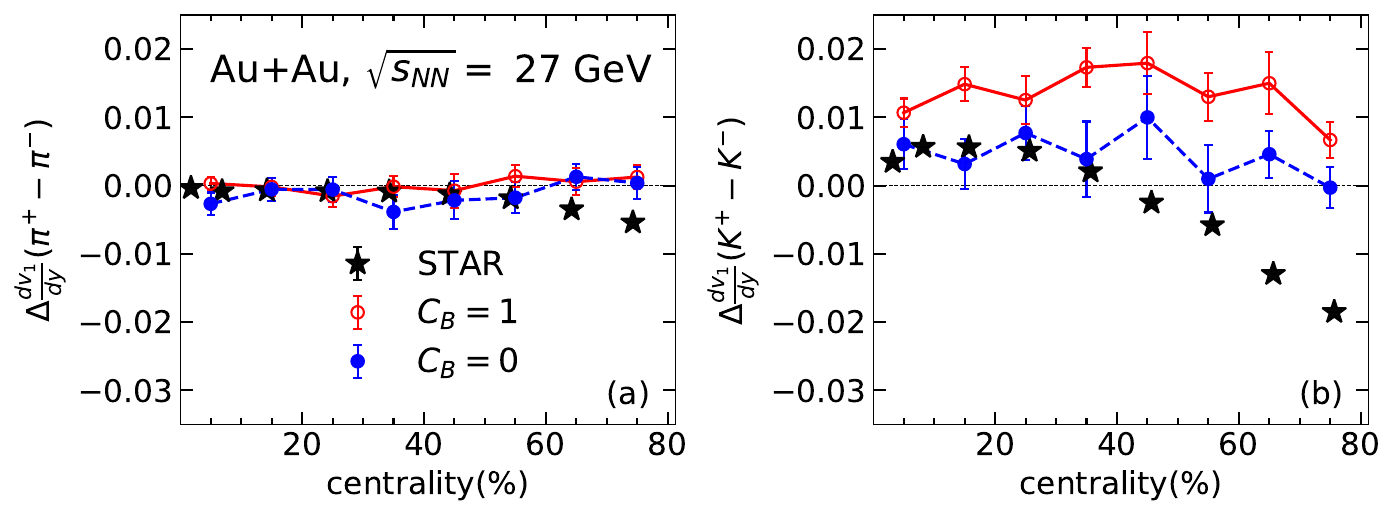} 
  \caption{Centrality dependence of the difference in mid-rapidity slope ($\Delta \frac{dv_1}{dy}$) between (a) $\pi^{+}$ and $\pi^{-}$, and (b) $K^{+}$ and $K^{-}$ in Au+Au collisions at $\sqrt{s_{\mathrm{NN}}} = 27$ GeV. The model calculations with baryon diffusion ($C_B = 1$) and without baryon diffusion ($C_B = 0$) are compared with the available STAR collaboration data \cite{STAR:2023jdd}.}
  \label{fig:Emback:dv1dy_cent_pion_kaon}
\end{figure}

Fig. \ref{fig:Emback:dv1dy_cent_pion_kaon} presents the results for the mesons. In this case, our model does not accurately reproduce the experimental data of the $\Delta \frac{dv_1}{dy}$ between $\pi^{+}-\pi^{-}$ and $K^{+}-K^{-}$. This discrepancy may arise because, in our framework, we do not independently initialize or evolve the strangeness and electric charge densities. Instead, these densities are constrained by the baryon distribution through the relations $n_S = 0$ and $n_{Q} = 0.4 n_B$. Moreover, we have not taken the diffusion of strangeness and electric charge densities in our model. The independent initialisation and evolution of $n_S$ and $n_Q$ could lead to different trends in $\Delta \frac{dv_1}{dy}(K^{+}-K^{-})$ or $\Delta \frac{dv_1}{dy}(\pi^{+}-\pi^{-})$ across centralities. Therefore, a more comprehensive investigation using a hydrodynamic framework that includes the evolution of all three conserved charges accounting their diffusion in the medium is essential to better understand these phenomena.

\section{Probing the sign change of $ \Delta d v_1 (p-\bar{p}) /dy$ from central to peripheral collisions in the presence of baryon diffusion}
\label{emfield:prob_sign_change}
As shown in Fig. \ref{fig:Emback:dv1dy_cent_proton_lambda}, the $\Delta dv_1/dy$ between $p$ and $\bar{p}$ exhibits a very different centrality trend when baryon diffusion is present ($C_B=1$) compared to when it is absent ($C_B=0$). In this section, we examine the reasons behind this behavior and how the initial baryon gradient in the transverse plane varies from central to peripheral collisions, ultimately influencing the splitting between the directed flow ($v_1$) of protons and anti-protons.

\begin{figure}
  \centering
  \includegraphics[width=0.5\textwidth]{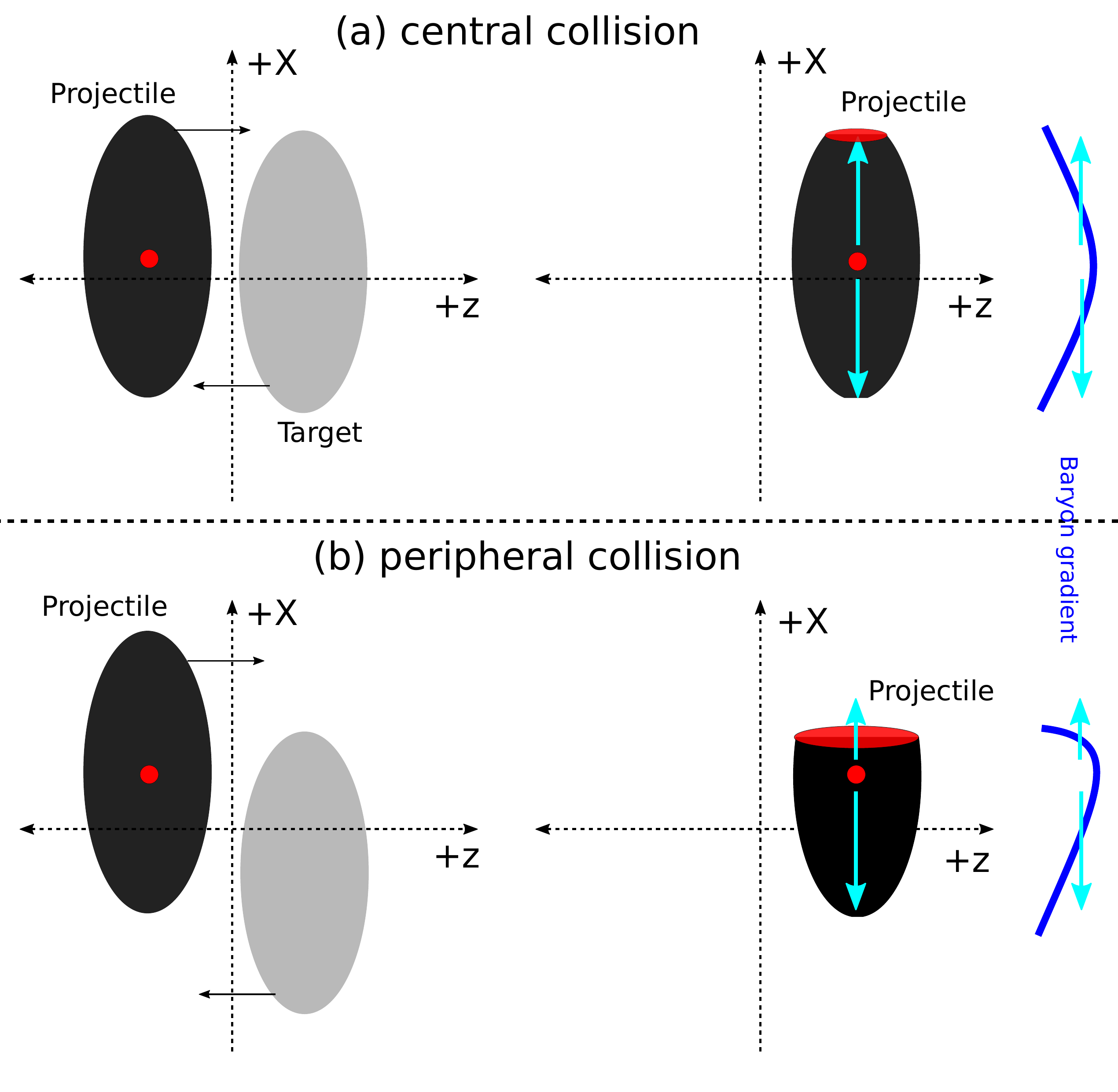} 
  \caption{Illustration of baryon gradient variations along $x$ direction at forward rapidity between central (top row) and peripheral (bottom row) Au+Au collisions. For each collision type, the left panel represents the pre-collision scenario, while the right panel depicts the post-collision scenario focusing on the forward rapidity region. At a forward rapidity slice, the transverse profile of baryon density is primarily governed by the participant distribution of the projectile nucleus. 
The dark blue lines represent baryon density distributions along the $x$-axis. These highlight the differences in initial baryon distribution between central and peripheral collisions. The baryon distributio is more symmetric in central collsion than the peripheral. Cyan arrows indicate the degree of asymmetries in baryon gradient along $x$ direction. This visualization highlights how the transverse baryon gradient changes from central to peripheral collisions, influencing baryon diffusion and subsequently impacting the directed flow splitting between baryons and anti-baryons. }
   \label{fig:draw_forward_assym}
\end{figure}

Fig. \ref{fig:draw_forward_assym} illustrates schematic representations (cartoons) of a central and a peripheral collision. Each scenario includes two stages: before the collision (left) and after the collision (right). For the post-collision scenario, we depict only the participants from the projectile nucleus positioned at forward rapidity. This is intentional, as our focus is on the forward rapidity region. According to the baryon profile defined in Eq. \ref{eq:two_component_baryon_profile}, the projectile nucleus—moving with positive rapidity—deposits a greater share of baryons along its direction of motion. Consequently, the baryon distribution at forward rapidity during the initial stage is primarily governed by the participant distribution of the projectile nucleus. Hence, this selective depiction highlights the crucial region under consideration.

The dark blue lines in the figure represent the baryon distribution along the $x$-axis at a positive rapidity slice for both central and peripheral collisions, shown adjacent to their respective post-collision diagrams. The baryon density deposited by the participant nucleus at any transverse position is governed by its transverse thickness, calculated by integrating the nuclear matter profile of the projectile nucleus along the $z$-direction (see Eq. \ref{eq:thickness_glau1}). Since the transverse thickness is maximal at the center of the nucleus and decreases radially outward, the baryon density is highest at the transverse position where the center is situated.  
In central collisions, where almost the entire nucleus participates, the baryon distribution is symmetric around the peak position. In contrast, peripheral collisions result in an asymmetric baryon distribution on either side of the baryon peak. This asymmetry in peripheral collisions generates asymmetric baryon gradients along the $x$-axis, represented by cyan arrows in the figure.

\begin{figure}
  \centering
  \includegraphics[width=0.5\textwidth]{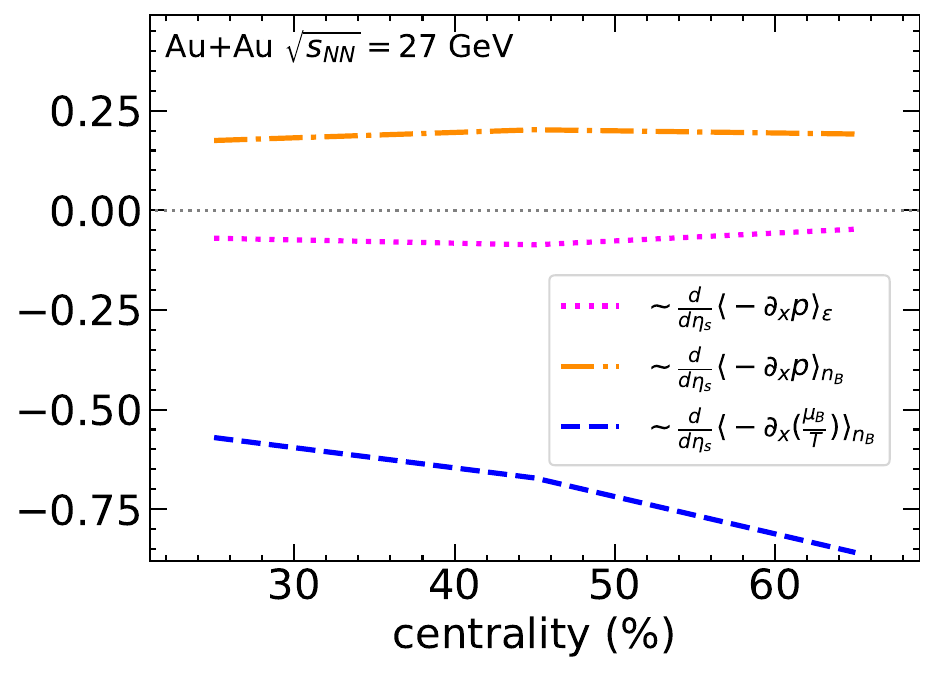} 
  \caption{Centrality dependence of mid-rapidity slopes for initial state estimators $\langle -\partial_x p \rangle_{\epsilon}$, $\langle -\partial_x p \rangle_{n_B}$, and $\langle - \partial_x \frac{\mu_B}{T} \rangle_{n_B}$ at $\sqrt{s_{\mathrm{NN}}} = 27$ GeV. These estimators quantify the initial state gradients of hydrodynamic field variables and illustrate the transition from central to peripheral collisions. The results are scaled for better visualization, demonstrating the significant role of baryon diffusion in centrality-dependent dynamics.}
  \label{fig:estim_cent}
\end{figure}

The baryon density distributions of the projectile nucleus at positive rapidity reveal a stark difference in baryon gradients along the $x$-axis between central and peripheral collisions. Consequently, the baryon diffusion, which is influenced by this gradient, also expected to differ, thereby affecting the baryonic flow. While the illustrations provide a qualitative understanding, a quantitative evaluation of the initial energy and baryon distribution gradients is essential. This quantification is presented in Fig. \ref{fig:estim_cent}, where the mid-rapidity slope of the initial state estimators, as discussed in Eqs. \ref{eq:estimator_1}, \ref{eq:estimator_2}, and \ref{eq:estimator_3}, is plotted as a function of centrality.

\begin{figure}
  \centering
  \includegraphics[width=1.0\textwidth]{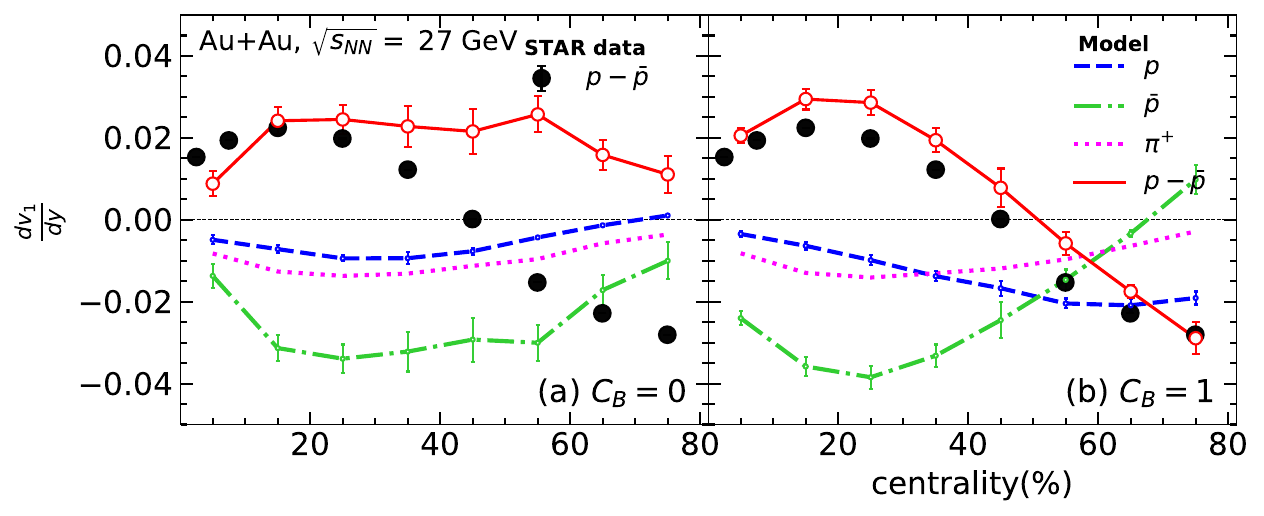} 
  \caption{Mid-rapidity slope of $v_1$ ($dv_1/dy$) for $\pi^{+}$, protons and antiprotons as a function of collision centrality at $\sqrt{s_{\mathrm{NN}}} = 27$ GeV. Panel (a) presents results for $C_B = 0$ (no baryon diffusion), while panel (b) shows results for $C_B = 1$ (baryon diffusion included). The comparison highlights the centrality-dependent effects of baryon diffusion, including the sign change in splitting between protons and antiprotons observed in peripheral collisions. The experimental data is taken from STAR collaboration \cite{STAR:2023jdd}. }
  \label{fig:Split_diff_CB}
\end{figure}

In the previous chapter, we demonstrated how effectively these predictors capture the flow behavior of identified hadrons, providing valuable insights into the system's dynamics. The mid-rapidity slope of $\langle -\partial_x p \rangle_{\epsilon}$ remains negative across all centralities. Consequently, in our model calculations, shown in Fig. \ref{fig:Split_diff_CB}, the $v_1$ of $\pi^{+}$ consistently remains negative. The relative distributions of baryon with respect to pressure, governed by $\langle -\partial_x p \rangle_{n_B}$, and energy with respect to pressure, governed by $\langle -\partial_x p \rangle_{\epsilon}$, show minimal centrality dependence. Their mid-rapidity slopes remain nearly constant and parallel with centrality variation, resulting in similar centrality trends for $v_1$ of protons, antiprotons, and $\pi^{+}$ in the absence of baryon diffusion $C_B=0$ (see Fig. \ref{fig:Split_diff_CB}(a)). Although there is a finite splitting between protons and antiprotons due to the presence of baryons in the medium, this splitting remains flat across centrality. In contrast, when baryon diffusion is included ($C_B=1$), the changing baryon gradient from central to peripheral collisions causes a changes in the $v_1$ slopes of protons and antiprotons, leading to a sign change in their splitting for peripheral collisions. This highlights the critical role of baryon diffusion in shaping the centrality-dependent behavior of $v_1$ splitting between baryons and anti-baryons.

Previously, we discussed that the estimator $\langle -\partial_x p \rangle_{n_B}$ governs the equilibrium component of the baryon flow along the $x$-direction, arising from advection due to $u^x$. Meanwhile, $\langle -\partial_x \frac{\mu_B}{T} \rangle_{n_B}$ characterizes baryon diffusion along the $x$-direction. Together, advection and diffusion determine the total baryon flow. 

Importantly, the centrality dependence of baryon diffusion is evident, as the mid-rapidity slope of $\langle -\partial_x \frac{\mu_B}{T} \rangle_{n_B}$ changes significantly from central to peripheral collisions, as shown in Fig. \ref{fig:estim_cent}. This variation causes the proton mid-rapidity slope to become increasingly negative with centrality, as shown in Fig. \ref{fig:Split_diff_CB}(b). In contrast, the slope of $v_1$ for $\bar{p}$ becomes progressively less negative with increasing centrality. This characteristic leads to a sign change in $\Delta dv_1/dy (p-\bar{p})$ from central to peripheral collsions.

\section{Investigation of baryon stopping effect on system size dependence of $\Delta dv_1/dy$}

Recently, at the Strangeness in Quark Matter (SQM) 2024 Conference in Strasbourg, the STAR collaboration presented centrality-dependent measurements of $\Delta dv_1/dy$ between oppositely charged hadrons ($\pi^+ - \pi^-$, $K^+ - K^-$, and $p - \bar{p}$) for $\sNN = 200$ GeV across different collision systems such as Au+Au, U+U, and Zr+Zr \cite{Taseer:2024sho,Taseer_talk_SQM_2024}. Upon comparison, it was observed that $\Delta dv_1/dy$ exhibits a notable dependence on the system size. This trend is most pronounced in the case of $p - \bar{p}$. The experimental measurements are plotted in Fig. \ref{fig:sys_split} and represented by symbols. 

\begin{figure}[htbp]
    \centering
    \begin{minipage}{0.45\textwidth} 
        \centering
        \includegraphics[width=\textwidth]{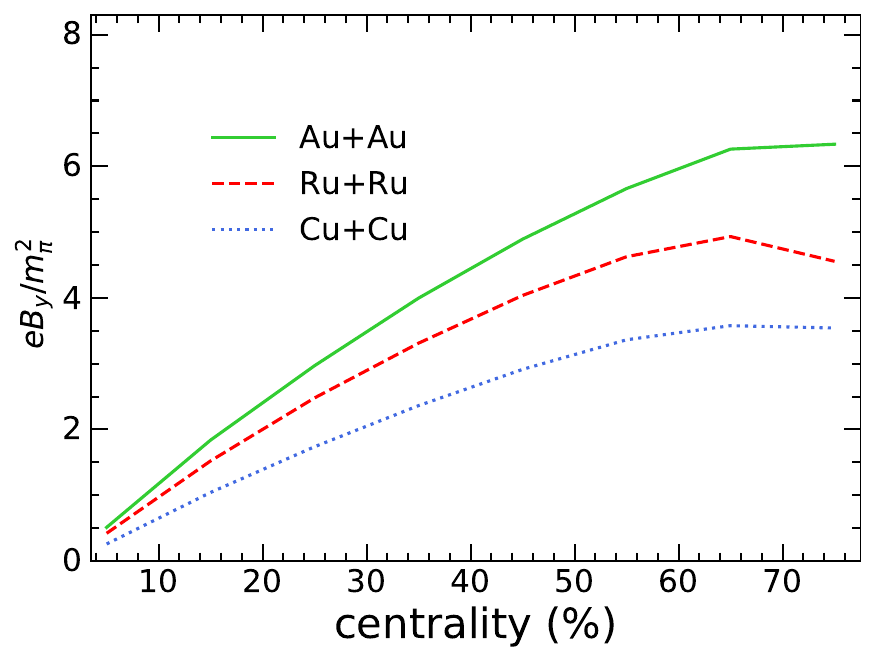}
        \caption{Dependence of the initial electromagnetic field strength $eB_y$ on system size and centrality. The magnetic field $eB_y$ is evaluated at $x_{\perp} = 0$ and $\eta_s = 0$ for Cu+Cu, Ru+Ru, and Au+Au collisions at $\sNN = 200$ GeV. The results show that $eB_y$ increases with system size and is stronger in more peripheral collisions. There is a clear system size dependence of strength of $eB_y$ in peripheral collisions whereas in central collsion the strength is very small.}
        \label{fig:eBy}
    \end{minipage}
    \hfill 
    \begin{minipage}{0.45\textwidth} 
        \centering
        \includegraphics[width=\textwidth]{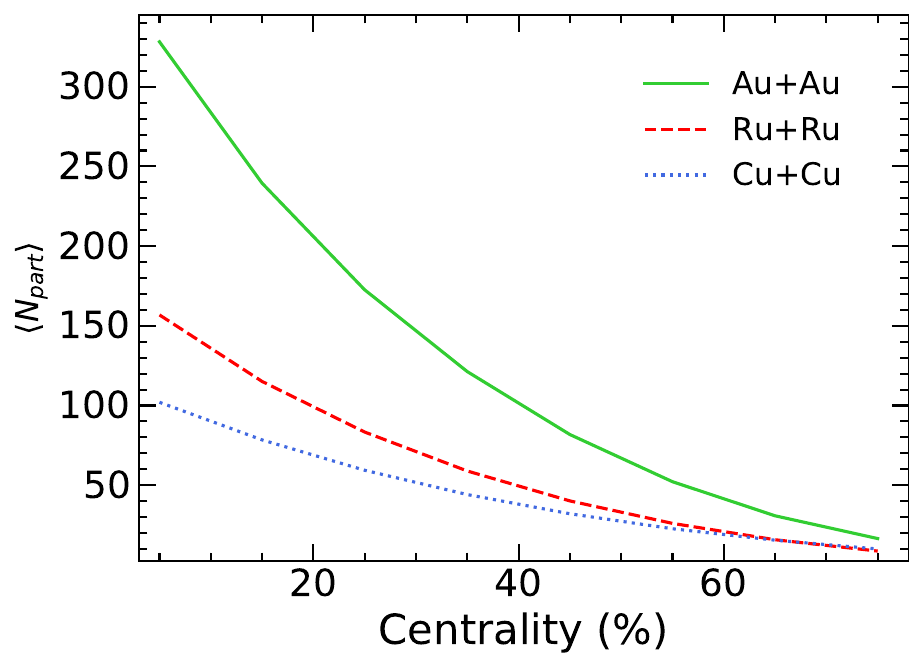}
        \caption{System size dependence of the average participant number ($\langle N_{\text{part}} \rangle$) as a function of centrality at $\sNN=200$ GeV. The plot illustrates how baryon stopping varies with system size, showing a larger average participant number for larger collision systems (e.g., Au+Au) at the same centrality, reflecting the increased baryon deposition in the medium for larger systems. There is a clear system size dependence of baryon stopping in central collisions.}
        \label{fig:sys_npart}
    \end{minipage}
    \label{fig:combined_figures}
\end{figure}

\begin{figure}
  \centering
  \includegraphics[width=1.0\textwidth]{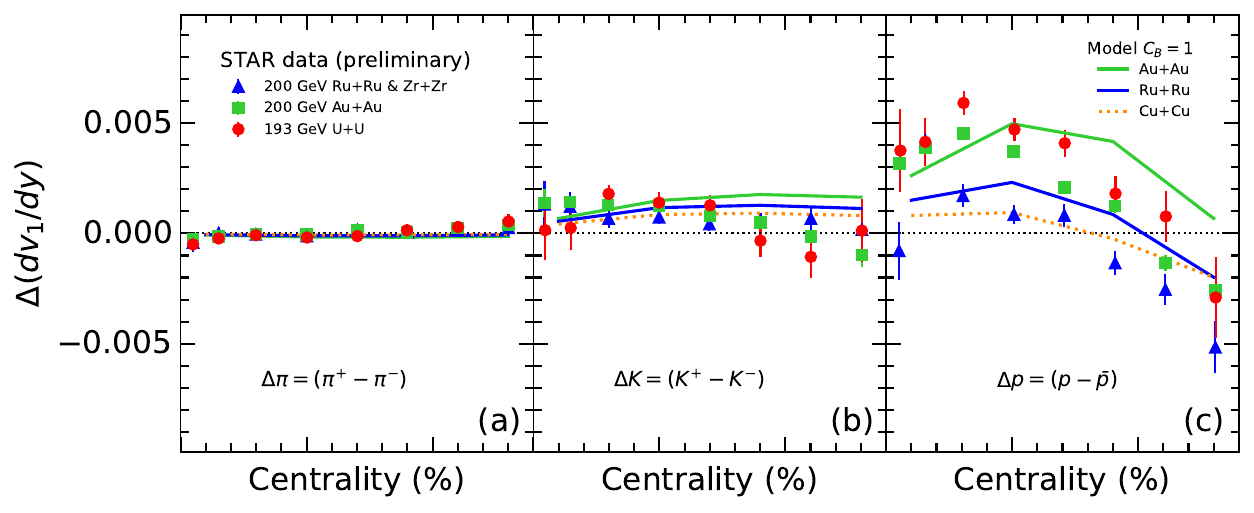} 
  \caption{Comparison of model calculations and experimental data for $\Delta dv_1/dy$ between (a) $\pi^{+}-\pi^{-}$, (b)$K^+-K^-$ (a) $p-\bar{p}$ across different systems. The model calculations are done for Cu+Cu, Ru+Ru, and Au+Au collisions at $\sNN = 200$ GeV whereas experimental data is available for Zr+Zr, Au+Au and U+U. Symbols represent preliminary STAR experimental data presented at SQM 2024 \cite{Taseer:2024sho,Taseer_talk_SQM_2024}, while model calculations, performed without incorporating electromagnetic field effects, are represented by lines. The model results indicate a system size dependence in $\Delta dv_1/dy(p-\bar{p})$ arising from baryon stopping.}
  \label{fig:sys_split}
\end{figure}

At a given centrality, the number of spectator nucleons differs across collision systems, which directly influences the strength of the generated electromagnetic (EM) field. To illustrate this, the model calculation of initial $eB_y$ produced by spectator nucleons at the position $(x,y,\eta_s)=(0,0,0)$ for Cu+Cu, Ru+Ru, and Au+Au collisions across different centralities is shown in Fig. \ref{fig:eBy}. The $eB_y$ calculation has been done using the procedure mentioned in Ref. \cite{Gursoy:2014aka}. The results reveal that the strength of the initial $eB_y$ increases with centrality, with peripheral collisions experiencing a stronger EM field. Furthermore, at a given centrality, larger collision systems exhibit a higher $eB_y$. This indicates a stronger EM field effect on the created fluid for larger systems. Consequently, the observed system size dependence of $\Delta dv_1/dy$ at $\sNN = 200$ GeV could be interpreted as a potential signature of the EM field.

However, it is essential to note that as the system size increases, the initial baryon deposition in the medium also grows. To demonstrate this, the average number of participants, $\langle N_{\mathrm{part}} \rangle$, is plotted as a function of centrality for different collision systems in Fig. \ref{fig:sys_npart}. The results show a clear system size dependence in the initial baryon stopping at a given centrality. This suggests that the observed system size dependence of $\Delta dv_1/dy$ could also arise from variations in the amount of initial baryon deposition, potentially complicating the interpretation as a signature of the EM field.

To examine the impact of baryon stopping on the system size dependence of $\Delta dv_1/dy$, we performed simulations using our model. These simulations excluded the electromagnetic (EM) field but incorporated baryon diffusion by setting $C_B = 1$. The calculations were conducted for Cu+Cu, Ru+Ru, and Au+Au collisions. Simulations for U+U collisions were not performed because uranium is a deformed nucleus. Our primary focus in this work is to explore the system size dependence of $v_1$ splitting arising from baryon stopping. Including U+U collisions would introduce potential additional contributions to $v_1$ splitting from nuclear deformation effects, which remain unexplored within our current framework. While the influence of deformation on $v_1$ splitting is a compelling avenue for future study, it is beyond the scope of this work.

Previous experimental measurements indicate that the $v_1$ of charged particles does not exhibit system size dependence \cite{STAR:2008jgm}. To ensure consistency with these observations, we independently adjusted the energy tilt parameter $\eta_m$ for each collision system. This adjustment was made to align the $v_1$ of charged hadrons across all systems for $10$-$40\%$ centrality, consistent with experimental data. The other parameter values used in these calculations at $\sNN = 200$ GeV are taken same as mentioned in Table \ref{param_for_model}.

Moreover, in this calculations, we did not perform hadronic transport simulations for the sampled hadrons from the hypersurface. Instead, we used the MUSIC particlization routine\footnote{The MUSIC particlization routine utilizes the Cooper-Frye formula to compute and store the invariant yield, $\frac{dN}{p_T dp_T dy d\phi}$, on a discretized grid of $y$, $p_T$, and $\phi$, while also accounting for contributions from resonance decays \cite{MUSIC_particlisation}. } to calculate the phase-space distribution of produced hadrons, which was then employed to determine the $v_1$ observables \cite{MUSIC_particlisation}. In the calculations, we included the effects of resonance decays. We verified that incorporating hadronic transport at the final stage has a negligible impact on the $v_1$ of identified hadrons, particularly at $\sNN = 200$ GeV. Consequently, the model results in this case are reliable and have the added advantage of eliminating statistical uncertainties associated with finite particle sampling.

The results from our model are shown in Fig. \ref{fig:sys_split}, where they are compared with experimental data. The splitting between $p$ and $\bar{p}$, presented in panel (c). The model demonstrates a clear system size dependence in $\Delta dv_1/dy(p-\bar{p})$ across all centralities, consistent with the observation that baryon density is higher in larger collision systems. This finding suggests that the extraction of the electromagnetic (EM) field signal via system size dependence is influenced by a background contribution from baryon stopping. However, it is important to note that the centrality trends of the EM field strength and the initial baryon deposition exhibit opposite behaviors, as seen in Fig. \ref{fig:eBy} and Fig. \ref{fig:sys_npart}. While the EM field effect is more prominent in peripheral collisions, baryon stopping dominates in central collisions. This observable, therefore, provides an opportunity to study the interplay between the EM field and baryon stopping, which could yield valuable insights. 

In our model calculations, we observed a pronounced system size dependence in $\Delta dv_1/dy(p-\bar{p})$ for central collisions, where the electromagnetic (EM) field strength is minimal. This suggests that the system size dependence of this splitting in central collisions primarily arises from the system size dependence of baryon stopping. Experimental data for $\Delta dv_1/dy(p-\bar{p})$ also indicates a potential system size dependence in central collisions, although the large experimental error bars limit a definitive conclusion. In contrast, for peripheral collisions, the experimental error bars are too significant to confirm the existence of a clear system size dependence.

The deposition and evolution of strangeness and electric charge density are expected to contribute to the $v_1$ splitting between $K^{+}-K^{-}$ and $\pi^{+}-\pi^{-}$. However, as previously discussed, our model does not incorporate the proper treatment of the other two conserved charges. The constriants $n_Q=0.4 n_B$ and $n_S=0$, imposed through NEoS-BQS EoS in the model calculation leads to non-zero $\mu_S$ and $\mu_Q$ in the medium which creates the splitting between $K^{+}-K^{-}$ and $\pi^{+}-\pi^{-}$ \cite{Monnai:2019hkn}. We observed that the splitting between $\pi^{+}-\pi^{-}$ is almost zero in our model whereas the $\Delta dv_1/dy(K^{+}-K^{-})$ shows a system size dependence. Since the initial nuclei do not carry any net strangeness, this observation in the kaon sector is particularly interesting. Future hydrodynamic simulations incorporating all conserved charges will be essential for understanding the system size dependence of $v_1$ splitting in $K^{+}-K^{-}$.

\section{Electric charge and strangeness-dependent directed flow splitting}

In another study by the STAR collaboration~\cite{STAR:2023wjl}, a new observable related to $v_1$ splitting was introduced and interpreted as being consistent with electromagnetic (EM) effects. This study focused on various hadrons composed entirely of produced quarks, such as $K^{-}(\bar{u}s)$, $\bar{p}(\bar{u}\bar{u}\bar{d})$, $\bar{\Lambda}(\bar{u}\bar{d}\bar{s})$, $\phi(s\bar{s})$, $\bar{\Xi}^{+}(\bar{d}\bar{s}\bar{s})$, $\Omega^{-}(sss)$, and $\bar{\Omega}^{+}(\bar{s}\bar{s}\bar{s})$.
Particle species are grouped into pairs as detailed Table~\ref{part_comb_table} having zero or close-to-zero total mass difference ($\Delta m$) at the constituent quark level. Notably, these pairs exhibit differences in net charge and net strangeness ($\Delta Q$ and $\Delta S$), providing an opportunity to investigate $v_1$ splitting as a function of these differences while minimizing the background contributions from mass differences~\cite{Sheikh:2021rew}. The slope of the $v_1$ difference of these combinations, denoted as $F_{\Delta} = d\Delta v_1/dy$, was observed to increase with $\Delta Q$, a trend attributed to the influence of the EM field~\cite{STAR:2023wjl}.

\begin{table}[ht]
\centering
\begin{tabular}{|p{0.9cm}|p{3.6cm}|p{0.9cm}|p{0.9cm}|p{0.9cm}|}
\hline  
Index & $\Delta v_1$ combinations & $\Delta Q$ & $\Delta S$ & $\Delta B$ \\
\hline  
1 & $\left[ \bar{p} + \phi \right] - \left[ K^{-} + \bar{\Lambda} \right]$ &  0 & 0 & 0 \\ 
2 & $\left[ \bar{\Lambda} \right] - \left[ \frac{1}{3} \Omega^{-} + \frac{2}{3} \bar{p} \right]$ & 1 & 2 & -2/3\\ 
3 & $\left[ \bar{\Lambda} \right] - \left[ K^{-} + \frac{1}{3} \bar{p} \right]$ & 4/3 & 2 &  -2/3\\ 
4 & $\left[ \bar{\Omega}^{+}  \right] - \left[  \Omega^{-} \right]$ & 2 & 6 & -2\\ 
5 & $\left[ \bar{\Xi}^{+} \right] - \left[ K^{-} + \frac{1}{3}  \Omega^{-}  \right]$ &  7/3 & 4 & -4/3\\ 
\hline
\end{tabular}
\caption{ Hadron combinations used to measure $v_1$ splitting as a function of $\Delta Q$, $\Delta S$, and $\Delta B$, as analyzed in the STAR collaboration study~\cite{STAR:2023wjl}. The table lists the indices, particle combinations, and their respective differences in net charge ($\Delta Q$), net strangeness ($\Delta S$), and net baryon number ($\Delta B$). }
\label{part_comb_table}
\end{table}

\begin{figure}
  \centering
  \includegraphics[width=1.0\textwidth]{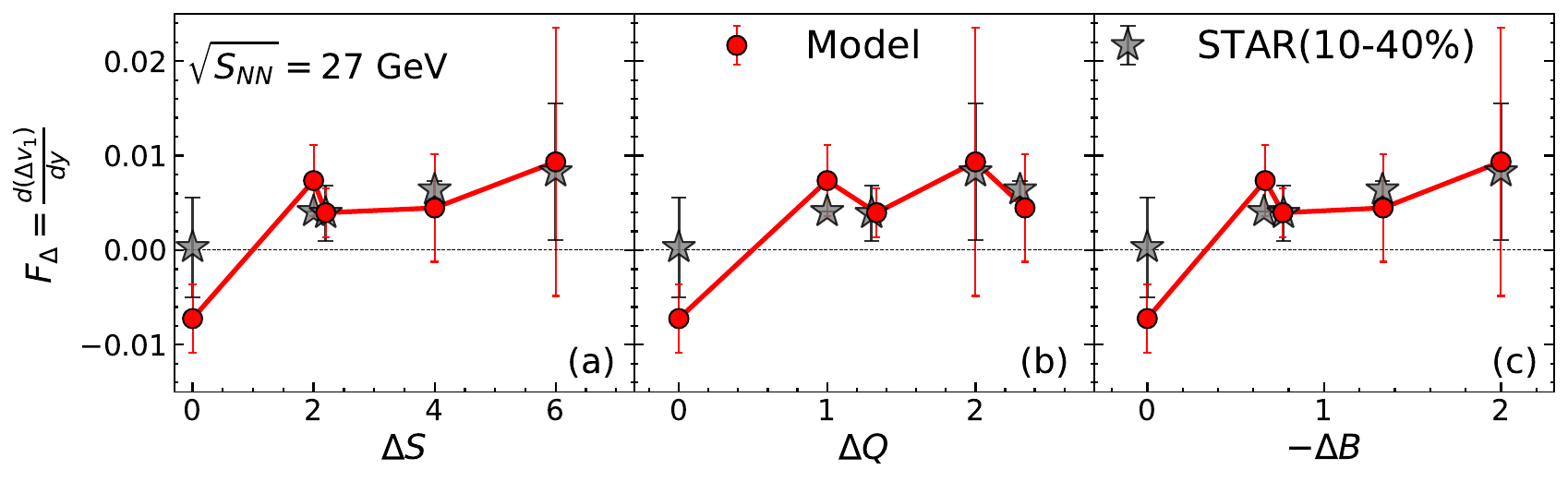} 
  \caption{Model calculations of $F_{\Delta} = d\Delta v_1/dy$ as a function of net charge difference ($\Delta Q$), net strangeness difference ($\Delta S$), and net baryon difference ($\Delta B$) for the hadron combinations listed in Table~\ref{part_comb_table}. The results incorporate baryon stopping effects with $C_B = 1$ and follow the same kinematic cuts as the experimental data. The experimental data is from STAR collaboration \cite{STAR:2023wjl}.}
  \label{fig:Emback:F_Delta_DeltaBQS}
\end{figure}

From our model calculations, we found that the considered particle pairs not only exhibit differences in $\Delta Q$ and $\Delta S$ but also possess a nonzero net baryon difference ($\Delta B$). To explore this further, we calculated $F_{\Delta}$ within our model and plotted as a function of $\Delta Q$, $\Delta S$, and $\Delta B$, as shown in Fig. \ref{fig:Emback:F_Delta_DeltaBQS}. The calculations were performed with nonzero baryon diffusion by setting $C_B = 1$. Moreover during $v_1$ calculation, we used the same kinematic cuts as in the experiment were applied. The model results show that $F_{\Delta}$ increases with $\Delta B$ and align closely with the experimental measurements. This observation indicates that the dependence of $d\Delta v_1/dy$ on $\Delta Q$ may not provide a definitive signature of EM fields. While the pair combinations effectively reduce the background contributions from mass differences, a significant background originating from baryon stopping physics persists, complicating the interpretation of the results.

\section{Chapter summary}
Recently, the STAR collaboration has presented multiple measurements focusing on charge-dependent directed flow splitting in the presence of electromagnetic (EM) fields \cite{STAR:2023jdd,Taseer:2024sho,STAR:2023wjl}. In Ref. \cite{STAR:2023jdd}, the mid-rapidity slope of the directed flow difference ($\Delta \frac{dv_1}{dy}$) between $\pi^{+}-\pi^{-}$, $K^{+}-K^{-}$, and $p-\bar{p}$ was measured as a function of centrality across various collision energies. These measurements reveal a sign change in $\Delta \frac{dv_1}{dy}$, transitioning from positive values in central collisions to negative values in peripheral collisions, with the most pronounced effect observed for $p-\bar{p}$. Additionaly, at the SQM 2024 conference, the same observable was presented for different collision systems at $\sNN=200$ GeV \cite{Taseer_talk_SQM_2024,Taseer:2024sho}, demonstrating a clear system size dependence, again with the strongest effect for $p-\bar{p}$. These trends in $\Delta \frac{dv_1}{dy}$, both with centrality and system size, have been interpreted as signatures of EM fields in heavy-ion collisions. The centrality dependence is attributed to the varying EM field strength from central to peripheral collisions, while the system size dependence is expected by the increase in EM field strength in larger collision systems, as the number of spectator nucleons—and thus the spectator charge—varies across collision systems.

The aim of our study is to demonstrate that the dependence of $\Delta \frac{dv_1}{dy}$ on centrality is not solely driven by the effects of electromagnetic fields on charged constituents but also includes a significant contribution from the physics of conserved charge stopping and dynamics. Using a hydrodynamic model that incorporates finite baryon density and baryon diffusion, we successfully reproduce the experimental measurements of the centrality and system size dependence of $\Delta \frac{dv_1}{dy}(p-\bar{p})$. Importantly, no electromagnetic field effects were included in our model calculations. This finding highlights the significant role of baryon stopping and its diffusion in influencing the directed flow splitting $\Delta \frac{dv_1}{dy}$ between $p$ and $\bar{p}$, complicating the extraction of electromagnetic field signals from this observable.

However, our model was unable to describe the $\Delta \frac{dv_1}{dy}$ observed between $\pi^{+}-\pi^{-}$ and $K^{+}-K^{-}$. This limitation arises from the absence of a proper treatment of net-strangeness and net-electric charge in our framework. While the model accounts for the initial deposition and evolution of net-baryon density, it assumes simplified relations for the other conserved charges: $n_S = 0$ and $n_Q = 0.4n_B$. Consequently, the interpretation of $\Delta \frac{dv_1}{dy}$ for mesons remains open. Similar to how baryon diffusion influences $p-\bar{p}$ splitting, the independent inhomogeneous distribution and diffusion of net-strangeness and net-electric charge could non-trivially affect the splitting between $\pi^{+}-\pi^{-}$ and $K^{+}-K^{-}$. Future investigations incorporating these conserved charges will be critical in studying the interplay between electromagnetic fields and conserved charge dynamics for these observables.

Moreover, in another study by the STAR collaboration \cite{STAR:2023wjl}, various hadron species were grouped into pairs, ensuring that each pair had a nearly zero net mass difference at the constituent quark level while maintaining distinct net charge differences ($\Delta Q$) between the pairs. It is observed that the slope of the directed flow difference, $F_\Delta = d\Delta v_1/dy$, increases with $\Delta Q$ for these pairs. This trend was interpreted as a signature of the electromagnetic fields present in heavy ion collisions. However, in our model calculations, we demonstrated that the particle combinations used in this analysis, which exhibit a non-zero $\Delta Q$, also have a non-zero $\Delta B$. Furthermore, our model successfully captures the $F_\Delta = d\Delta v_1/dy$ dependence on $\Delta Q$ without including any electromagnetic field effects. This finding suggests that the observed increase in $F_\Delta$ may not be solely attributed to the charge difference. Instead, it appears to be predominantly driven by the difference in baryon number, $\Delta B$.

\def \la{\langle}
\def \ra{\rangle}
\chapter{Baryon diffusion coefficient of the strongly interacting medium}
\label{ch:baryondiffusion}

The relativistic hydrodynamic model has proven highly effective in describing the dynamics of the strongly interacting matter produced in relativistic heavy-ion collisions at the highest RHIC and LHC energies \cite{Romatschke:2007mq,Schenke:2011bn,Denicol:2018wdp,Karpenko:2015xea,Shen:2012vn}. Three essential components are required for modeling heavy-ion collisions using relativistic dissipative hydrodynamics: the initial conditions of the conserved quantities, the equation of state of the QCD medium, and the transport coefficients of the medium, which characterize its dissipative currents. At LHC energies and the highest RHIC energies, the central rapidity region exhibits a negligible net conserved charge density \cite{ALICE:2013mez}. Consequently, it is customary to focus exclusively on the evolution of the energy-momentum tensor ($T^{\mu \nu}$) within the framework of relativistic dissipative hydrodynamics, while neglecting other conserved charges such as net baryon density, net electric charge, and net strangeness \cite{Gale:2013da,Schenke:2011bn,Alver:2010dn,Gale:2012rq,Song:2008hj,Song:2010mg,Heinz:2005bw,Dusling:2007gi,Bozek:2011ua,Bernhard:2019bmu}. In this context, shear viscosity ($\eta$) and bulk viscosity ($\zeta$) emerge as the primary transport coefficients governing dissipation in the medium. These coefficients drive the generation of viscous quantities, such as the shear stress tensor ($\pi^{\mu \nu}$) and bulk viscous pressure ($\Pi$), as hydrodynamic responses to gradients of $T^{\mu \nu}$.

Reliable first-principles estimates of $\eta$ and $\zeta$ are still lacking, making it standard practice to treat these transport coefficients as free parameters in hydrodynamic models. They are typically constrained by comparing model predictions with experimental data once a sensitive observable is identified. Notably, one of the hallmark achievements of RHIC phenomenology has been the extraction of $\eta$ for the QCD medium by comparing model predictions with measurements of flow coefficients \cite{Romatschke:2007mq,Karpenko:2015xea, Gotz:2022naz, Niemi:2015qia,Schenke:2011bn,Denicol:2015nhu,DerradideSouza:2015kpt,Gale:2013da,Heinz:2013th,Bernhard:2019bmu,Roy:2018rnb}. Efforts have also been made to estimate the bulk viscosity of the QCD medium through similar approaches \cite{Rose:2014fba,Ryu:2015vwa,Bernhard:2019bmu,Roy:2011pk}.

As the collision energy decreases in the Beam Energy Scan (BES) program at RHIC, the net proton yield indicates that the matter produced in the central rapidity region becomes increasingly enriched with net baryon density \cite{STAR:2017sal,BRAHMS:2003wwg,BRAHMS:2009wlg,NA49:2010lhg}. Unlike at higher beam energies, this observation highlights the necessity of accounting for the evolution of baryon density \cite{Denicol:2018wdp,Wu:2021fjf,Cimerman:2023hjw,De:2022yxq,Monnai:2012jc}. Consequently, the relativistic dissipative hydrodynamics framework must incorporate the dynamics of baryon-conserved charges. This adaptation requires several additional components: the QCD equation of state (EoS) at non-zero baryon densities, an initial condition for baryon density, and, in addition to shear viscosity ($\eta$) and bulk viscosity ($\zeta$), a new transport coefficient—the baryon diffusion coefficient ($\kappa_B$) \cite{Rougemont:2015ona,Denicol:2018wdp,Rougemont:2017tlu}—to characterize the medium's response to baryon density inhomogeneities.

The QCD EoS for the relevant ranges of temperature ($T$) and baryon chemical potential ($\mu_B$), applicable to Au+Au collisions across $\sqrt{s_{\text{NN}}}$ = 7.7–200 GeV, is now accessible through state-of-the-art lattice QCD computations \cite{HotQCD:2014kol,HotQCD:2012fhj, Ding:2015fca,Bazavov:2017dus,Borsanyi:2013bia,Borsanyi:2018grb,Bazavov:2020bjn,Noronha-Hostler:2019ayj,Monnai:2019hkn}. Additionally, a suitable initial condition for baryon density at BES energies has been established, as discussed in the previous chapter. This initial condition successfully captures data trends in the directed flow ($v_1$) of identified hadrons, particularly the observed $v_1$ splitting between protons ($p$) and antiprotons ($\bar{p}$) \cite{Parida:2022zse,Parida:2022ppj}. The availability of these critical inputs—the EoS and initial conditions—paves the way to focus on extracting $\kappa_B$ by comparing model predictions with experimental data for observables that are sensitive to this transport coefficient.

In our work, we employ the hydrodynamic framework MUSIC that considers the baryon diffusion current \cite{Denicol:2018wdp,Li:2018fow}. Our goal is to extract the baryon diffusion coefficient by comparing model predictions with experimentally measured observables.

\section{Model calculation details}

\begin{wrapfigure}{l}{0.5\textwidth} 
    \centering
    \includegraphics[width=0.45\textwidth]{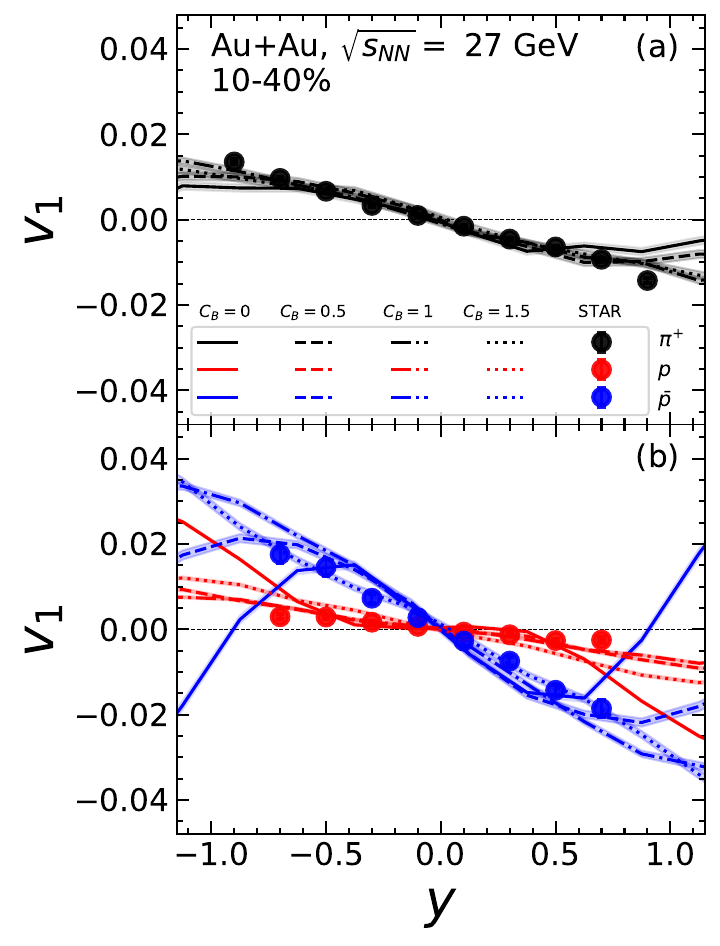}
    \caption{Rapidity dependence of directed flow ($v_1$) for (a) $\pi^{+}$ ,(b) $p$ and $\bar{p}$ in 10-40\% centrality Au+Au collisions at $\sqrt{s_{\text{NN}}}=27$ GeV. The model calculations, performed using different values of the baryon diffusion strength parameter ($C_B$), are compared with experimental measurements from the STAR collaboration \cite{STAR:2014clz}.}
    \label{fig:v1_ydiffCB}
\end{wrapfigure}

In this work, we study Au+Au collisions at $\sqrt{s_{\text{NN}}} = 27$ GeV using the hybrid framework described in Chapter \ref{ch:framework}. For the initial conditions of the hydrodynamic model, we employ a tilted energy density profile and the net baryon deposition profile outlined in Chapter \ref{ch:baryon_tilt}. For baryon diffusion coefficient $\kappa_B$, we use the ansatz based on kinetic
theory approach \cite{Denicol:2018wdp}:
\begin{equation}
\kappa_B = \frac{C_B}{T} n_B \left( \frac{1}{3} \coth{\left(\frac{\mu_B}{T}\right)} - \frac{n_B T}{\epsilon + p} \right),
\label{eq:kappaB_form2}
\end{equation}
where $n_B$, $\epsilon$, $P$, $T$, and $\mu_B$ represent the baryon number density, energy density, pressure, temperature, and baryon chemical potential, respectively. The parameter $C_B$ is an arbitrary constant that determines the strength of baryon diffusion and remains largely unknown for the Quantum Chromodynamics (QCD) medium.

To investigate the effects of $C_B$, we perform relativistic dissipative hydrodynamic simulations for four different values: $C_B = 0, 0.5, 1, \text{and } 1.5$. For each value of $C_B$, the remaining model parameters are independently tuned to reproduce the experimental measurement of the mid-rapidity yield of identified hadrons including proton, anti-proton and net-proton as well as the rapidity-differential $v_1$ of identified hadrons in the central rapidity region. The specific parameter values used for each $C_B$ are detailed in Table \ref{param_for_model_diffCB_27GeV}.

Fig. \ref{fig:v1_ydiffCB} shows the model calculations of the rapidity dependence of $v_1$ for $\pi^{+}$ in panel (a) and for $p$, $\bar{p}$ in panel (b), obtained using different values of $C_B$ in the simulations. In the central rapidity region, where experimental data are available, the $v_1(y)$ curves for different $C_B$ values align closely with each other. However, at larger rapidities, noticeable differences emerge. If experimental measurements were extended to these larger rapidity regions, $v_1(y)$ could potentially constrain $\kappa_B$.

With the currently available data \cite{STAR:2014clz}, the rapidity-dependent $v_1$ measurements do not distinguish between different $C_B$ values and, therefore, cannot effectively constrain $\kappa_B$. Moving forward, we focus on identifying other observables that are highly sensitive to $\kappa_B$, which could provide a robust means of constraining its value.

\begin{center}
\begin{table}[h!]
\centering
\begin{tabular}{|p{0.4cm}|p{0.7cm}|p{1.1cm}|p{0.55cm}|p{0.4cm}|p{0.4cm}|p{0.5cm}|p{0.5cm}|p{0.6cm}|p{0.6cm}|p{0.45cm}|p{0.55cm}|}
\hline 
 $C_B$ & $\tau_0$\tiny{(fm)} &$\epsilon_{0}$ \tiny{(GeV/fm$^{3}$)} & $\alpha$ &  $\eta_{0}$ & $\sigma_{\eta}$ & $\eta/s$ & $\eta_{0}^{n_{B}}$ & $\sigma_{B,-}$ & $\sigma_{B,+}$ & $\eta_m$ & $\omega$ \\ \hline
0.0 & 1.2  &  2.4 & 0.11  &  1.3  &  0.7  & 0.08 &  1.9  &  1.1   &  0.2  & 1.0 & 0.17  \\
\hline 
0.5 & 1.2  &  2.4 & 0.11  &  1.3  &  0.7  & 0.08 &  2.1  &  1.1   &  0.2  & 1.0 & 0.14  \\ 
\hline
1.0 & 1.2  &  2.4 & 0.11  &  1.3  &  0.7  & 0.08 &  2.3  &  1.1   &  0.2  & 1.1 & 0.11  \\ 
\hline
1.5 & 1.2  &  2.4 & 0.11  &  1.3  &  0.7  & 0.08 &  2.5  &  1.1   &  0.2  & 1.2 & 0.10  \\ 
\hline
\end{tabular}
\caption{ Model parameters used in the simulations with different $C_B$ value at $\sqrt{S_{NN}}=27$ GeV . }
\label{param_for_model_diffCB_27GeV}
\end{table}
\end{center}

\section{Constraining the baryon diffusion coefficient}

\begin{figure}[htbp]
    \centering
    \begin{minipage}{0.47\textwidth} 
        \centering
        \includegraphics[width=\textwidth]{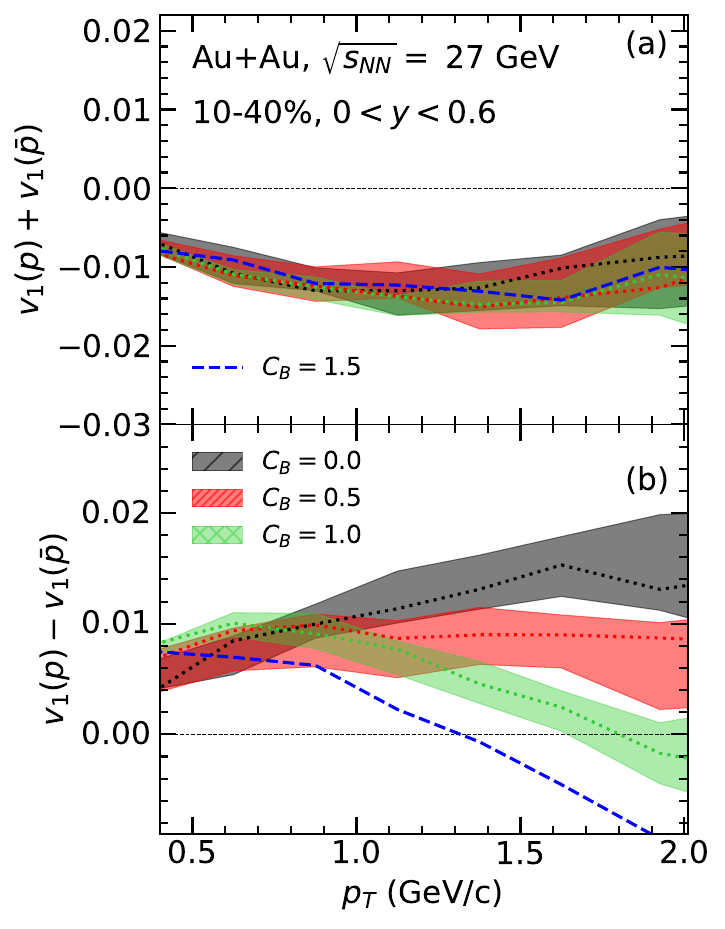}
        \caption{Model calculations of the transverse momentum dependence of (a) the sum of the directed flow  and (b) the difference in directed flow between proton ($p$) and anti-proton ($\bar{p}$) in Au+Au collisions of 10-40\% centrality at $\sqrt{s_{\text{NN}}}=27$ GeV. The results, calculated with varying values of the baryon diffusion coefficient ($C_B$), are plotted. The shaded bands represent the systematic uncertainties arising from variations in model parameters.}
        \label{fig:Deltav1_ppbar_pT}
    \end{minipage}
    \hfill 
    \begin{minipage}{0.47\textwidth} 
        \centering
        \includegraphics[width=\textwidth]{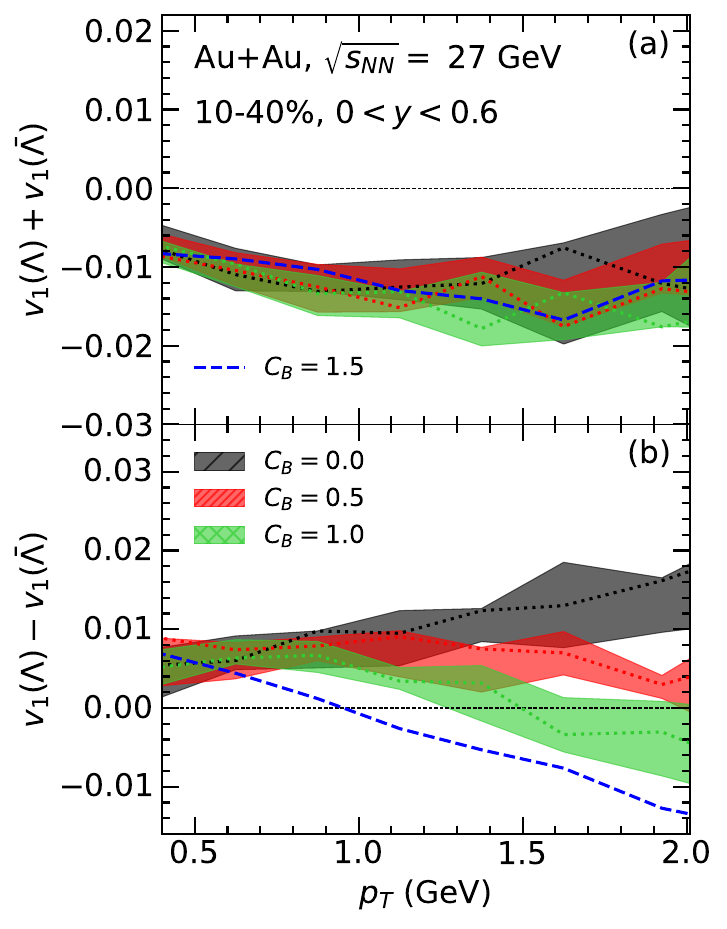}
        \caption{Model calculations of the transverse momentum dependence of (a) the sum of the directed flow and (b) the difference in directed flow between lambda ($\Lambda$) and anti-lambda ($\bar{\Lambda}$) in Au+Au collisions of 10-40\% centrality at $\sqrt{s_{\text{NN}}}=27$ GeV. The results, calculated with varying values of the baryon diffusion coefficient ($C_B$), are plotted. The shaded bands represent the systematic uncertainties arising from variations in model parameters.}
        \label{fig:Deltav1_LLbar_pT}
    \end{minipage}
    \label{fig:combined_figures}
\end{figure}

After calibrating our model parameters to reproduce the experimental data for the rapidity-dependent $v_1$ splitting between baryons and anti-baryons at 10–40\% centrality, we calculated their $p_T$ dependence. While calculating the $p_T$ differential $v_1$, we consider the hadrons within rapidity range $0<y<0.6$. We made this choice because in this region the $v_1(y)$ of identified hadrons calculated using different $C_B$ values are almost same. Fig. \ref{fig:Deltav1_ppbar_pT}(a) shows the sum of the $p_T$-differential $v_1$ for $p$ and $\bar{p}$ and Fig. \ref{fig:Deltav1_LLbar_pT}(a) presents the corresponding sum for $\Lambda$ and $\bar{\Lambda}$. These observables appear to be independent of the chosen $C_B$ value and, therefore, cannot be used to constrain $\kappa_B$. However, comparison of this observable with future experimental data would be a non-trivial check of our model, independent of $\kappa_B$.

In panel (b) of both Fig. \ref{fig:Deltav1_ppbar_pT} and Fig. \ref{fig:Deltav1_LLbar_pT}, we plot the $p_T$-differential $v_1$ difference for $p - \bar{p}$ and $\Lambda - \bar{\Lambda}$. This observable is notably sensitive to variations in $C_B$ and, consequently, $\kappa_B$. A comparison between the model and experimental data for this observable could enable the determination of $\kappa_B$, a fundamental transport coefficient that characterizes the QCD medium at non-zero baryon density. This finding represents the main result of our study.

We systematically studied the uncertainties arising from variations in other model parameters. The goal was to determine whether the observed sensitivity of this observable to $C_B$ remains robust under these variations. The shaded bands in Fig. \ref{fig:Deltav1_ppbar_pT} and Fig. \ref{fig:Deltav1_LLbar_pT} represent the systematic uncertainties introduced by varying key model parameters. Specifically, we varied the initial hydrodynamic starting time by 40\%, adjusted $\eta_0^{n_B}$ by 10\%, modified $\eta/s$ by 25\%, from their default value and also considered a temperature- and $\mu_B$-dependent $\eta/s$ following the prescription in Ref. \cite{Shen:2020jwv}. Additionally, we relaxed the initial Bjorken flow assumption using the approach described in Sec. \ref{baryon_tilt:Bjorken_flow_break} of Chapter \ref{ch:baryon_tilt}. These systematics were performed by varying one parameter at a time, with the remaining parameters readjusted as needed to maintain agreement with experimental data on rapidity-dependent charged particle yields, mid-rapidity net-proton yields, and the rapidity-dependent $v_1(y)$ of identified hadrons.

For $C_B = 1.5$, we found no suitable parameter space that simultaneously captures these basic observables when parameters are varied, so calculations for this value of $C_B$ were performed using only the default parameter set listed in Table \ref{param_for_model_diffCB_27GeV}. Despite these systematic variations, the sensitivity of $p_T$ differential $\Delta v_1$ to different $C_B$ values persists. Therefore, future experimental measurements of this observable will allow us performing a Bayesian analysis to put strong constraint on the value of $C_B$.

In Section \ref{emfield:prob_sign_change} of the previous chapter, we discussed how the baryon gradient along the $x$-direction evolves from central to peripheral collisions, leading to significant variations in the baryon diffusion current ($q^{\mu}$). This current, which arises as a response to the baryon gradient, differs significantly across centrality classes. Consequently, the centrality dependence of the splitting in directed flow ($\Delta \frac{dv_1}{dy}$) between baryons and anti-baryons exhibits a sign change. Since the baryon diffusion current  $q^{\mu} \sim \kappa_B \nabla_{\mu} \frac{\mu_B}{T}$, for a given centrality—or equivalently, for a specific baryon density profile—the strength of the diffusion current is directly influenced by the baryon diffusion coefficient $\kappa_B$. Therefore, variations in $\kappa_B$ can produce different magnitudes of baryon diffusion, which, in turn, could manifest as distinct trends in the centrality dependence of $\Delta \frac{dv_1}{dy}$. This behavior is precisely what we observe in our model calculations.

\begin{figure}[htbp]
  \centering
  \includegraphics[width=0.6\textwidth]{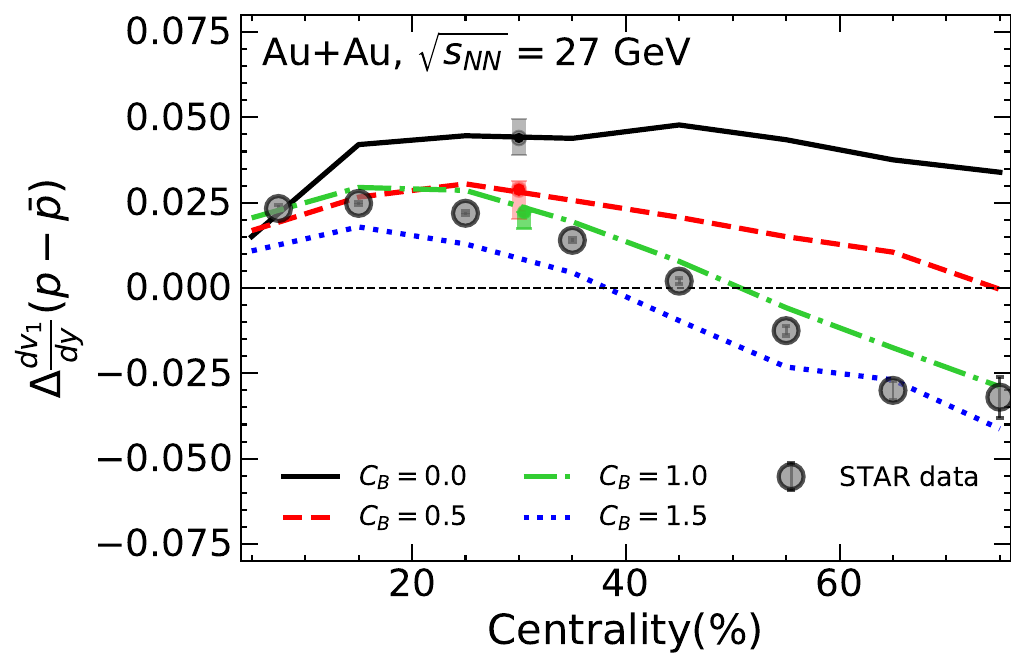} 
  \caption{Centrality dependence of the splitting of mid-rapidity directed flow slope ($\Delta \frac{dv_1}{dy}$) between protons ($p$) and anti-protons ($\bar{p}$) in Au+Au collisions at $\sqrt{s_{\text{NN}}}=27$ GeV. Model calculations are shown for four values of the baryon diffusion strength parameter $C_B$ (0, 0.5, 1.0, and 1.5). Error bars represent systematic uncertainties estimated from variations in model parameters. The systematic uncertainty calculations have been done only for 10-40\% centrality. The experimental data represented by symbols is from STAR collaboration \cite{STAR:2023jdd}. }
  \label{fig:v1y_4CB}
\end{figure}

We have presented the model calculations for the centrality dependence of $\Delta \frac{dv_1}{dy}(p - \bar{p})$ for different values of $C_B$ in Fig. \ref{fig:v1y_4CB}. To estimate the systematic uncertainties arising from variations in model parameters, we performed these variations for the $10$–$40\%$ centrality class and assumed these uncertainties to be approximately representative for all centralities. The parameter variations are identical to those described in the context of Figs. \ref{fig:Deltav1_ppbar_pT} and \ref{fig:Deltav1_LLbar_pT}, and the resulting uncertainties are indicated as error bars in the plot. The results reveal clear differences among the predictions for different $C_B$ values, which enable us to put constraints on $C_B$. Specifically, the $\Delta \frac{dv_1}{dy}(p - \bar{p})$  for $C_B = 0$ and $C_B = 0.5$ remain positive across all centralities, whereas those for $C_B = 1$ and $C_B = 1.5$ exhibit a sign change in mid-central collisions. The sign change occurs at increasingly central collisions for larger values of $C_B$. Focusing on centralities below $40\%$, where contributions from the electromagnetic field to this observable are expected to be minimal \cite{Gursoy:2018yai,STAR:2023jdd} and the hydrodynamic model framework is more reliable, our analysis suggests a preference for $0.5 < C_B < 1.5$. However, a more quantitative constraint on $\kappa_B$ would require a Bayesian analysis, which we leave for future work. Nonetheless, this study identifies relevant observables that can be employed in Bayesian analyses to impose robust constraints on $\kappa_B$.

An important point to note is that the constituents (quarks) in the system produced during heavy-ion collisions carry multiple conserved charges: baryon number (B), strangeness (S), and electric charge (Q). As a result, the diffusion processes of these conserved charges are interrelated \cite{Fotakis:2021diq,Rose:2020sjv,Fotakis:2019nbq,Greif:2017byw,Das:2021bkz,Dey:2024hhc}. A non-zero gradient in the density of any one conserved charge can induce diffusion currents ($q^{\mu}$) for all three charges. This interdependence implies that the evolution of these conserved charges is inherently coupled, as expressed in Eq. \ref{eq:kappa_matrix}. The resulting diffusion coefficient matrix includes diagonal terms ($\kappa_{BB}$, $\kappa_{SS}$, $\kappa_{QQ}$), representing the standard diffusion coefficients for each charge, and off-diagonal terms that quantify the coupling between the diffusion of different conserved charges:

\beq
\begin{pmatrix}
q^{\mu}_B \\ q^{\mu}_S \\ q^{\mu}_Q 
\end{pmatrix}
=
\begin{pmatrix}
\kappa_{BB} &  \kappa_{BS} & \kappa_{BQ} \\
\kappa_{SB} &  \kappa_{SS} & \kappa_{SQ} \\
\kappa_{QB} &  \kappa_{QS} & \kappa_{QQ} \\
\end{pmatrix}
\begin{pmatrix}
\nabla^{\mu} \left( \mu_B / T \right) \\
\nabla^{\mu} \left( \mu_S / T \right) \\
\nabla^{\mu} \left( \mu_Q / T \right) \\
\end{pmatrix}
\label{eq:kappa_matrix}
\eeq

Recent studies indicate that the off-diagonal elements of the diffusion matrix are comparable in magnitude to the diagonal terms \cite{Fotakis:2021diq,Rose:2020sjv,Fotakis:2019nbq,Greif:2017byw,Das:2021bkz}. This finding underscores the significance of coupling between the diffusion currents of conserved charges, particularly in simulations of low-energy heavy-ion collisions. Therefore, a comprehensive study of conserved charge diffusion requires estimating the full diffusion matrix, with all its elements extracted simultaneously through model-to-data comparisons. Such an approach necessitates the development of a hydrodynamic framework capable of evolving all three conserved charges alongside the energy-momentum tensor, starting from appropriate initial conditions and employing a suitable equation of state. Significant progress in this direction has been made by Fotakis et al. \cite{Fotakis:2022usk}, who derived second-order equations of motion for the dissipative quantities within the $(10 + 4N)$-moment approximation.

Research in the area of hydrodynamic evolution of all three conserved charges remains limited \cite{Pihan:2024lxw,Plumberg:2024leb}, with most studies focusing only on the net-baryon dynamics—the dominant conserved charge in the system produced during heavy-ion collisions \cite{Denicol:2018wdp,Li:2018fow,Denicol:2018wdp, Du:2022yok, Bozek:2022svy,Shen:2020jwv, Shen:2023awv,Shen:2022oyg,Shen:2017bsr,Du:2018mpf}. Recent progress has aimed to understand the dynamics of a single conserved charge density, laying the groundwork for future studies that could generalize these findings to encompass the dynamics of all conserved charges.

\section{Chapter summary}
In this chapter, we utilized a relativistic dissipative hydrodynamics framework with a Glauber model-based initial condition for energy and baryon deposition. This setup was carefully calibrated to reproduce the rapidity dependence of charged particle multiplicity, net proton yield, and the directed flow splitting ($v_1$) between protons and anti-protons. To model baryon diffusion, we adopted an ansatz derived from kinetic theory \cite{Denicol:2018wdp}: $\kappa_B = \frac{C_B}{T} n_B \left( \frac{1}{3} \coth{\left(\frac{\mu_B}{T}\right)} - \frac{n_B T}{\epsilon + p } \right)$, where $C_B$ is an arbitrary constant that remains largely unconstrained for the QCD  medium.
Our study revealed that the transverse momentum ($p_T$) dependence of the directed flow splitting between baryons and anti-baryons is highly sensitive to the variations in $C_B$. Based on this, we proposed the $p_T$-differential $v_1$ splitting between protons and anti-protons as a robust observable for probing the baryon diffusion coefficient or, equivalently, the parameter $C_B$ in the QCD medium. Additionally, we utilized recent measurements from the STAR collaboration of the mid-rapidity directed flow slope splitting ($\Delta \frac{dv_1}{dy}$) between protons and anti-protons as a function of collision centrality \cite{STAR:2023jdd}. These data allowed us to estimate the range of $C_B$, finding it to be constrained within $0.5 < C_B < 1.5$ under the current parametrization of $\kappa_B$. This study paves the way for further investigations into the complete $3 \times 3$ diffusion matrix of the QCD medium, using similar observables constructed from hadrons with relevant quantum numbers. Our work opens up a novel portal to probe the dynamics of the conserved charges in QCD medium by enabling theory to data comparison.

\def \la{\langle}
\def \ra{\rangle}
\def \v1even{v_1^{\text{even}}}
\chapter{Rapidity even $v_1$ of identified hadrons}
\label{ch:evenv1}

Hydrodynamic modeling of heavy-ion collisions at finite baryon density requires an initial baryon density profile as input. This initial baryon density is closely related to the baryon stopping mechanisim. However, no first-principle calculation currently exists to describe baryon stopping, making phenomenological modeling a  reliable approach \cite{Denicol:2018wdp,Pihan:2024lxw, Du:2022yok, Bozek:2022svy,Shen:2020jwv, Shen:2023awv,Shen:2022oyg,Shen:2017bsr,Du:2018mpf}. Recent phenomenological models for initial baryon deposition, inspired by the baryon junction picture, have proven effective in reproducing experimental data related to baryons, such as the rapidity-differential net-proton yield and the rapidity-differential directed flow of baryons \cite{Du:2022yok,Pihan:2024lxw}. These successes not only validate the utility of such models in probing baryon stopping but also provide an avenue for gaining deeper phenomenological insights into the baryon junction conjecture through model-to-data comparisons.

In the baryon junction picture, the valence quarks are connected by gluons in a Y-shaped topological structure, with the junction itself carrying the baryon number. Recently, efforts in relativistic heavy-ion collisions have focused on investigating potential signatures of this baryon junction \cite{STAR:2024lvy,Lewis:2022arg,Pihan:2023dsb,Pihan:2024lxw} to track the actual baryon number carriers in a hadron. It has been suggested that nucleon-nucleon collisions can involve both single- and double-junction stopping mechanisms \cite{Kharzeev:1996sq,Lewis:2022arg}.
For a nucleon-nucleon collision at beam rapidity $Y_{\text{beam}}$, the rapidity-dependent cross-sections for single-junction stopping are expressed as $\sim \exp{(\alpha_J (y-Y_{\text{beam}}))}$ for the target and $\sim \exp{(\alpha_J (y+Y_{\text{beam}}))}$ for the projectile, with $\alpha_J=0.42$ \cite{Kharzeev:1996sq}. In contrast, the cross-section for double-junction stopping exhibits no rapidity dependence, scaling as $\sim \exp{(Y_{\text{beam}})}$ \cite{Kharzeev:1996sq,Lewis:2022arg}.

In the phenomenologically successful baryon deposition model proposed in previous chapters, we introduced a two-component baryon deposition scheme in which the initial baryon profile receives contributions from both participant nucleons and binary collisions. This model could be interpreted in a baryon juction picture, where the baryon deposition by participant sources is associated with single-junction stopping, while double-junction stopping is linked to binary collisions. Drawing insights from the single-junction stopping cross-section, a rapidity-asymmetric baryon deposition is employed for the projectile and target participants \cite{Du:2022yok}. In contrast, for the binary collisions associated with double-junction stopping, we adopt a rapidity-symmetric deposition profile that generalises the rapidity independent double junction stopping cross-section. The baryons stopped immediately after the collision are not necessarily in thermal or chemical equilibrium \cite{Bialas:2016epd}. However, the initial profile used as input for hydrodynamic simulations represents the thermalized distribution of energy and net baryon density at a constant proper time $\tau_0$. Pre-equilibrium dynamics preceding thermalization can alter the spatial distribution of baryon density, offering flexibility to deviate from a strictly flat rapidity distribution for the symmetric component.

Notably, in a recent study (Ref.~\cite{Du:2022yok}), along with rapidity-asymmetric deposition for participants, a rapidity-flat baryon deposition profile is combined to account for double-junction stopping (though this rapidity flat deposition profile was not explicitly associated with binary collisions). This approach successfully reproduces the directed flow ($v_1$) of protons over a wide range of collision energies.

In this work, we investigate this rapidity-symmetric deposition profile that could be associated with double-junction stopping. We explore two different types of initial net-baryon deposition model, which differ only in the modeling of the symmetric component of the baryon deposition. However, the asymmetric deposition component, associated with participant nucleons and single-junction stopping, is kept identical in both profiles. 

In model calculation, we show that the choice of the rapidity-symmetric profile is sensitive to the rapidity-even $v_1$ ($\v1even$) splitting between protons and anti-protons. Future experimental measurements of this $\v1even$ splitting can provide strong constraints on the rapidity-symmetric deposition profile of net-baryons. This could offer valuable phenomenological insights into the baryon junction picture.

\section{Rapidity even profiles}

\begin{figure}[htbp]
  \centering
  \includegraphics[width=0.9\textwidth]{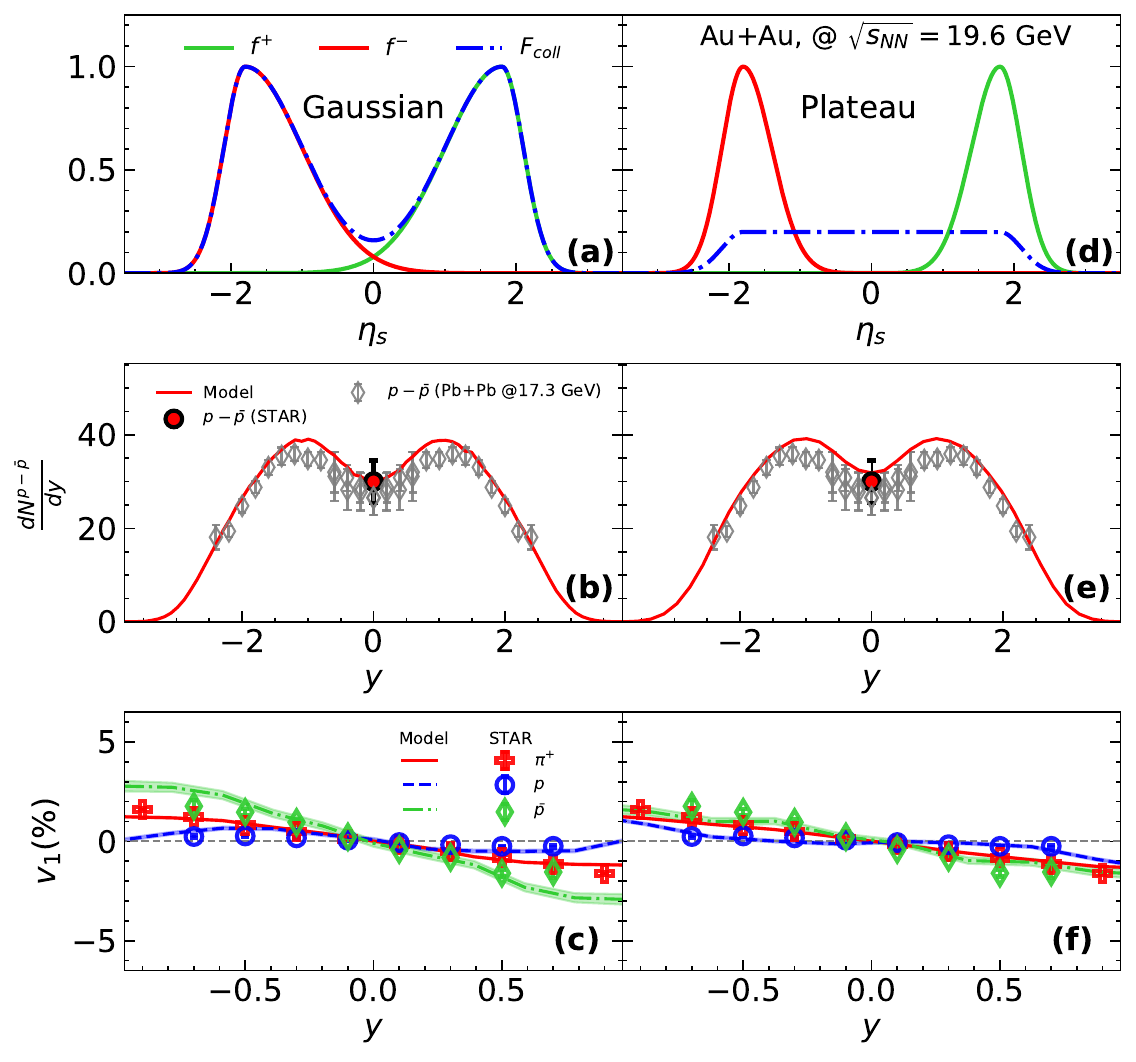} 
  \caption{Demonstration of the Gaussian and Plateau baryon deposition profiles used in our study, shown in panels (a) and (d), respectively. Using these profiles, we performed hybrid simulations of Au+Au collisions at $\sNN=19.6$ GeV. The rapidity-differential net proton yields calculated with the Gaussian and Plateau profiles are presented in panels (b) and (e), while panels (c) and (f) display the rapidity-odd $v_1$ for $\pi^{+}$, $p$, and $\bar{p}$ obtained from the Gaussian and Plateau profiles, respectively. Model calculations (lines) are compared with experimental data (markers). The rapidity-differential net proton measurements from Pb+Pb collisions at $\sNN=17.3$ GeV (NA49 Collaboration) \cite{NA49:2010lhg} are included for comparison, alongside mid-rapidity net proton yields from the STAR Collaboration \cite{STAR:2017sal}, shown as red circles in panels (b) and (e). Additionally, the rapidity-odd $v_1$ of identified hadrons is compared to measurements from the STAR Collaboration \cite{STAR:2014clz}. }
  \label{fig:Gauss_plat_netp_v1_smooth}
\end{figure}

The forward-backward symmetric profile taken in our default initial baryon deposition model is outlined in Eq. \ref{eq:fcollnb_default} and shown in Fig. \ref{fig:Gauss_plat_netp_v1_smooth}(a). Additionally, in this chapter, we adopt a plateau-type form, as used in Ref. \cite{Du:2022yok}, for the deposition by binary collision sources. The initial three-dimensional distribution of net-baryon density with a plateau-type symmetric envelope profile is constructed similarly to Eq. \ref{eq:two_component_baryon_profile}, but with the $f_{coll}^{n_B}$ component taking the following form:
\begin{equation}
    f_{coll}^{n_B} =  \exp \left(  -\frac{ \left( \vert \eta_{s} \vert - \eta_{0}^{n_B} \right)^2}{2 \sigma_{B,+}^2}   
    \theta (\vert \eta_{s} \vert - \eta_{0}^{n_B} ) \right)
\end{equation}

This plateau-type baryon deposition scheme is illustrated in Fig. \ref{fig:Gauss_plat_netp_v1_smooth}(d). Moving forward, we will refer to our default profile, as outlined in Eq. \ref{eq:fcollnb_default}, as the Gaussian profile, and this new type of profile will be called the plateau profile.

Our goal now is to study the observable consequences of these two different profiles. We aim to demonstrate observables that, upon measurement and comparison with our model calculations, could distinguish between these two types of baryon deposition profiles. Such comparisons would provide crucial phenomenological insights into the baryon stopping mechanism in heavy-ion collisions, as well as into the conjectured baryon junction picture.

\begin{table}[ht]
\centering
\begin{tabular}{|p{1.3cm}|p{1.3cm}|p{1.3cm}|p{1.3cm}|p{1.3cm}|p{1.3cm}|}
\hline
\multicolumn{6}{|c|}{Gaussian Profile} \\
\hline 
$C_B$ &  $\eta_{0}^{n_{B}}$ & $\sigma_{B,-}$ & $\sigma_{B,+}$ & $\omega$ & $\eta_m$ \\
\hline
1.0 & 1.8 & 0.8 & 0.3 & 0.15 & 0.8 \\
\hline
\multicolumn{6}{|c|}{Plateau Profile} \\
\hline 
$C_B$ & $\eta_{0}^{n_{B}}$ & $\sigma_{B,-}$ & $\sigma_{B,+}$ & $\omega$ & $\eta_m$ \\
\hline
1.0 & 1.8 & 0.4 & 0.3 & 0.06 & 1.0 \\
\hline

\hline 
\end{tabular}
\caption{Parameter values for the two types of initial baryon deposition profiles, Gaussian and plateau, used in the simulation at $\sqrt{s_{\text{NN}}}=19.6$ GeV. The Gaussian and plateau profiles are depicted in Fig. \ref{fig:Gauss_plat_netp_v1_smooth}(a) and Fig. \ref{fig:Gauss_plat_netp_v1_smooth}(d), respectively. }
\label{tab:param_for_diff_prof}
\end{table}

We have performed hydrodynamic + hadronic transport simulations using both types of initial net-baryon profiles. For consistency, we take the same tilted initial condition for energy distribution, as detailed in Sec.\ref{Sec:TIC}, with both the baryon profiles. For the simulation, we prepared event-averaged smooth initial conditions, as described in Sec.\ref{sec2}, and carried out single-shot hydrodynamic evolutions of these initial conditions. The simulation results for Au+Au collisions at $\sNN=19.6$ GeV are presented in Fig.~\ref{fig:Gauss_plat_netp_v1_smooth}. In Figs.\ref{fig:Gauss_plat_netp_v1_smooth}(b) and \ref{fig:Gauss_plat_netp_v1_smooth}(e), we show the rapidity distributions of net-protons calculated using the Gaussian and plateau profiles, respectively, along with the corresponding experimental data. Additionally, Figs.\ref{fig:Gauss_plat_netp_v1_smooth}(c) and \ref{fig:Gauss_plat_netp_v1_smooth}(f) present the $v_1(y)$  for $\pi^{+},p$ and $\bar{p}$ calculated with these two profiles. The specific parameter values assigned to each profile are listed in Table~\ref{tab:param_for_diff_prof}. We have already studied the Gaussian profile in detail in previous chapters, and as expected, it explains the experimental data well. Interestingly, with the plateau-type deposition scheme, we identified parameter sets that effectively reproduce both the experimental data for the rapidity distribution of net-protons and the $v_1(y)$ of $\pi^{+},p$ and $\bar{p}$ simultaneously.

Although not shown here, it is worth mentioning that we also found parameter spaces at $\sNN=27$ and $\sNN=200$ GeV that successfully captures the experimental data on the rapidity distribution of net-protons and the $v_1(y)$ of $\pi^{+},p$ and $\bar{p}$ for both profiles. This observation suggests that the rapidity-odd $v_1$ cannot distinguish between the symmetric profile forms, as both the plateau and Gaussian profiles explain the experimental data equally well. However, since the rapidity-even profile differs between the two cases, we should calculate and investigate the rapidity-even component of $v_1$. As discussed in Chapter \ref{ch:intro}, the rapidity-even $v_1$ arises due to event-by-event fluctuations in nucleon positions, and therefore it requires event-by-event hydrodynamic simulations, which we have not yet performed. Instead, we have used smooth initial conditions for the hydrodynamic evolution.

\section{Effect of event-by-event fluctuation on rapidity-odd $v_1$}

Since the rapidity-odd $v_1$ is generated by the forward-backward symmetry breaking in the initial condition, it is primarily driven by the geometry of the collsion and expected to be less sensitive to fluctuations. Consequently, in our previous calculations, we used smooth initial conditions as input to the hydrodynamics, neglecting event-by-event fluctuations. Moreover, in hydrodynamic model study of rapidity odd $v_1$, it is always a common practice to take smooth initial condition as input \cite{Du:2022yok, Bozek:2022svy, Shen:2020jwv,Jiang:2021foj,Bozek:2010bi,Jiang:2021ajc,Jiang:2023fad}. However, for the study of rapidity-even $v_1$, it is necessary to perform event-by-event simulations.

\begin{figure}[htbp]
  \centering
  \includegraphics[width=0.5\textwidth]{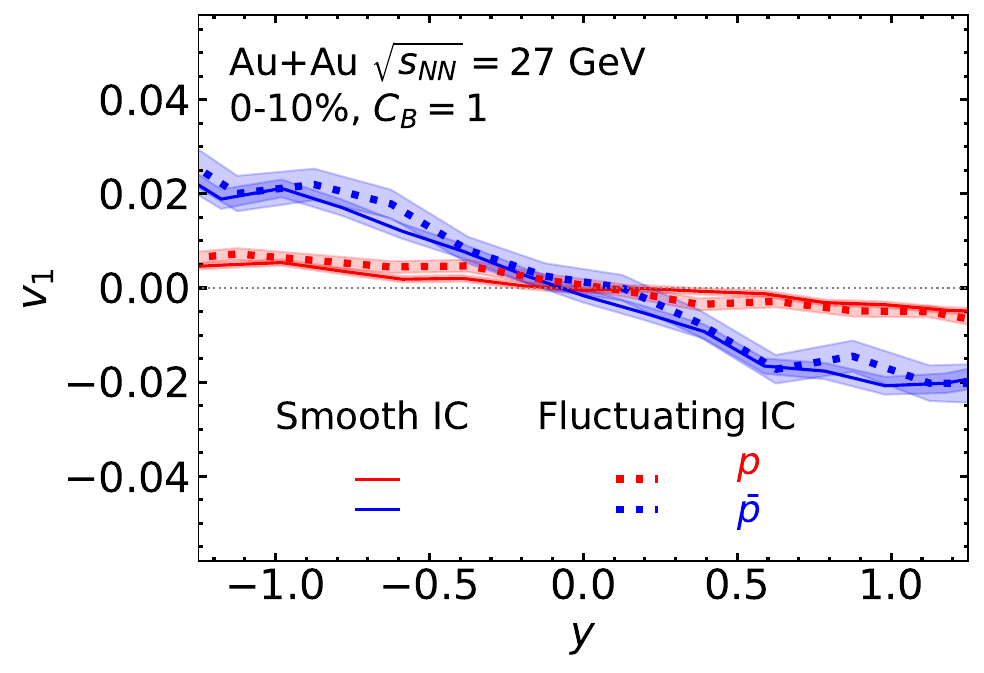} 
  \caption{The rapidity-odd $v_1$ for $p$, and $\bar{p}$ in 0-10\% centrality Au+Au collisions at $\sNN = 27$ GeV. The simulation results calculated using smooth and fluctuating initial conditions are compared. The results highlight that, there is negligible impact of event-by-event fluctuations on the rapidity-odd component of directed flow for identified hadrons.}
  \label{fig:v1smoothfluc}
\end{figure}

By performing event-by-event simulations at $\sNN=27$ GeV for 0-10\% centrality, using the same parameter values obtained by matching with experimental data in the smooth initial condition case, we observed that the rapidity-odd $v_1$ of identified hadrons is minimally affected by the event-by-event nucleon position fluctuations in the model. To demonstrate the effect of these fluctuations, in Fig.~\ref{fig:v1smoothfluc}, we compare the $v_1(y)$ of protons and anti-protons obtained from simulations with smooth initial conditions and with event-by-event simulations.

Our model calculations also show that the rapidity distribution of net-protons, as well as the rapidity distribution of identified hadron yields, does not change when performing event-by-event simulations, provided the same parameter values are used as in the smooth initial condition case.

These observations allow us to perform event-by-event hydrodynamic simulations with the same parameter values used in the smooth initial condition case, enabling us to reliably describe experimental data and make predictions for the rapidity-even $v_1$ of identified hadrons.

\section{Rapidity-even $v_1$ of charged hadrons}

To study the rapidity-even $v_1$, we performed event-by-event model simulations for Au+Au collisions in 0-10\% centrality at $\sNN=$ 200, 27, and 19.6 GeV. The hydrodynamic evolution were conducted with a non-zero baryon diffusion coefficient $C_B=1$ and using the NEoS-BQS equation of state \cite{Monnai:2019hkn}. The parameter values used in the simulations\footnote{It is important to note that, with these model parameters, we are able to capture the experimental measurements of various bulk observables, such as the rapidity distribution of charged particle yields, the rapidity distribution of net proton yields, $v_2(p_T)$ of charged hadrons, and rapidity-odd $v_1(y)$ of identified hadrons simultaneously.} with the Gaussian profile are as detailed in Table \ref{param_for_model} of Chapter \ref{ch:baryon_tilt}, whereas for the plateau profile the paramters are mentioned in Table \ref{param_for_plateau_profile}.

\begin{center}
\begin{table}[h!]
\centering
\begin{tabular}{|p{0.9cm}|p{0.4cm}|p{0.7cm}|p{1.1cm}|p{0.55cm}|p{0.4cm}|p{0.4cm}|p{0.5cm}|p{0.6cm}|p{0.6cm}|p{0.45cm}|p{0.55cm}|}
\hline 
$\sqrt{S_{NN}}$ \tiny{(GeV)} & $C_B$ & $\tau_0$\tiny{(fm)} &$\epsilon_{0}$ \tiny{(GeV/fm$^{3}$)} & $\alpha$ &  $\eta_{0}$ & $\sigma_{\eta}$ & $\eta_{0}^{n_{B}}$ & $\sigma_{B,-}$ & $\sigma_{B,+}$ & $\eta_m$ & $\omega$ \\ \hline
200  & 1.0 & 0.6  &  8.0 & 0.14  &  1.3  &  1.5  &  5.0  &  1.6   &  0.1  & 2.5 & 0.03  \\ 
\hline
27   & 1.0 & 1.2  &  2.4 & 0.11  &  1.3  &  0.7  &  2.3  &  0.8   &  0.2  & 1.0 & 0.06  \\ 
\hline
19.6 & 1.0 & 1.8  &  1.55 & 0.1  &  1.3  &  0.4  &  1.8  &  0.4   &  0.3 & 1.0 & 0.06  \\ 
\hline
\end{tabular}
\caption{ Model parameters used in the simulations employing the plateau profile. }
\label{param_for_plateau_profile}
\end{table}
\end{center}

For each collision energy, we performed hydrodynamic evolution for 200 initial configurations, and after evolving each configuration, we obtained the hypersurface of $\epsilon_f = 0.26$ GeV/fm$^3$. From each hypersurface, we performed 1000 particleization events. Subsequently, the particles sampled from these hypersurfaces were put event by event into the transport code for late-stage evolution. Taking the final hadron obtained from these $200 \times 1000$ events, we evaluated the rapidity-even $v_1$ using the prescription proposed by Luzum and Ollitrault \cite{Luzum:2010fb}. In this method, the event plane 
$Q$ vector is defined with a weight $w_i=p_T - \frac{\la p_T^2 \ra}{\la p_T \ra}$.
\beq
Q \exp(i\Psi_1) = \la w_i \exp(i\phi_i) \ra
\eeq
The $Q$-weighted value of the rapidity even component of the directed flow coefficient is calculated as:
\beq
\v1even = \frac{ \la Q \cos(\phi_i - \Psi_1)  \ra }{\sqrt{\la Q^2 \ra}} 
\eeq
where $\la...\ra$ represents the average over all tracks within the selected kinematic cut.

\begin{figure}[htbp]
  \centering
  \includegraphics[width=0.9\textwidth]{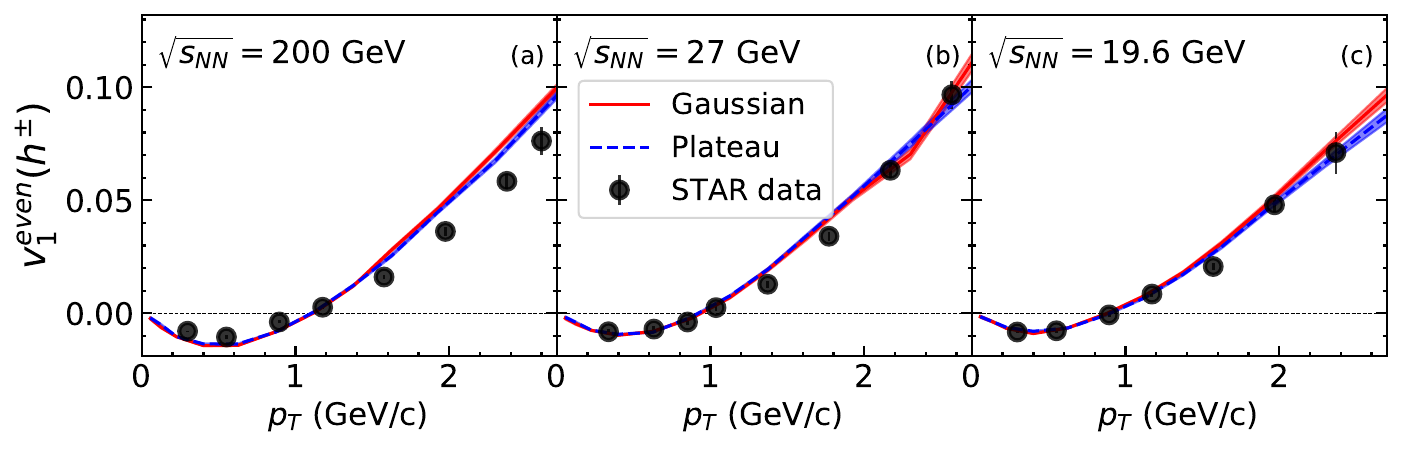} 
  \caption{Transverse momentum ($p_T$) dependence of the rapidity-even directed flow ($v_1^{\text{even}}$) for charged hadrons in Au+Au collisions at (a) $\sNN=200$ GeV, (b) $\sNN=27$ GeV, and (c) $\sNN=19.6$ GeV of 0-10\% centrality. Model calculations using Gaussian and plateau profiles are represented by solid and dashed lines respectively, while experimental data points from the STAR collaboration are depicted as markers for comparison \cite{STAR:2018gji}. }
  \label{fig:v1evenptch}
\end{figure}

\begin{figure}[htbp]
  \centering
  \includegraphics[width=0.5\textwidth]{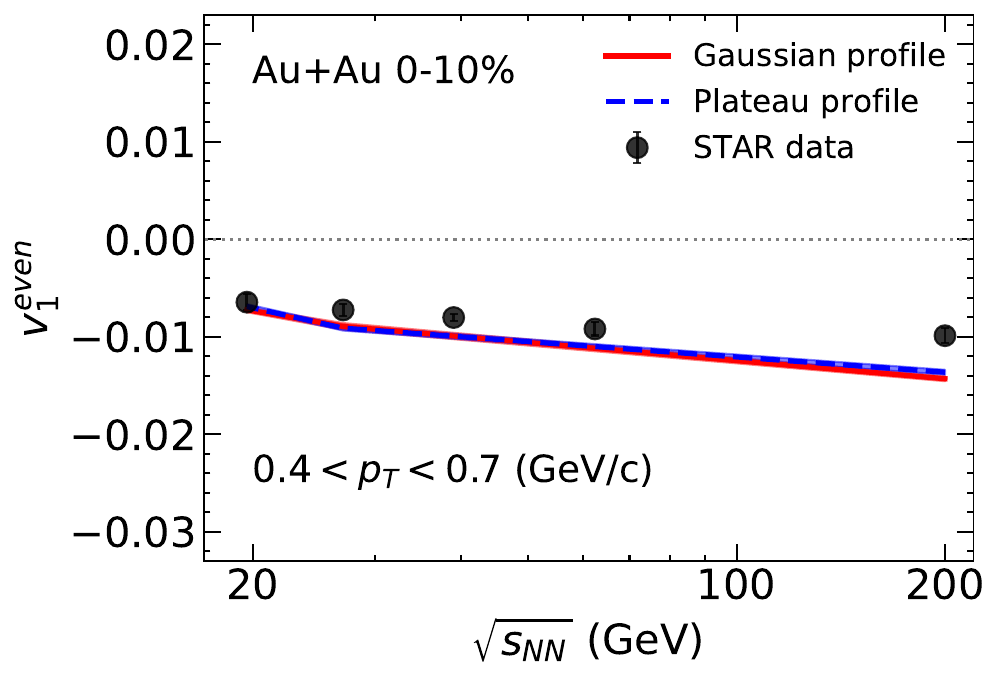} 
  \caption{Beam energy ($\sNN$) dependence of the rapidity-even directed flow ($v_1^{\text{even}}$) for charged hadrons in Au+Au collisions of 0-10\% centrality. Model calculations employing Gaussian and plateau profiles are shown as solid and dashed lines, respectively, while experimental data from the STAR collaboration are represented by markers for comparison \cite{STAR:2018gji}.}
  \label{fig:v1evensnnch}
\end{figure}

First, we studied our model ability to capture the already measured $p_T$ and $\sNN$ dependence of the rapidity-even $v_1$ of charged hadrons. In Fig.\ref{fig:v1evenptch}, we plot the $p_T$-differential $\v1even$ of charged hadrons in the mid-rapidity region for $\sNN=$ 200, 27, and 19.6 GeV. Then, in Fig.\ref{fig:v1evensnnch}, we plot the $\sNN$ dependence of $\v1even$ and compare it with experimental measurements from the STAR collaboration \cite{STAR:2018gji}. We observed a very small beam energy dependence in the charged hadron $\v1even$, which is also observed in the experimental data. Most importantly, our model calculations, using both the Gaussian and plateau-type initial baryon deposition profiles, exhibit the same $\v1even$ of charged hadrons and show a very good agreement with the experimental data. As expected, the charged hadron $\v1even$, primarily dominated by the contribution of pions, is mainly generated by energy density fluctuations within the fireball and is insensitive to the baryon profile.

\section{Rapidity-even $v_1$ of identified hadrons}

\begin{figure}[htbp]
  \centering
  \includegraphics[width=0.9\textwidth]{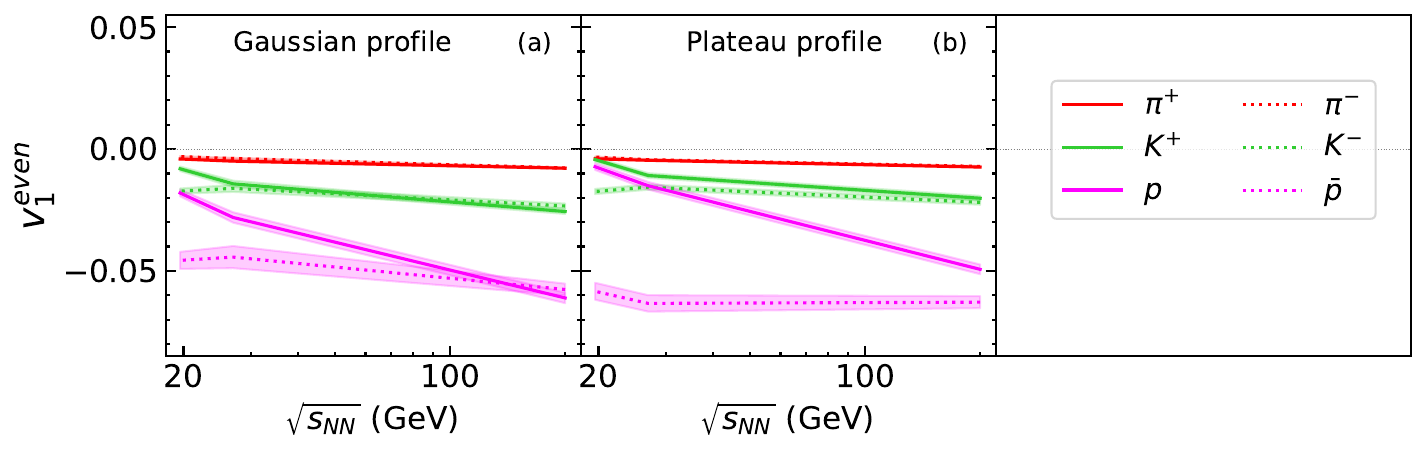} 
  \caption{Beam energy ($\sNN$) dependence of the rapidity-even directed flow ($v_1^{\text{even}}$) for identified particles ($\pi^+$, $\pi^-$, $K^+$, $K^-$, $p$, and $\bar{p}$) in Au+Au collisions of 0-10\% centrality . This $\v1even$ measurement has been done in the mid-rapidity region ($\vert y \vert<0.5$). Model calculations using Gaussian and plateau baryon deposition profiles are shown in panel (a) and (b), respectively.}
  \label{fig:v1evensnniden}
\end{figure}

\begin{figure}[htbp]
  \centering
  \includegraphics[width=0.5\textwidth]{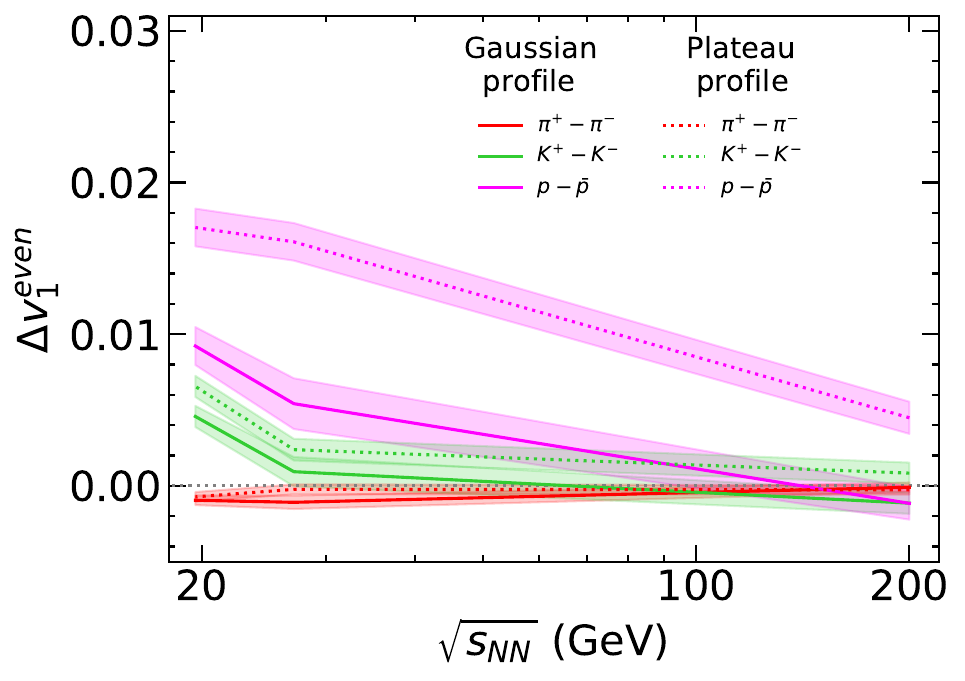} 
  \caption{Beam energy ($\sNN$) dependence of the rapidity-even directed flow ($v_1^{\text{even}}$) splitting between $(\pi^{+}-\pi^{-}), (K^{+}-K^{-})$ and $(p-\bar{p})$ in Au+Au collisions of 0-10\% centrality. This $\v1even$ measurement has been done in the mid-rapidity region ($\vert y \vert<0.5$). The model calculations, using both the Gaussian and plateau profiles, are represented by solid and dotted lines, respectively.}
  \label{fig:Deltav1eveniden}
\end{figure}

In our model, we calculated the beam energy dependence of the rapidity-even $v_1$ for identified hadrons ($\pi^+,\pi^-,K^+,K^-,p,\bar{p}$) in the mid-rapidity region, as shown in Fig. \ref{fig:v1evensnniden}. The results obtained using the Gaussian profile and the plateau profile are presented separately in panel (a) and panel (b), respectively. We observed that the beam energy dependence of $\v1even$ for pions, kaons, and protons differs significantly where proton $\v1even$ shows the strongest dependence on $\sNN$. There is no noticeable difference between $\v1even$ for $\pi^{+}-\pi^{-}$, while a significant difference is observed between $K^{+}$ and $K^{-}$ at lower $\sNN$. Most notably, the $\v1even$ for protons and anti-protons, as well as the splitting between them, exhibits a substantial difference between the Gaussian and plateau profiles.

To quantify the splitting of $\v1even$, we plotted $\Delta \v1even$ for $\pi^{+}-\pi^{-}$, $K^{+}-K^{-}$ and $p-\bar{p}$ as a function of collision energy in Fig. \ref{fig:Deltav1eveniden}. We found that there is a significant splitting between $\v1even$ for protons and anti-protons in both the plateau and Gaussian profiles. In event-by-event simulation, the positions of the $N_{part}$ and $N_{coll}$ sources fluctuate from event to event, which in turn causes fluctuations in the constructed baryon density profile. This fluctuation leads to the splitting of $\v1even$ between protons and anti-protons. As the baryon stopping effect becomes more pronounced at lower collision energies, the splitting increases as the collision energy decreases. Additionally, we observed a finite splitting between $K^{+}$ and $K^{-}$, which can be attributed to the non-zero strangeness chemical potential $\mu_S$ of the medium, generated by the NEoS-BQS EoS \cite{Monnai:2019hkn}. This EoS imposes the constraint that the strangeness density $n_S=0$ and the electric charge density $n_Q=0.4 n_B$. As the strangeness density is coupled to the baryon density in our model, $\Delta \v1even (K^{+}-K^{-})$ follows a similar trend to $\Delta \v1even (p-\bar{p})$ as a function of collision energy.  Notably, our model calculations do not account for the diffusion of strangeness or electric charge. Future improvements in our model, incorporating independent initialization and the evolution of all three conserved charges (baryon number, electric charge, and strangeness) along with their diffusion-could alter the predicted $\v1even$ splittings between $\pi^{+} - \pi^{-}$ and $K^{+} - K^{-}$.

The Gaussian profile shows nearly half the magnitude of $\Delta \v1even (p-\bar{p})$ at all considered $\sNN$ compared to the plateau profile. Therefore, future measurements of the rapidity-differential $\v1even$ for identified hadrons, especially the $\v1even$ splitting between protons and anti-protons, could serve to discriminate between different initial baryon profiles. These measurements could provide stringent constraints on modeling the initial baryon profile.

\section{Rapidity-differential splitting of $\v1even$ between proton and anti-proton}

\begin{figure}[htbp]
    \centering
    \begin{minipage}{0.45\textwidth} 
   \centering
  \includegraphics[width=0.99\textwidth]{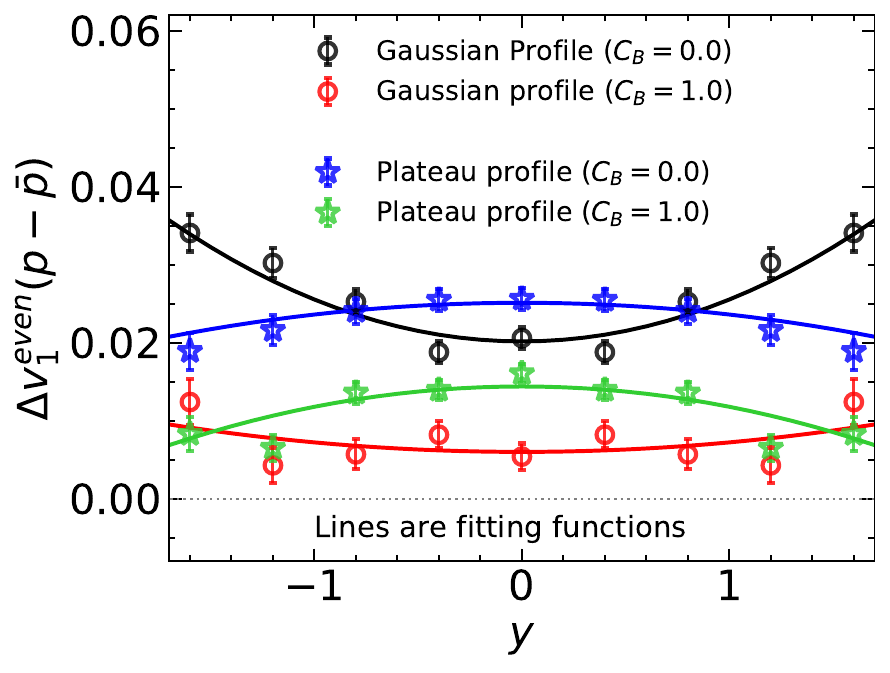} 
  \caption{Rapidity differential $v_1^{\text{even}}$ splitting between protons and anti-protons, $\Delta v_1^{\text{even}} (p-\bar{p})$ in 0-10\% centrality Au+Au collisions at $\sNN=27$ GeV. The model  results obtained using both Gaussian and plateau profiles, for both $C_B=0$ and $C_B=1$ are presented for comparison.}
  \label{fig:Deltav1y_ppbar_27GeV}
    \end{minipage}
    \hfill 
    \begin{minipage}{0.45\textwidth} 
  \centering
  \includegraphics[width=0.99\textwidth]{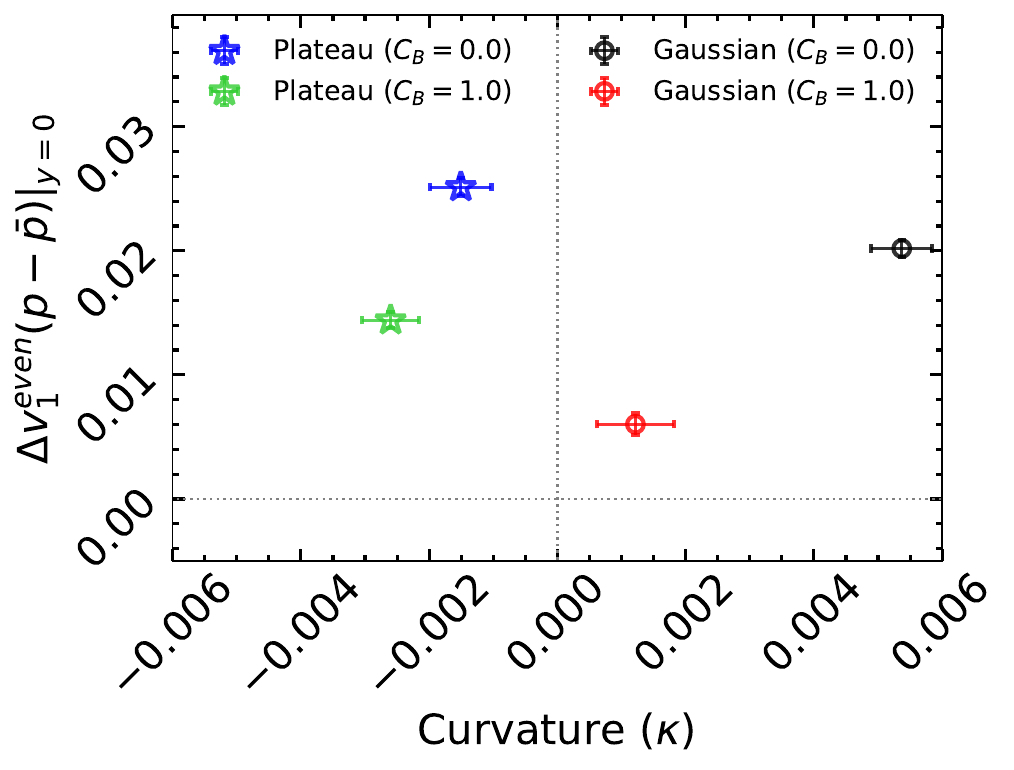} 
  \caption{Mid-rapidity $\Delta v_1^{\text{even}} (p-\bar{p})$ magnitude and the second-order derivative of the rapidity differential $v_1^{\text{even}}$ splitting between protons and anti-protons at mid-rapidity (denoted as curvature $\kappa$) in 0-10\% centrality Au+Au collisions at $\sNN=27$ GeV. Model calculation results using both Gaussian and plateau profiles, along with simulations for $C_B=0$ and $C_B=1$, are presented for comparison. }
  \label{fig:v1even_curvature_2D}
    \end{minipage}
    \label{fig:combined_figures}
\end{figure}

We have plotted the rapidity-differential $\Delta \v1even (p-\bar{p})$ for 0-10\% centrality Au+Au collisions at $\sNN=$ 27 GeV in Fig. \ref{fig:Deltav1y_ppbar_27GeV}. The model results obtained from the calculations with the Gaussian and plateau profiles are compared. In previous sections of this chapter, we presented the $\v1even$ results calculated using a non-zero baryon diffusion coefficient ($C_B=1$). For comparison, in Fig. \ref{fig:Deltav1y_ppbar_27GeV}, we also show the results from our model calculations with $C_B=0$. We observed that the shape of the rapidity dependence of $\Delta \v1even (p-\bar{p})$ differs significantly between the Gaussian and plateau profiles. Specifically, the rapidity dependence is concave in the Gaussian profile case, whereas it is convex in the plateau profile case. Furthermore, there is a strong dependence of $\Delta \v1even (p-\bar{p})$ on $C_B$. In the $C_B=0$ case, the splitting is more pronounced at all rapidities compared to the $C_B=1$ case for both profiles.

To present the results more compactly and facilitate comparison of the model calculations with different baryon profiles and $C_B$ values, we have plotted the magnitude of $\Delta \v1even (y)$ at mid-rapidity and the second-order derivative of rapidity differential $\Delta \v1even (p-\bar{p})$ at mid-rapidity (referred to as the curvature, $\kappa$) together in Fig. \ref{fig:v1even_curvature_2D}. From this plot, it is evident that the curvature for the Gaussian profile is positive, while it is negative for the plateau profile. Additionally, there is a difference in the magnitude of the mid-rapidity 
$\Delta \v1even (p-\bar{p})$ between the $C_B=0$ and $C_B=1$ cases for both profiles. The calculations with different baryon profiles and $C_B$ values are distinct in the $\Delta \v1even(y)\vert_{y=0}$ and $\frac{d^2 \Delta \v1even(y)}{dy^2}$ (curvature) space.

Our observations suggest that simultaneous experimental measurements of the mid-rapidity value of $\Delta \v1even (p-\bar{p})$ and its curvature at mid-rapidity, followed by comparison with model calculations, could provide constraints on both the baryon diffusion coefficient and the initial baryon profile of the medium.

As discussed earlier, the Gaussian and plateau profiles differ in the symmetric component of the baryon deposition profile. This symmetric component is associated with double junction stopping in the baryon junction picture. The sensitivity of the rapidity-differential $\v1even$ splitting between protons and anti-protons to these two types of baryon deposition profiles highlights the importance of this observable as a crucial experimental tool to get valuable phenomenological insights into the baryon stopping mechanism in heavy ion collisions and to put constraints on the baryon diffusion coefficient of the medium.

\section{Chapter summary}

Our two-component baryon deposition model can be interpreted within the baryon junction picture. In this Glauber geometry-based phenomenological model of initial baryon stopping, the baryon deposition along rapidity consists of both a forward-backward asymmetric component and a symmetric component. The forward-backward asymmetric deposition, associated with participant sources, follows the single-junction stopping picture, while the symmetric component, arising from binary collisions, is linked to double-junction stopping. 

In the double-junction stopping, the stopping cross-section is independent of rapidity \cite{Kharzeev:1996sq}. Generalizing this and preserving the symmetry of the double-junction stopping cross-section, we introduced a rapidity-dependent but symmetric baryon deposition profile for the binary collision sources in our model. Alternatively, in this chapter, we considered a rapidity-independent flat baryon deposition profile for the binary collision sources, following exactly the double-junction stopping cross-section form. These two profiles, differing only in the symmetric component of baryon deposition, are referred to as the Gaussian and Plateau profiles in our study.

We performed hydrodynamic evolution with both these initial baryon deposition profiles. Both profiles were able to capture important bulk observables, such as the rapidity-odd $v_1$ of identified hadrons and the rapidity-even $v_1$ of charged hadrons. However, we found that the rapidity-even $v_1$ measurement of identified particles is particularly sensitive to the symmetric component of the initial baryon deposition, especially the $\v1even$ splitting between proton and anti-proton. This sensitivity allows for discrimination between the two profiles. Furthermore, we showed that the rapidity-even $v_1$ splitting could also provide constraints on baryon diffusion.

Our key proposal is that future experimental measurements of the rapidity-differential $v_1^{\text{even}}$ splitting between protons and anti-protons will offer crucial phenomenological insights into the baryon junction picture. These measurements could also provide stringent constraints in modelling the initial baryon distribution profile, and help constrain the baryon diffusion coefficient of the medium.

\backmatter
\chapter{Thesis summary}

The Beam Energy Scan (BES) program at the Relativistic Heavy Ion Collider (RHIC) is dedicated to probe the Quantum Chromodynamics (QCD) phase diagram in detail. A key objective is to identify experimental signatures of the conjectured QCD critical point, which is predicted to occur at finite baryon chemical potential. To this end, event-by-event fluctuation measurements of net-baryon numbers have been performed across a broad range of collision energies. However, these fluctuation measurements are performed within specific rapidity windows. Fluctuations arising from event-by-event fluctuations in initial baryon stopping and its subsequent evolution can introduce non-critical contributions to the relevant observables in the measured rapidity window. These contributions, being unrelated to critical phenomena, necessitate the establishment of a robust baseline for non-critical fluctuations, against which critical fluctuations can be identified. .

Additionally, recent measurements by the STAR collaboration have investigated the directed flow splitting between oppositely charged hadrons as a potential signature of electromagnetic (EM) fields generated in heavy-ion collisions. The chosen pairs of oppositely charged hadrons, such as protons and anti-protons, not only have opposite electric charges but also carry opposite baryon numbers. Consequently, these observables, intended for the study of EM fields, are also expected to be significantly influenced by background effects arising from baryon stopping dynamics. This underscores the necessity of a precise understanding of the baseline behavior of relevant observables in the absence of both critical phenomena and EM fields at BES energies. Such baselines are essential for isolating and identifying novel physics signals.

Achieving this goal requires comprehensive dynamical simulations of the entire system at finite baryon density. These simulations must begin with proper initial conditions, particularly for baryon deposition, and incorporate accurately constrained transport coefficients. In this context, the focus of this thesis is a phenomenological investigation of initial baryon stopping and its diffusion during the system's evolution within a hydrodynamic framework. Specifically, we explore net-baryon-related physics at BES energies by studying the directed flow ($v_1$) of identified hadrons, with particular attention to baryons and anti-baryons.

The thesis begins with a brief introduction to heavy ion physics and outlines the motivation for a detailed study of baryon dynamics at BES in Chapter 1. Chapter 2 provides an overview of the hybrid framework—integrating hydrodynamic evolution and hadronic transport—established in Ref. \cite{Denicol:2018wdp}, which serves as the foundation for this study.

In Chapter 3, we examine the directed flow in the absence of baryon density. Subsequently, we revisit the phenomenologically successful tilted energy deposition model proposed by Bożek and Wyskiel, comparing it with the more recent tilted source model introduced by the CCNU group. Our analysis reveals that the CCNU model is essentially a generalization of the Bożek-Wyskiel model. We investigate in detail how the initial state tilted geometry induces asymmetric sideward flow, reproducing the rapidity-differential directed flow of charged hadrons.

Despite these advancements, existing models have consistently struggled to reproduce the directed flow of identified hadrons, particularly protons and anti-protons, across the collision energy range explored in the BES program. This failure can be attributed to the improper modeling of the initial baryon distribution in the thermalized fireball, which is closely linked to baryon stopping. To address this challenge, Chapter 4 introduces a novel initial baryon deposition model based on Glauber geometry. This model employs a two-component deposition scheme in which baryon deposition within the thermalized fireball is influenced by both participant nucleons and binary collision sources.

The two-component model is inspired by insights from microscopic models such as LEXUS, which suggest that baryon stopping depends on the number of binary collisions. Additionally, the model could be interpreted within the baryon junction picture. In this framework, single junction stopping involves one unit of baryon number and produces a forward-backward asymmetry in the rapidity deposition profile, depending on whether the junction originates from the forward- or backward-going nucleus. Conversely, double junction stopping, formed by pair of junctions from the two colliding nucleons, deposits two units of baryon number and yield a symmetric rapidity deposition profile. Accordingly, in our model, single junction stopping could be interpreted to associate with participant nucleons, while double junction stopping linked with binary collisions.

When coupled with a tilted energy density profile, our baryon deposition model successfully describes the rapidity-differential $v_1$ of identified hadrons across the collision energy range $\sqrt{s_{NN}} = 7.7 - 200$ GeV, as well as the rapidity-differential net-proton yield. Notably, this approach captures the elusive $v_1$ splitting between protons and anti-protons, along with the double sign change in the directed flow of net-protons and net-lambdas observed between $\sqrt{s_{NN}} = 39$ GeV and $11.5$ GeV. Furthermore, our model calculations indicate that baryon diffusion plays a critical role in the observed sign change of the mid-rapidity slope of proton directed flow between $11.5$ GeV and $7.7$ GeV—a phenomenon interpreted as a potential signature of a first-order phase transition.

In the final two sections of Chapter 4, we investigate two interesting phenomena related to directed flow. The first is the impact of the hadronic stage on the directed flow of $K^{*0}$ resonances, and the second is the splitting of elliptic flow between hadrons originating from two distinct regions of phase space: one with $p_x > 0$ and the other with $p_x < 0$, where the $x$-axis aligns with the direction of the impact parameter. Both studies incorporate the effects of baryon density in the medium.

Our analysis demonstrates that the $v_1$ of the $K^{*0}$ is significantly influenced during the hadronic stage due to asymmetric signal loss on opposite sides of the $p_x$-axis in momentum space, a consequence of the tilted fireball geometry. We find that the late-stage hadronic rescattering phase introduces substantial qualitative changes to the $v_1$ of $K^{*0}$, resulting in $dv_1/dy(K^{*0}) - dv_1/dy(K^+)$ and $dv_1/dy(\phi) - dv_1/dy(K^+)$ having opposite signs. This effect is more pronounced in central collisions compared to peripheral ones, due to higher multiplicity and a longer duration of the hadronic phase. Additionally, the effect is further enhanced in low-energy collisions, where the stronger breaking of boost invariance amplifies the asymmetry.

The splitting of elliptic flow between different momentum space regions of produced hadrons has been proposed as a sensitive probe of the angular momentum carried by the fireball. Our study reveals that this elliptic flow splitting is primarily driven by contributions from directed and triangular flows. As such, it provides a valuable constraint on models of initial-state rapidity distribution within the fireball.

With an appropriate initial condition established, our focus shifts to constraining the baryon diffusion coefficient ($\kappa_B$), as detailed in Chapter 5. To achieve this, we employ an ansatz derived from kinetic theory:
\begin{equation}
\kappa_B = \frac{C_B}{T} n_B \left( \frac{1}{3} \coth{\left(\frac{\mu_B}{T}\right)} - \frac{n_B T}{\epsilon + p} \right),
\label{eq:kappaB_form2}
\end{equation}
where $\epsilon$, $P$, $T$, $n_B$, and $\mu_B$ denote the energy density, pressure, temperature, baryon number density, and baryon chemical potential, respectively. In this formulation, $C_B$ is a parameter that regulates the strength of baryon diffusion. Consequently, determining $\kappa_B$ becomes a matter of extracting the value of $C_B$.

By comparing model calculations for various $C_B$ values with experimental measurements of the centrality dependence of the directed flow slope splitting between protons and anti-protons, we provide the first estimation of $C_B$, finding it to lie within the range $0.5 < C_B < 1.5$. Furthermore, our model calculations reveal that the transverse momentum differential splitting of directed flow between protons and anti-protons exhibits strong sensitivity to $C_B$. This makes it an excellent observable for further constraining $\kappa_B$.

In Chapter 6, we investigate the centrality-dependent splitting of the $v_1(y)$ slope ($\Delta dv_1/dy$) between oppositely charged hadrons, recently measured by the STAR collaboration. The STAR measurement argued that the observed sign change of $\Delta dv_1/dy$ with centrality could be a signature of the electromagnetic field generated in heavy-ion collisions. However, our model calculations, which do not incorporate any electromagnetic field effects but consider only baryon diffusion, successfully reproduce the centrality trend of the $v_1$ slope splitting between protons and anti-protons.

Through this analysis, we demonstrate the pivotal role of baryon diffusion in shaping the centrality dependence of $\Delta dv_1/dy$. Specifically, our findings highlight that baryon stopping in the initial stages of the collision introduces a significant background that can influence signals attributed to the electromagnetic field. This underscores the necessity of accounting for such non-electromagnetic effects when interpreting experimental results.

In the final chapter, we explore rapidity-even directed flow in a baryon-rich medium. Building on the initial condition model discussed earlier, we provide predictions for the rapidity-even $v_1$ of identified hadrons across different collision energies—observables that have yet to be measured experimentally.

Our study focuses on the rapidity-symmetric deposition profile associated with double-junction stopping in the model. The calculations reveal that this rapidity-symmetric profile strongly influences the rapidity-even $v_1$ splitting between protons and anti-protons. We propose that future experimental measurements of this $v_1^{\text{even}}$ splitting could offer significant constraints on the rapidity-symmetric component of the baryon deposition profile, providing valuable phenomenological insights into the baryon junction picture.

Furthermore, our model calculations demonstrate that the rapidity-differential $v_1$ splitting between protons and anti-protons is simultaneously sensitive to the initial baryon deposition profile and the baryon diffusion coefficient. This dual sensitivity makes it an ideal observable for constraining both the baryon deposition profile and $\kappa_B$, enhancing our understanding of baryon dynamics in heavy-ion collisions.

\appendix
\chapter{Appendix}
\label{ch:appendix1}
\def \la{\langle}
\def \ra{\rangle}

This appendix discusses the importance of determining the direction of $\vec{b}$ (or the positive x-axis) in a collision event when measuring or calculating $v_1^{\text{odd}}$.

\begin{figure}
  \centering
  \includegraphics[width=1\textwidth]{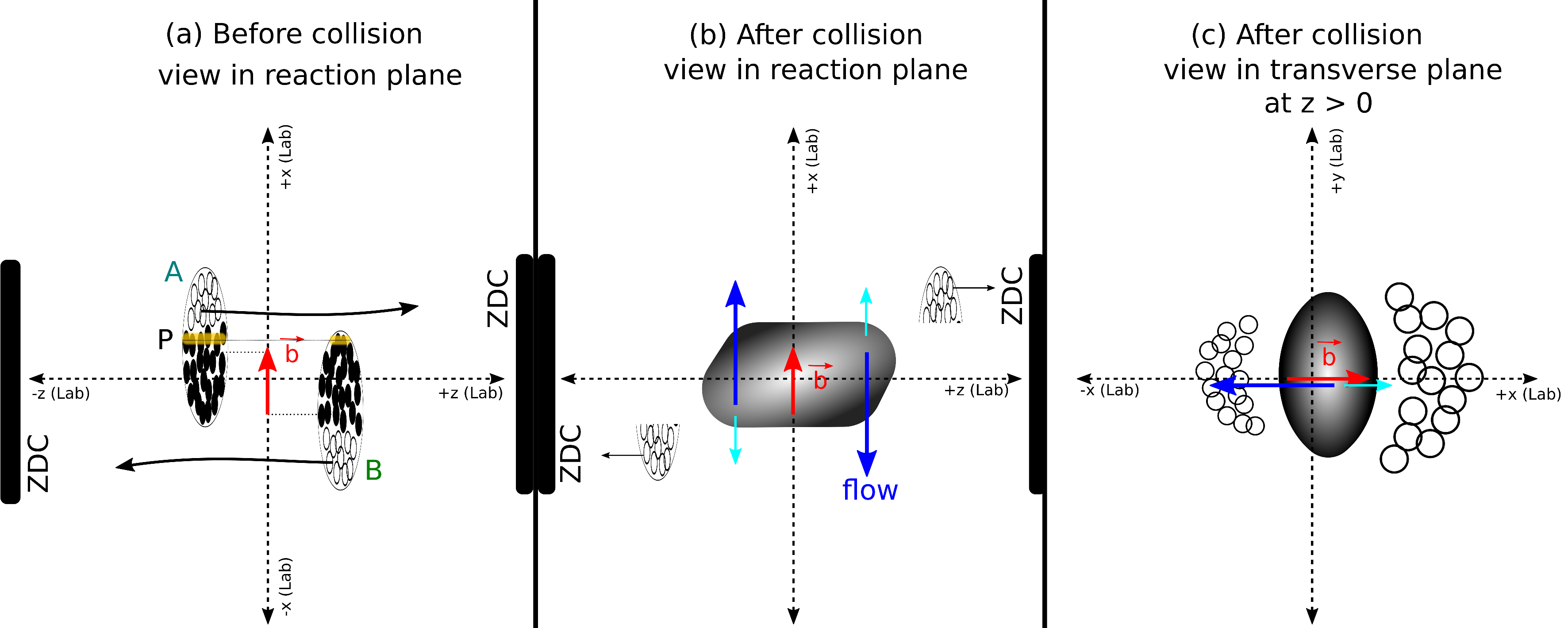} 
  \caption{A diagram demonstrating the breaking of forward-backward symmetry along the longitudinal direction and the generation of rapidity-odd directed flow in a non-central collision. Panel (a) shows the pre-collision scenario in the xz plane of the laboratory coordinate system, panel (b) depicts the post-collision scenario in the xz plane, and panel (c) illustrates the post-collision scenario in the transverse plane for the z > 0 region. Additional details are provided in the text.    }
  \label{fig:v1odd_gen}
\end{figure}

Consider a non-central symmetric nucleus collision event as depicted in Fig. \ref{fig:v1odd_gen}(a). The event is shown in the $xz$-plane of the laboratory coordinate system (LCS). In this scenario, the projectile nucleus, labeled as A, moves along the positive $z$-axis of the LCS, while the target nucleus, labeled as B, moves in the opposite direction, towards the negative $z$-axis. Before the collision, the center of the projectile nucleus is positioned on the positive side of the laboratory $x$-axis, and the center of the target nucleus is on the negative side. For simplicity, assume that the $y$-coordinate of both nuclei centers is at y=0 in the LCS. The impact parameter vector $\vec{b}$ (defined as the vector joining the center of the target nucleus to the center of the projectile nucleus) is aligned with the positive x-axis of the LCS. The dark-colored nucleons in both nuclei represent the participants in the collision, while the open symbols represent the spectators. After the collision, the spectators fly away and hit the Zero Degree Calorimeter (ZDC) detectors located far from the origin, near the beam axis. In experiments like STAR at RHIC \cite{STAR:2008jgm} or ALICE at LHC \cite{ALICE:2013xri}, the ZDC is used to provide information about the transverse position of the spectators on both sides \cite{Florkowski:2010zz}.

Before the collision, consider a transverse position labeled P on the positive side of the laboratory $x$-axis. At this position, the local matter density (or participant density) of the projectile nucleus is higher than that of the target nucleus. During the collision, participants from both nuclei at transverse position P will overlap and deposit their energy. However, due to this local asymmetry in participant density, there will be an asymmetry in longitudinal momentum between the forward and backward moving sources. Consequently, the forward-backward symmetry is broken, resulting in a net longitudinal momentum along the positive $z$-direction of the laboratory coordinate system (LCS). As a result, the matter in the $x>0$ region of the LCS will be pushed toward the positive $z$-direction, while in the $x<0$ region, it will be pushed toward the negative $z$-direction. This process leads to the generation of a deformed matter distribution or an asymmetric longitudinal flow. A representation of such a deformed matter distribution following the collision is depicted in Fig. \ref{fig:v1odd_gen} (b). In this deformed profile, at any non-zero $z$ position, there is a top-down dipolar asymmetry in energy deposition is observed across the x-axis.
\begin{figure}
  \centering
  \includegraphics[width=0.7\textwidth]{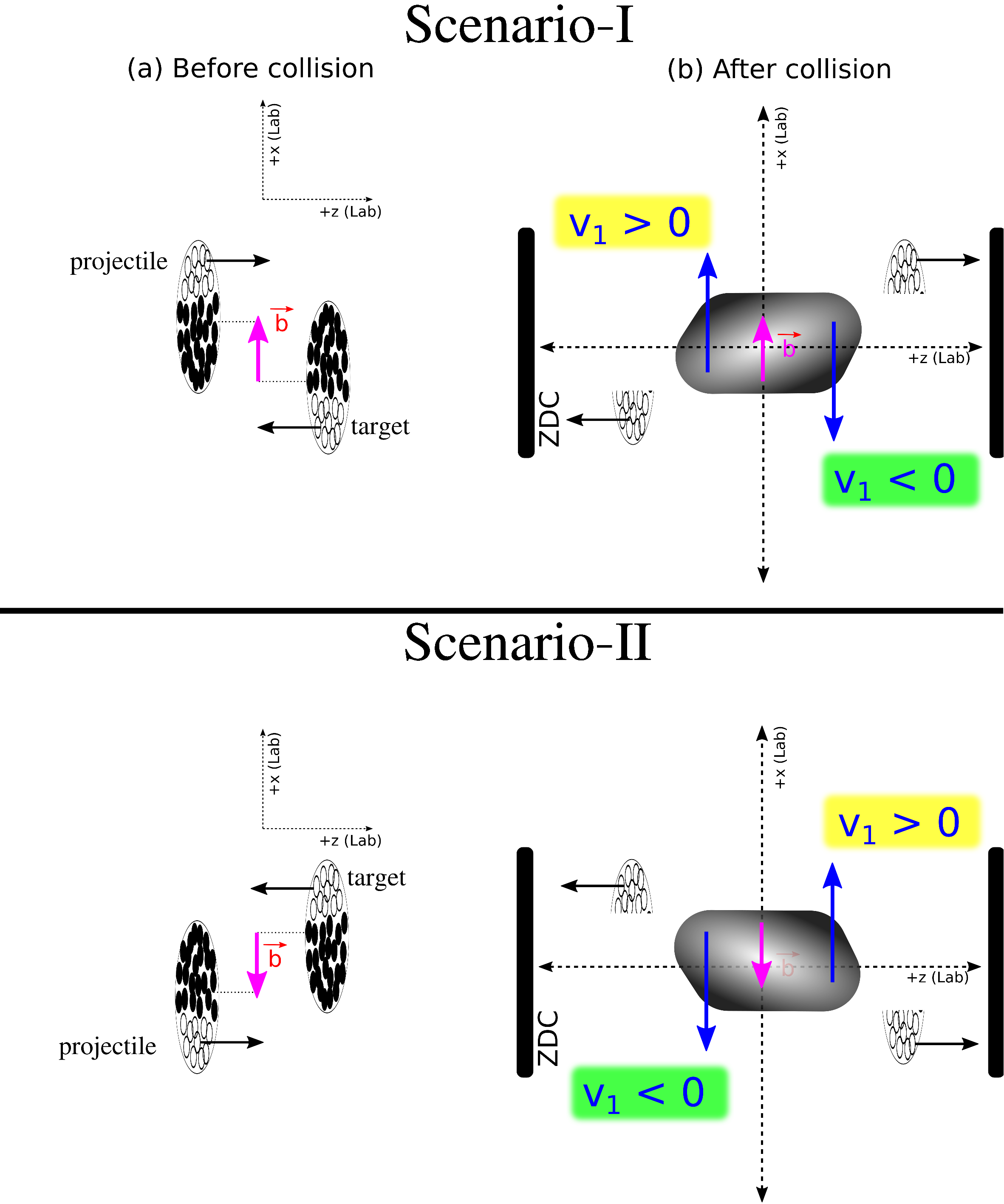} 
  \caption{A diagram showing the correlation between the sign of $v_1^{\text{odd}}$ and the direction of the impact parameter vector ($\vec{b}$) in both the forward and backward rapidity regions. }
  \label{fig:v1_b_corr}
\end{figure}

If we assume a smooth distribution of nuclear matter within the nucleus, ignoring fluctuations due to nucleon positions, there is no top-down asymmetry at $z = 0$. However, this asymmetry emerges at non-zero values of $z$.  Similar to how the azimuthal asymmetry in non-central collisions leads to a non-zero initial eccentricity in the transverse plane, producing collective elliptic flow, the evolution of this initial dipole deformation generates a directed flow along the impact parameter direction in the $z<0$ region and anti-parallel to it in the $z>0$ region. This flow is an odd function of the longitudinal coordinate and manifests as a function of rapidity ($y$) in the momentum space of produced particles. Notably, in central collisions ($\vert \vec{b} \vert = 0 $), this symmetry breaking is absent, leading to zero $v_1^{\text{odd}}$.

Rapidity-odd directed flow arises due to the collision geometry, developing parallel to the impact parameter in the negative space-time rapidity region ($z<0$) and anti-parallel in the positive space-time rapidity region ($z>0$). Therefore, it is crucial to determine the direction of $\vec{b}$ in each collision event. To illustrate the significance of determining the direction of $\vec{b}$ in measuring $v_1^{\text{odd}}$, Fig. \ref{fig:v1_b_corr} presents two collision scenarios.
In both scenarios, the lab frame's x-axis is fixed in the same direction, but the difference lies in the collision geometry. In the first scenario, $\vec{b}$ aligns with the positive x-axis of the lab coordinate system (LCS), while in the second scenario, $\vec{b}$ aligns with the negative x-axis of the LCS. If we use the lab x-axis as the reference for the collision event instead of the direction of $\vec{b}$, averaging over these configurations leads to $v_1^{\text{odd}} = \langle \cos \phi \rangle = 0$ across all rapidities.

Since the direction of $\vec{b}$ is random in experiments, if we consider the positive $x$-axis of the LCS as the positive $p_x$ direction (or $\phi = 0$) and calculate the event-averaged $v_1 = \langle \cos{\phi} \rangle$, we would get a zero magnitude across all rapidities. Instead, to accurately measure $v_1^{\text{odd}}$, we should determine the direction of $\vec{b}$ for each event, assign it as the positive $x$-direction for that event, and then calculate $v_1 = \langle \cos \phi \rangle = \langle p_x / p_T \rangle$ by averaging over tracks within each event, followed by averaging over all events.

In experiments, the direction of $\vec{b}$ cannot be determined directly. However, the direction of the perpendicular vector extending from the center of target spectators detected at the ZDC on the negative $z$ side to the center of projectile spectators detected at the ZDC on the positive $z$ side can serve as a good proxy for the direction of $\vec{b}$. In experiments the angle between $\vec{b}$ and the positive $x$-axis of the LCS is approximated through the first-order spectator plane angle ($\Psi^{\text{SP}}_{1}$). The experimental techniques for determining $\Psi^{\text{SP}}_{1}$ are detailed in ref. \cite{STAR:2008jgm,ALICE:2013xri}. $v_1^{\text{odd}}$ is measured by rotating the azimuthal angle of each track by $\Psi^{\text{SP}}_{1}$ of the respective event, then averaging over all tracks and ultimately over all events:
\beq
v_1^{\text{odd}} = \langle \langle  \cos(\phi-\Psi^{\text{SP}}_{1}) \rangle \rangle.
\eeq

{\raggedright \printbibliography[heading=bibintoc, title={References}]}

\end{document}